\documentclass[aps,pre,reprint,groupedaddress]{revtex4-2}

\usepackage{makecell}
\usepackage{graphicx}
\usepackage{amsmath}
\usepackage[colorlinks=true, linkcolor=blue, citecolor=blue,urlcolor=blue]{hyperref}

\begin{document}

%\preprint{}

\title{From Convex to Non-convex: Evolution of Rarefaction-dispersive shock interactions and the Influence of Non-convexity}


\author{Jia-Xue Niu}
%\thanks{}
\affiliation{School of Mathematics, Taiyuan University  of Technology, Taiyuan 030024, China}
\author{Rui Guo}
\email{gr81@sina.com}
\affiliation{School of Mathematics, Taiyuan University  of Technology, Taiyuan 030024, China}

\author{Hua-Ying Ren}
\affiliation{School of Mathematics, Taiyuan University  of Technology, Taiyuan 030024, China}

\author{Ya-Hui Huang}
\affiliation{School of Mathematics, Taiyuan University  of Technology, Taiyuan 030024, China}


\date{\today}

\begin{abstract}
In this paper, we focus on the analytical description of the interaction between a rarefaction and dispersive shock wave across both convex and non-convex cases, with particular attention to the effect of the non-convexity, within the framework of the Gardner equation. For convex structures, internal oscillations degenerate into small amplitude harmonic waves or a modulated soliton train as $t\rightarrow \infty$, accompanied by either a retained rarefaction part, or a new dispersive shock wave emanating from it. Taking into account the non-convexity, when $\alpha>0$, we find that kinks either remain non-participating in the interaction at all, or only act to switch polarities of convex structures. As for $\alpha<0$, we solve the Gardner--Whitham equations with three varying Riemann invariants, to analyze the rarefaction-contact dispersive shock interaction where the only possible configuration is that the rarefaction wave is on the left. It is demonstrated that the rarefaction wave will be completely drawn into the interaction region, with a changed contact dispersive shock wave escaping from the left. And internal oscillations eventually degenerate into an asymptotic algebraic soliton train as $t\rightarrow \infty$. In addition, we study the interaction between a rarefaction wave and composite structure consisting of the contact and classical dispersive shock parts under two distinct situations: (\romannumeral1) For the composite structure-rarefaction interaction, the contact part remains inactive in the interaction, and internal oscillations ultimately degenerate into a contact dispersive shock wave. (\romannumeral2) For the rarefaction-composite structure case, the entire composite structure participates in the interaction, during which the region concerned undergoing a transition from one subregion to two and then back to one.
\end{abstract}

%\keywords{}

\maketitle

\section{INTRODUCTION}

The competition and balance between nonlinearity and dispersion effects are widely present in nature and various physical systems, including fluid mechanics, plasmas, nonlinear optics and Bose-Einstein condensates, and can give rise to various nonlinear wave phenomena~\cite{Nlw1,Nlw2,Nlw3,Nlw4,Nlw5}. Among them, the dispersive shock wave (DSW), serving as the counterpart of the classical viscous shock wave in the conservative continuous media, has attracted widespread attention in recent years~\cite{Scholarpedia2009,physicad2016,pre2026mkdv,pre2026FPU,prl2025MR}. Owing to the dispersive regularization of a gradient catastrophe, the DSW is generated, which appears as an expanding oscillatory wavetrain connecting two distinct hydrodynamic states. At its two edges, it reduces to solitons and small-amplitude harmonic waves, respectively. The first analytic description of a DSW solution was given by Gurevich and Pitaevski\v{i} in 1974~\cite{GP1974}, by applying the Whitham modulation theory proposed in 1965~\cite{whitham1965} to the Korteweg--de Vries (KdV) equation. It is point out that, for the step initial data, the discontinuity either expands self-similarly into a smooth expansion fan, i.e., the rarefaction wave (RW), or evolves into a DSW. Subsequently, the analytic description of DSWs and RWs has been extended to a wide range of integrable and non-integrable equations~\cite{physicad1995,2000,chaos2005nonintegrable,pre2019nonintegrable,siam2017,pre2012,jpc2018,pra2020,pre2023}.

In fact, the DSW and RW are commonly observed to arise side by side in many systems. For example, the decay of the photonic dam into a pair of oppositely propagating RW and DSW has been observed~\cite{PRL2017}, and such coexistence is also reported in a one-dimensional droplet-bearing environment~\cite{pradbe}. In addition, the so-called dispersive rarefaction shocks have been realized in three-dimensional printed soft-chain systems, in both experiments and numerical simulations~\cite{DRS}. This naturally leads to a more intricate and interesting question: how does the system evolve when the above expanding waves propagate and collide in space? Such interactions are no longer simple superpositions, but involve nonlinear interactions with dynamical behaviors far more complex, such as emission and absorption of solitons~\cite{pradsoliton,pradbsoliton}, and dynamic reorganization of wave patterns\cite{physicad2007,physicad2012,pre2009,pre2013,pla2013,chaos2002}. Among them, a prototypical one is the interaction between DSWs, and it usually triggers multiphase modulation theory~\cite{cpam1980,jns2006}, which is typically addressed by means of the long-time asymptotic analysis, and numerical methods~\cite{pre2009,pre2013,pla2013,physicad2007}. By contrast, the interaction between RWs and DSWs involves solely single-phase dynamics, which raises the possibility of a fairly straightforward analytic description, and this work focuses exclusively on such interaction.

In recent years, this type of interaction has attracted increasing attention. It should be noted that the classical Riemann problem is incapable of capturing this interaction, which is instead replaced by the double-step initial data, characterized by two initial discontinuities. By introducing two independent jump interfaces, this formulation describes how two distinct macroscopic dispersive fluid states are connected through a dynamically evolving interaction region. The emergence of the interaction region destroys the self-similarity of solutions of the Whitham equations, rendering the hodograph transform indispensable: by mapping the space-time plane to the Riemann invariants plane, it significantly simplifies the solution process~\cite{2000}. For the KdV equation, the box and well initial-value problems and six typical scenarios in the double-step initial value problem have been studied~\cite{chaos2002,pre2009}. As time approaches infinity, the unidirectional interaction between the RW and DSW continues without their separation, leading ultimately to a RW or a DSW accompanied by either a small-amplitude oscillatory wavetrain or a soliton train. In contrast, it can be bidirectional within the framework of the defocusing NLS equation, and the refraction problem, namely the dispersive analogue of the shock-rarefaction collision in classical gas dynamics, has been investigated~\cite{physicad2012}. It is indicated by numerical simulations and Whitham modulation theory that the two waves completely separate after interaction, and this process alters the parameters of both waves. In addition, the DSW fitting method applicable to non-integrable dispersive systems has been extended to this problem for the NLS equation with saturable nonlinearity~\cite{physicad2012}.

It is worth remarking that researches on this type of interaction have mostly been limited to the convex case. Non-convex equations, on the other hand, display far more intricate dispersive hydrodynamic features.~To be more specific, for equations featuring non-convex dispersion, there exsit the radiating DSW, cross-over DSW, and traveling DSW~\cite{siamj2017,waterwave2025}; for integrable equations featuring non-convex flux, such as the modified KdV (mKdV), Gardner, and derivative nonlinear Schr\"{o}dinger-type equations, the kink and contact DSW (CDSW) emerge for particular initial data~\cite{siam2017,pre2012,jpc2018,pra2020,pre2023}. Although the box and well problems have been examined for the mKdV equation, the non-convex nature was not adequately captured~\cite{CPL2026}. Consequently, the investigation of the classification and evolution of RW-DSW interactions influenced by non-convex effects is still unclear.

To address this, we adopt the Gardner equation
\begin{equation}
u_t+f(u)_x+u_{xxx}=0
\end{equation}
with the hydrodynamic flux function $f(u)=3u^2-2\alpha u^3$ in which the coefficient $\alpha$ can be positive or negative depending on the considered physical problem, as a prototypical system to investigate the interaction between a RW and DSW. Equation (1) has been proven to be integrable~\cite{jpsj}, and can describe internal waves in a stratified medium as well as nonlinear wave phenomena in a number of physical contexts~\cite{bg1,bg2,bg3,bg4}. The Gardner--Whitham modulation system in a Riemann invariant form has been derived~\cite{pre2012}, and there exist two sets of the traveling wave parameters corresponding to the same set of the Riemann invariants due to the invariance of Eq.~(1) with respect to the transformation $u\rightarrow\frac{1}{\alpha}-u$. Because the flux function $f(u)$ is non-convex, which is reflected in the violation of either strict hyperbolicity or genuine nonlinearity of the Gardner--Whitham equations~\cite{Hyperbolicsystems}, kinks ($\alpha>0$) and CDSWs ($\alpha<0$) are allowed when the initial step data cross the inflection point $u=\frac{1}{2\alpha}$, relative to the convex KdV case. Although the Riemann problem of Eq.~(1) has been studied, the interaction between RWs and DSWs has not yet been reported. Moreover, owing to its non-convexity, it is reasonable and meaningful to choose it as the subject of this paper. Notably, the authors have demonstrated based on the solitonic modulation system that the interaction between the kink and convex RW or DSW leads to polarity reversal of the latter~\cite{sol-mean}, which is helpful to analyze the impact of kinks on the interaction in this paper.

The organization of this paper is as follows: In Sec.~\uppercase\expandafter{\romannumeral2}, we will recall the Riemann problem for Eq.~(1) firstly, and then provide a classification of the double-step initial value problem in Tables~1 and 2. In Sec.~\uppercase\expandafter{\romannumeral3}, the interaction between convex RWs and DSWs will be studied when the initial value does not cross the inflection point of the flux function, and the analytical expression of the interaction region as well as the asymptotic behavior at large time will be analyzed. In Sec.~\uppercase\expandafter{\romannumeral4}, we will discuss the influence of non-convex flux on the interaction through the generation of kinks when $\alpha>0$. In Sec.~\uppercase\expandafter{\romannumeral5}, we will analyze, for $\alpha<0$, the interaction between the classical RW and the special non-convex DSW, i.e., the pure CDSW. Furthermore, the case of composite structures containing the CDSW will be discussed in Sec.~\uppercase\expandafter{\romannumeral6}. Finally, Section \uppercase\expandafter{\romannumeral7} will present conclusions of this paper. In addition, we will explain the periodic solutions and Gardner--Whitham equations, the hodograph transform, and the modulated phase in Appendixes A-C, respectively.

\section{INITIAL VALUE PROBLEM}

As $\alpha=0$, Equation (1) reduces to the KdV equation whose discontinuous initial value problem has been explored~\cite{GP1974}. If the left plateau is higher, one shall find the DSW with the bright soliton front taking into account the dispersion effect. However, the situation becomes even more complex for Eq.~(1), owing to its non-convexity. When $\alpha$ takes different signs, the discontinuous initial value
\begin{equation}
u\left( x,0 \right) =\left\{ \begin{array}{l}
	u_-,\ \ \ x<0,\\
	u_+,\ \ \ x>0,\\
\end{array} \right.
\end{equation}
evolves into distinct nonlinear wave structures~\cite{pre2012}, including RWs, DSWs, kinks and CDSWs, as shown in Fig.~1. The commonality of evolution lies in the fact that RWs and DSWs have dual polarity for both $\alpha>0$ and $\alpha<0$. When the initial condition crosses the inflection point $u=\frac{1}{2\alpha}$ of the hydrodynamic flux function in Eq.~(1), the initial value (2) evolves into a composite structure: a RW or DSW combines with a kink for $\alpha>0$, or a CDSW for $\alpha<0$.
 \begin{figure*}
 \includegraphics[scale=0.7]{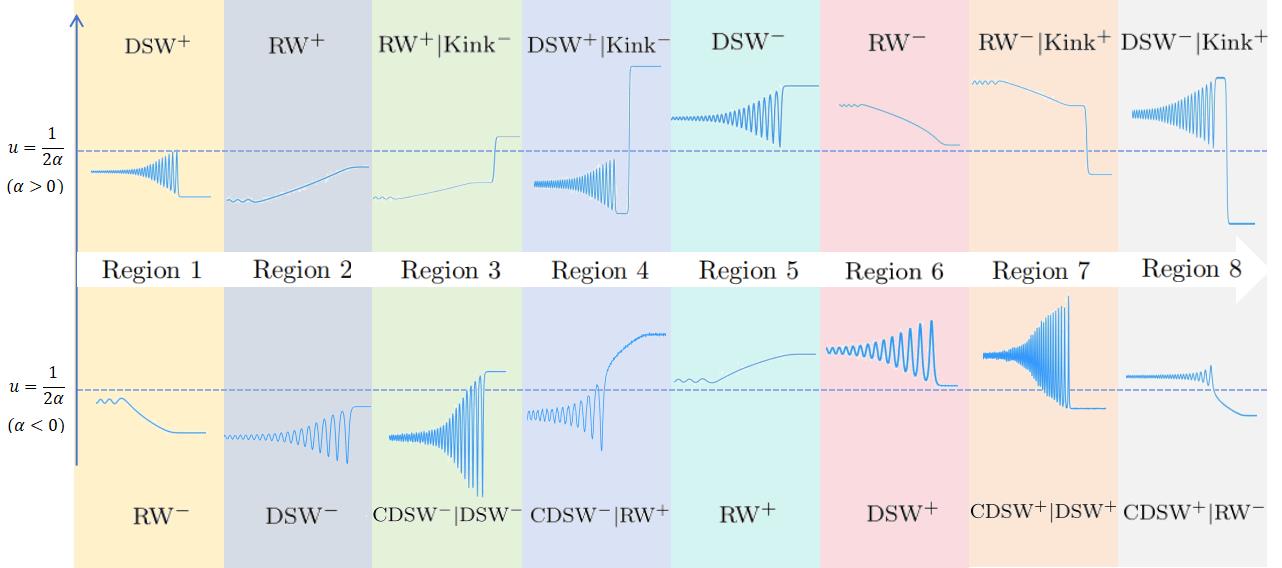}%
 \caption{ Complete classification of solutions to the discontinuous initial value problem of Eq.~(1). The superscripts $+/-$ denote normal or reversed ones.}
 \label{Fig.~$1$.}
 \end{figure*}
Based on the above, we consider the double-step initial data
\begin{equation}
u(x,0)=\left\{ \begin{array}{l}
	u_l,\ \ \ x<0,\\
	u_m,\ \ 0<x<D,\\
	u_r,\ \ \ x>D,\\
\end{array} \right.
\end{equation}
for Eq.~(1). The relative ordering of the three parameters yields six distinct cases, namely,  \uppercase\expandafter{\romannumeral1}: $u_l>u_m>u_r$, \uppercase\expandafter{\romannumeral2}: $u_l<u_m<u_r$, \uppercase\expandafter{\romannumeral3}: $u_m<u_l<u_r$, \uppercase\expandafter{\romannumeral4}: $u_m<u_r<u_l$, \uppercase\expandafter{\romannumeral5}: $u_r<u_l<u_m$, \uppercase\expandafter{\romannumeral6}: $u_l<u_r<u_m$. Taking into account the inflection point $u=\frac{1}{2\alpha}$, the evolving results are shown in Tables~1-2. In this paper, we only focus on evolutions involving collisions between RWs and DSWs. Interactions among RWs alone or among DSWs alone are beyond the scope of this paper and will be considered in the future.
 \begin{table*}%[H] %add [H] placement to break table across pages
 \caption{\label{Table. 1}Results of the double-step initial data (3) for $\alpha>0$.}
% \begin{ruledtabular}
{\fontsize{8pt}{12pt}\selectfont
 \begin{tabular}{|c|c|c|c|}
  \hline
  \multicolumn{2}{|c|}{Case \uppercase\expandafter{\romannumeral1}: $u_l>u_m>u_r$} & \multicolumn{2}{c|}{Case \uppercase\expandafter{\romannumeral2}: $u_l<u_m<u_r$}\\
  \hline
    Wave structure & Step condition &Wave structure & Step condition \\
   \hline
    $\text{DSW}^++\text{DSW}^+$ & $u_l<\frac{1}{2\alpha}$& $\text{RW}^++\text{RW}^+$ & $u_r<\frac{1}{2\alpha}$ \\
   \hline
    $\text{RW}^-+\text{RW}^-|\text{Kink}^+$ & \makecell{$u_r<\frac{1}{2\alpha}<u_m$\\$ u_m+u_r>\frac{1}{\alpha}$} & $\text{RW}^++\text{RW}^+|\text{Kink}^-$ & \makecell{$u_m<\frac{1}{2\alpha}<u_r$\\$u_m+u_r<\frac{1}{\alpha}$} \\
   \hline
    $\text{RW}^-+\text{RW}^-$ & $\frac{1}{2\alpha}<u_r$& $\text{RW}^++\text{DSW}^+|\text{Kink}^-$ & \makecell{$u_m<\frac{1}{2\alpha}<u_r$\\$u_m+u_r>\frac{1}{\alpha}$} \\
   \hline
    $\text{RW}^-|\text{Kink}^++\text{DSW}^+$ & \makecell{$u_m<\frac{1}{2\alpha}<u_l$\\$ u_l+u_m>\frac{1}{\alpha}$}& $\text{RW}^+|\text{Kink}^-+\text{DSW}^-$ & \makecell{$u_l<\frac{1}{2\alpha}<u_m$\\$u_l+u_m<\frac{1}{\alpha}$} \\
   \hline
    $\text{RW}^-+\text{DSW}^-|\text{Kink}^+$ & \makecell{$u_r<\frac{1}{2\alpha}<u_m$\\$ u_m+u_r<\frac{1}{\alpha}$}& $\text{DSW}^+|\text{Kink}^-+\text{DSW}^-$ & \makecell{$u_l<\frac{1}{2\alpha}<u_m$\\$u_l+u_m>\frac{1}{\alpha}$} \\
   \hline
    $\text{DSW}^+|\text{Kink}^++\text{DSW}^+$ & \makecell{$u_m<\frac{1}{2\alpha}<u_l$\\$ u_l+u_m<\frac{1}{\alpha}$}& $\text{DSW}^-+\text{DSW}^-$ & $\frac{1}{2\alpha}<u_l$ \\
  \hline
  \hline
  \multicolumn{2}{|c|}{Case \uppercase\expandafter{\romannumeral3}: $u_m<u_l<u_r$} & \multicolumn{2}{c|}{Case \uppercase\expandafter{\romannumeral4}: $u_m<u_r<u_l$}\\
  \hline
   Wave structure & Step condition &Wave structure & Step condition \\
   \hline
  $\text{DSW}^++\text{RW}^+$ & $u_r<\frac{1}{2\alpha}$ & $\text{DSW}^++\text{RW}^+$ & $u_l<\frac{1}{2\alpha}$\\
  \hline
  $\text{DSW}^++\text{RW}^+|\text{Kink}^-$ & \makecell{$u_l<\frac{1}{2\alpha}<u_r$\\$u_m+u_r<\frac{1}{\alpha}$} & $\text{RW}^-+\text{DSW}^-$ & $\frac{1}{2\alpha}<u_m$\\
  \hline
  $\text{DSW}^++\text{DSW}^+|\text{Kink}^-$ & \makecell{$u_l<\frac{1}{2\alpha}<u_r$\\$u_m+u_r<\frac{1}{\alpha}$} & $\text{RW}^-|\text{Kink}^++\text{RW}^+|\text{Kink}^-$ & \makecell{$u_m<\frac{1}{2\alpha}<u_r$\\$u_m+u_r<\frac{1}{\alpha}<u_l+u_m$}\\
  \hline
  $\text{RW}^-+\text{DSW}^-$ & $\frac{1}{2\alpha}<u_m$ & $\text{RW}^-|\text{Kink}^++\text{RW}^+$ & \makecell{$u_r<\frac{1}{2\alpha}<u_l$\\$u_l+u_m>\frac{1}{\alpha}$}\\
  \hline
  $\text{RW}^-|\text{Kink}^++\text{DSW}^+|\text{Kink}^-$ & \makecell{$u_m<\frac{1}{2\alpha}<u_l$\\$\frac{1}{\alpha}<u_l+u_m$} & $\text{RW}^-|\text{Kink}^++\text{DSW}^+|\text{Kink}^-$ & \makecell{$u_m<\frac{1}{2\alpha}<u_r$\\$\frac{1}{\alpha}<u_m+u_r$}\\
  \hline
  $\text{DSW}^-|\text{Kink}^++\text{RW}^+|\text{Kink}^-$ & \makecell{$u_m<\frac{1}{2\alpha}<u_l$\\$u_m+u_r<\frac{1}{\alpha}$} & $\text{DSW}^-|\text{Kink}^++\text{RW}^+$ & \makecell{$u_r<\frac{1}{2\alpha}<u_l$\\$u_l+u_m<\frac{1}{\alpha}$}\\
  \hline
  $\text{DSW}^-|\text{Kink}^++\text{DSW}^+|\text{Kink}^-$ & \makecell{$u_m<\frac{1}{2\alpha}<u_l$\\$u_l+u_m<\frac{1}{\alpha}<u_m+u_r$} & $\text{DSW}^-|\text{Kink}^++\text{RW}^+|\text{Kink}^-$ & \makecell{$u_m<\frac{1}{2\alpha}<u_r$\\$u_l+u_m<\frac{1}{\alpha}$}\\
  \hline
  \hline
  \multicolumn{2}{|c|}{Case \uppercase\expandafter{\romannumeral5}: $u_r<u_l<u_m$} & \multicolumn{2}{c|}{Case \uppercase\expandafter{\romannumeral6}: $u_l<u_r<u_m$}\\
  \hline
   Wave structure & Step condition &Wave structure & Step condition \\
   \hline
  $\text{RW}^++\text{DSW}^+$ & $u_m<\frac{1}{2\alpha}$ & $\text{RW}^++\text{DSW}^+$ & $u_m<\frac{1}{2\alpha}$\\
  \hline
  $\text{RW}^+|\text{Kink}^-+\text{DSW}^-|\text{Kink}^+$ & \makecell{$u_l<\frac{1}{2\alpha}<u_m$\\$u_1+u_m<\frac{1}{\alpha}$} & $\text{RW}^+|\text{Kink}^-+\text{RW}^-$ & \makecell{$u_l<\frac{1}{2\alpha}<u_r$\\$u_l+u_m<\frac{1}{\alpha}$}\\
  \hline
  $\text{DSW}^+|\text{Kink}^-+\text{RW}^-|\text{Kink}^+$ & \makecell{$u_l<\frac{1}{2\alpha}<u_m$\\$\frac{1}{\alpha}<u_m+u_r$} & $\text{RW}^+|\text{Kink}^-+\text{RW}^-|\text{Kink}^+$ & \makecell{$u_r<\frac{1}{2\alpha}<u_m$\\$u_1+u_m<\frac{1}{\alpha}<u_m+u_r$}\\
  \hline
  $\text{DSW}^+|\text{Kink}^-+\text{DSW}^-|\text{Kink}^+$ & \makecell{$u_l<\frac{1}{2\alpha}<u_m$\\$u_m+u_r<\frac{1}{\alpha}<u_l+u_m$} & $\text{DSW}^+|\text{Kink}^-+\text{RW}^-$ & \makecell{$u_l<\frac{1}{2\alpha}<u_r$\\$u_l+u_m>\frac{1}{\alpha}$}\\
  \hline
  $\text{DSW}^-+\text{RW}^-$ & $\frac{1}{2\alpha}<u_r$ & $\text{DSW}^+|\text{Kink}^-+\text{RW}^-|\text{Kink}^+$ & \makecell{$u_r<\frac{1}{2\alpha}<u_m$\\$\frac{1}{\alpha}<u_l+u_m$}\\
  \hline
  $\text{DSW}^-+\text{RW}^-|\text{Kink}^+$ & \makecell{$u_r<\frac{1}{2\alpha}<u_l$\\$u_m+u_r>\frac{1}{\alpha}$} & $\text{RW}^+|\text{Kink}^-+\text{DSW}^-|\text{Kink}^+$ & \makecell{$u_r<\frac{1}{2\alpha}<u_m$\\$u_m+u_r<\frac{1}{\alpha}$}\\
  \hline
  $\text{DSW}^-+\text{DSW}^-|\text{Kink}^+$ & \makecell{$u_r<\frac{1}{2\alpha}<u_l$\\$u_m+u_r<\frac{1}{\alpha}$} & $\text{DSW}^-+\text{RW}^-$ & $\frac{1}{2\alpha}<u_l$\\
  \hline
 \end{tabular}}
% \end{ruledtabular}
 \end{table*}
  \begin{table*}%[H] %add [H] placement to break table across pages
 \caption{\label{Table. 2}Results of the double-step initial data (3) for $\alpha<0$.}
% \begin{ruledtabular}
{\fontsize{8pt}{12pt}\selectfont
 \begin{tabular}{|c|c|c|c|}
  \hline
  \multicolumn{2}{|c|}{Case \uppercase\expandafter{\romannumeral1}: $u_r>u_m>u_l$} & \multicolumn{2}{c|}{Case \uppercase\expandafter{\romannumeral2}: $u_l<u_m<u_r$}\\
  \hline
    Wave structure & Step condition &Wave structure & Step condition \\
   \hline
    $\text{RW}^-+\text{RW}^-$ & $u_l<\frac{1}{2\alpha}$& $\text{DSW}^-+\text{DSW}^-$ & $u_r<\frac{1}{2\alpha}$ \\
   \hline
    $\text{DSW}^++\text{CDSW}^+|\text{DSW}^+$ & \makecell{$u_r<\frac{1}{2\alpha}<u_m$\\$ u_m+u_r>\frac{1}{\alpha}$} & $\text{DSW}^-+\text{CDSW}^-|\text{DSW}^-$ & \makecell{$u_m<\frac{1}{2\alpha}<u_r$\\$u_m+u_r<\frac{1}{\alpha}$} \\
   \hline
    $\text{DSW}^++\text{DSW}^+$ & $\frac{1}{2\alpha}<u_r$& $\text{DSW}^-+\text{CDSW}^-|\text{RW}^+$ & \makecell{$u_m<\frac{1}{2\alpha}<u_r$\\$u_m+u_r>\frac{1}{\alpha}$} \\
   \hline
    $\text{CDSW}^+|\text{DSW}^++\text{RW}^-$ & \makecell{$u_m<\frac{1}{2\alpha}<u_l$\\$ u_l+u_m>\frac{1}{\alpha}$}& $\text{DSW}^-|\text{DSW}^-+\text{RW}^+$ & \makecell{$u_l<\frac{1}{2\alpha}<u_m$\\$u_l+u_m<\frac{1}{\alpha}$} \\
   \hline
    $\text{DSW}^++\text{CDSW}^+|\text{RW}^-$ & \makecell{$u_r<\frac{1}{2\alpha}<u_m$\\$ u_m+u_r<\frac{1}{\alpha}$}& $\text{CDSW}^-|\text{RW}^++\text{RW}^+$ & \makecell{$u_l<\frac{1}{2\alpha}<u_m$\\$u_l+u_m>\frac{1}{\alpha}$} \\
   \hline
    $\text{CDSW}^+|\text{RW}^-+\text{RW}^-$ & \makecell{$u_m<\frac{1}{2\alpha}<u_l$\\$ u_l+u_m<\frac{1}{\alpha}$}& $\text{RW}^++\text{RW}^+$ & $\frac{1}{2\alpha}<u_l$ \\
  \hline
  \hline
  \multicolumn{2}{|c|}{Case \uppercase\expandafter{\romannumeral3}: $u_m<u_l<u_r$} & \multicolumn{2}{c|}{Case \uppercase\expandafter{\romannumeral4}: $u_m<u_r<u_l$}\\
  \hline
   Wave structure & Step condition &Wave structure & Step condition \\
   \hline
  $\text{RW}^-+\text{DSW}^-$ & $u_r<\frac{1}{2\alpha}$ & $\text{RW}^-+\text{DSW}^-$ & $u_l<\frac{1}{2\alpha}$\\
  \hline
  $\text{RW}^-+\text{CDSW}^-|\text{DSW}^-$ & \makecell{$u_l<\frac{1}{2\alpha}<u_r$\\$u_m+u_r<\frac{1}{\alpha}$} & $\text{DSW}^++\text{RW}^+$ & $\frac{1}{2\alpha}<u_m$\\
  \hline
  $\text{RW}^-+\text{CDSW}^-|\text{RW}^+$ & \makecell{$u_l<\frac{1}{2\alpha}<u_r$\\$u_m+u_r>\frac{1}{\alpha}$} & $\text{CDSW}^+|\text{DSW}^++\text{CDSW}^-|\text{DSW}^-$ & \makecell{$u_m<\frac{1}{2\alpha}<u_r$\\$u_m+u_r<\frac{1}{\alpha}<u_l+u_m$}\\
  \hline
  $\text{DSW}^++\text{RW}^+$ & $\frac{1}{2\alpha}<u_m$ & $\text{CSDW}^+|\text{DSW}^++\text{DSW}^-$ & \makecell{$u_r<\frac{1}{2\alpha}<u_l$\\$u_l+u_m>\frac{1}{\alpha}$}\\
  \hline
  $\text{DSW}^++\text{CDSW}^-|\text{RW}^+$ & \makecell{$u_m<\frac{1}{2\alpha}<u_l$\\$\frac{1}{\alpha}<u_l+u_m$} & $\text{CDSW}^+|\text{DSW}^++\text{CDSW}^-|\text{RW}^+$ & \makecell{$u_m<\frac{1}{2\alpha}<u_r$\\$\frac{1}{\alpha}<u_m+u_r$}\\
  \hline
  $\text{CDSW}^+|\text{RW}^-+\text{CDSW}^-|\text{DSW}^-$ & \makecell{$u_m<\frac{1}{2\alpha}<u_l$\\$u_m+u_r<\frac{1}{\alpha}$} & $\text{CDSW}^+|\text{RW}^-+\text{DSW}^-$ & \makecell{$u_r<\frac{1}{2\alpha}<u_l$\\$u_l+u_m<\frac{1}{\alpha}$}\\
  \hline
  $\text{CDSW}^+|\text{RW}^-+\text{CDSW}^-|\text{RW}^+$ & \makecell{$u_m<\frac{1}{2\alpha}<u_l$\\$u_l+u_m<\frac{1}{\alpha}<u_m+u_r$} & $\text{CDSW}^+|\text{RW}^-+\text{CDSW}^-|\text{DSW}^-$ & \makecell{$u_m<\frac{1}{2\alpha}<u_r$\\$u_l+u_m<\frac{1}{\alpha}$}\\
  \hline
  \hline
  \multicolumn{2}{|c|}{Case \uppercase\expandafter{\romannumeral5}: $u_r<u_l<u_m$} & \multicolumn{2}{c|}{Case \uppercase\expandafter{\romannumeral6}: $u_l<u_r<u_m$}\\
  \hline
   Wave structure & Step condition &Wave structure & Step condition \\
   \hline
  $\text{DSW}^-+\text{RW}^-$ & $u_m<\frac{1}{2\alpha}$ & $\text{DSW}^-+\text{RW}^-$ & $u_m<\frac{1}{2\alpha}$\\
  \hline
  $\text{CDSW}^-|\text{DSW}^-+\text{CDSW}^+|\text{RW}^-$ & \makecell{$u_l<\frac{1}{2\alpha}<u_m$\\$u_1+u_m<\frac{1}{\alpha}$} & $\text{CDSW}^-|\text{DSW}^-+\text{DSW}^+$ & \makecell{$u_l<\frac{1}{2\alpha}<u_r$\\$u_l+u_m<\frac{1}{\alpha}$}\\
  \hline
  $\text{CDSW}^-|\text{RW}^++\text{CDSW}^+|\text{DSW}^+$ & \makecell{$u_l<\frac{1}{2\alpha}<u_m$\\$\frac{1}{\alpha}<u_m+u_r$} & $\text{CDSW}^-|\text{DSW}^-+\text{CDSW}^+|\text{DSW}^+$ & \makecell{$u_r<\frac{1}{2\alpha}<u_m$\\$u_1+u_m<\frac{1}{\alpha}<u_m+u_r$}\\
  \hline
  $\text{CDSW}^-|\text{RW}^++\text{CDSW}^+|\text{RW}^-$ & \makecell{$u_l<\frac{1}{2\alpha}<u_m$\\$u_m+u_r<\frac{1}{\alpha}<u_l+u_m$} & $\text{CDSW}^-|\text{RW}^++\text{DSW}^+$ & \makecell{$u_l<\frac{1}{2\alpha}<u_r$\\$u_l+u_m>\frac{1}{\alpha}$}\\
  \hline
  $\text{RW}^++\text{DSW}^+$ & $\frac{1}{2\alpha}<u_r$ & $\text{CDSW}^-|\text{RW}^++\text{CDSW}^+|\text{DSW}^+$ & \makecell{$u_r<\frac{1}{2\alpha}<u_m$\\$\frac{1}{\alpha}<u_l+u_m$}\\
  \hline
  $\text{RW}^++\text{CDSW}^+|\text{DSW}^+$ & \makecell{$u_r<\frac{1}{2\alpha}<u_l$\\$u_m+u_r>\frac{1}{\alpha}$} & $\text{CDSW}^-|\text{DSW}^-+\text{CDSW}^+|\text{RW}^-$ & \makecell{$u_r<\frac{1}{2\alpha}<u_m$\\$u_m+u_r<\frac{1}{\alpha}$}\\
  \hline
  $\text{RW}^++\text{CDSW}^+|\text{RW}^-$ & \makecell{$u_r<\frac{1}{2\alpha}<u_l$\\$u_m+u_r<\frac{1}{\alpha}$} & $\text{RW}^++\text{DSW}^+$ & $\frac{1}{2\alpha}<u_l$\\
  \hline
 \end{tabular}}
% \end{ruledtabular}
 \end{table*}

In order to make the first two members of the Gardner--Whitham hierarchy (see Appendix A) match each other on the phase transition boundaries, we introduce the Riemann invariant $r=\chi(u)=u(1-\alpha u)$ with the following initial state
\begin{equation}
r(x,0)=\left\{ \begin{array}{l}
	\chi_l,\ \ \ x<0,\\
	\chi_m,\ \ 0<x<D,\\
	\chi_r,\ \ \ x>D,\\
\end{array} \right.
\end{equation}
where $\chi_l=\chi(u_l)$, $\chi_m=\chi(u_m)$, and $\chi_r=\chi(u_r)$.

After wave breaking, the Riemann invariant, originally depending on a single parameter, is extended to a three-parameter family, and their evolutions as functions of space and time variables are governed by Eq.~(A6).
\section{INTERACTION BETWEEN CONVEX STRUCTURES}
In this section we consider a simple scenario: the interaction between pure RWs and DSWs, which only occurs in Case \uppercase\expandafter{\romannumeral3} and \uppercase\expandafter{\romannumeral6} for both $\alpha>0$ and $\alpha<0$. At this point, the initial data (3) don't cross the inflection point, and the function $\chi(u)$ is strictly monotonic. We distinguish the two interaction types by the left-to-right ordering: the RW-DSW interaction (RW left of DSW) and the DSW-RW interaction (DSW left of RW).
\subsection{DSW-RW interaction}
When $\text{min}(\chi_l,\chi_r)>\chi_m$, the initial data (3) give rise to a DSW generated at the discontinuity $x=0$ and a RW generated at $x=D$, in the early stage of the evolution.
\subsubsection{Partial retention of the rarefaction wave}
Taking the case where $u_m<u_l<u_r<\frac{1}{2\alpha}\Rightarrow \chi_m<\chi_l<\chi_r$ as an example, as time evolves, the $(x,t)$ plane is divided into five regions (see Figs.~2 and 4). There are two expanding waves, a DSW and a RW, connected by a plateau. Noting that the RW propagating between the following boundaries
\begin{equation}
x_{\text{RW}}^-=6\chi_mt+D,\ \ x_{\text{RW}}^+=6\chi_rt+D,
\end{equation}
can be expressed as
\begin{equation}
u(\tau)=\frac{1}{2\alpha}\left( 1-\sqrt{1-\frac{2\alpha}{3}\left(\tau-\frac{D}{t}\right)} \right),
\end{equation}
where $\tau=\frac{x}{t}$ is the self-similar variable. And the DSW oscillating between two boundaries
\begin{align}
&x_{\text{DSW}}^-=s_{\text{DSW}}^-t=(12\chi_m-6\chi_l)t,\ \ m\rightarrow 0,\notag\\
&x_{\text{DSW}}^+=s_{\text{DSW}}^+t=(2\chi_m+4\chi_l)t,\ \ m\rightarrow 1,
\end{align}
can be described by the single-phase solution (A3) with the following three Riemann invariants in terms of the self-similar form
\begin{equation}
r_1=\chi_m,\ \ \tau=v_2(\chi_m,r_2,\chi_l),\ \ r_3=\chi_l.
\end{equation}
With the evolution of time, the width of the plateau between them gradually decreases and eventully disappears, then the soliton edge of DSW and the trailing edge of RW intersect at $(x_1,t_1)$ satisfying
$x_1=x_{\text{DSW}}^+(t_1)=x_{\text{RW}}^-(t_1)$, i.e.,
\begin{equation}
x_1=\frac{D(2\chi_l+\chi_m)}{2(\chi_l-\chi_m)},\ \ t_1=\frac{D}{4(\chi_l-\chi_m)}.
\end{equation}
If $\alpha(u_l+u_m)>0$, one can see that it takes more time from initial state until the disappearance of the plateau compared to the KdV equation, for which one has $t_1=\frac{D}{4(u_l-u_m)}$. This can be also interpreted by the fact that the DSW of Eq.~(1) is narrower~\cite{pre2012}, and the result is opposite as $\alpha(u_l+u_m)<0$.

\begin{figure}
\includegraphics[scale=0.21]{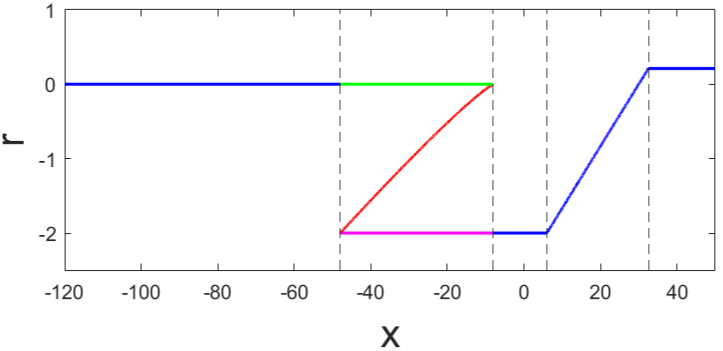}\hfill
\includegraphics[scale=0.31]{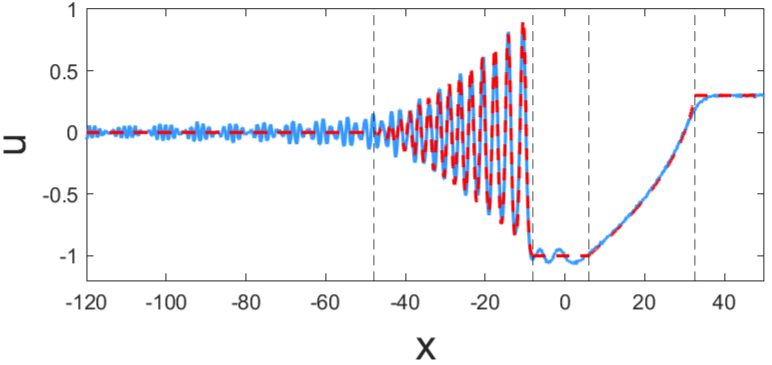}\\
\includegraphics[scale=0.31]{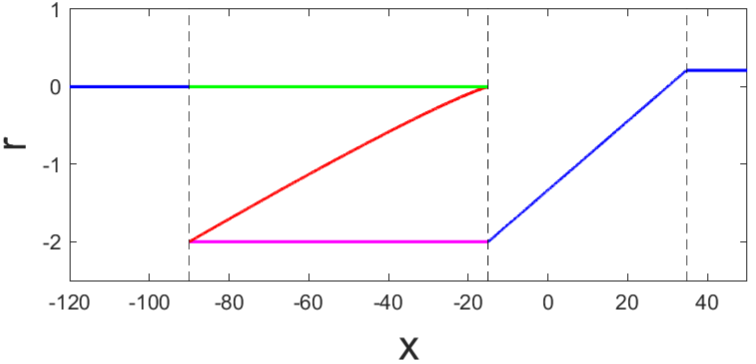}\hfill
\includegraphics[scale=0.31]{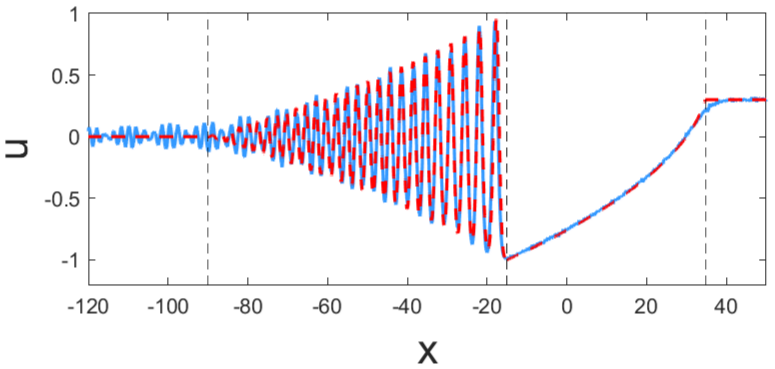}\\
\includegraphics[scale=0.31]{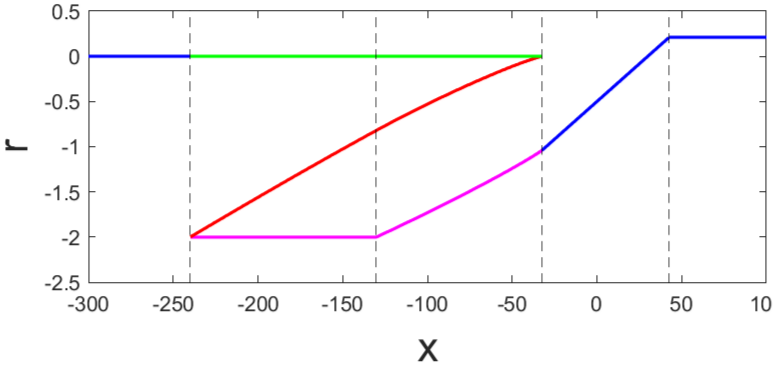}\hfill
\includegraphics[scale=0.31]{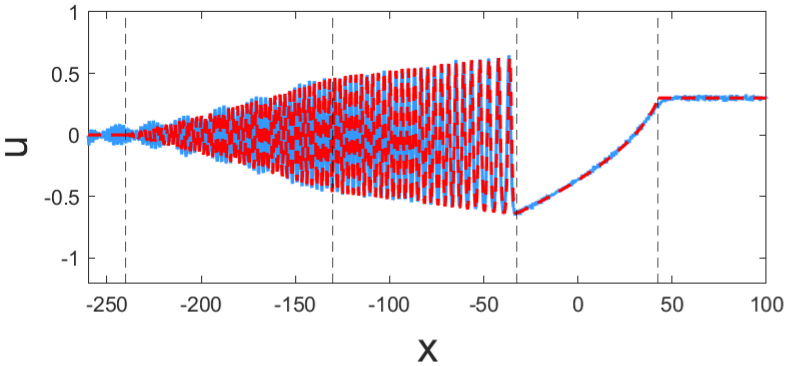}\\

\vspace{0cm}{\footnotesize\hspace{0.0cm}(a)\hspace{4cm}(b)}
\caption{Evolution of DSW-RW interaction with retained RW, at $t=2$, $t=t_1$, and $t=10$, respectively. The parameters are $u_l=0$, $u_m=-1$, $u_r=0.3$, $D=30$, $\alpha=1$. (a) Riemann invariants. In the multi-value region, the pink line represents $r_1$, the red one represents $r_2$, and the green one corresponds to $r_3$. The same applies below. (b) Solution of Eq.~(1). The red dashed line represents the analytical results based on the Whitham modulation theory, while the blue solid line represents the numerical results.}
\label{Figs.~2.}
\end{figure}

After the collision, Riemann invariants are mixed, leading to the emergence of a new region, as depicted in the third row of Figs.~2 and 4. This region is still described by the single-phase periodic solution (A3) with a family of three Riemann invariants, among which $r_3=\chi_l$ is always a constant but the remaining Riemann invariants vary with spatiotemporal variables. Therefore, the Gardner--Whitham equations are no longer solvable by self-similar solutions, necessitating a re-solving of Eq.~(A6).

In the interaction region, two varying Riemann invariants have boundary conditions as follows
\begin{align}
r_1=\chi_m,\ \ \ r_3=\chi_l,\ \ \ \ &\text{on the left boundary},\notag\\
r_2=r_3=\chi_l,\ \ \ \ \ \ \ \ \ &\text{on the right boundary}.
\end{align}
After the hodograph transform, the $(x,t)$ plane is transformed to the hodograph $(r_1,r_2)$ plane shown in Fig.~3(a), and implicit solutions of Eq.~(A6) can be determined by the single function $\hat{g}(r_1,r_2)$ satisfying Euler-Poisson-Darboux (EPD) equation (B4). According to the general solution (B5), we need to derive the boundary conditions for the phase function $\hat{g}$.

At the left boundary of the interaction region, we consider Eq.~(B2) for $i=2$ as
\begin{equation}
x-v_2(\chi_m,r_2,\chi_l)t=\hat{w}_2(\chi_m,r_2),
\end{equation}
taking into account the first condition in Eq.~(10). Because the left boundary is connected to the DSW, we obtain the left boundary condition for $\hat{w}_2(r_1,r_2)$
\begin{equation}
\hat{w}_2(\chi_m,r_2)=\hat{g}(\chi_m,r_2)-\frac{L(\chi_m,r_2,\chi_l)}{\partial_2L(\chi_m,r_2,\chi_l)}\partial_2\hat{g}(\chi_m,r_2)=0,
\end{equation}
comparing with Eq.~(8). Integrating the above equation yields
\begin{equation}
\hat{g}(\chi_m,r_2)=\hat{c}_1L(\chi_m,r_2,\chi_l),
\end{equation}
with constant $\hat{c}_1$. Similarly, according to the second condition in Eq.~(10), the Riemann invariant $r_1$ matches the solution of the dispersionless limit of Eq.~(1) and $r_2\rightarrow r_3$, at the right boundary of the interaction region. Comparing Eq.~(B2) for $j=1$ with the Riemann invariant $r=\frac{x-D}{t}$, we have the right boundary condition for $\hat{w}_1(r_1,r_2)$ as
\begin{align}
\hat{w}_1(r_1,\chi_l)&=\hat{g}(r_1,\chi_l)-\frac{L(r_1,\chi_l,\chi_l)}{\partial_1L(r_1,\chi_l,\chi_l)}\partial_1\hat{g}(r_1,\chi_l)\notag\\
&=D.
\end{align}
Upon integration, we obtain
\begin{equation}
\hat{g}(r_1,\chi_l)=\frac{\hat{c}_2}{\sqrt{\chi_l-r_1}}+D,
\end{equation}
with constant $\hat{c}_2$.

Next, we need to determine unknown functions $\hat{\phi}_1(r)$ and $\hat{\phi}_2(r)$ in the general solution (B5) for $i=1,\ j=2$, using boundary conditions Eqs.~(13) and (15). Without loss of generality, we assume that $a_1=\chi_m,\ a_2=\chi_l$, and the second integral converges to 0 as $r_2\rightarrow\chi_l$, then one yields
\begin{equation}
\int_{\chi_m}^{r_1}\frac{\hat{\phi}_1(r)}{\sqrt{r_1-r}\sqrt{\hat{\chi}_l-r}}dr=\frac{\hat{c}_2}{\sqrt{\chi_l-r_1}}+D,
\end{equation}
noting that we have modified the form to ensure the formulas under square roots are always positive within the integral interval. The inverse Abel transform gives
\begin{align}
\hat{\phi}_1(r_1)&=\frac{\sqrt{\chi_l-r_1}}{\pi}\frac{d}{dr_1}\int_{\chi_m}^{r_1}\frac{\hat{c}_2+D\sqrt{\chi_l-r}}{\sqrt{r_1-r}}dr\notag\\
&=\frac{\hat{c}_2\sqrt{\chi_l-\chi_m}+D(\chi_l-r_1)}{\pi\sqrt{r_1-\chi_m}\sqrt{\chi_l-r_1}}.
\end{align}
Let $\hat{c}_1=0$ and $\hat{\phi}_2\equiv0$, then the phase function $\hat{g}(r_1,r_2)$ satisfies
\begin{equation}
\hat{g}(\chi_m,r_2)=\underset{r_1\rightarrow \chi_m}{\lim}\int_{\chi_m}^{r_1}\frac{\hat{\phi}_1(r)}{\sqrt{r_1-r}\sqrt{r_2-r}}dr=0,
\end{equation}
which requires$\underset{r\rightarrow \chi_m}{\lim}\hat{\phi}_1(r)=0$, i.e.,
\begin{equation}
\hat{c}_2=-D\sqrt{\chi_l-\chi_m},
\end{equation}
then
\begin{equation}
\hat{\phi}_1(r)=-\frac{D}{\pi}\frac{\sqrt{r-\chi_m}}{\sqrt{\chi_l-r}}.
\end{equation}
Finally,
\begin{align}
\hat{g}(r_1,r_2)&=-\frac{D}{\pi}\int_{\chi_m}^{r_1}\frac{\sqrt{r-\chi_m}}{\sqrt{r_1-r}\sqrt{r_2-r}\sqrt{\chi_l-r}}dr\notag\\
&=\frac{2D}{\pi}\frac{(\chi_l-\chi_m)\big(\Pi(\hat{n},\hat{k})-\text{K}(\hat{k})\big)}{\sqrt{r_2-\chi_m}\sqrt{\chi_l-r_1}},
\end{align}
where $\text{K}(\hat{k})$ and $\Pi(\hat{n},\hat{k})$ is the complete elliptic integral of the first and third kind with
$$
\hat{n}=\frac{\chi_m-r_1}{\chi_l-r_1},\ \ \hat{k}=\frac{(r_1-\chi_m)(\chi_l-r_2)}{(r_2-\chi_m)(\chi_l-r_1)}.
$$
Now the oscillation in the interaction region can be analytically described by the single-phase periodic wave (A3) in which Riemann invariants are governed by Eqs.~(B2), (B3) and (21). Its two boundaries are
\begin{align}
x^-&=v_1(\chi_m,r_2,\chi_l)t+\hat{w}_1(\chi_m,r_2)\notag\\
&=v_2(\chi_m,r_2,\chi_l)t+\hat{w}_2(\chi_m,r_2),
\end{align}
and
\begin{align}
x^+&=v_1(r_1,\chi_l,\chi_l)t+\hat{w}_1(r_1,\chi_l)\notag\\
&=v_2(r_1,\chi_l,\chi_l)t+\hat{w}_2(r_1,\chi_l).
\end{align}
Figs.~2 show the evolution of the Riemann invariants and the solution of Eq.~(1). Our analytical results agree well with the numerical simulations. When $\chi_l<\chi_r$, Fig.~3(b) illustrates the evolution on the $(x,t)$ plane, and we can see the retained RW from it.

\begin{figure*}
\includegraphics[scale=0.5]{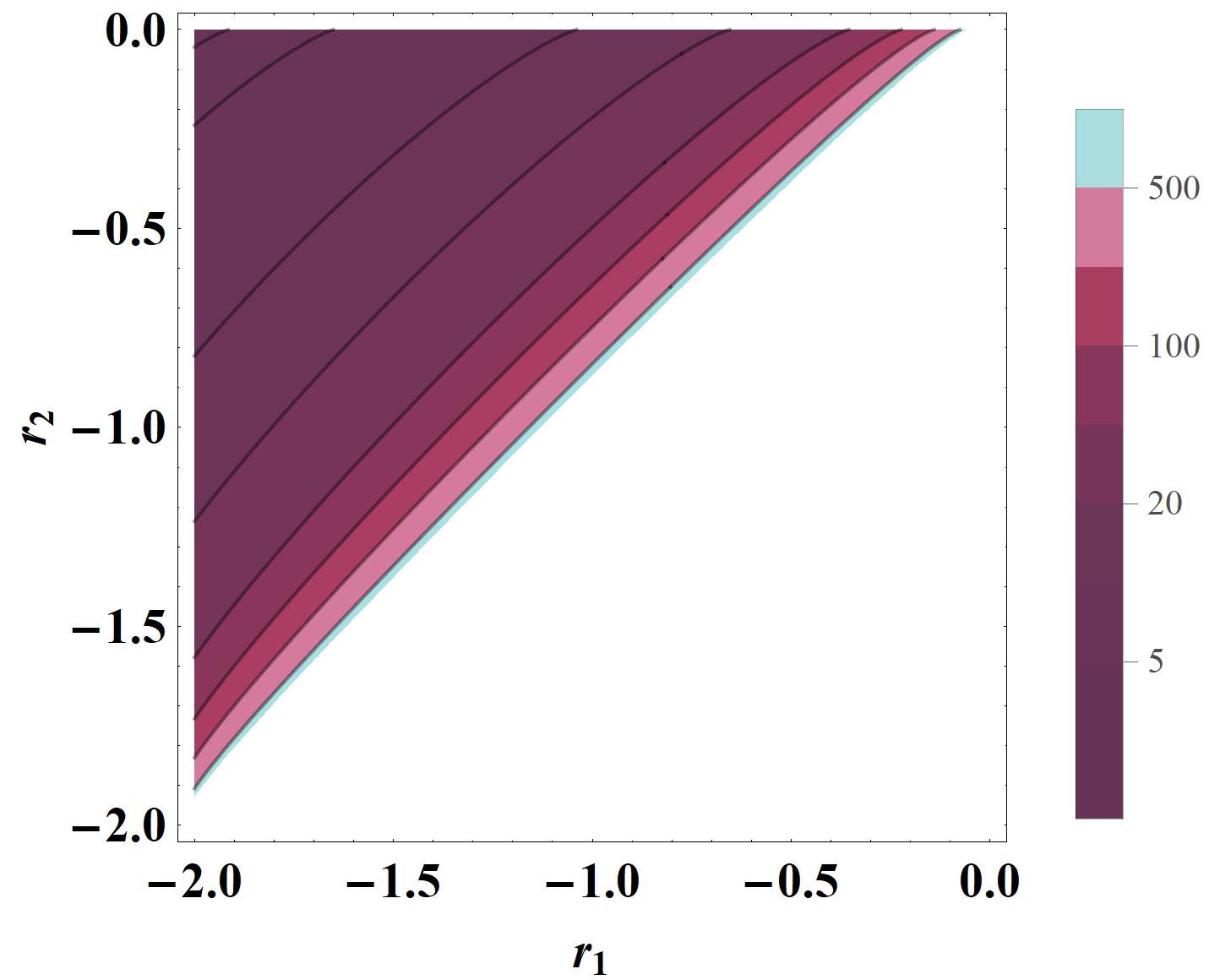}\hfill
\includegraphics[scale=0.38]{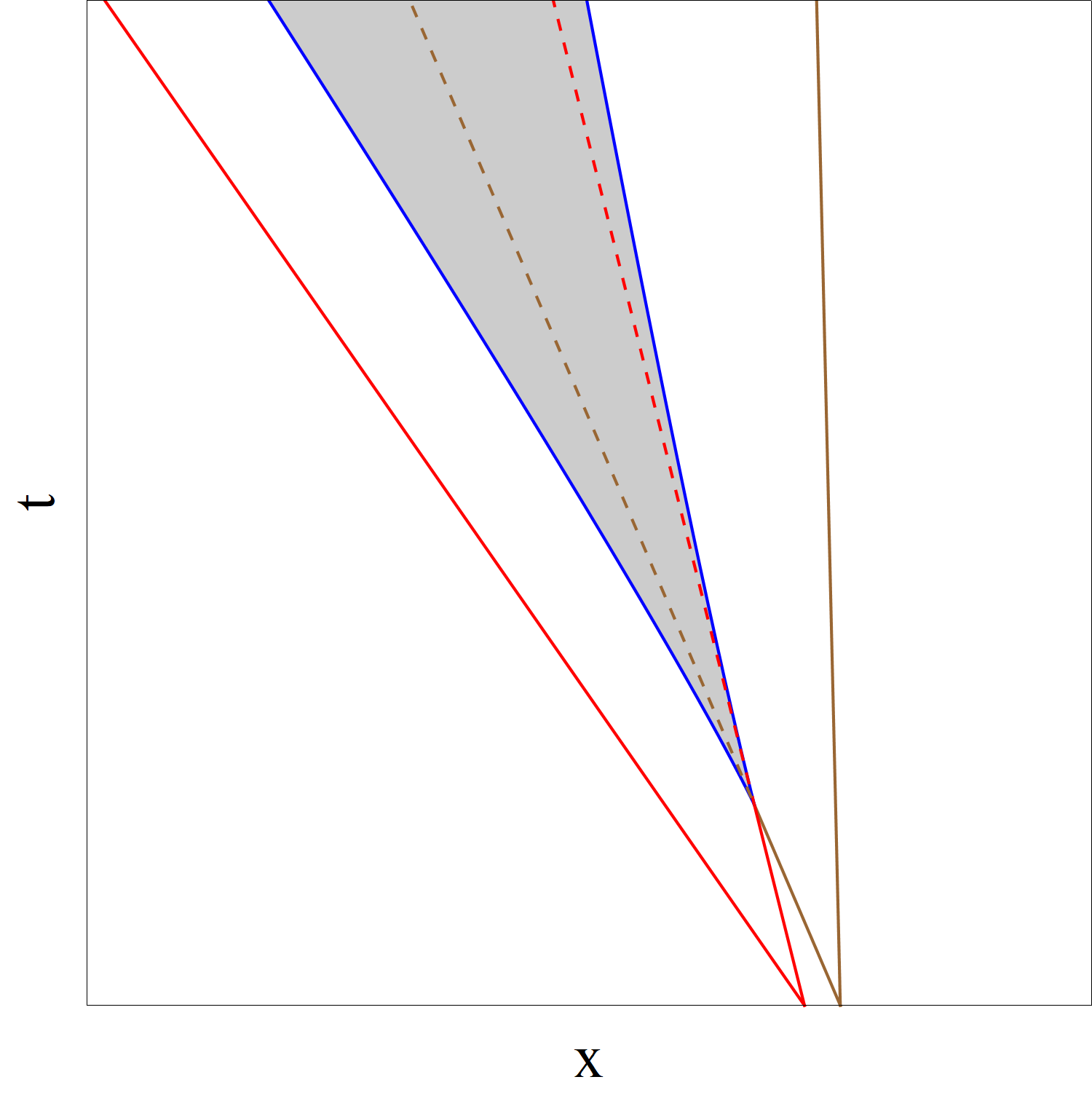}\hfill
\includegraphics[scale=0.38]{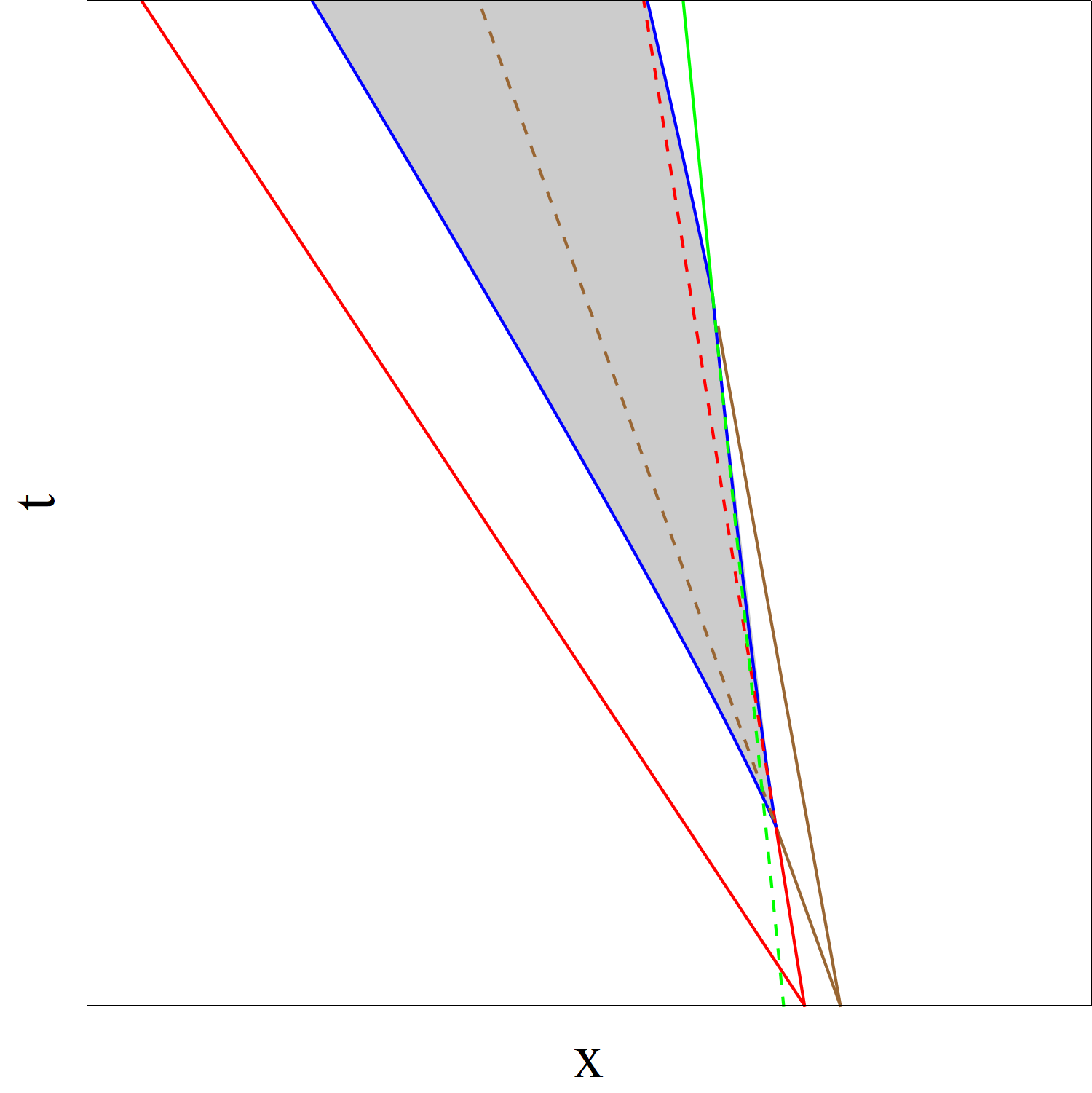}\hfill\\
\vspace{0cm}{\footnotesize\hspace{1.0cm}(a)\hspace{5.3cm}(b)\hspace{5.3cm}(c)}
\caption{(a) Contour plot of $\frac{\hat{w}_2(r_1,r_2)-\hat{w}_1(r_1,r_2)}{v_1(r_1,r_2,\chi_l)-v_2(r_1,r_2,\chi_l)}$ in the hodograph $(r_1,r_2)$ plane with parameters $u_l=0$, $u_m=-1$, $u_r=0.3$; (b) Evolution with retained RW on the $(x,t)$ plane; (c) Evolution with escaping DSW on the $(x,t)$ plane. The brown, red and blue lines are edges of the RW, DSW and interaction region, respectively, and the dashed lines represents original trajectories if there had been no collision. The green line is the leading edge of the DSW existing from the interaction region.}
\label{Figs.~3.}
\end{figure*}

From above analyses, the interaction region gradually expands, so we need to discuss its asymptotic state when $t\rightarrow \infty$ by finding the asymptotic boundaries.

On the right boundary, the second Whitham velocity represents the boundary expansion speed as follows
\begin{equation}
\frac{dx^+}{dt}=v_2(r_1,\chi_l,\chi_l)=2(r_1+2\chi_l).
\end{equation}
Due to $x^+$ separates the interaction and RW region, Riemann invariant $r_1$ equals to $\frac{x^+-D}{6t}$, then we have
\begin{equation}
x^+=D+6\chi_lt+\hat{c}t^{\frac{1}{3}},
\end{equation}
where $\hat{c}=-32\big(D^2(\chi_l-\chi_m)\big)^\frac{1}{3}$ determined by the key point (9). Then the right boundary can be expressed in terms of $t$ as
\begin{equation}
x^+=D+6\chi_lt-32\big(D^2(\chi_l-\chi_m)t\big)^\frac{1}{3},
\end{equation}
the first two term of above equation is the same as the trailing edge of a RW evolving from the initial data $u\left( x,t \right) =\left\{ \begin{array}{l}
	u_l,\ \ \ x<D\\
	u_r,\ \ \ x>D\\
\end{array} \right. $.

On the left boundary,
\begin{equation}
\hat{w}_1(\chi_m,r_2)=\frac{D\sqrt{r_2-\chi_m}}{(1-\mu)\sqrt{\chi_l-\chi_m}},\ \   \hat{w}_2(\chi_m,r_2)=0,
\end{equation}
where $\mu=\text{E}(m)/\text{K}(m)$ omitted from the independent variable $m$. From Eq.~(B2), varying Riemann invariants $r_1$ and $r_2$ can be connected by the following formula
\begin{equation}
t=\frac{\hat{w}_2(r_1,r_2)-\hat{w}_1(r_1,r_2)}{v_1(r_1,r_2,\chi_l)-v_2(r_1,r_2,\chi_l)}
\end{equation}
for fixed $t$. Substituting Eqs.~(A13) and (21) into (22) and (28), the left boundary can be determined by modulus $m$ as follows
\begin{align}
t^-=&\frac{D}{4\sqrt{m}(\chi_l-\chi_m)}\Bigg(1+\frac{(1-m)(1-\mu)}{(2-m)\mu-2(1-m)}\Bigg),\notag\\
x^-=&\frac{D}{2\sqrt{m}(\chi_l-\chi_m)}\Bigg(\chi_m+\frac{\chi_l\big((1+m)\mu-(1-m)\big)}{(2-m)\mu-2(1-m)}\Bigg.\notag\\
&\Bigg.-\frac{3(\chi_l-\chi_m)m(1-m)}{(2-m)\mu-2(1-m)}\Bigg).
\end{align}

Fig.~3(a) shows the trajectories of $r_1$ and $r_2$ for different time in the hodograph plane, and implies the fact that two varying Riemann invariants are asymptotically equal as $t\rightarrow \infty$, i.e., $m\rightarrow 0$, so the modulated single-phase periodic wave in the interaction region degenerates into a small amplitude wave train. Expanding Eqs.~(29) near $m=0$, one yields
\begin{align}
t^-&=\frac{D}{3(\chi_l-\chi_m})m^{-\frac{3}{2}}-\frac{D}{24(\chi_l-\chi_m)\sqrt{m}}+O(\sqrt{m}),\notag\\
x^-&=\frac{2D(2\chi_m-\chi_l)}{(\chi_l-\chi_m)m^{\frac{3}{2}}}+\frac{D(13\chi_l-14\chi_m)}{4(\chi_l-\chi_m)\sqrt{m}}+O(\sqrt{m}).
\end{align}
\subsubsection {Partial transmission of the dispersive shock wave}
In addition, we study a special case in which the DSW propagates out of the interaction region.  This evolution is shown in Figs.~4 for the interaction between a RW and a DSW with a dark soliton front. If $\chi_l>\chi_r$, there is another critical instant $t_2$ that needs to be considered, that is when the whole RW is absorbed into the interaction region, i.e., the interaction region is directly connected with the right plateau. At this point,
$$
r_1=\chi_r,\ \ r_2=r_3=\chi_l,
$$
on the right boundary of the interaction region, which is also the leading edge of the RW, so
\begin{equation}
x^+=v_2(\chi_r,\chi_l,\chi_l)t_2+\hat{w}_2(\chi_r,\chi_l)=x_{\text{RW}}^+=6\chi_rt_2+D,\notag
\end{equation}
which gives
\begin{align}
&t_2=\frac{D-\hat{w}_2(\chi_r,\chi_l)}{4(\chi_l-\chi_r)}=\frac{D\sqrt{\chi_l-\chi_m}}{4(\chi_l-\chi_r)^\frac{3}{2}},\notag\\
&x_2=D\left(1+\frac{3\chi_r\sqrt{\chi_l-\chi_m}}{2(\chi_l-\chi_r)^\frac{3}{2}}\right).
\end{align}

When $t>t_2$, both sides of the interaction region are connected directly to DSWs, as the soliton edge of the right DSW emerges first. For the right DSW, the modulation solution is given by two constant Riemann invariants
\begin{equation}
r_1=\chi_r,\ \ r_3=\chi_l,
\end{equation}
and the remaining one being a simple wave modulation solution
\begin{align}
x&=v_2(\chi_r,r_2,\chi_l)t+\hat{P}(r_2)\notag\\
&=\left(2(\chi_r+r_2+\chi_l)-\frac{4(r_2-\chi_r)(1-m)}{\mu(m)-(1-m)}\right)t+\hat{P}(r_2),
\end{align}
where modulus in elliptic integral is
\begin{equation}
m=\frac{r_2-\chi_r}{\chi_l-\chi_r},
\end{equation}
and the function $\hat{P}(r_2)$ is given by
\begin{align}
\hat{P}(r)=\hat{w}_2(r,\chi_l)
=\frac{2D\big( (\chi_l-\chi_m)\Pi(\hat{p},\hat{q})+\hat{Y}\big)}{\pi\sqrt{r-\chi_m}\sqrt{\chi_l-\chi_r}},
\end{align}
with
\begin{align}
&\hat{Y}=\frac{(\chi_l-\chi_r)\Big((\chi_l-\chi_m)\big(1-\mu(\hat{y})\big)\text{K}(\hat{q})-(r-\chi_m)\text{E}(\hat{q})\Big)}{(\chi_l-\chi_r)\mu(\hat{y})-(\chi_l-r)},\notag\\
&\hat{p}=\frac{\chi_m-\chi_r}{\chi_l-\chi_r},\ \ \hat{q}=\frac{(\chi_l-r)(\chi_r-\chi_m)}{(r-\chi_m)(\chi_l-\chi_r)},\ \ \hat{y}=\frac{r-\chi_r}{\chi_l-\chi_r}.
\end{align}
This DSW is a simple wave solution corresponding to the following initial data in terms of Riemann invariant for Eq.~(1)
\begin{equation}
r(x,0)=\hat{P}^{-1}(x),
\end{equation}
where $\hat{P}^{-1}(x)$ being the inverse of $x=\hat{P}(r)$. Then, the leading edge of the transmitted DSW are found by Eq.~(33) as $2(\chi_l+2\chi_r)t+\hat{P}(\chi_l)$ shown by the green line in Figs.~3(c). The phase shift is $\hat{P}(\chi_l)$, compared with the original DSW.
\begin{figure}
\includegraphics[scale=0.32]{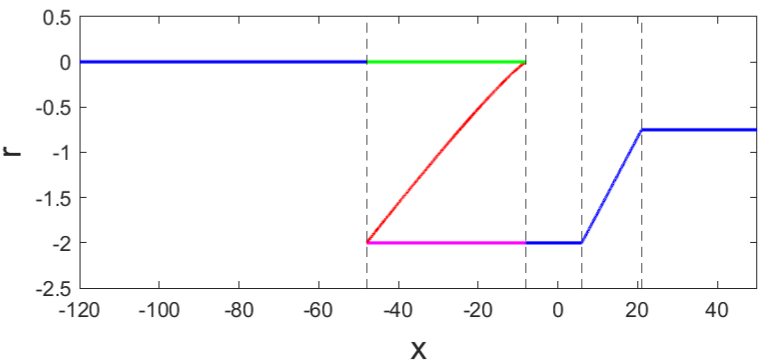}\hfill
\includegraphics[scale=0.32]{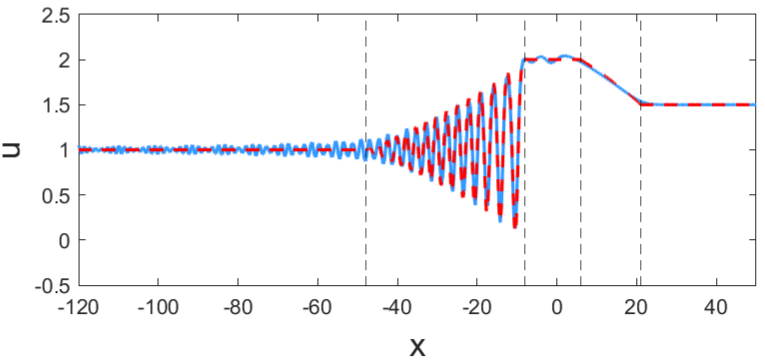}\\
\includegraphics[scale=0.32]{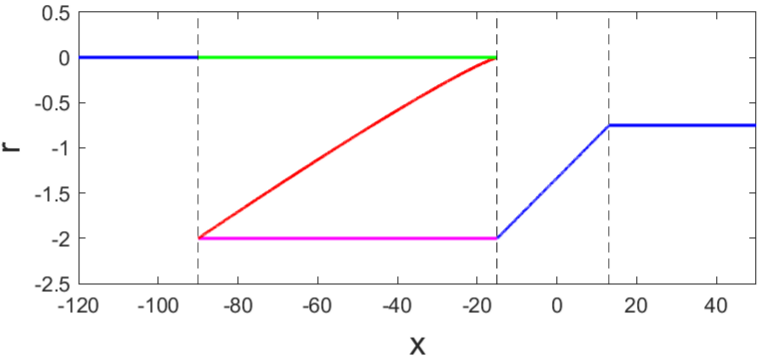}\hfill
\includegraphics[scale=0.32]{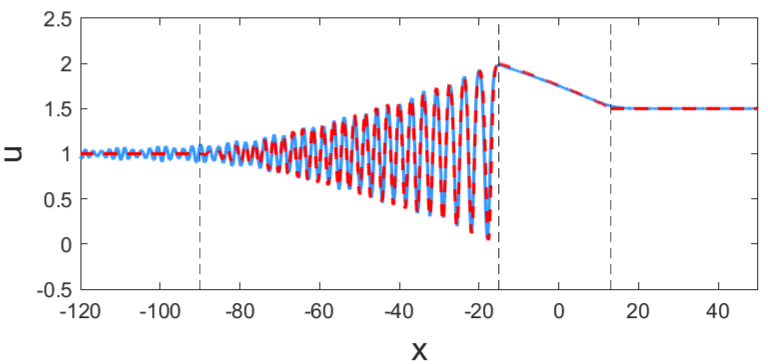}\\
\includegraphics[scale=0.32]{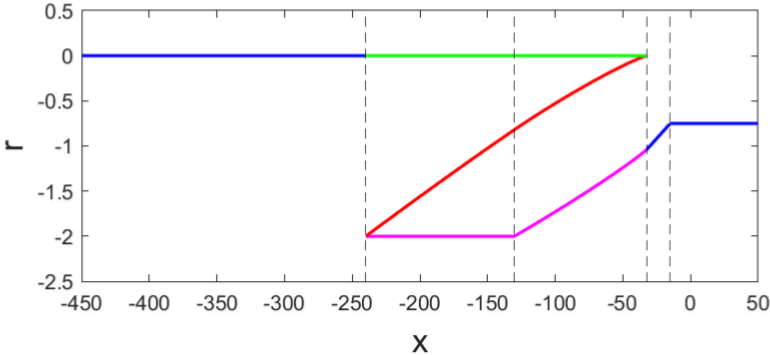}\hfill
\includegraphics[scale=0.32]{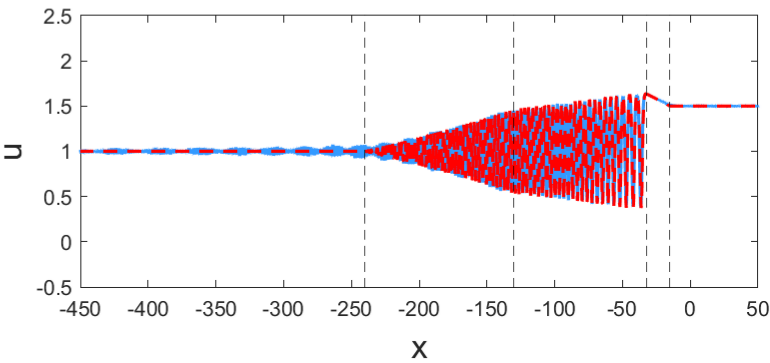}\\
\includegraphics[scale=0.32]{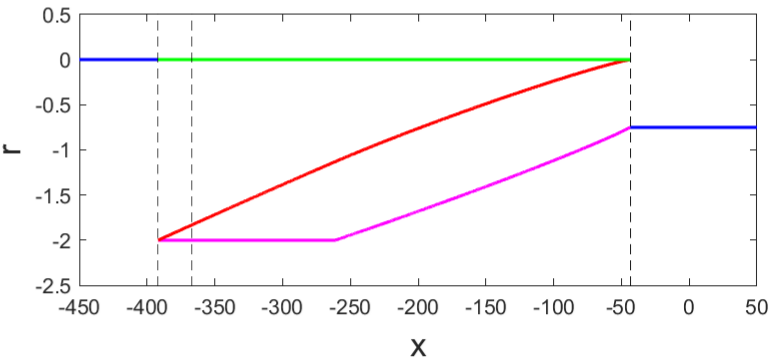}\hfill
\includegraphics[scale=0.32]{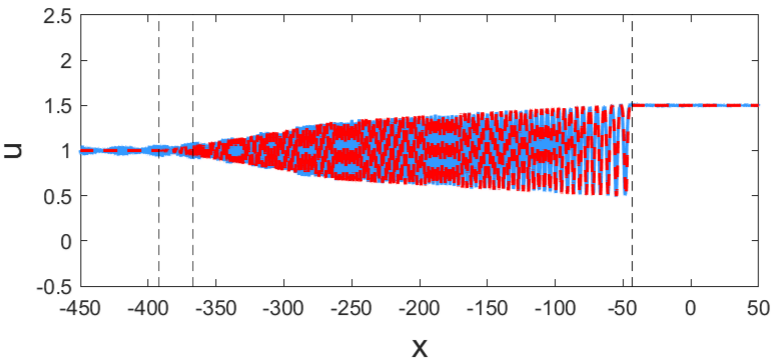}\\
\includegraphics[scale=0.32]{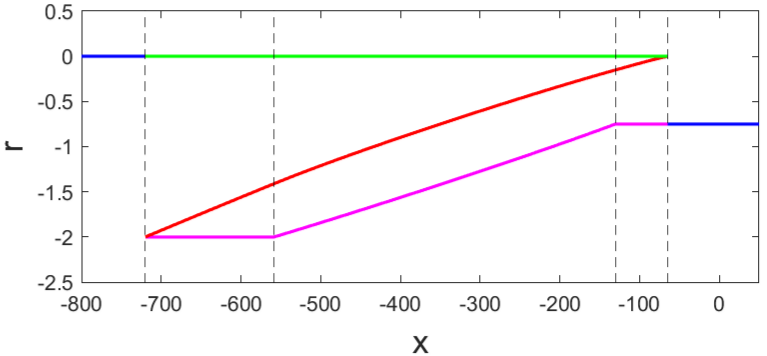}\hfill
\includegraphics[scale=0.32]{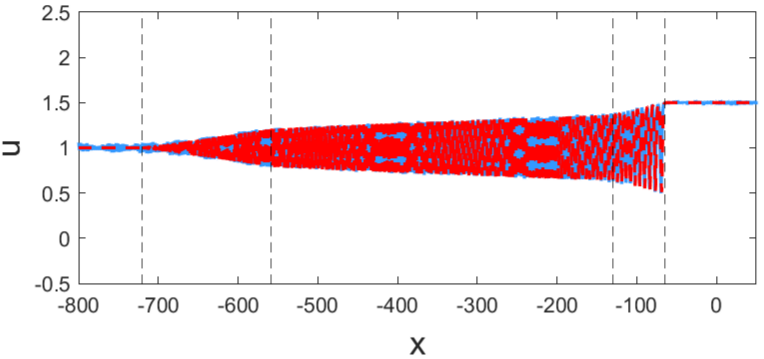}\\

{\footnotesize\hspace{0.0cm}(a)\hspace{4cm}(b)}
\caption {Evolution of DSW-RW interaction with transmitted DSW, at $t=2$, $t=t_1$, $t=10$, $t=t_2$, and $t=30$, respectively. The parameters are $u_l=1$, $u_m=2$, $u_r=1.5$, $D=30$, $\alpha=1$. (a) Riemann invariants; (b) Solution of Eq.~(1). The red dashed line represents the analytical results based on the Whitham modulation theory, while the blue solid line represents the numerical results.}
\label{Figs.~4.}
\end{figure}
\subsection{RW-DSW interaction}
In this section, we focus on an initial value satisfying $\chi_m>\text{max} (\chi_l, \chi_r)$, which is different from Sec.~\uppercase\expandafter{\romannumeral3} A. This causes the previous structure to switch left and right, resulting in a RW on the left and a DSW on the right.

\begin{figure}
\includegraphics[scale=0.33]{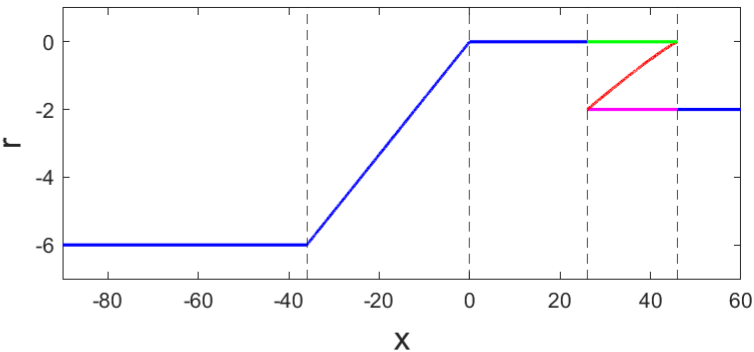}\hfill
\includegraphics[scale=0.33]{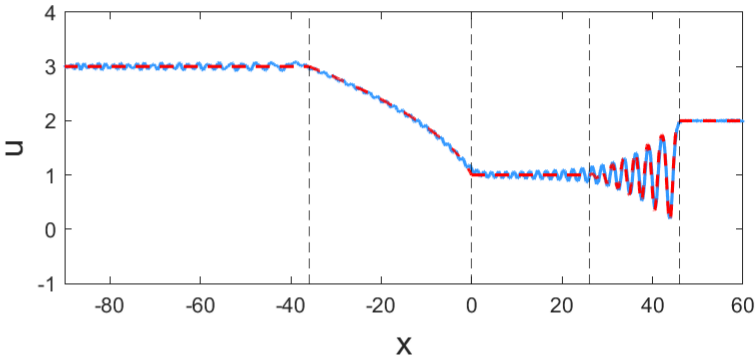}\\
\includegraphics[scale=0.33]{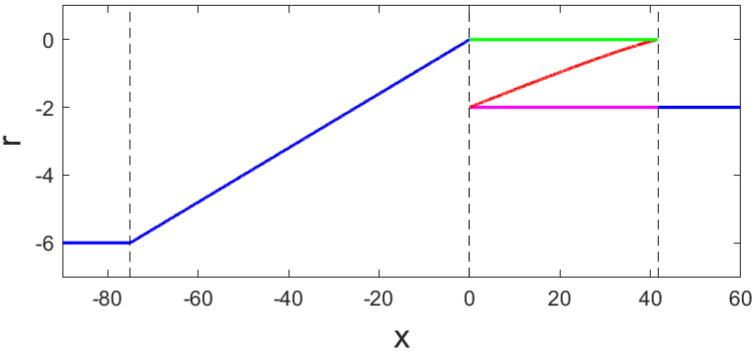}\hfill
\includegraphics[scale=0.33]{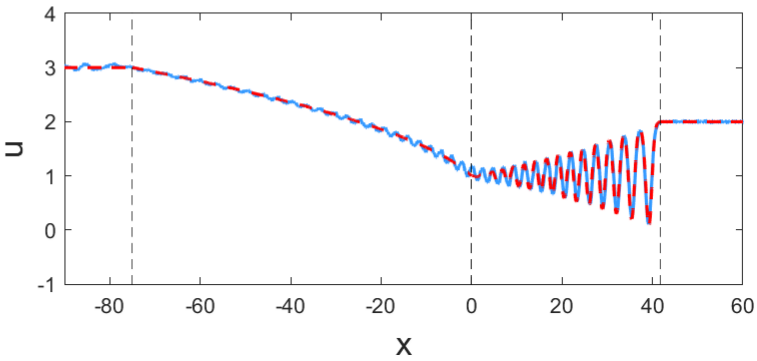}\\
\includegraphics[scale=0.33]{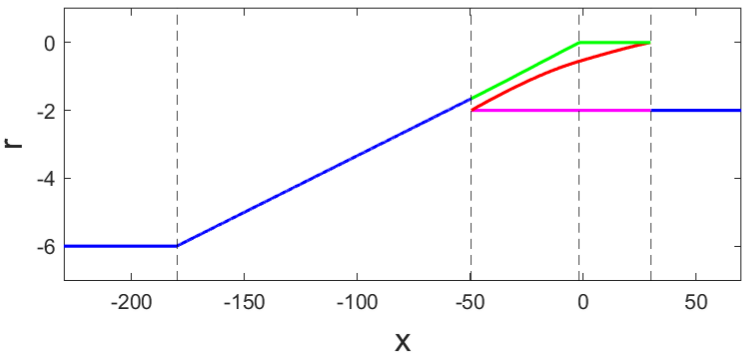}\hfill
\includegraphics[scale=0.33]{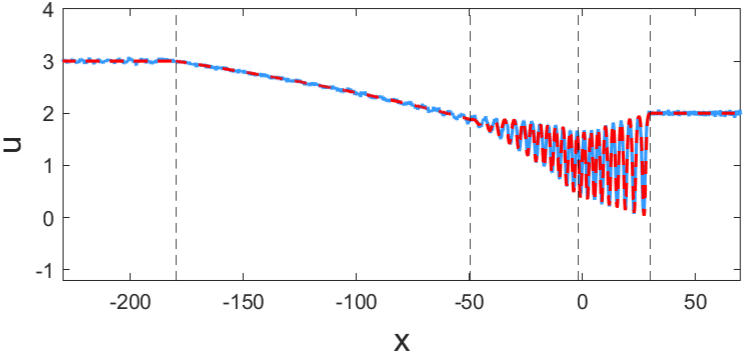}\\

{\footnotesize\hspace{0.0cm}(a)\hspace{4cm}(b)}
\caption{Evolution of RW-DSW interaction with retained RW, at $t=1$, $t=t_1$, and $t=5$, respectively. The parameters are $u_l=3$, $u_m=1$, $u_r=2$, $D=50$, $\alpha=1$. (a) Riemann invariants; (b) Solution of Eq.~(1). The red dashed line represents the analytical results based on the Whitham modulation theory, while the blue solid line represents the numerical results.}
\label{Figs.~5.}
\end{figure}
\subsubsection{Partial retention of the rarefaction wave}
For initial data (3) with $\frac{1}{2\alpha}<u_m<u_r<u_l$, we have $\chi_l<\chi_r<\chi_m$, resulting wave breaking at $x=D$. Subsequently, the $(x,t)$ plane is divided into five regions (see the first row in Figs.~5 and 7). For the centered RW $u(\tau)=\frac{1}{2\alpha}\left( 1+\sqrt{1-\frac{2\alpha}{3}\tau} \right)$ with boundaries
\begin{equation}
x_{\text{RW}}^-=6\chi_lt,\ \ x_{\text{RW}}^+=6\chi_mt,
\end{equation}
the dispersionless Riemann invariant is $r=\frac{\tau}{6}$. The DSW oscillates between the trailing edge
\begin{equation}
x_{\text{DSW}}^-=(12\chi_r-6\chi_m)t+D,
\end{equation}
and the leading edge
\begin{equation}
x_{\text{DSW}}^+=(2\chi_r+4\chi_m)t+D.
\end{equation}
Obviously, the middle plateau gradually decreases due to $\chi_m>\chi_r$, and disappears if $x_{\text{RW}}^+=x_{\text{DSW}}^-$, implying that the centered RW will intersects the DSW at $(x_1,t_1)$ as follows
\begin{equation}
x_1=\frac{\chi_mD}{2(\chi_m-\chi_r)},\ \ t_1=\frac{D}{12(\chi_m-\chi_r)}.
\end{equation}
The collision here has no counterpart in the KdV equation due to it only allows normal RWs and DSWs, we can not compare the time required for the collision to occur like Sec.~\uppercase\expandafter{\romannumeral3} A.

After the leading edge of the RW colliding with the trailing edge of the DSW, there occurs a new single-phase region (see the third row in Figs.~5 and 7) in which two Riemann invariants vary and self-similar solutions are no longer applicable.

The interaction region separates the remaining parts of the RW and DSW, so the varying Riemann invariants $r_2$ and $r_3$ satisfy
\begin{align}
r_1=r_2=\chi_r,\ \ \ \ &\text{on the left boundary},\notag\\
r_1=\chi_r,\ \ \ r_3=\chi_m,\ \ \ \ &\text{on the right boundary}.
\end{align}
With the help of the hodograph transform, the solutions of two non-trivial Whitham modulation equations are implicitly determined by
\begin{equation}
x-v_2t=\check{w}_2(r_2,r_3),\ \ x-v_3t=\check{w}_3(r_2,r_3).
\end{equation}
Next, we will transform the boundary conditions of Riemann invariants $r_i$ into the counterpart of $\check{w}_i$ and the solution $\check{g}$ of the EPD equation. At the left boundary, combining Eq.~(43) with (42), we have
\begin{equation}
\check{w}_3(\chi_r,r_3)=\check{g}(\chi_r,r_3)-\frac{L(\chi_r,\chi_r,r_3)}{\partial_3L(\chi_r,\chi_r,r_3)}\partial_3\check{g}(\chi_r,r_3)=0,
\end{equation}
considering the limit (A13). Integrating the above equation yields
\begin{equation}
\check{g}(\chi_r,r_3)=\frac{\check{c}_1}{\sqrt{r_3-\chi_r}},
\end{equation}
with the constant $\check{c}_1$. At the right boundary of the interaction region, $r_2$ is self-similar through $\frac{x-D}{t}=v_2(\chi_r,r_2,\chi_m)$, then one has
\begin{equation}
\check{w}_2(r_2,\chi_m)=\check{g}(r_2,\chi_m)-\frac{L(\chi_r,r_2,\chi_m)}{\partial_2L(\chi_r,r_2,\chi_m)}\partial_2\check{g}(r_2,\chi_m)=D,
\end{equation}
which gives
\begin{equation}
\check{g}(r_2,\chi_m)=\check{c}_2\text{K}(\frac{r_2-\chi_r}{\chi_m-\chi_r})+D.
\end{equation}

Next, we need to find $\check{g}(r_2,r_3)$ possessing boundary conditions (45) and (47). From Eq.~(B5), phase $\check{g}(r_2,r_3)$ has the following general form
\begin{align}
\check{g}(r_2,r_3)=&\int_{\chi_r}^{r_2}\frac{\check{\phi}_2(r)}{\sqrt{r_2-r}\sqrt{r_3-r}}dr\notag\\
&+\int_{r_3}^{\chi_m}\frac{\check{\phi}_3(r)}{\sqrt{r-r_2}\sqrt{r-r_3}}dr.
\end{align}
Without loss of generality, we assume $\check{c}_2=0$ and the second integral converges to zero as $r_3\rightarrow \chi_m$, then Eq.~(47) is rewritten as
\begin{equation}
\check{g}(r_2,\chi_m)=\int_{\chi_r}^{r_2}\frac{\check{\phi}_2(r)}{\sqrt{r_2-r}\sqrt{\chi_m-r}}dr=D.
\end{equation}
By means of the inverse Abel transform, the unknown function $\check{\phi}_2$ can be solved as
\begin{align}
\check{\phi}_2(r_2)&=\frac{\sqrt{\chi_m-r_2}}{\pi}\frac{d}{dr_2}\int_{\chi_r}^{r_2}\frac{D}{\sqrt{r_2-r}}dr\notag\\
&=\frac{D\sqrt{\chi_m-r}}{\pi\sqrt{r-\chi_r}},
\end{align}
so
\begin{align}
\check{g}(r_2,r_3)=&\int_{\chi_r}^{r_2}\frac{D\sqrt{\chi_m-r}}{\pi\sqrt{r_2-r}\sqrt{r_3-r}\sqrt{r-\chi_r}}dr\notag\\
&+\int_{r_3}^{\chi_m}\frac{\check{\phi}_3(r)}{\sqrt{r-r_2}\sqrt{r-r_3}}dr.
\end{align}
We turn to another boundary condition (45) for phase $\check{g}(r_2,r_3)$ on the left boundary of the interaction region. Considering the limit of the first term in Eq.~(51) as $r_2\rightarrow r_1=\chi_r$,
\begin{equation}
\underset{r_2\rightarrow \chi _r}{\lim}\int_{\chi _r}^{r_2}{\frac{\check{\phi} _2\left( r \right)}{\sqrt{r_2-r}\sqrt{r_3-r}}dr}=\frac{D\sqrt{\chi _m-\chi _r}}{\sqrt{r_3-\chi _r}}.
\end{equation}
Then Eq.~(45) is satisfied when $\check{\phi}_3=0$ if we set $\check{c}_1=D\sqrt{\chi_m-\chi_r}$. Finally, the phase $\check{g}(r_2,r_3)$ can be expressed in term of the third complete elliptic integral as follows
\begin{align}
\check{g}(r_2,r_3)&=\int_{\chi_r}^{r_2}\frac{D\sqrt{\chi_m-r}}{\pi\sqrt{r_2-r}\sqrt{r_3-r}\sqrt{r-\chi_r}}dr\notag\\
&=\frac{2D}{\pi}\frac{\chi_m-\chi_r}{\sqrt{r_3-\chi_r}\sqrt{\chi_m-r_2}}\Pi(\check{n},\check{k}),
\end{align}
with
\begin{equation}
\check{n}=-\frac{r_2-\chi_r}{\chi_m-r_2},\ \ \check{k}=\frac{(r_2-\chi_r)(\chi_m-r_3)}{(r_3-\chi_r)(\chi_m-r_2)}.
\end{equation}

Above all, the oscillation in the interaction region can be analytically described by the single-phase periodic wave (A3) with Riemann invariants given by Eqs.~(43), (B3) and (53). The left boundary of the interaction region is
\begin{align}
x^-=&v_2(\chi_r,\chi_r,r_3)t+\check{w}_2(\chi_r,r_3)\notag\\
=&v_3(\chi_r,\chi_r,r_3)t+\check{w}_3(\chi_r,r_3),\notag
\end{align}
and the right one is
\begin{align}
x^+&=v_2(\chi_r,r_2,\chi_m)t+\check{w}_2(r_2,\chi_m)\notag\\
&=v_3(\chi_r,r_2,\chi_m)t+\check{w}_3(r_2,\chi_m).\notag
\end{align} The numerical and analytical results evolving from initial data (3), shown in Figs.~5, demonstrate excellent consistency. If $\chi_l<\chi_r$, the retained RW can be observed in Fig.~6(b) which shows the evolution on the $(x,t)$ plane.
\begin{figure*}
\includegraphics[scale=0.5]{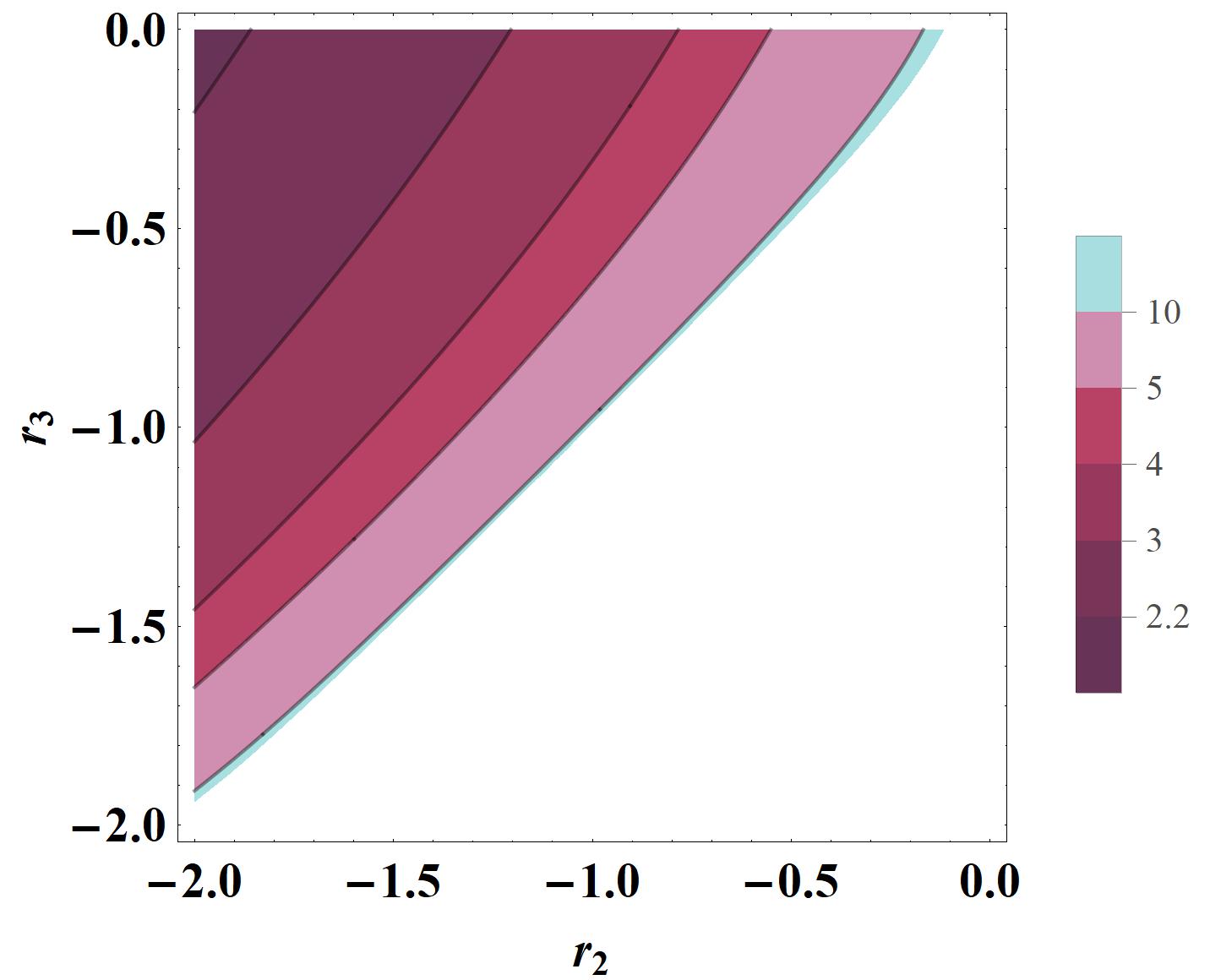}\hfill
\includegraphics[scale=0.38]{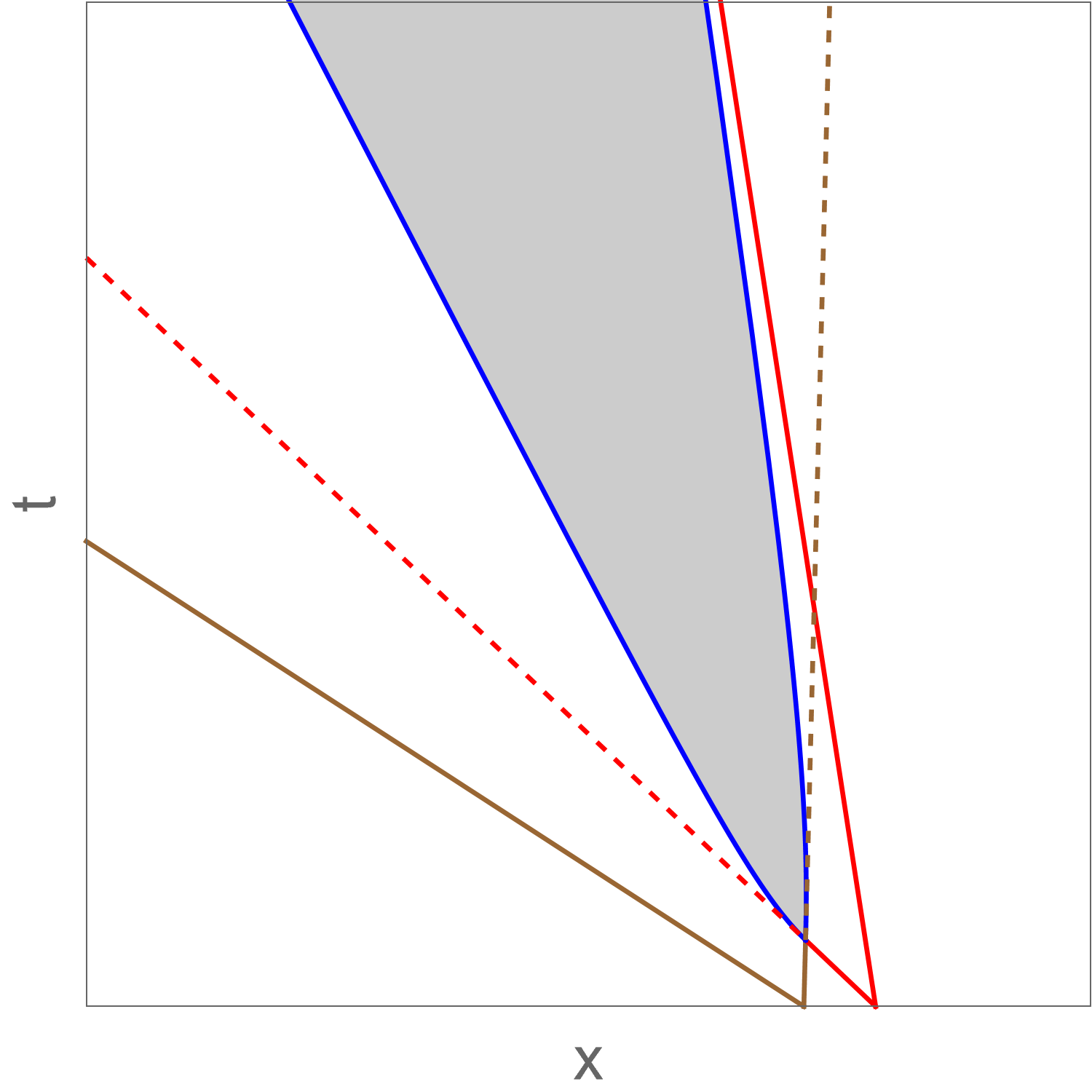}\hfill
\includegraphics[scale=0.38]{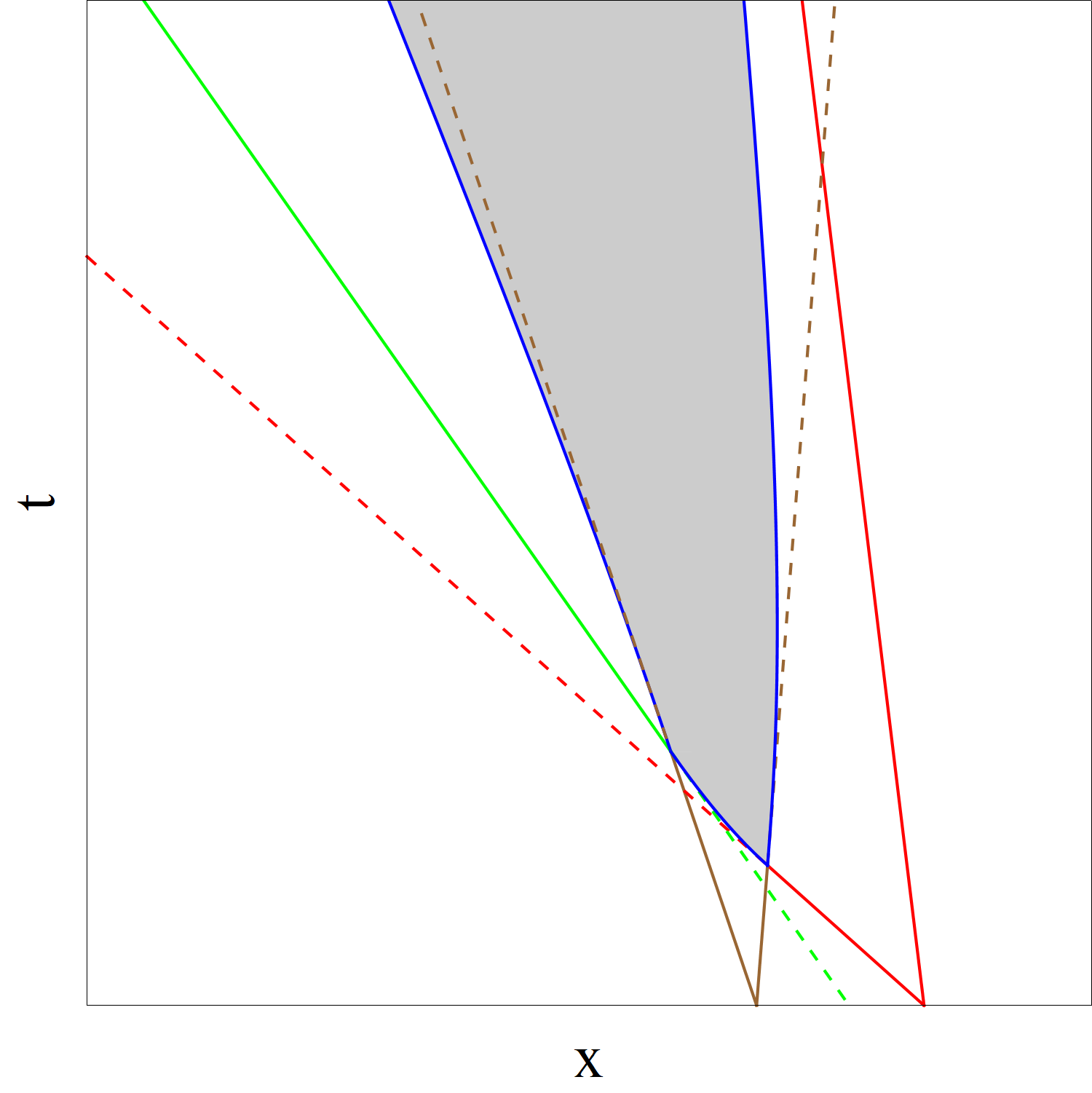}\hfill\\
{\footnotesize\hspace{1.0cm}(a)\hspace{5.3cm}(b)\hspace{5.3cm}(c)}
\caption{(a) Contour plot of $\frac{\check{w}_3(r_2,r_3)-\check{w}_2(r_2,r_3)}{v_2(\chi_r,r_2,r_3)-v_3(\chi_r,r_2,r_3)}$ on the hodograph $(r_2,r_3)$ plane with parameters $u_l=3$, $u_m=1$, $u_r=2$; (b) Evolution with retained RW on the $(x,t)$ plane; (c) Evolution with escaping DSW on the $(x,t)$ plane. The brown, red and blue lines are edges of the RW, DSW and interaction region, respectively, and the dashed lines represents original trajectories if there had been no collision. The green line is the leading edge of the DSW existing from the interaction region.}
\label{Figs.~6.}
\end{figure*}

Similar to Sec.~\uppercase\expandafter{\romannumeral3} A, the interaction region continues to expand, and the asymptotic boundaries at infinite time is worth discussing.

On the left boundary adjacent to the RW, $r_2\rightarrow r_1$, then we have
\begin{equation}
\frac{dx^-}{dt}=v_2(\chi_r,\chi_r,r_3)=12\chi_r-6r_3,
\end{equation}
taking account to Eq.~(A13). Due to $r_3$ equals to $\frac{x^-}{6t}$, one yields
\begin{equation}
x^-=6\chi_rt+\frac{\check{c}}{t},
\end{equation}
where $\check{c}=\frac{D^2}{24(\chi_m-\chi_r)}$ determined by Eq.~(41) in the $(x,t)$ plane. Finally, the left boundary can be expressed in terms of $t$ as
\begin{equation}
x^-=6\chi_rt+\frac{D^2}{24(\chi_m-\chi_r)t},
\end{equation}
the first term of above equation is the same as the leading edge of a RW evolving from the initial data $u\left( x,t \right) =\left\{ \begin{array}{l}
	u_l,\ \ \ x<0\\
	u_r,\ \ \ x>0\\
\end{array} \right. $.

On the right side,
\begin{align}
&\check{w}_2(r_2,\chi_m)=D,\notag\\ &\check{w}_3(r_2,\chi_m)=D\big(1-\frac{\sqrt{\chi_m-r_2}}{\mu\sqrt{\chi_m-\chi_3}}\big).
\end{align}
At a fixed time, if one of the varying Riemann invariants is selected within $(\chi_r,\chi_m)$, then the other can be uniquely determined by the following equation
\begin{equation}
t=\frac{\check{w}_3(r_2,r_3)-\check{w}_2(r_2,r_3)}{v_2(\chi_r,r_2,r_3)-v_3(\chi_r,r_2,r_3)}.
\end{equation}
Fig.~6(a) shows the contour plot of $\frac{\check{w}_3(r_2,r_3)-\check{w}_2(r_2,r_3)}{v_2(\chi_r,r_2,r_3)-v_3(\chi_r,r_2,r_3)}$, demonstrating the evolution trajectory of $r_2$ and $r_3$ on the hodograph plane at different times. It can be seen that as time approaches $t_1$, the trajectory degenerates into a point $(\chi_r,\chi_m)$, which is consistent with our analysis and represents the initial formation of the interaction region. Taking into account Eqs.~(B3) and (53), as well as the Whitham characteristic velocities $v_i$, the trajectory of the right boundary of the interaction region can be described by the following parameter equation
\begin{align}
t^+=&\frac{D(\mu+m-1)}{4\sqrt{1-m}(\chi_m-\chi_r)\big((1+m)\mu+m-1\big)},\notag\\
x^+=&D\Bigg(1+\frac{(\chi_m+2\chi_r)(\mu+m-1)}{2(\chi_m-\chi_r)\sqrt{1-m}\big((1+m)\mu+m-1\big)}\big.\notag\\
&\big.-\frac{m\mu+3m(1-m)}{2\sqrt{1-m}\big((1+m)\mu+m-1\big)} \Bigg).
\end{align}
Fig.~6(a) implies the fact that two varying Riemann invariants are asymptotically equal as $t\rightarrow \infty$, i.e., $m\rightarrow 1$, so the oscillation degenerates into a dark soliton train. Expanding Eq.~(60) near $m=1$, one yields
\begin{widetext}
\begin{align}
t^+=&-\frac{D}{8(\chi_m-\chi_r)\sqrt{1-m}}+\frac{D(\text{ln}\frac{16}{1-m}-2)}{32(\chi_m-\chi_r)}\sqrt{1-m}+O\big((1-m)^\frac{3}{2}\big),\notag\\
x^+=&-\frac{D(2\chi_m+\chi_r)}{4(\chi_m-\chi_r)\sqrt{1-m}}+D-3D\frac{\big(2\chi_r-(2\chi_m-\chi_r)\text{ln}\frac{16}{1-m} \big)}{16(\chi_m-\chi_r)}\sqrt{1-m}+O\big((1-m)^\frac{3}{2}\big).
\end{align}
\end{widetext}
\subsubsection{Partial transmission of the dispersive shock wave}
In particular, if $\chi_l>\chi_r$ (see Figs.~7), we need to consider $t_2$ when the whole RW is absorbed into the interaction region. On the left boundary of the interaction region,
$$
r_1=r_2=\chi_r,\ \ r_3=\chi_l,
$$
which is also the trailing edge of the RW, so
\begin{equation}
x^-=v_2(\chi_r,\chi_r,\chi_l)t_2+\check{w}_2(\chi_r,\chi_l)=x_{\text{RW}}^-=6\chi_lt_2,\notag
\end{equation}
which gives\\
\begin{equation}
t_2=\frac{\check{w}_2(\chi_r,\chi_l)}{12(\chi_l-\chi_r)}.
\end{equation}

When $t>t_2$, there occurs a new DSW (see the green line in Fig.~6(c)) on the left side of the interaction region, as the small amplitude edge emerges first. And Riemann invariants are governed by
\begin{equation}
r_1=\chi_r,\ \ r_3=\chi_l,
\end{equation}
and
\begin{align}
x&=v_2(\chi_r,r_2,\chi_l)t+\check{P}(r_2)\notag\\
&=\left(2(\chi_r+r_2+\chi_l)-\frac{4(r_2-\chi_r)(1-m)}{\mu-(1-m)}\right)t+\check{P}(r_2),
\end{align}
where modulus in elliptic integral is $
m=\frac{r_2-\chi_r}{\chi_l-\chi_r}$ and the function $\check{P}(r_2)$ is given by
\begin{align}
\check{P}(r)=&\check{g}(r,\chi_l)-\frac{L(\chi_r,r_,\chi_l)}{\partial_2L(\chi_r,r,\chi_l)}\partial_2\check{g}(r,\chi_l)\notag\\
=&\frac{2D\big( (\chi_m-\chi_r)\Pi(\check{p},\check{q})+\check{Y}\big)}{\pi\sqrt{\chi_m-r}\sqrt{\chi_l-\chi_r}},
\end{align}
with
\begin{align}
&\check{Y}=\frac{(\chi_l-r)(\chi_m-\chi_r)\text{K}(\check{q})-(\chi_m-r)(\chi_l-\chi_r)\text{E}(\check{q})}{(\chi_l-\chi_r)\mu(\check{y})-(\chi_l-r)},\notag\\
&\check{p}=\frac{r-\chi_r}{r-\chi_m},\ \ \check{q}=\frac{(\chi_m-\chi_l)(r-\chi_r)}{(\chi_m-r)(\chi_l-\chi_r)},\ \ \check{y}=\frac{r-\chi_r}{\chi_l-\chi_r}.
\end{align}
This DSW is a simple wave solution corresponding to the initial data $r(x,0)=\check{P}^{-1}(x)$ for Eq.~(1). Then the edges of the DSW can be found similar to Sec.~\uppercase\expandafter{\romannumeral3} A, and the phase shift is $\check{P}(\chi_r)-D$, compared with the original DSW.
\begin{figure}
\includegraphics[scale=0.33]{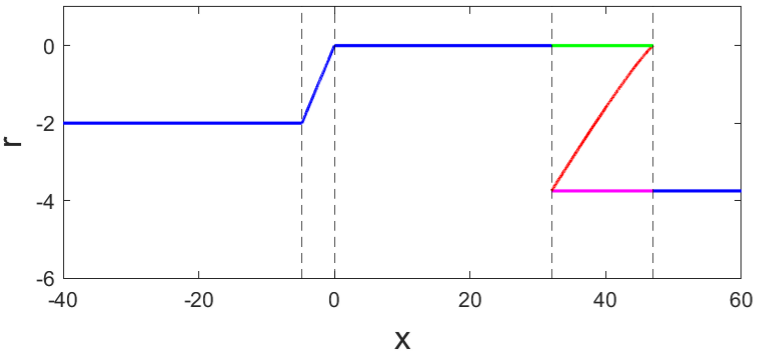}\hfill
\includegraphics[scale=0.33]{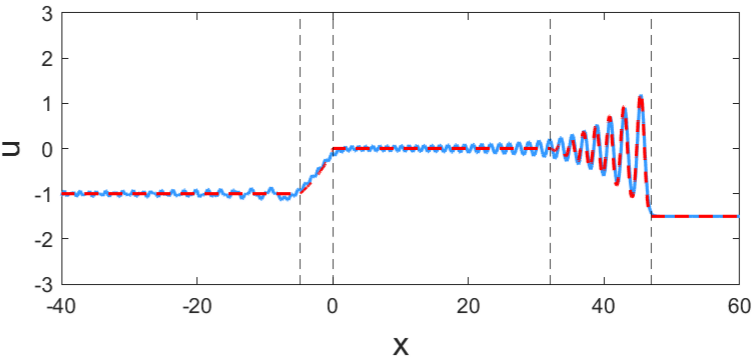}\\
\includegraphics[scale=0.33]{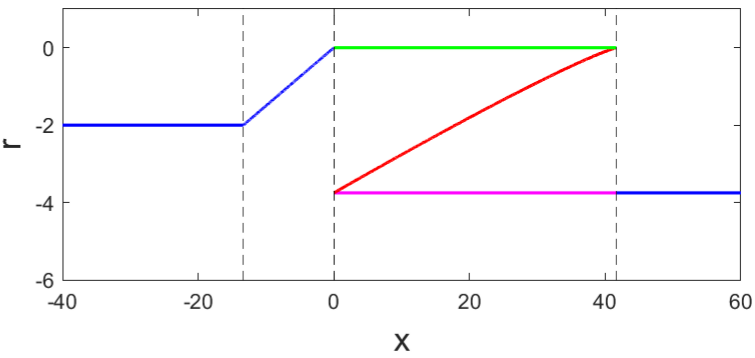}\hfill
\includegraphics[scale=0.33]{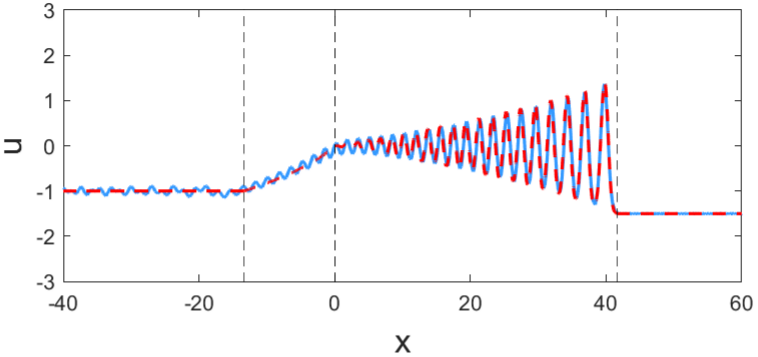}\\
\includegraphics[scale=0.33]{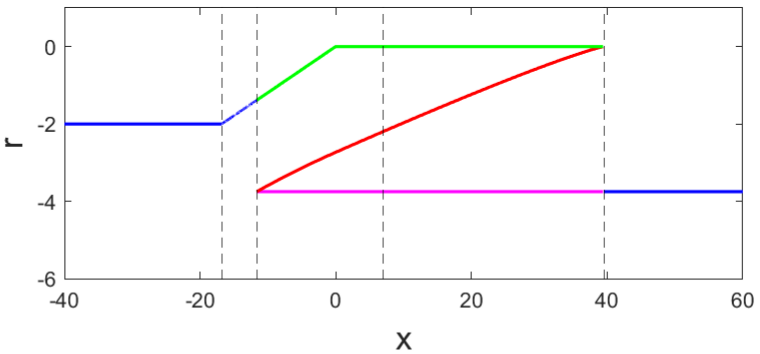}\hfill
\includegraphics[scale=0.33]{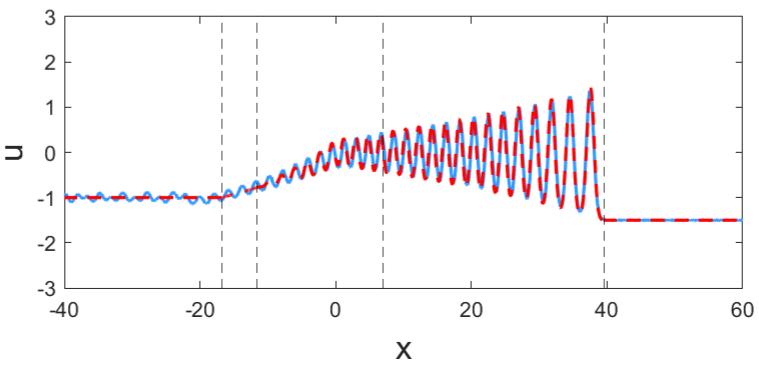}\\
\includegraphics[scale=0.33]{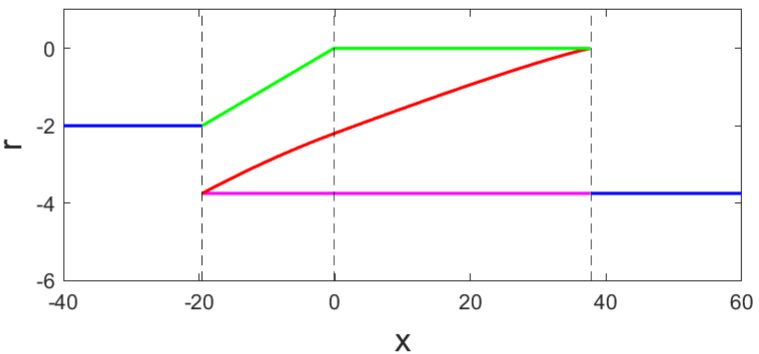}\hfill
\includegraphics[scale=0.33]{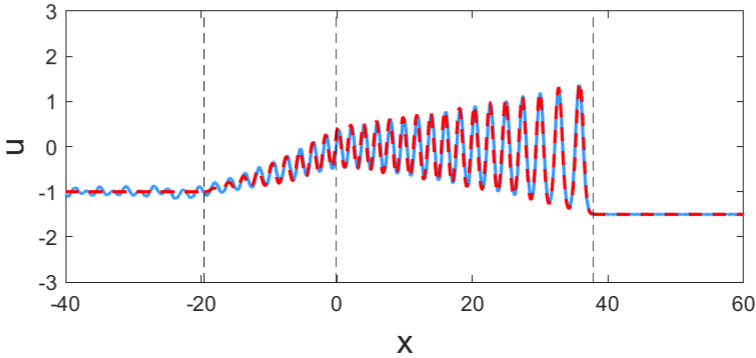}\\
\includegraphics[scale=0.33]{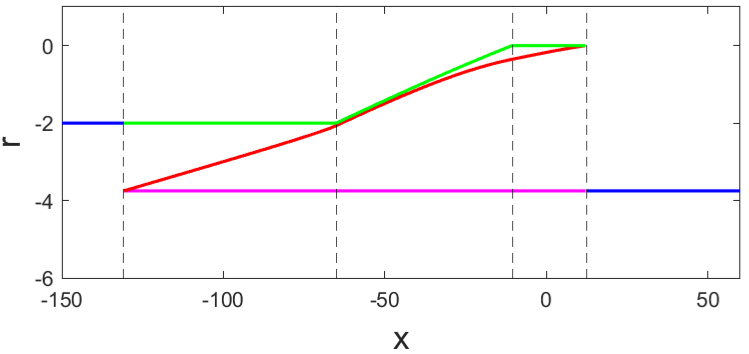}\hfill
\includegraphics[scale=0.33]{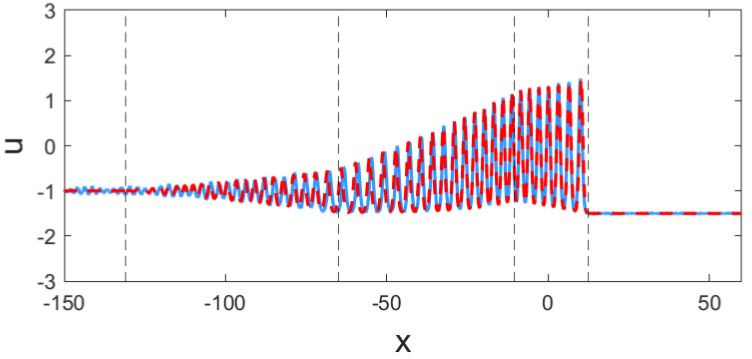}\\

{\footnotesize\hspace{0.0cm}(a)\hspace{4cm}(b)}
\caption{ Evolution of RW-DSW interaction with transmitted DSW, at $t=0.4$, $t=t_1$, $t=1.4$, $t=t_2$, and $t=5$, respectively. The parameters are $u_l=-1$, $u_m=0$, $u_r=-1.5$, $D=50$, $\alpha=1$. (a) Riemann invariants; (b) Solution of Eq.~(1). The red dashed line represents the analytical results based on the Whitham modulation theory, while the blue solid line represents the numerical results.}
\label{Figs.~$7$.}
\end{figure}
\subsection{General scenario}
Now we turn our attention to other double-step initial data evolving into a RW and a DSW in the early stage of evolution for both $\alpha>0$ and $\alpha<0$. The corresponding initial Riemann invariant satisfies either $\text{min}(\chi_l,\chi_r)>\chi_m$ or $\text{max}(\chi_l,\chi_r)<\chi_m$, and the modulation solution and large time asymptotic state in the interaction region can be attributed to results of Secs.~\uppercase\expandafter{\romannumeral3} A and \uppercase\expandafter{\romannumeral3} B.

As $t\rightarrow\infty$, a new DSW appears if $\chi_l>\chi_r$, while part of the original RW is retained if $\chi_l<\chi_r$. Furthermore, the monotonicity of the function $\chi(u)$ on $\big(\text{min}(\chi_l,\chi_r),\text{max}(\chi_l,\chi_r) \big)$ will affect the polarity of the structure on the right side of the interaction at large time, namely, they are normal when monotonically increasing, and reversed in the opposite case, according to Eqs.~(A9)-(A10). This conclusion is consistent with the following step initial value problem
\begin{equation}
u\left( x,0 \right) =\left\{ \begin{array}{l}
	u_l,\ \ x<0,\\
	u_r,\ \ x>0,\\
\end{array} \right. \ \ \text{if}\ \ u_l>u_r,
$$
$$
u\left( x,0 \right) =\left\{ \begin{array}{l}
	u_l,\,\,\,\,x<D,\\
	u_r,\,\,\,\,x>D,\\
\end{array} \right. \,\,\,\,\text{if}\ \ u_l<u_r,
\end{equation}
excepting that the additional small amplitude harmonic wave or soliton train.

For $\alpha>0$, the small amplitude harmonic wave corresponds to $r_2\rightarrow r_1$, implying that
\begin{align}
u_2&=u_3=\frac{1}{2\alpha}(1-\sqrt{1-4\alpha r_3}),\notag\\
u_1&=\frac{1}{2\alpha}(1-2\sqrt{1-4\alpha r_2}+\sqrt{1-4\alpha r_3}),\notag\\
u_4&=\frac{1}{2\alpha}(1+2\sqrt{1-4\alpha r_2}+\sqrt{1-4\alpha r_3}),
\end{align}
where $r_3$ is a constant, and $r_2$ is determined by Eqs.(21), (53) and (B2)-(B3). On the other hand, the soliton train corresponds to $r_2\rightarrow r_3$, whose polarity is also determined by the monotonicity of $\chi(u)$, and velocity is $c=2(r_1+2r_2)$. One yileds
\begin{align}
u_1&=u_2=\frac{1}{2\alpha}(1-\sqrt{1-4\alpha r_1}),\notag\\
u_3&=\frac{1}{2\alpha}(1+\sqrt{1-4\alpha r_1}-2\sqrt{1-4\alpha r_2}),\notag\\
u_4&=\frac{1}{2\alpha}(1+\sqrt{1-4\alpha r_1}+2\sqrt{1-4\alpha r_2}),
\end{align}
for the bright soliton train with amplitude $a=\frac{1}{\alpha}(\sqrt{1-4\alpha r_1}-\sqrt{1-4\alpha r_2})$, and
\begin{align}
u_1&=\frac{1}{2\alpha}(1-\sqrt{1-4\alpha r_1}-2\sqrt{1-4\alpha r_2}),\notag\\
u_2&=\frac{1}{2\alpha}(1-\sqrt{1-4\alpha r_1}+2\sqrt{1-4\alpha r_2}),\notag\\
u_3&=u_4=\frac{1}{2\alpha}(1-\sqrt{1-4\alpha r_1}),
\end{align}
for the dark soliton train with amplitude $a=\frac{1}{\alpha}(\sqrt{1-4\alpha r_1}+\sqrt{1-4\alpha r_2})$, where $r_1$ is a constant.

Similarly, for $\alpha<0$, the small amplitude harmonic wave corresponds to
\begin{align}
u_1&=u_2=\frac{1}{2\alpha}(1+\sqrt{1-4\alpha r_3}),\notag\\
u_3&=\frac{1}{2\alpha}(1+2\sqrt{1-4\alpha r_2}-\sqrt{1-4\alpha r_3}),\notag\\
u_4&=\frac{1}{2\alpha}(1-2\sqrt{1-4\alpha r_2}-\sqrt{1-4\alpha r_3}),
\end{align}
when oscillating in $u_1\leq u\leq u_2$, and
\begin{align}
u_1&=\frac{1}{2\alpha}(1+2\sqrt{1-4\alpha r_2}+\sqrt{1-4\alpha r_3}),\notag\\
u_2&=\frac{1}{2\alpha}(1-2\sqrt{1-4\alpha r_2}+\sqrt{1-4\alpha r_3}),\notag\\
u_3&=u_4=\frac{1}{2\alpha}(1-\sqrt{1-4\alpha r_3}),
\end{align}
when oscillating in $u_3\leq u\leq u_4$. In addition, both the bright and dark soliton train moving at velocity $c=2(r_1+2r_2)$ have amplitude $a=\frac{1}{2\alpha}(1+2\sqrt{1-4\alpha r_1}-2\sqrt{1-4\alpha r_2})$.

Obviously, the velocity and frequency of the small amplitude harmonic wave, as well as the amplitude and velocity of the soliton train are modulated by the Gardner--Whitham equations.
\section{EFFECT FROM KINKS}
In Ref.~\cite{sol-mean}, the authors have discussed the interaction between convex mean flows and kinks for Eq.~(1), showing that kinks impart polarity reversal to RWs and DSWs. Here, we will analyze the impact of kinks on the interaction, i.e., the initial data crosses the inflection point $u=\frac{1}{2\alpha}$.

For the case that the DSW first passes through kink, and then interacts with the RW, there is a composite structure RW$|$Kink emanating from $x=0$ and a DSW emanating from $x=D$. After passing through the kink, the polarity of DSW is reversed, and the analysis of interaction with RW is similar to Sec.~\uppercase\expandafter{\romannumeral3} B. The leading edge velocity of the DSW is $2\chi_r+4\chi_m$. Alternatively, focusing on the case that the RW first passes through kink, and then interacts with the DSW, there is a composite structure DSW$|$Kink emanating from $x=0$ and a RW emanating from $x=D$. After passing through the kink, the polarity of RW is reversed, and the analysis of interaction with DSW is similar to Sec.~\uppercase\expandafter{\romannumeral3} A. The leading edge velocity is $6\chi_r$ for the right RW, and $2\chi_r+4\chi_l$ for the new DSW. Considering the velocity of the kink $\frac{1}{\alpha}+2\chi_r$, it only plays a role in reversing polarity and does not participate in RW-DSW interactions.

If a RW interacts with a DSW first, regardless of whether the single-phase periodic wave in the interaction region degenerates into a soliton train or a small amplitude harmonic wave as $t\rightarrow\infty$, this region and the convex RW or DSW on the left or right side as a whole have a lower velocity on the leading edge than the kink. As for the interaction between two composite structures, it can be seen as having an additional kink on the far right side, which will not affect the entire interaction process. The final asymptotic state has two kinks with opposite polarity moving at the same speed on the far right side.
\section{INTERACTION BETWEEN RWS AND CDSWS}
For $\alpha<0$, Equation (1) admits special DSWs connecting platforms $u_l$ and $u_r$ such that $u_l+u_r=\frac{1}{\alpha}$, namely CDSWs in which $r_1\equiv r_2$. It is necessary to introduce
the classical Whitham combination
\begin{align}\notag
R_1=\frac{1}{2}(u_2+u_3)=\frac{1}{2\alpha}(1+\sqrt{1-4\alpha r_1}),\\
R_2=\frac{1}{2}(u_2+u_4)=\frac{1}{2\alpha}(1-\sqrt{1-4\alpha r_2}),\\
R_3=\frac{1}{2}(u_3+u_4)=\frac{1}{2\alpha}(1-\sqrt{1-4\alpha r_3}),\notag
\end{align}
$(R_1\leq R_2\leq R_3)$ for the normal CDSW and
\begin{align}\notag
R_1=\frac{1}{2}(u_2+u_3)=\frac{1}{2\alpha}(1-\sqrt{1-4\alpha r_1}),\\
R_2=\frac{1}{2}(u_1+u_3)=\frac{1}{2\alpha}(1+\sqrt{1+4\alpha r_2}),\\
R_3=\frac{1}{2}(u_1+u_2)=\frac{1}{2\alpha}(1+\sqrt{1-4\alpha r_3}),\notag
\end{align}
$(R_3\leq R_2\leq R_1)$ for the reversed CDSW. $R_i$ can be expressed in terms of $u_l$ and $u_r$
$$
\left\{ \begin{array}{l}
	R_1=\frac{1}{2\alpha}\left( 1+\sqrt{2\alpha ^2\left( u_l-\frac{1}{2\alpha} \right) ^2+\frac{1}{2}\left( 1-\frac{2\alpha}{3}\tau \right)} \right) ,\\
	R_2=\frac{1}{2\alpha}\left( 1-\sqrt{2\alpha ^2\left( u_l-\frac{1}{2\alpha} \right) ^2+\frac{1}{2}\left( 1-\frac{2\alpha}{3}\tau \right)} \right) ,\\
	R_3=u_l,\\
\end{array} \right.
$$
for the normal CDSW and
$$
\left\{ \begin{array}{l}
	R_1=\frac{1}{2\alpha}\left( 1-\sqrt{2\alpha ^2\left( u_l-\frac{1}{2\alpha} \right) ^2+\frac{1}{2}\left( 1-\frac{2\alpha}{3}\tau \right)} \right) ,\\
	R_2=\frac{1}{2\alpha}\left( 1+\sqrt{2\alpha ^2\left( u_l-\frac{1}{2\alpha} \right) ^2+\frac{1}{2}\left( 1-\frac{2\alpha}{3}\tau \right)} \right) ,\\
	R_3=u_l,\\
\end{array} \right.
$$
for the reversed CDSW. And Riemann invariants satisfy
$$
\left\{ \begin{array}{l}
	r_1=r_2=\frac{1}{12}\left( \tau +6\chi _l \right) ,\\
	r_3=\chi _l,\\
\end{array} \right.
$$
for both CDSWs with boundary velocities $
s^-=\frac{3}{\alpha}-6\chi_l,\ s^+=6\chi_l=6\chi_r$.

Obviously, when combined with DSWs, the soliton front edge cannot be fully displayed, accompanied by the velocity $s_*=12\chi_r-6\chi_l$ at the connection. However, the leading edge of the CDSW can be connected with a RW, implying that the collision occurs only when the RW is located on the left of the CDSW.
\begin{figure*}
\includegraphics[scale=0.3]{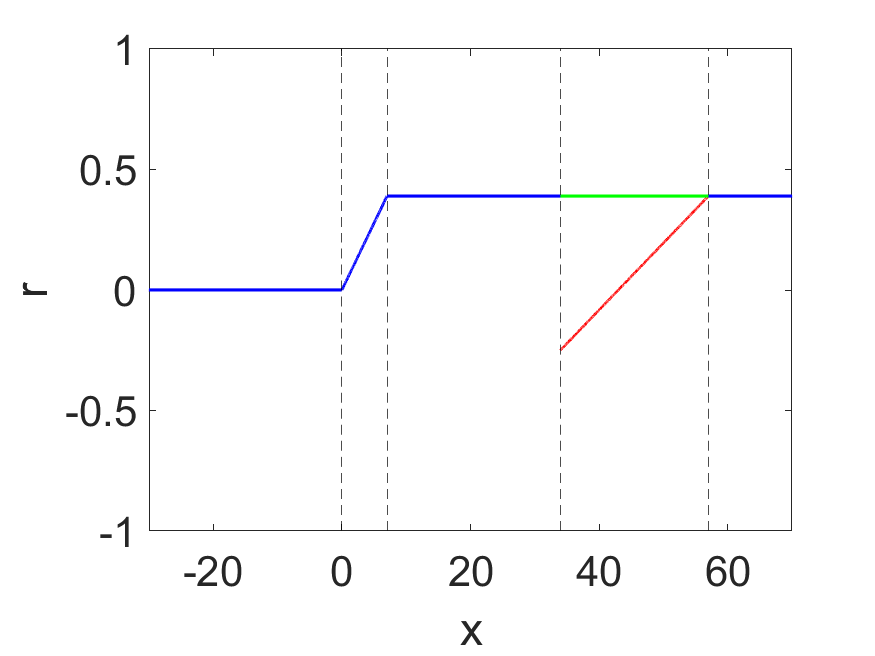}\hfill
\includegraphics[scale=0.3]{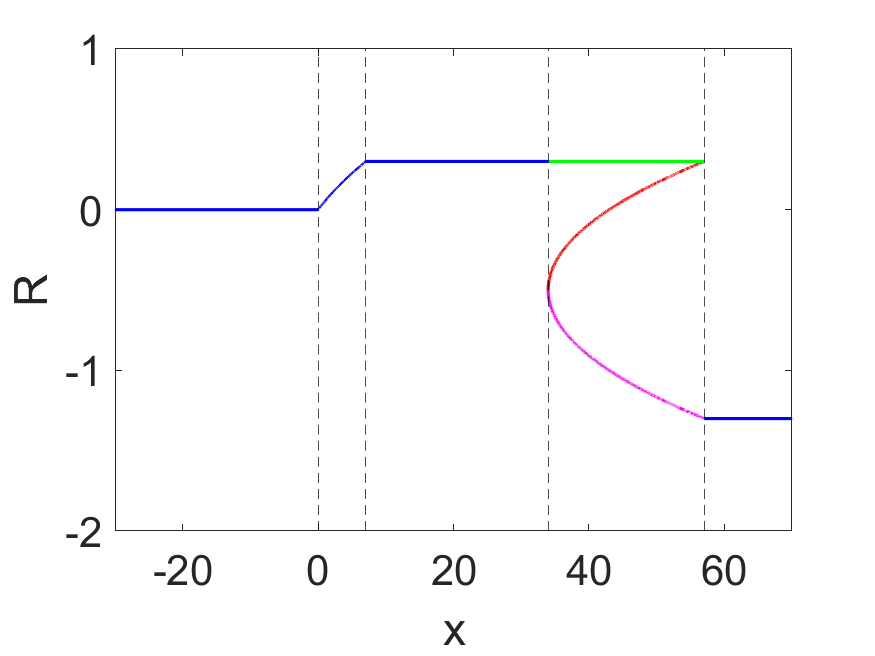}\hfill
\includegraphics[scale=0.3]{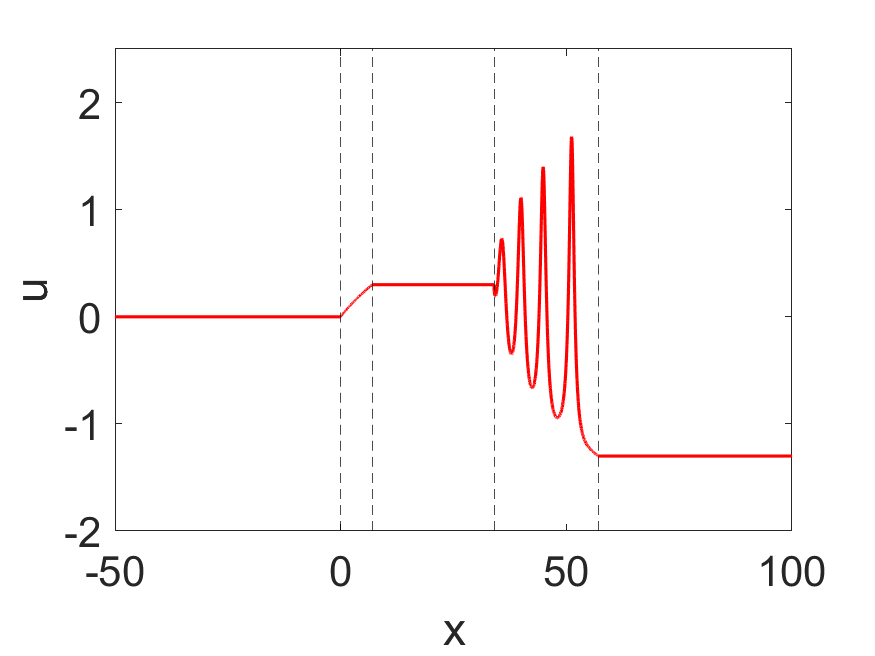}\hfill
\includegraphics[scale=0.3]{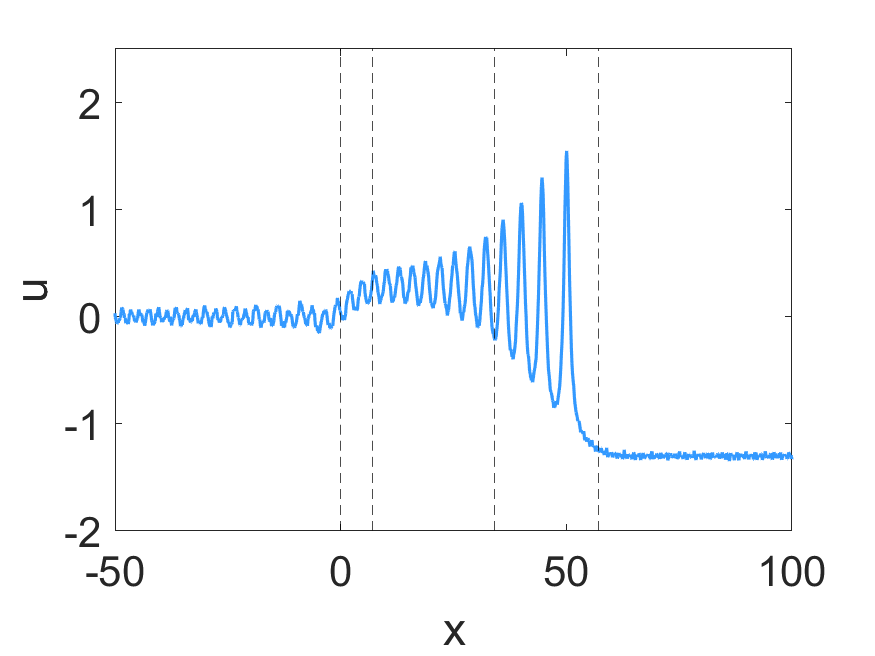}\hfill\\
\includegraphics[scale=0.3]{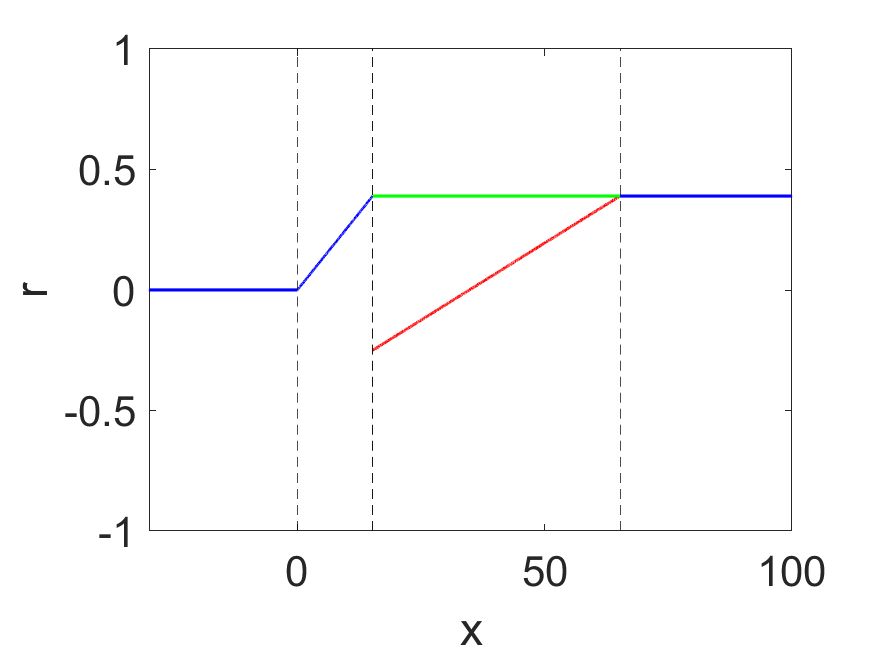}\hfill
\includegraphics[scale=0.3]{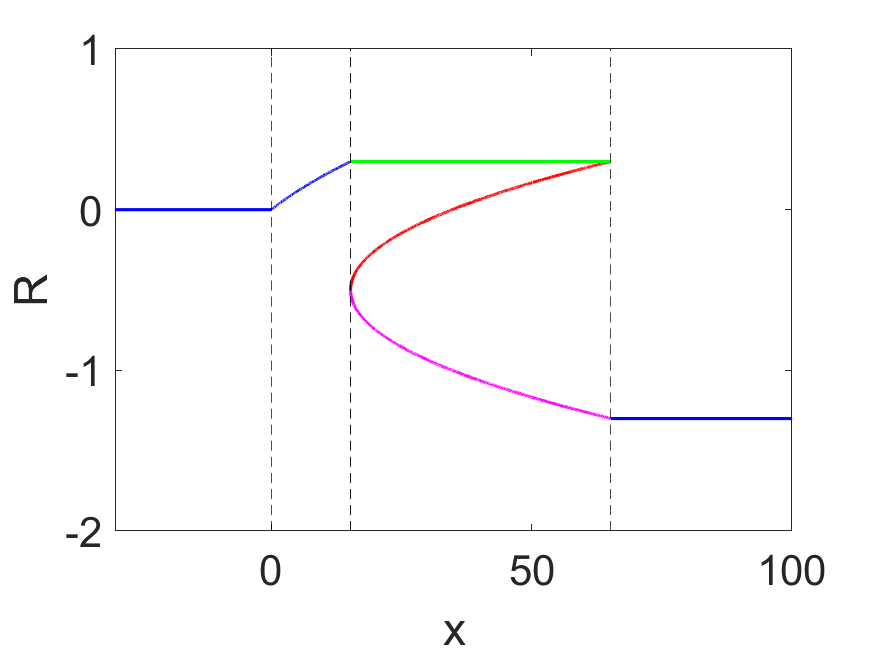}\hfill
\includegraphics[scale=0.3]{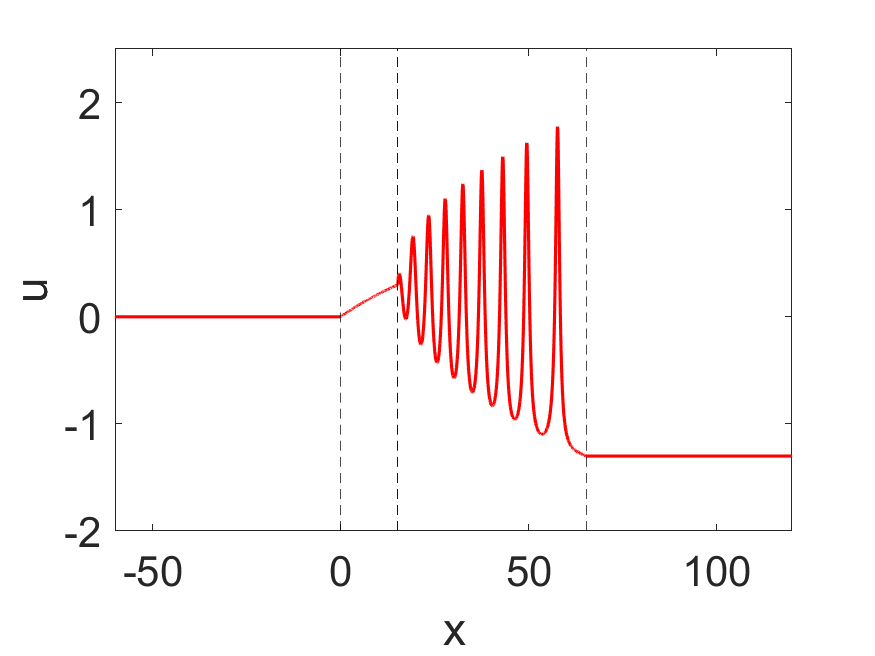}\hfill
\includegraphics[scale=0.3]{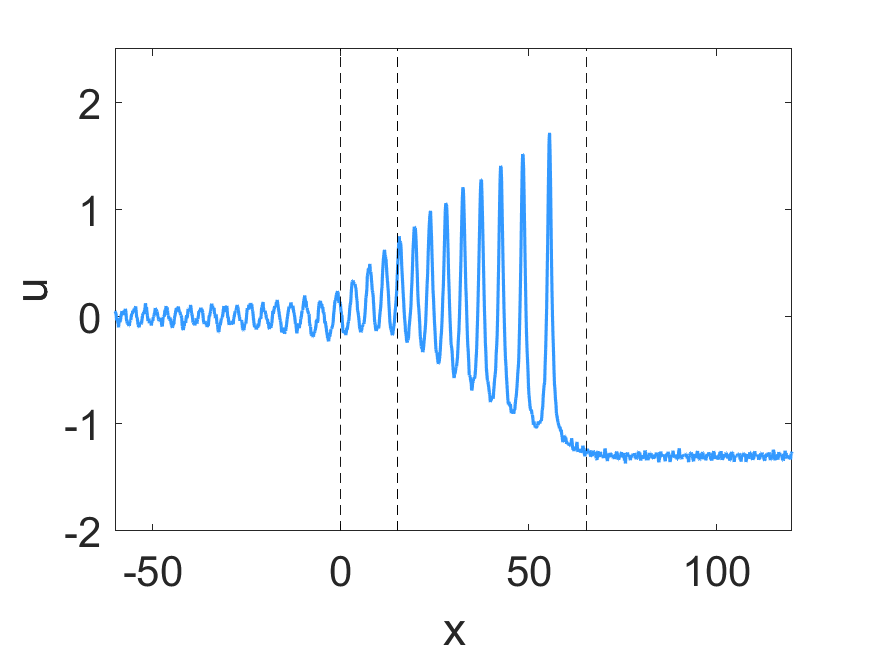}\hfill\\
\includegraphics[scale=0.3]{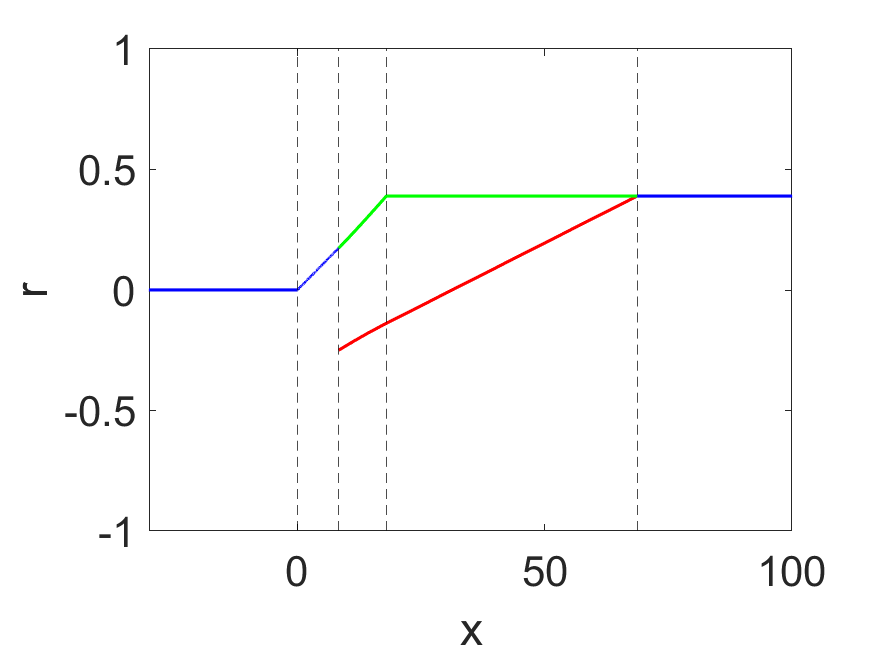}\hfill
\includegraphics[scale=0.3]{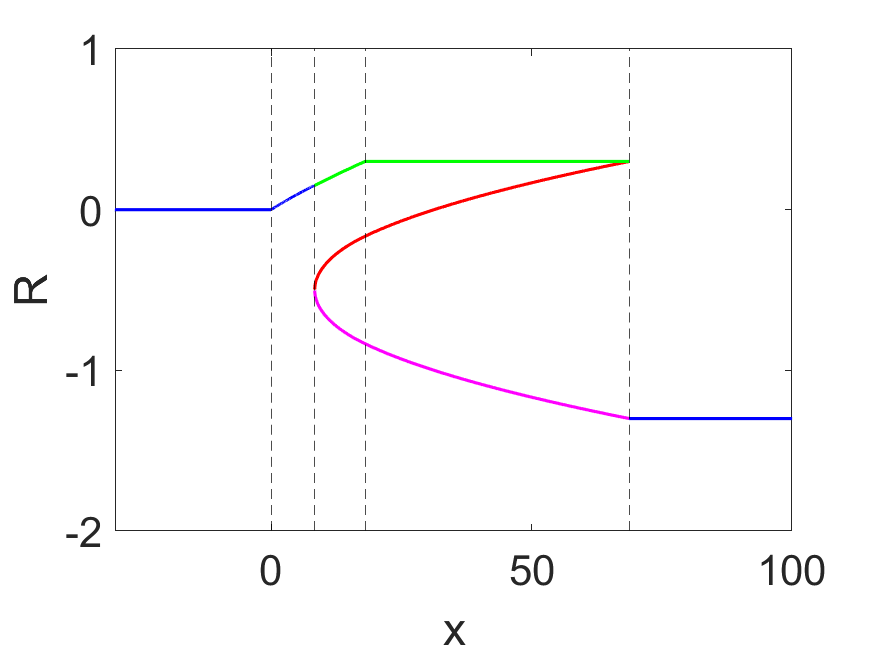}\hfill
\includegraphics[scale=0.3]{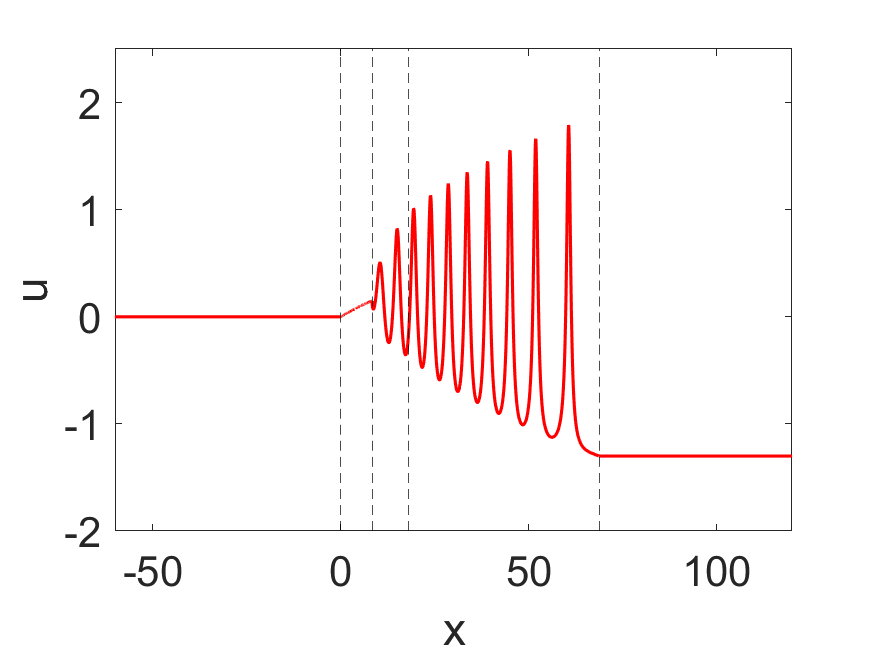}\hfill
\includegraphics[scale=0.3]{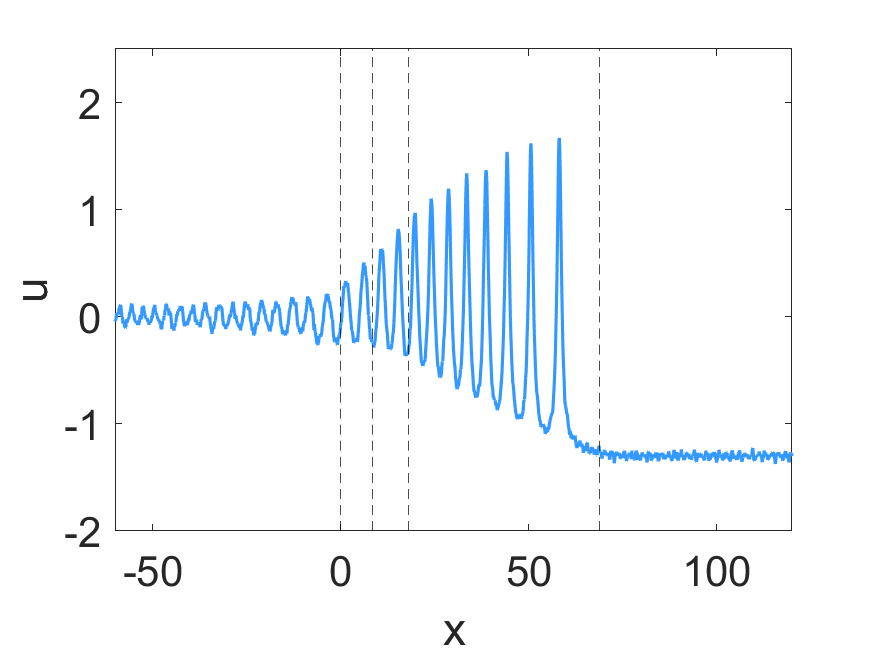}\hfill\\
\includegraphics[scale=0.3]{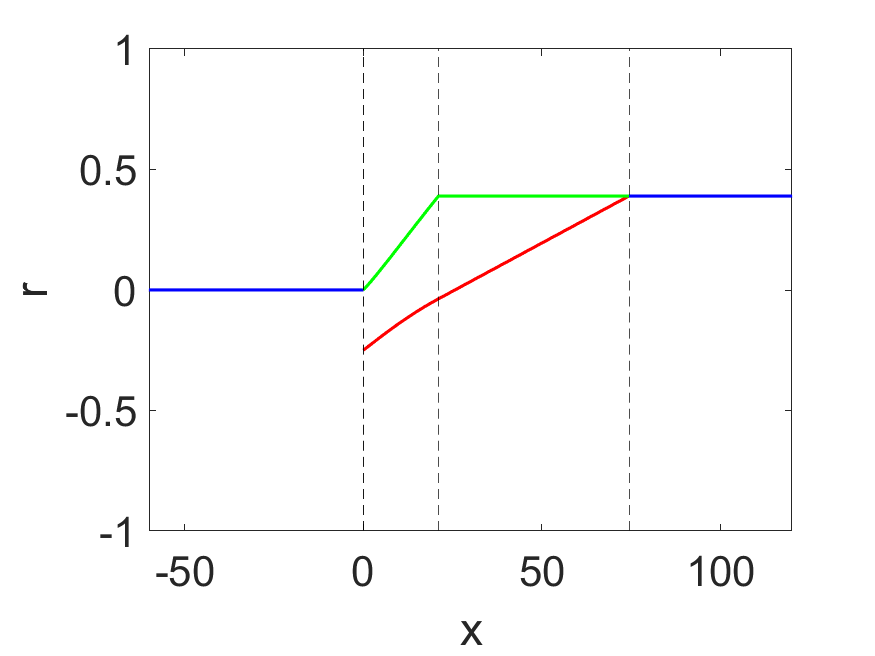}\hfill
\includegraphics[scale=0.3]{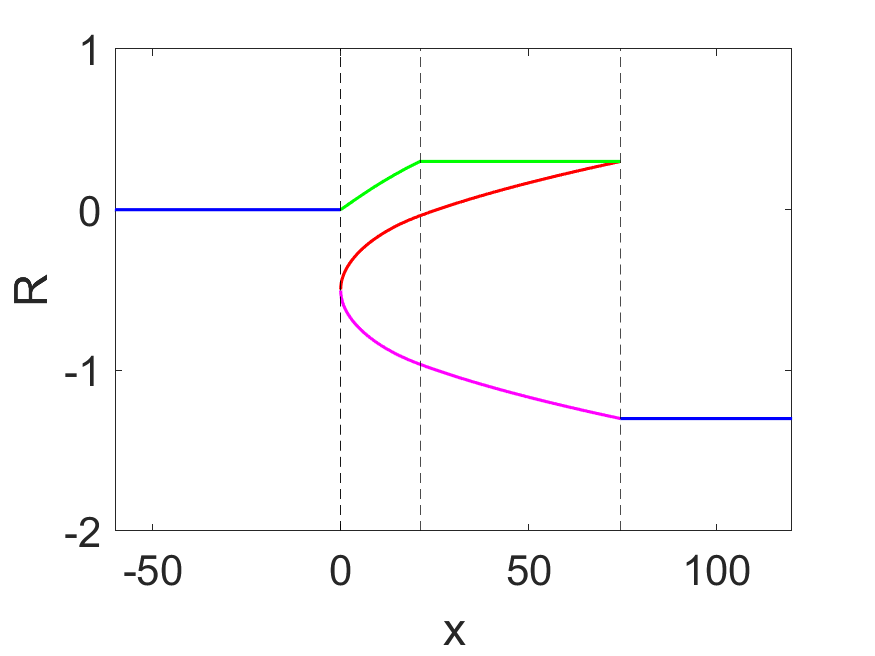}\hfill
\includegraphics[scale=0.3]{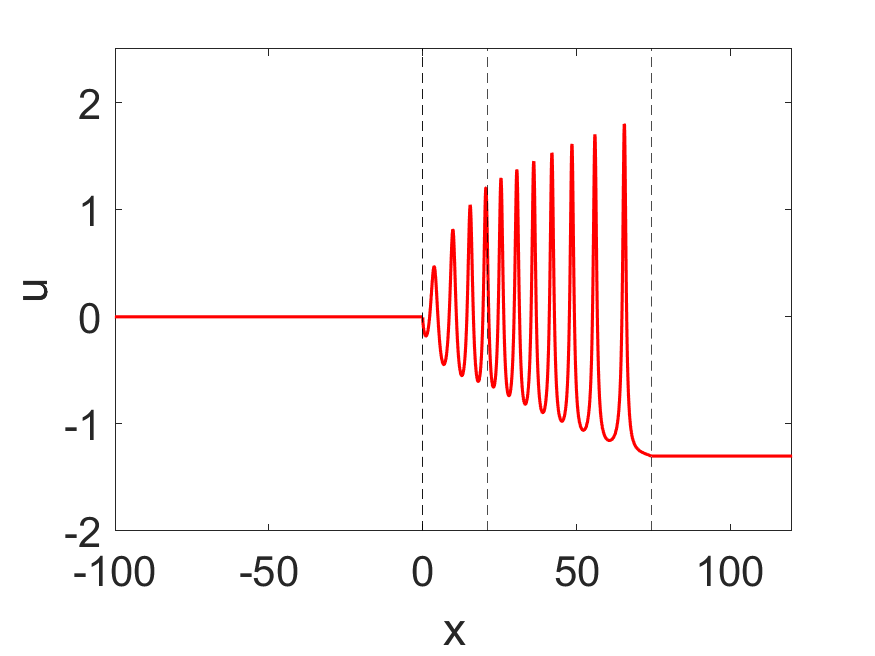}\hfill
\includegraphics[scale=0.3]{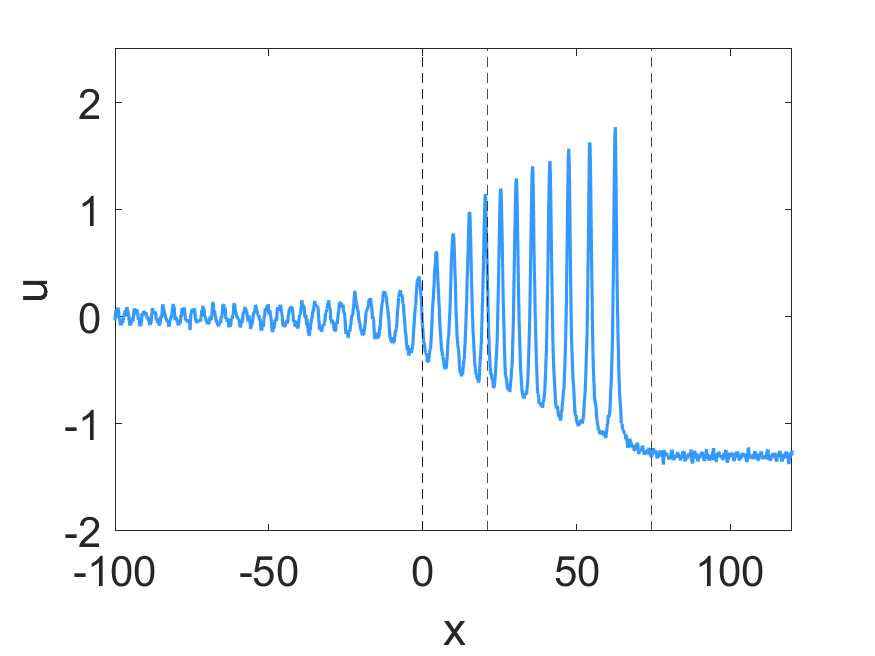}\hfill\\
\includegraphics[scale=0.3]{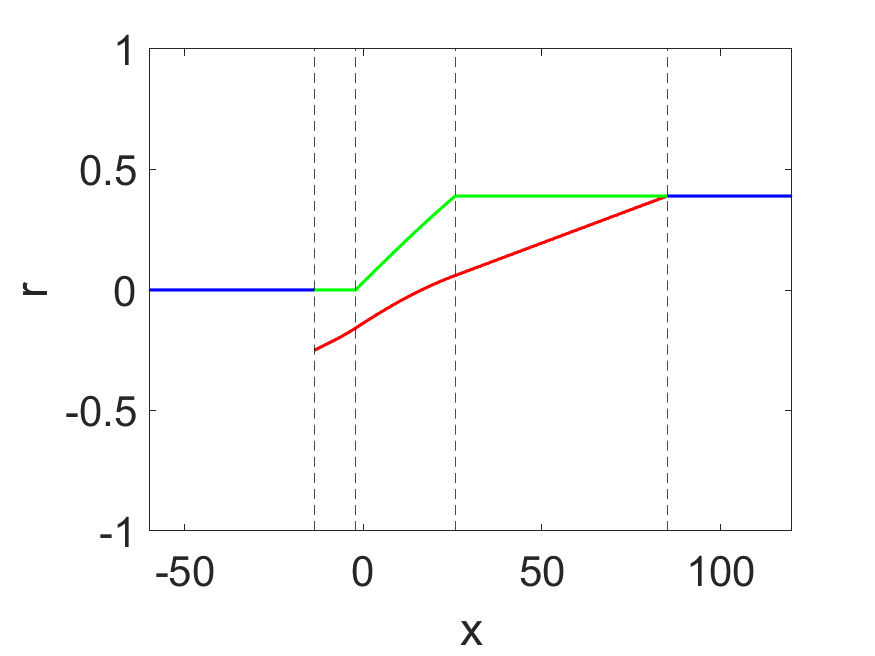}\hfill
\includegraphics[scale=0.3]{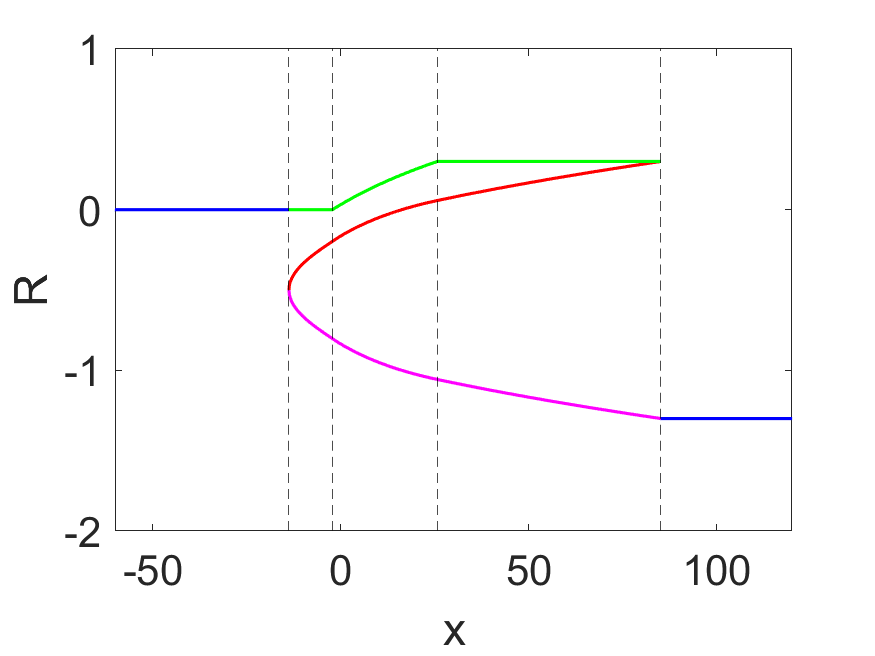}\hfill
\includegraphics[scale=0.3]{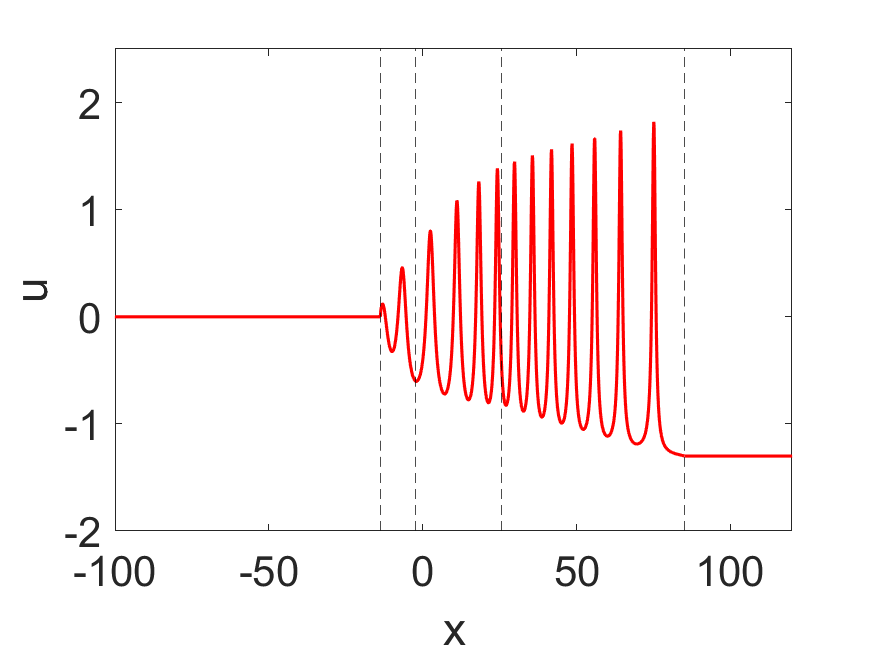}\hfill
\includegraphics[scale=0.3]{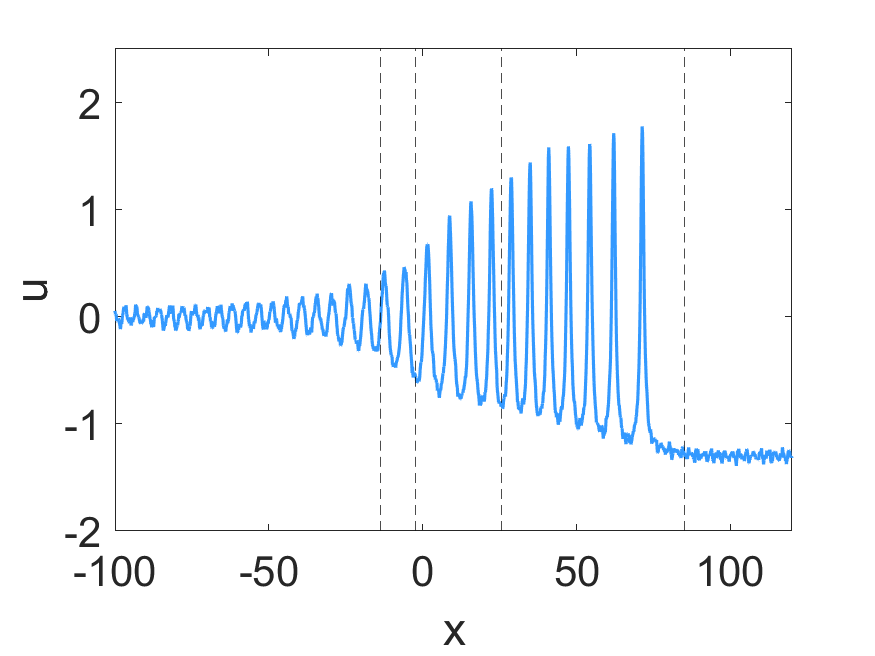}\hfill\\

{\footnotesize\hspace{0.0cm}(a)\hspace{4cm}(b)\hspace{4cm}(c)\hspace{4cm}(d)}
\caption{Evolution of RW-normal CDSW interaction, at $t=3$, $t=t_1$, $t=8$, $t=t_2$, and $t=15$, respectively. The parameters are $u_l=0$, $u_m=0.3$, $u_r=-1.3$, $D=50$, $\alpha=-1$. (a) Riemann invariants $r$; (b) Combination of Riemann invariants $R$; (c) Analytical solutions of Eq.~(1) based on the Whitham modulation theory; (d) Numerical solutions of Eq.~(1).}
\label{Figs.~8.}
\end{figure*}
\begin{figure*}
\includegraphics[scale=0.3]{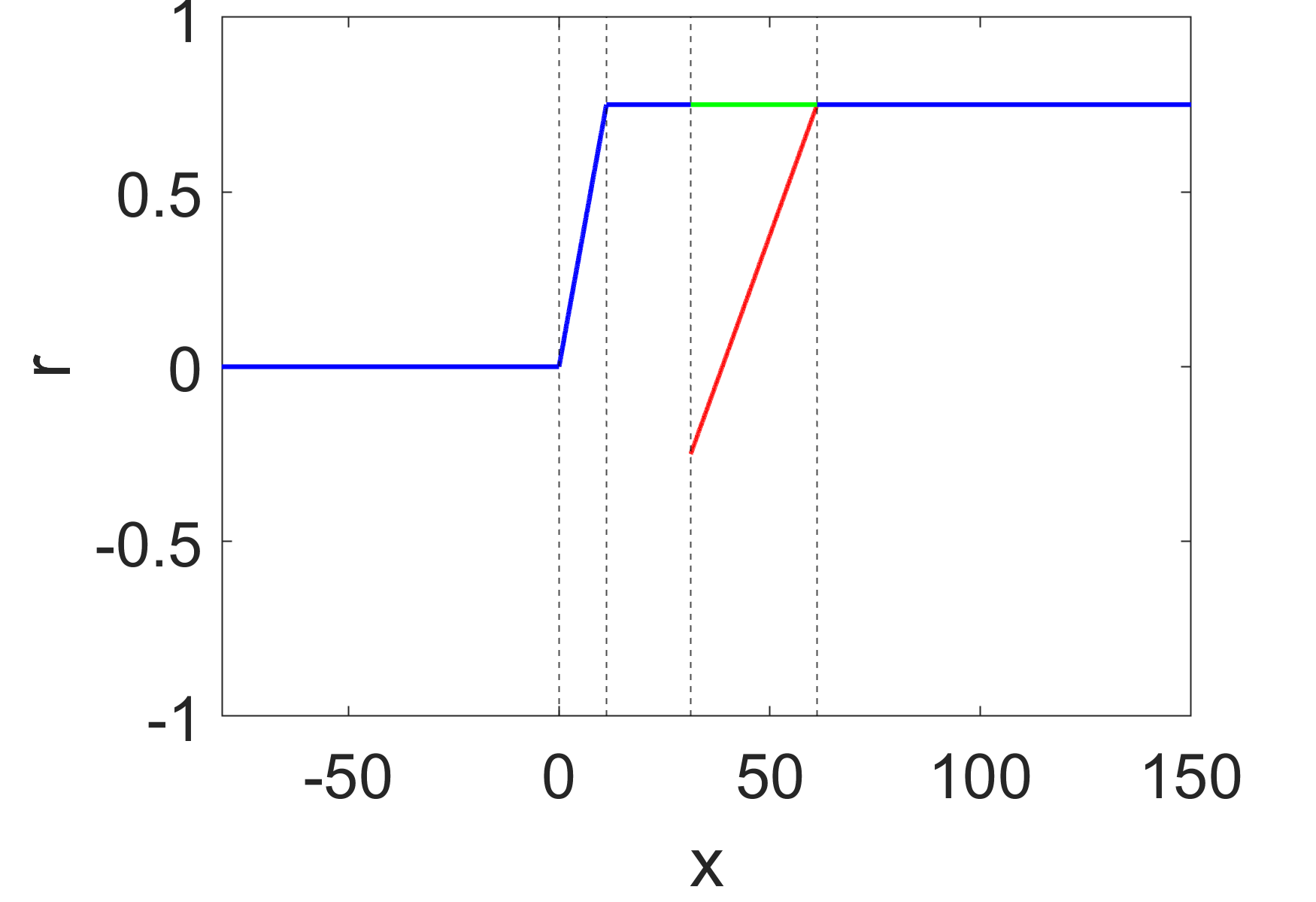}\hfill
\includegraphics[scale=0.3]{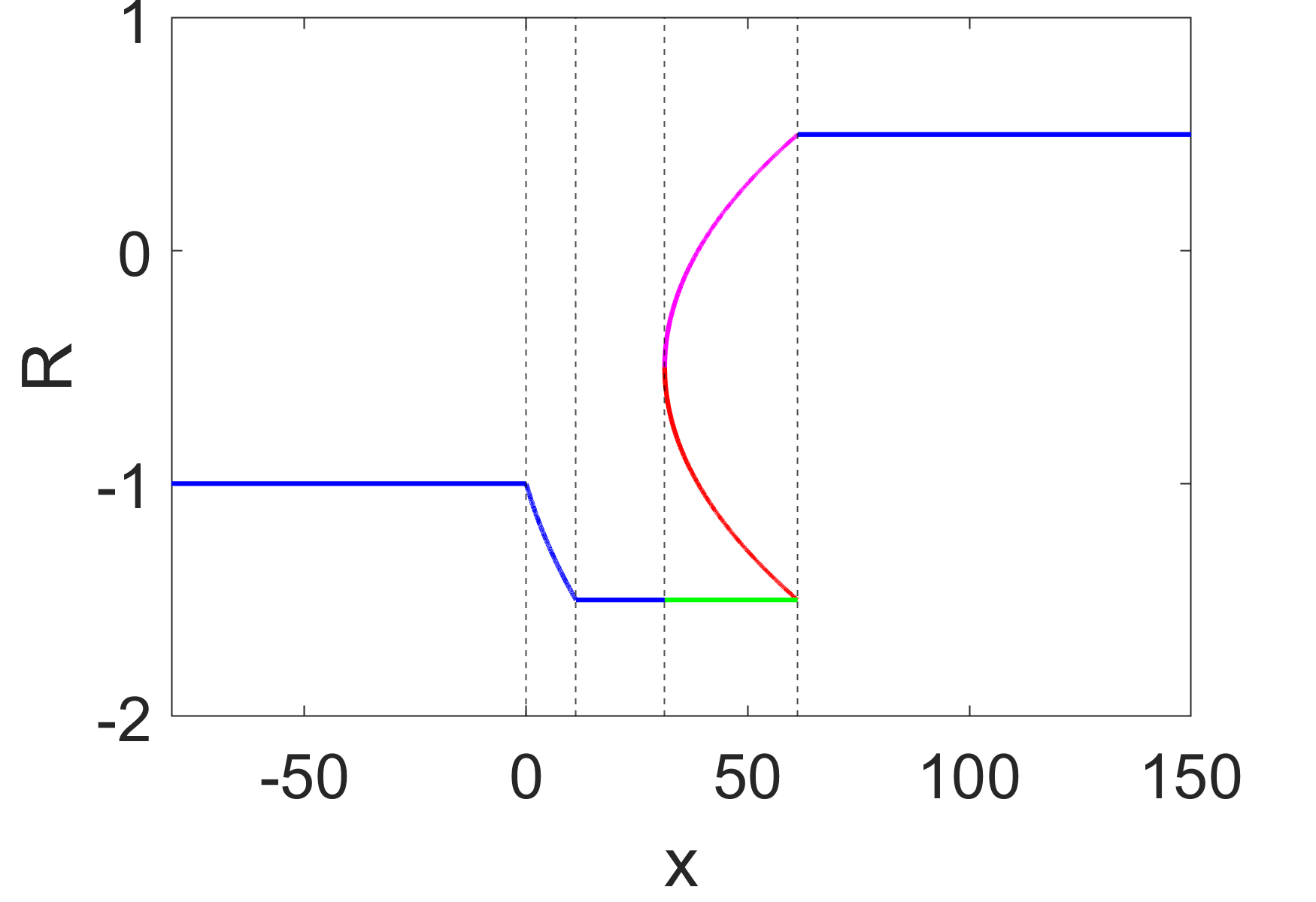}\hfill
\includegraphics[scale=0.3]{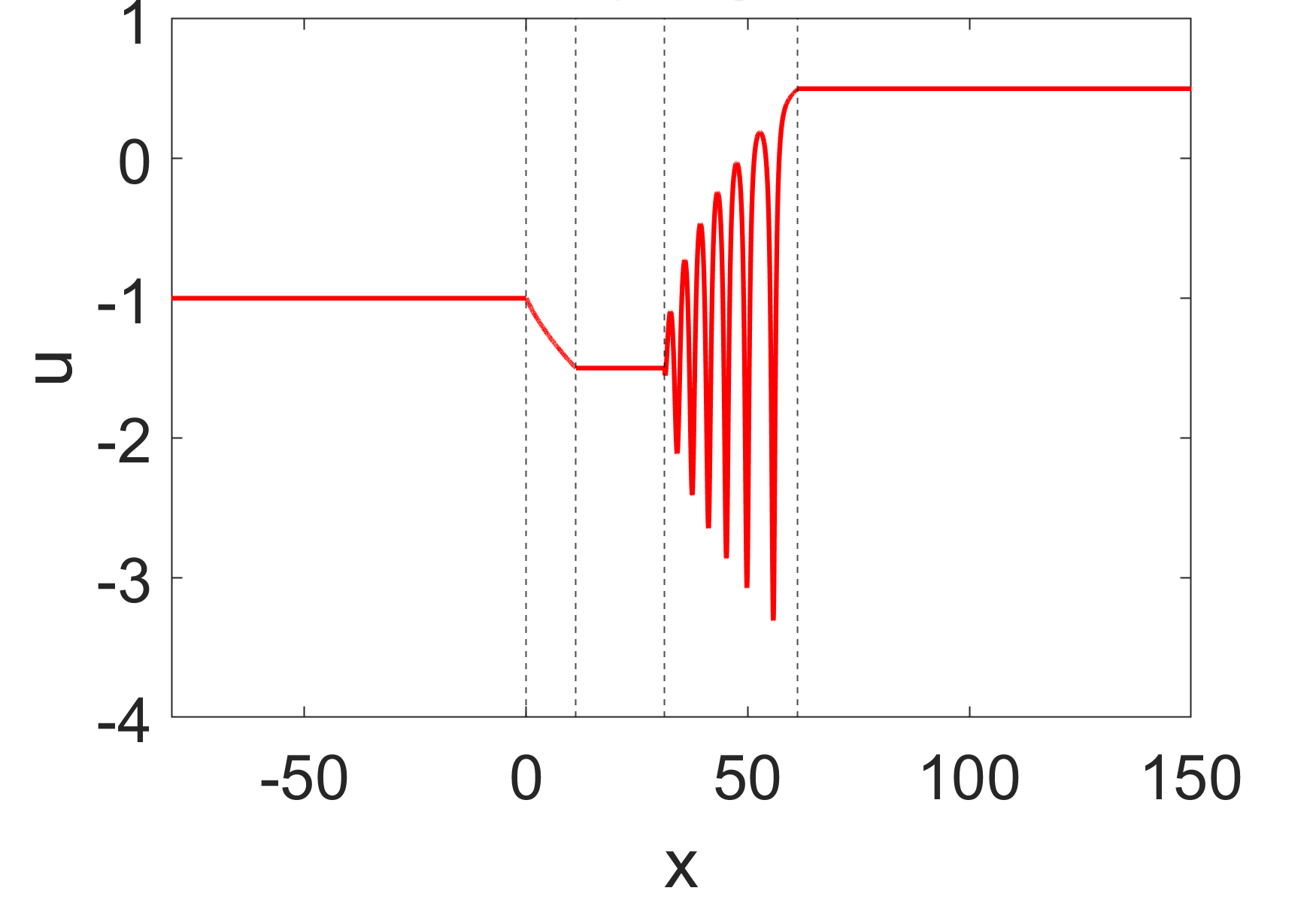}\hfill
\includegraphics[scale=0.3]{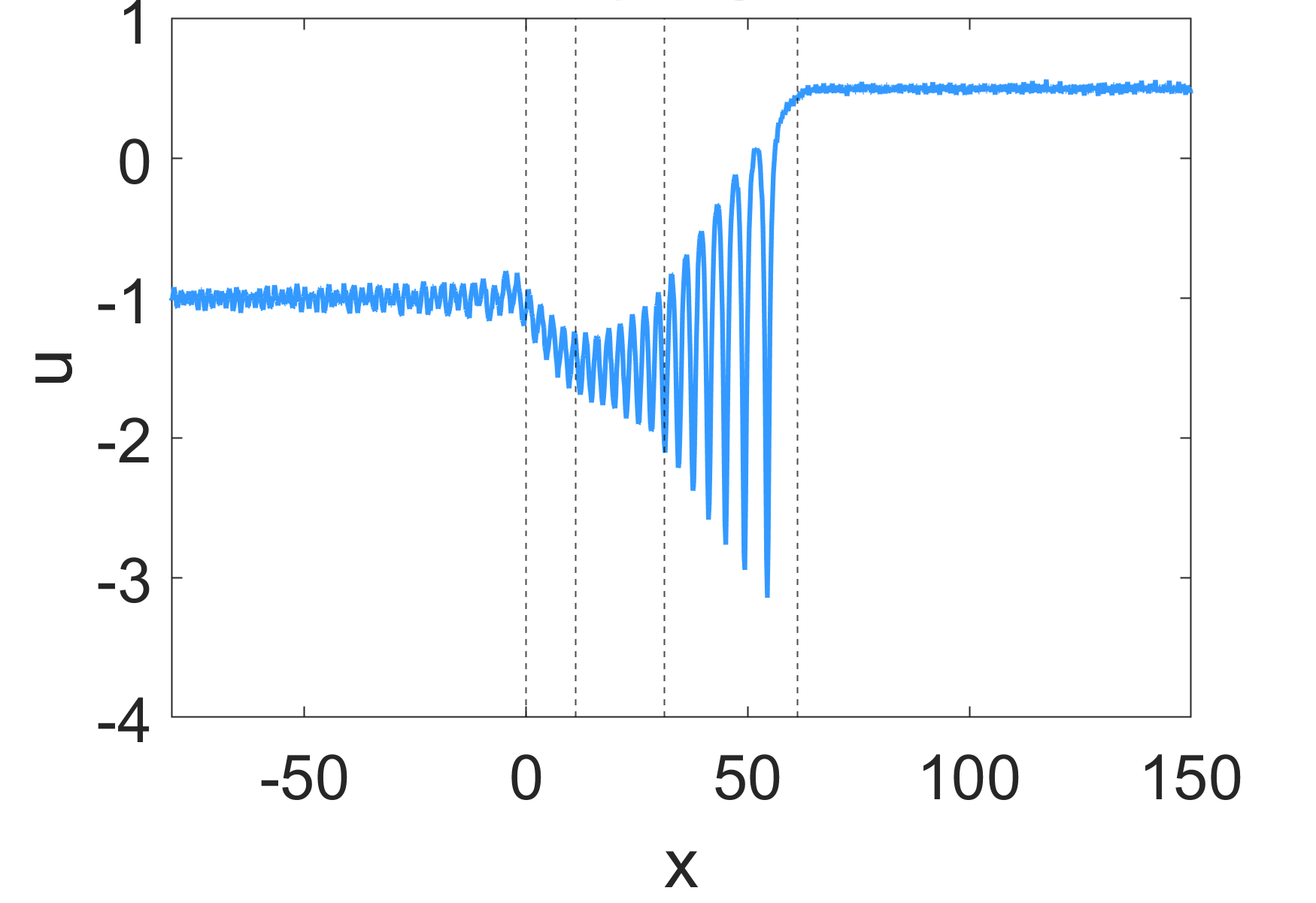}\hfill\\
\includegraphics[scale=0.3]{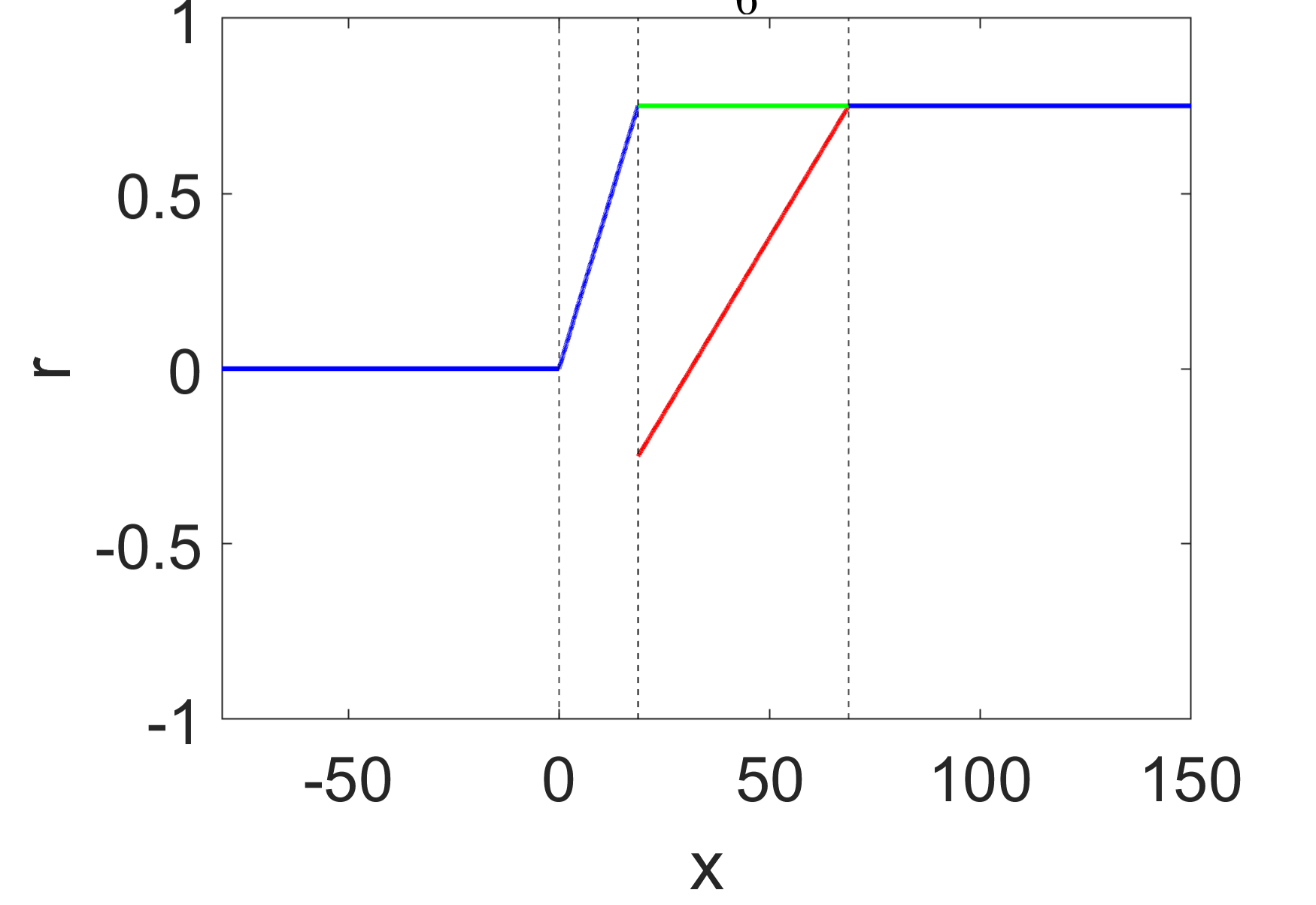}\hfill
\includegraphics[scale=0.3]{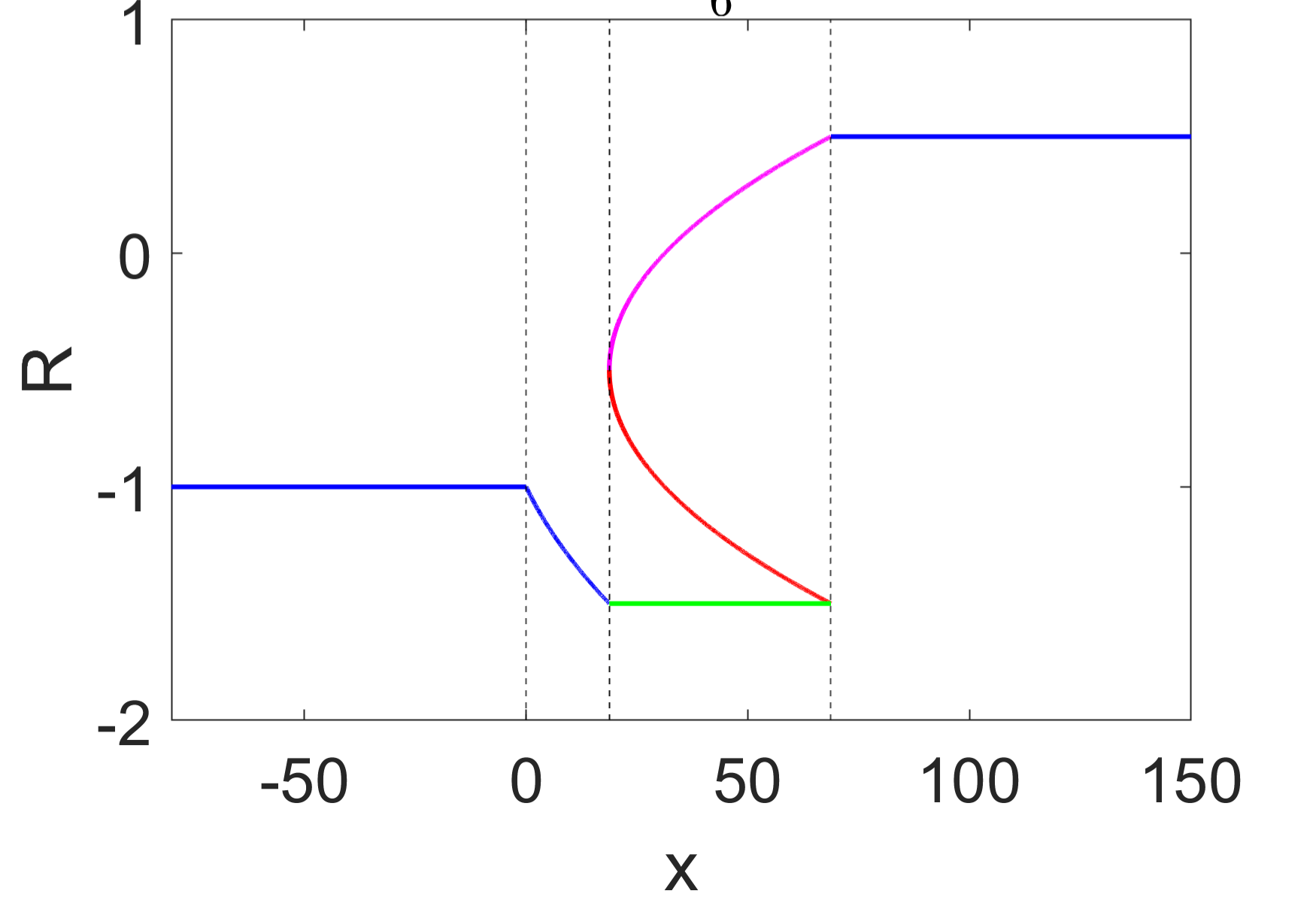}\hfill
\includegraphics[scale=0.3]{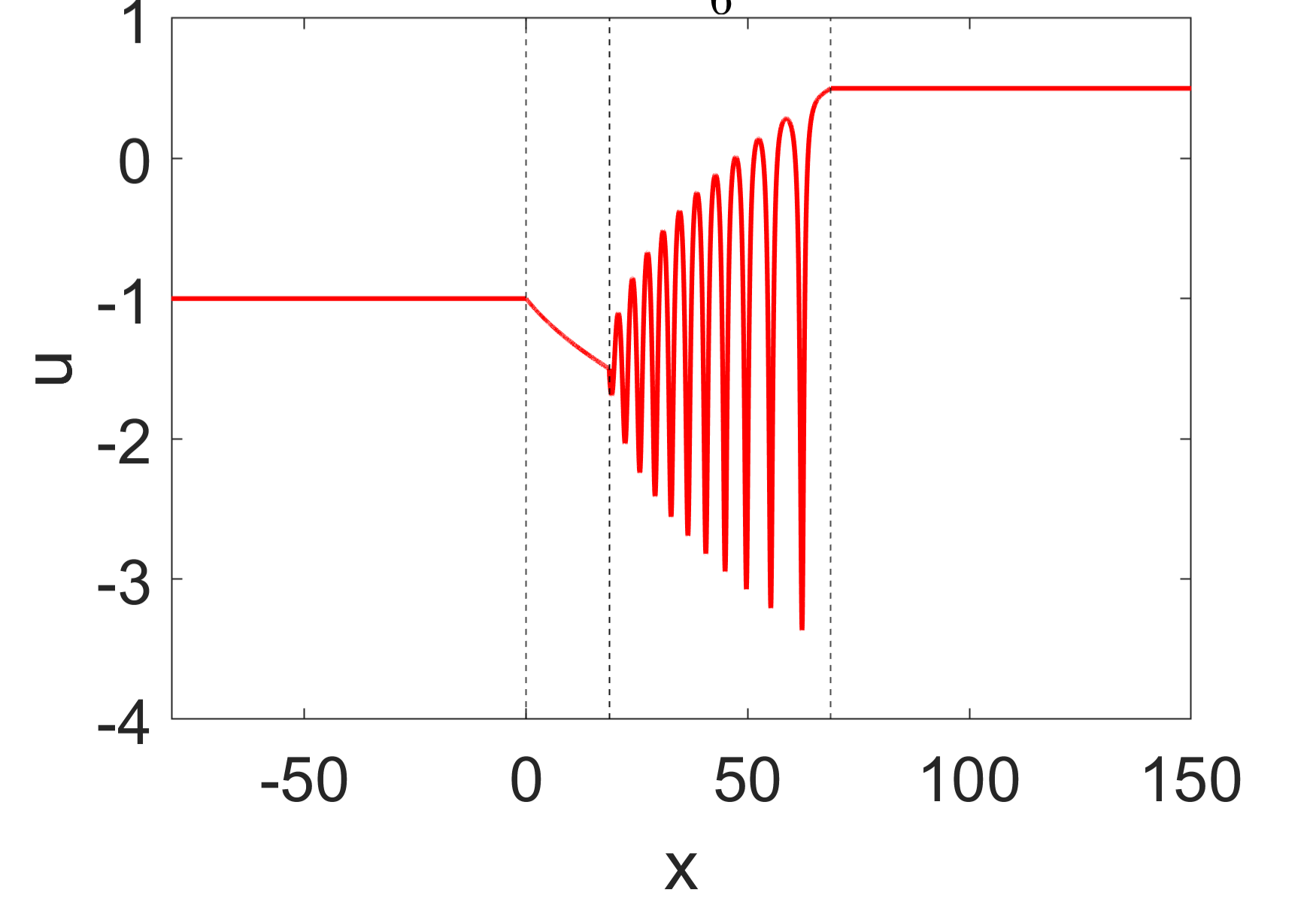}\hfill
\includegraphics[scale=0.3]{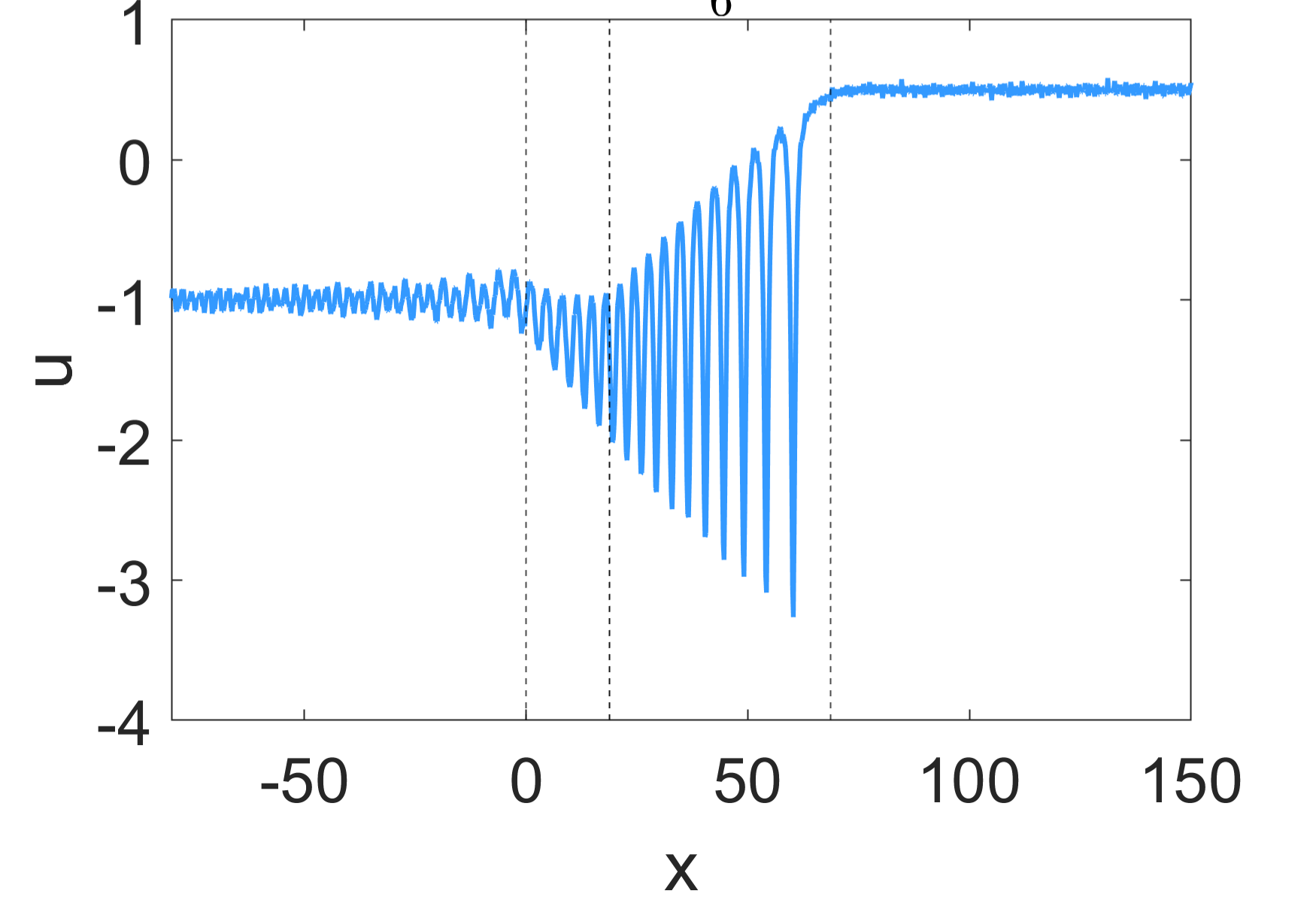}\\
\includegraphics[scale=0.3]{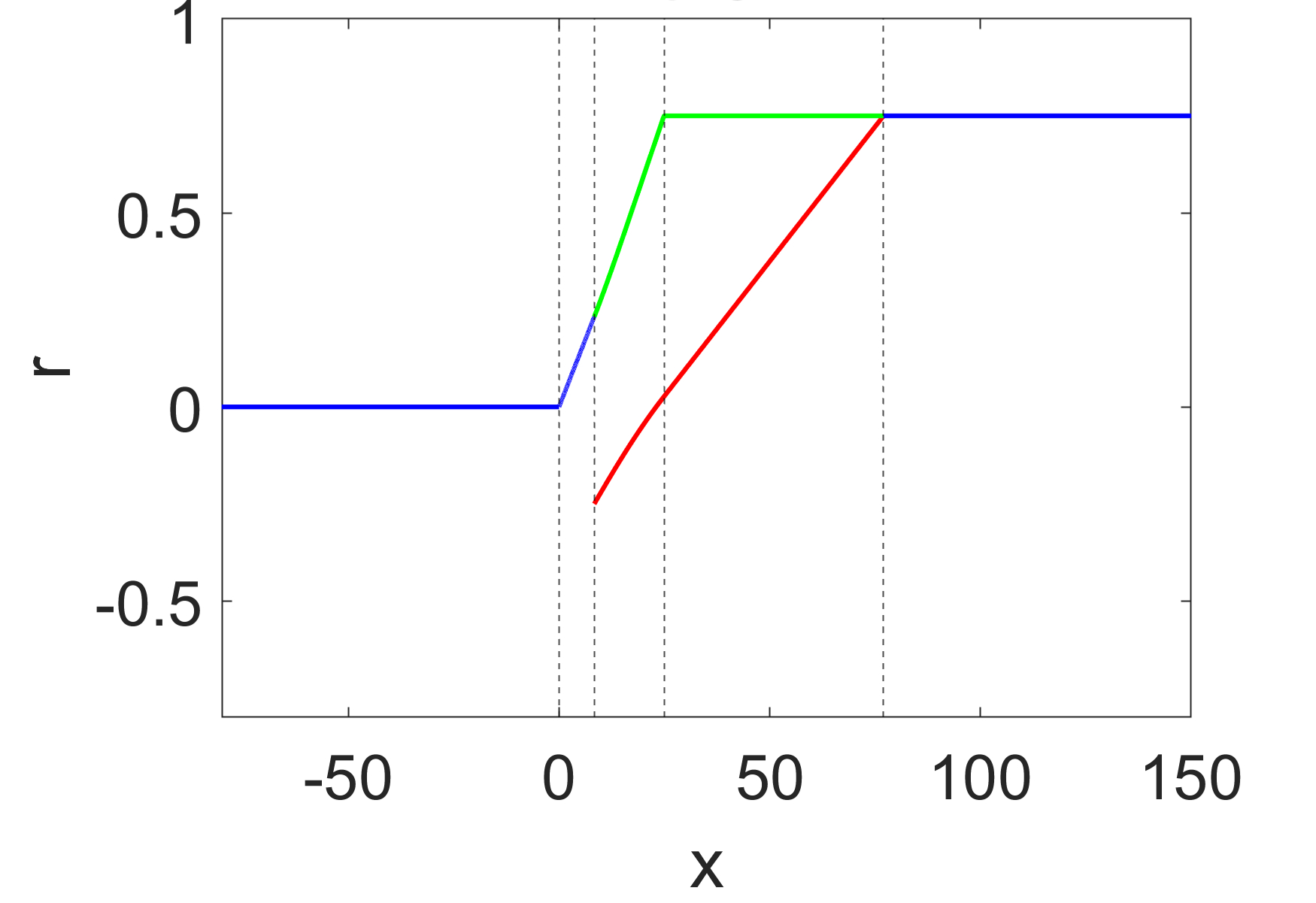}\hfill
\includegraphics[scale=0.3]{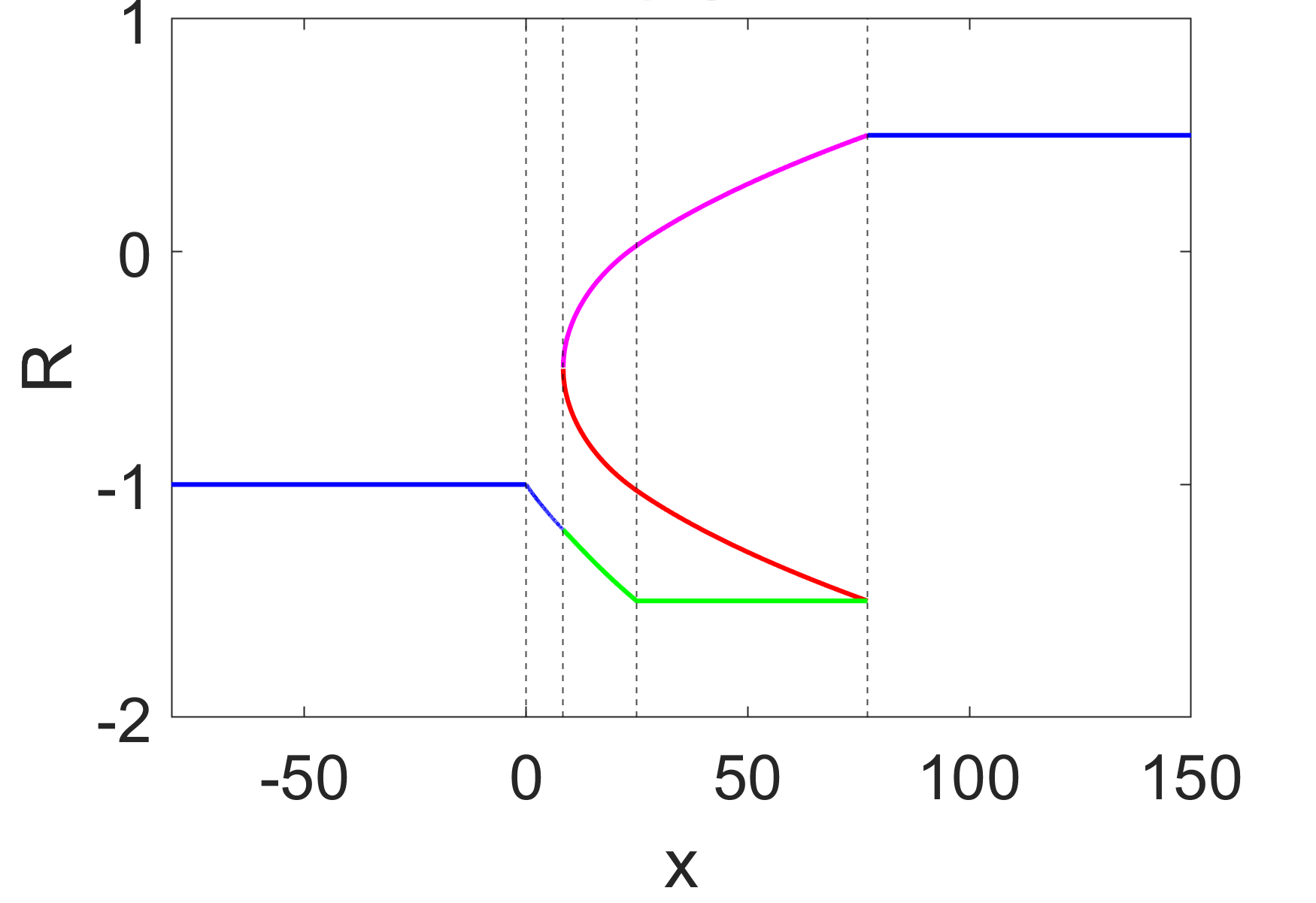}\hfill
\includegraphics[scale=0.3]{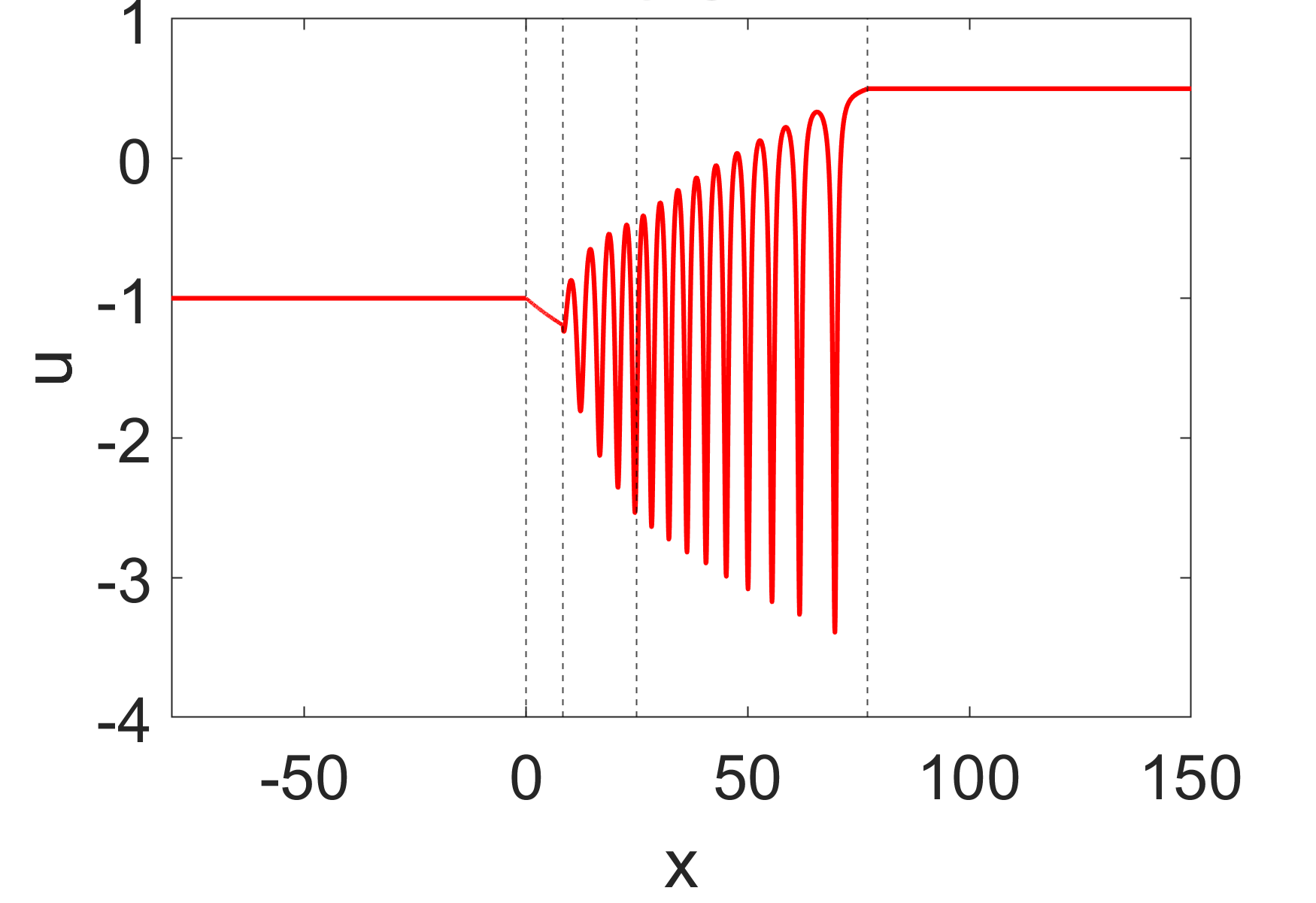}\hfill
\includegraphics[scale=0.3]{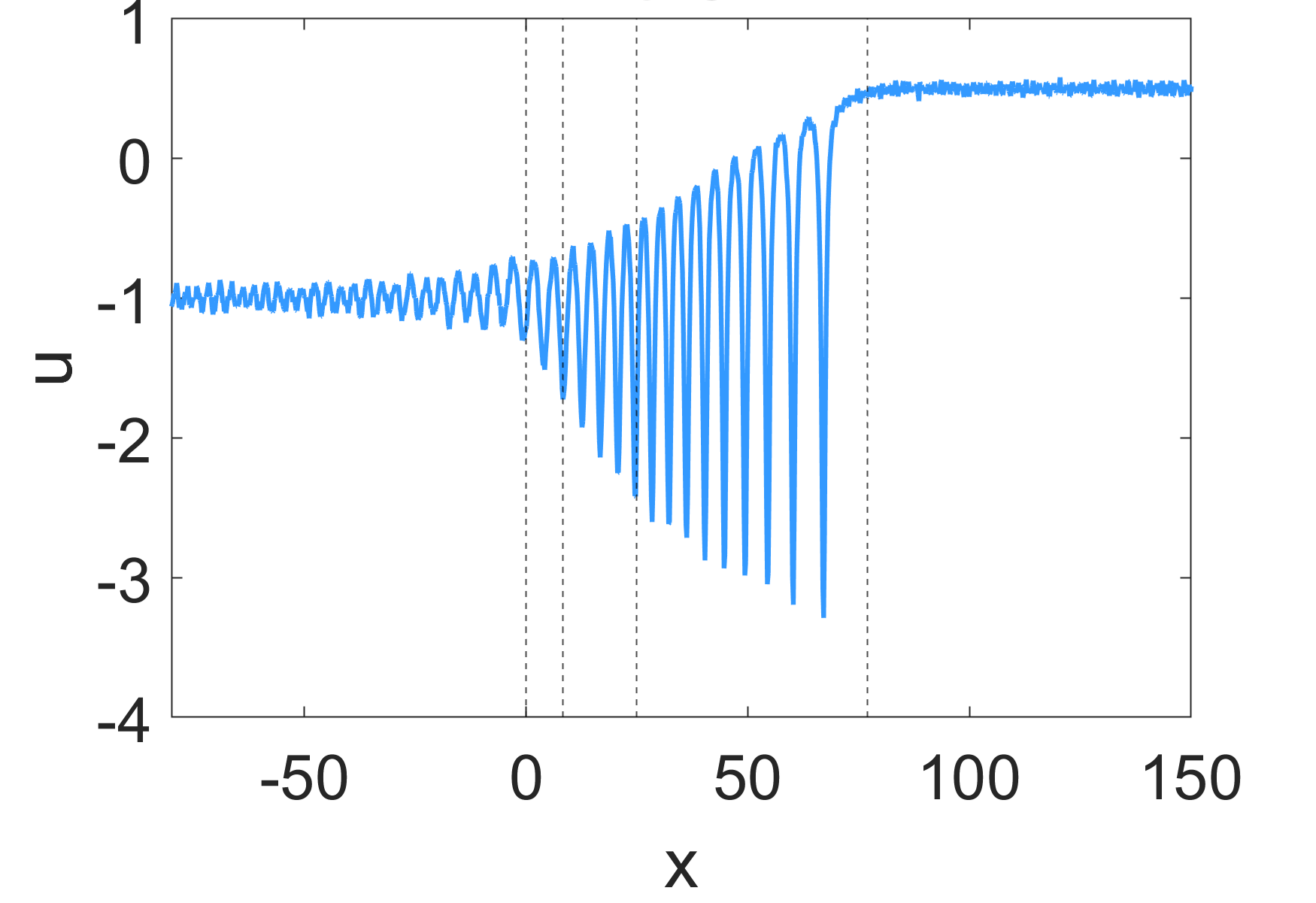}\\
\includegraphics[scale=0.3]{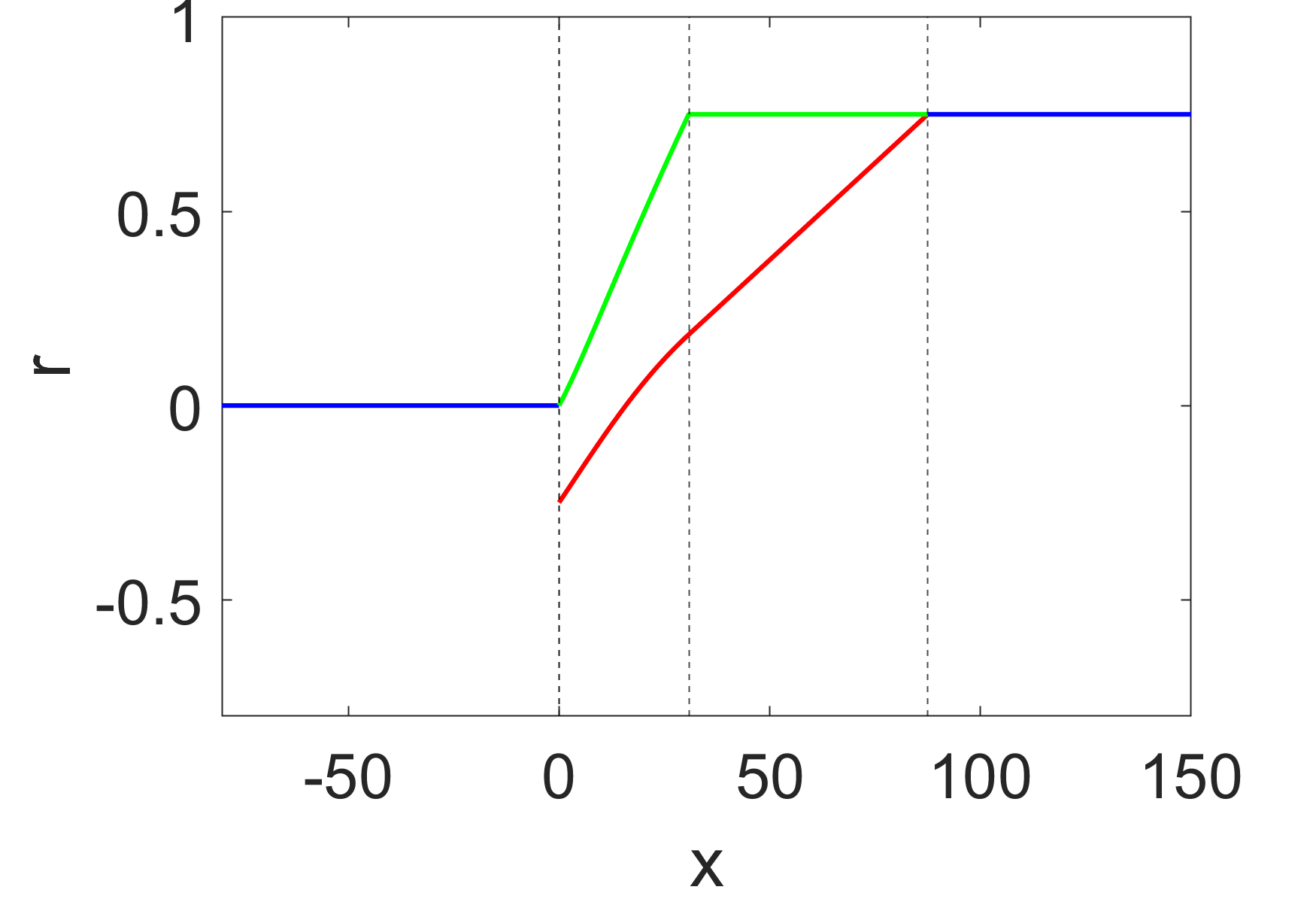}\hfill
\includegraphics[scale=0.3]{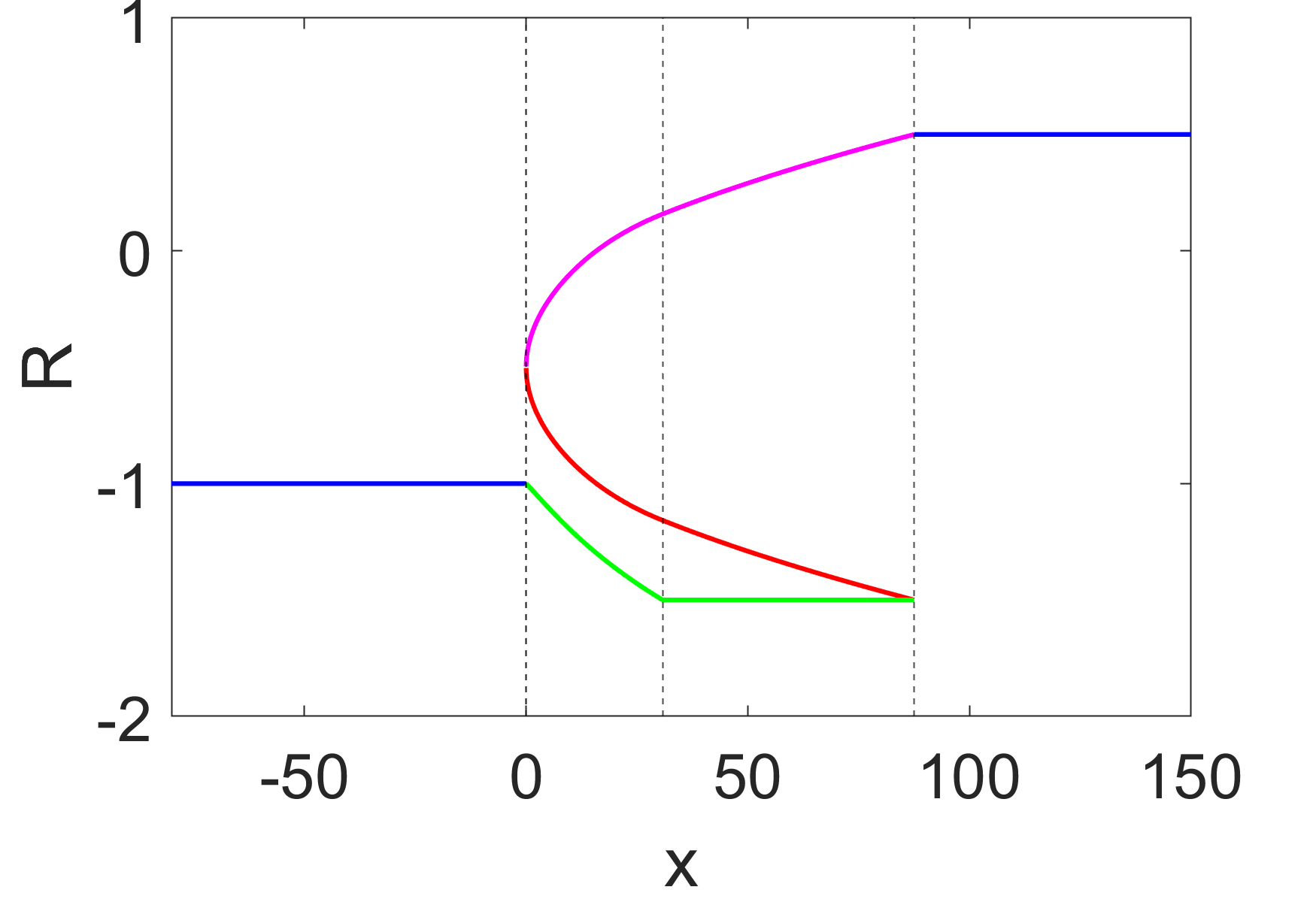}\hfill
\includegraphics[scale=0.3]{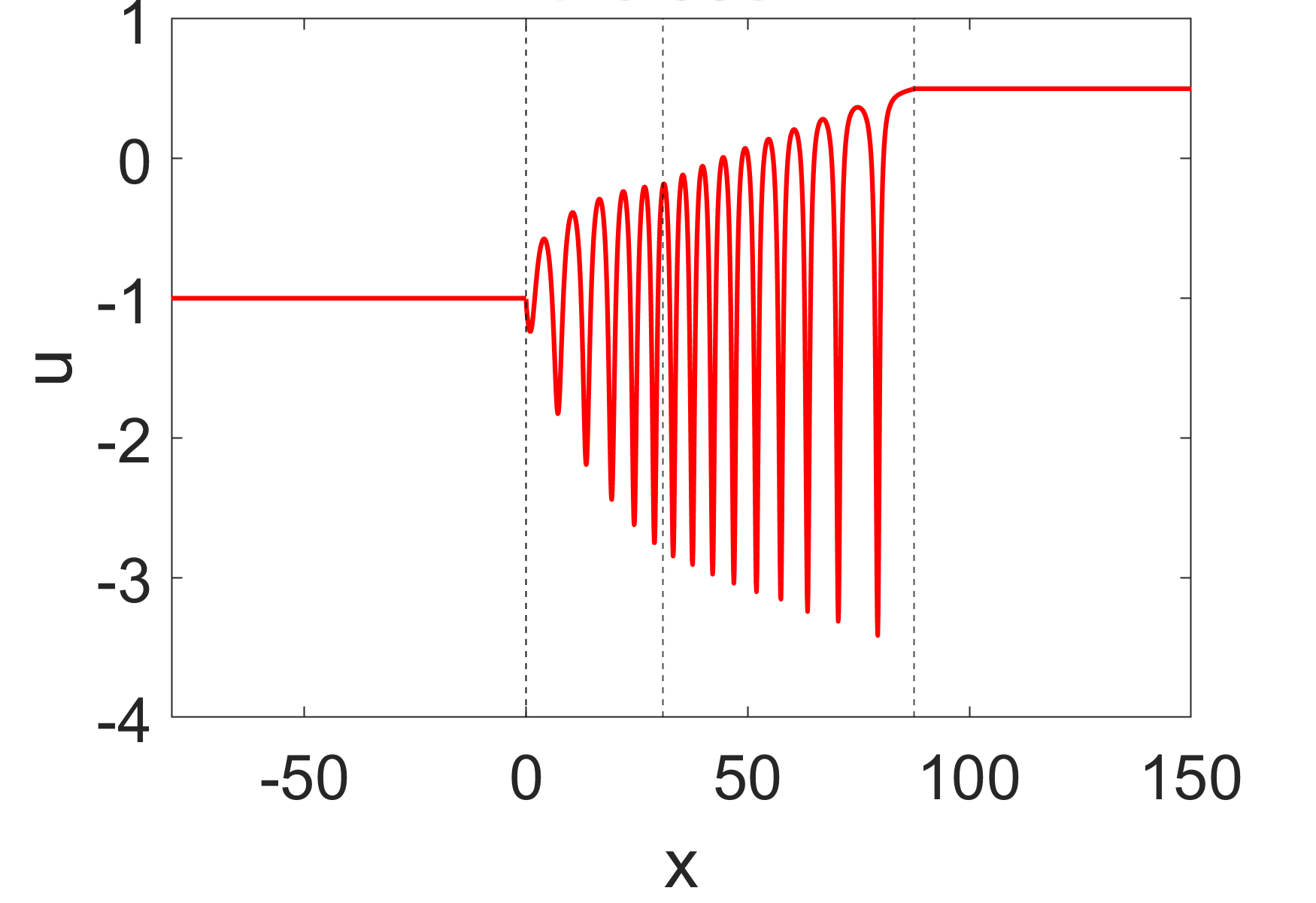}\hfill
\includegraphics[scale=0.3]{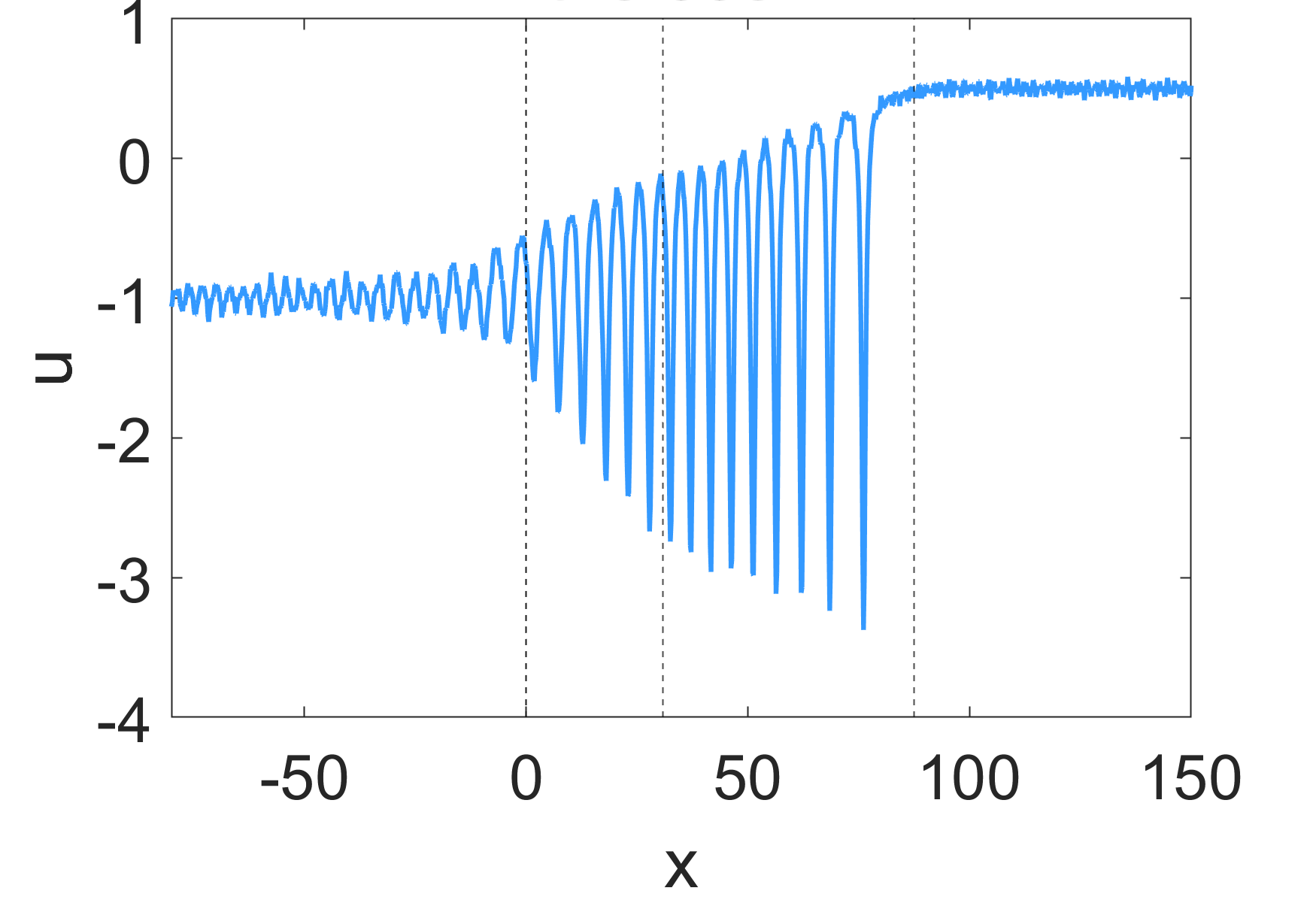}\hfill\\
\includegraphics[scale=0.3]{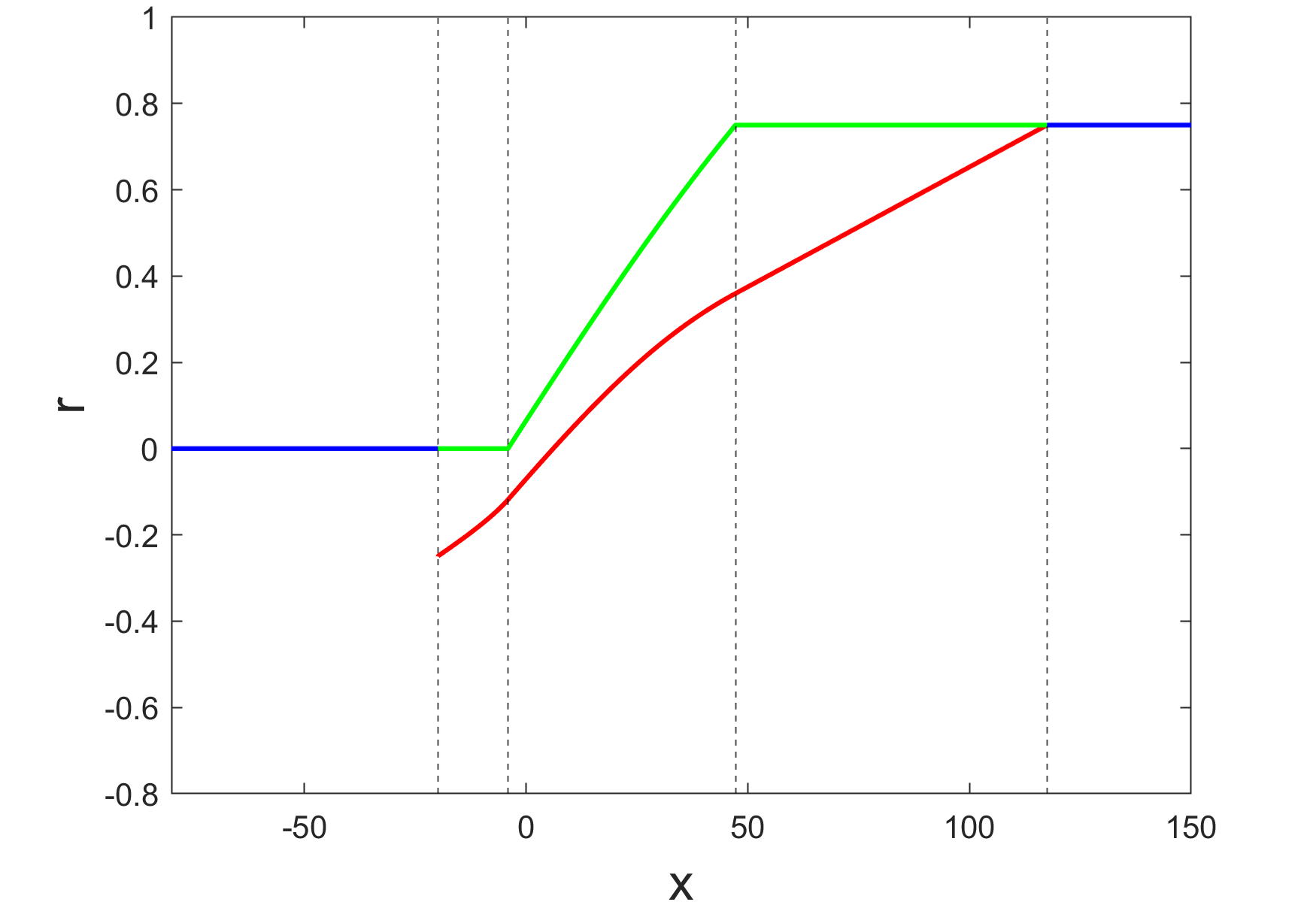}\hfill
\includegraphics[scale=0.3]{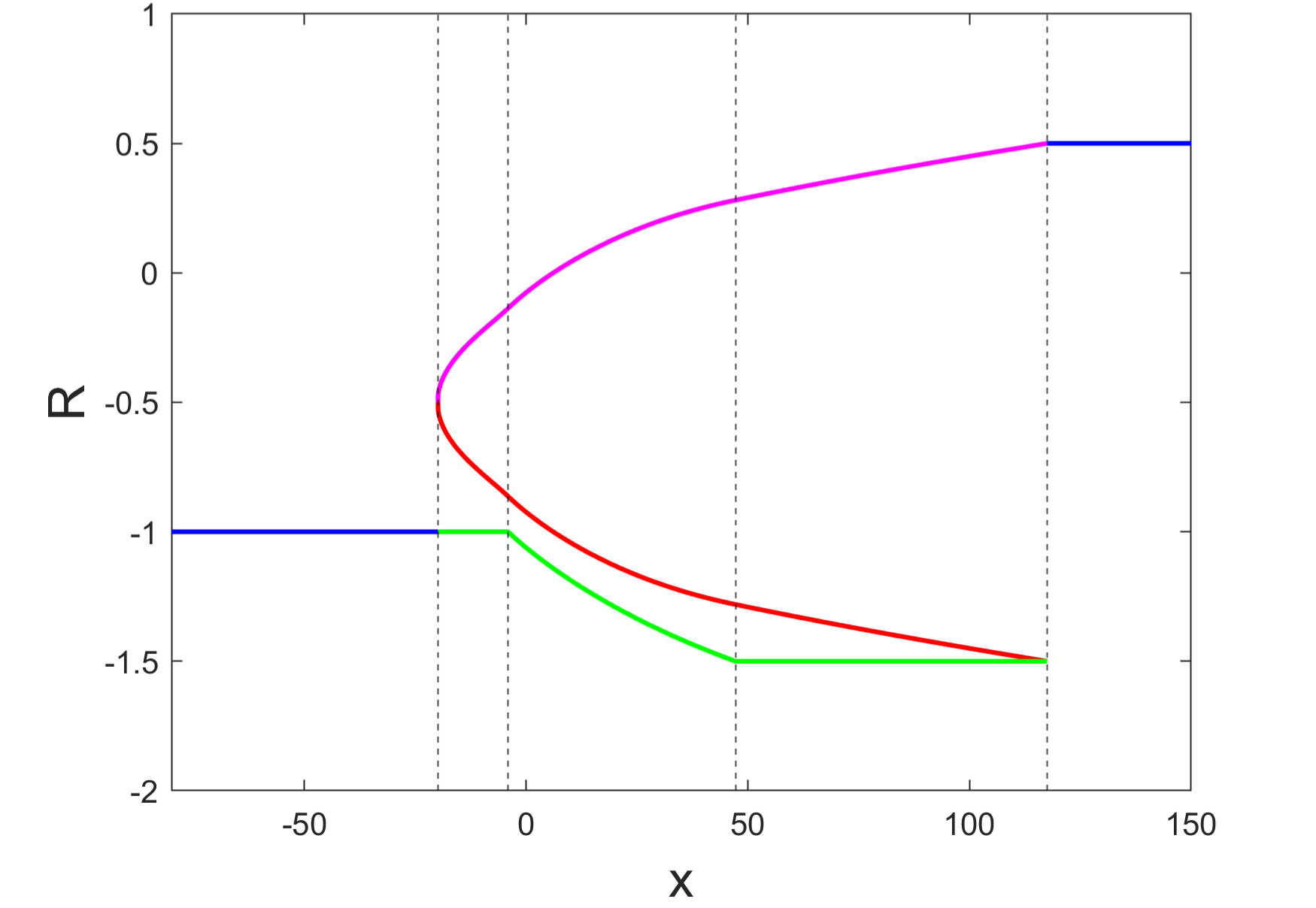}\hfill
\includegraphics[scale=0.3]{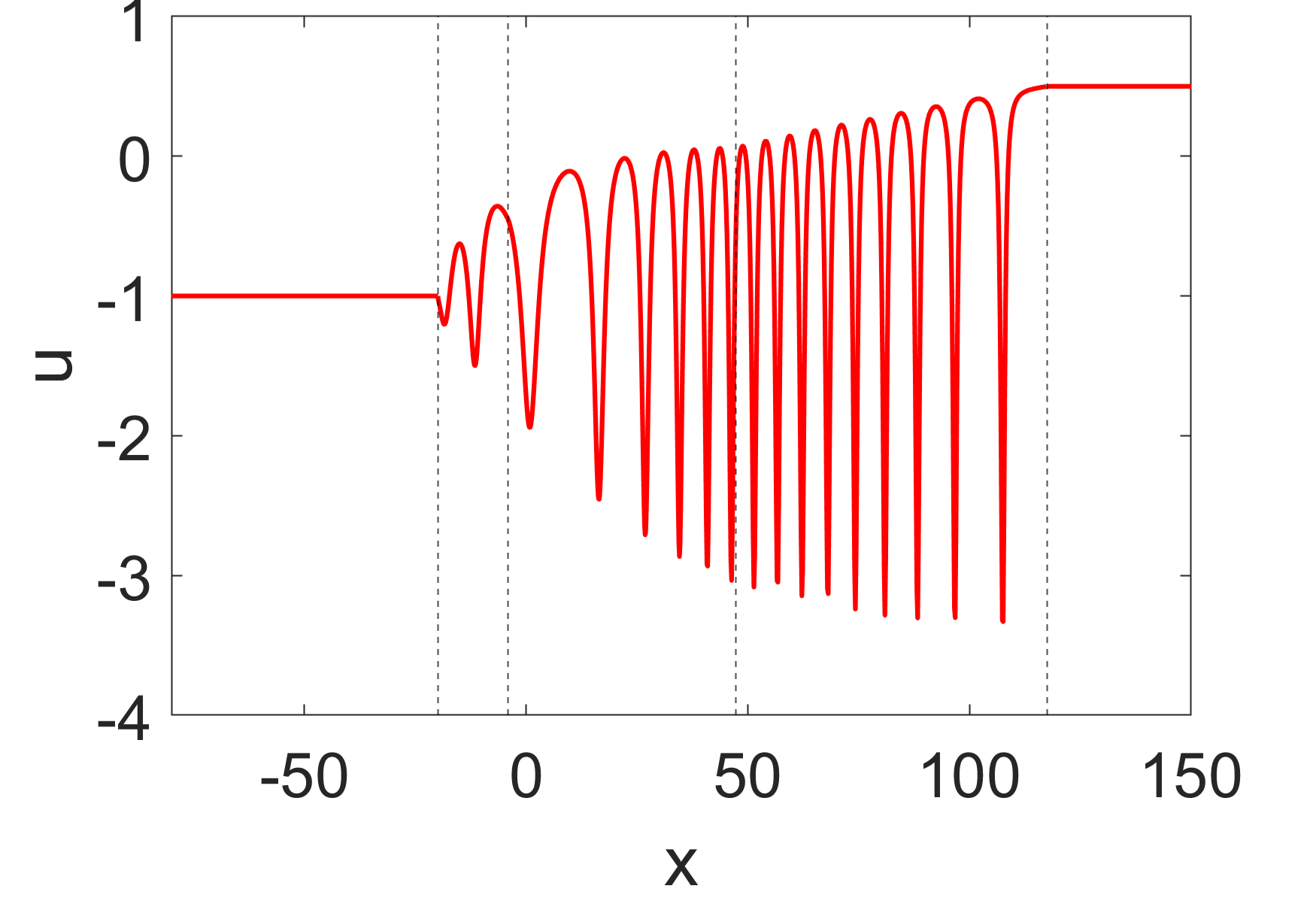}\hfill
\includegraphics[scale=0.3]{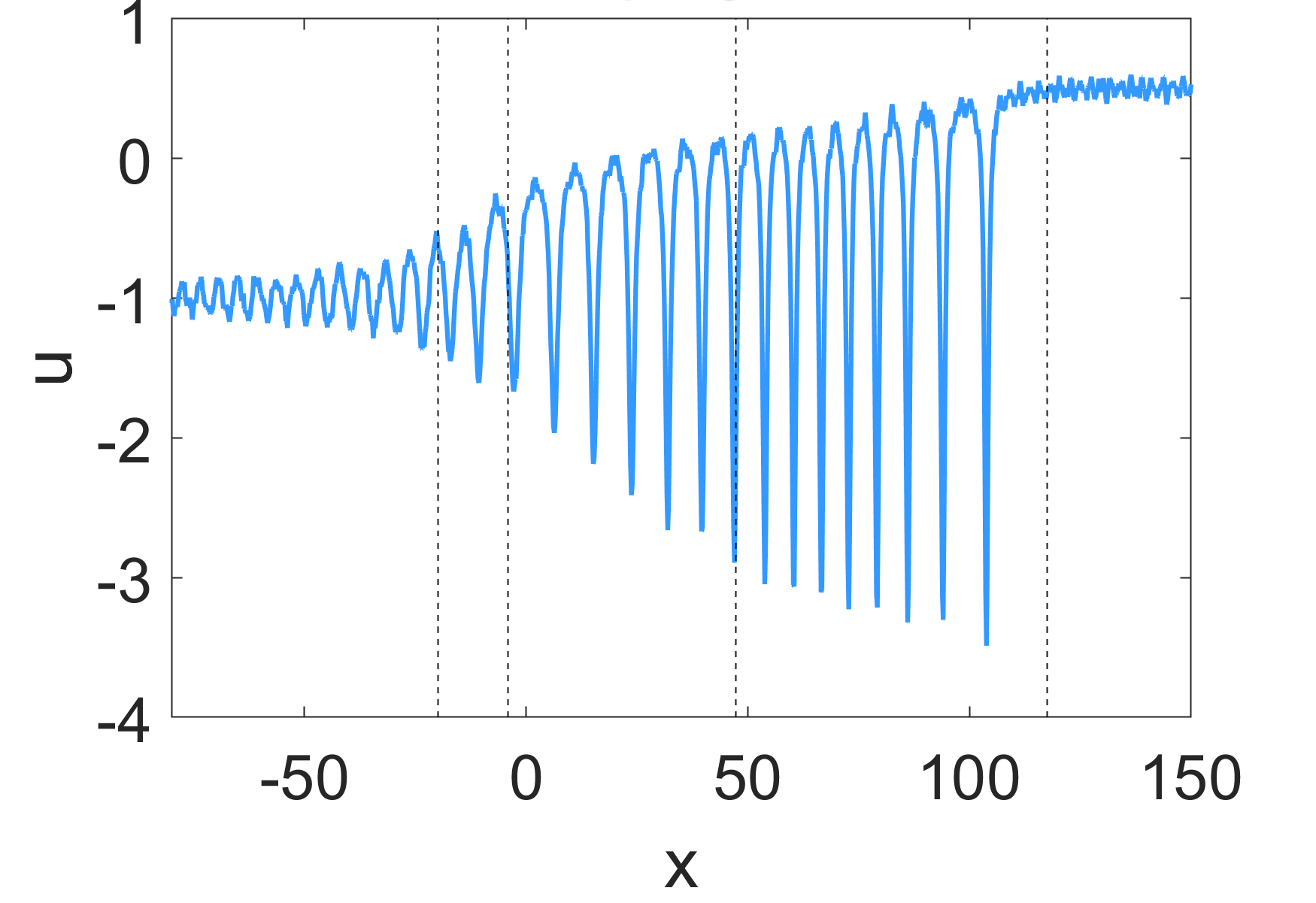}\hfill\\

{\footnotesize\hspace{0.0cm}(a)\hspace{4cm}(b)\hspace{4cm}(c)\hspace{4cm}(d)}
\caption{Evolution of RW-reversed CDSW interaction, at $t=2.5$, $t=t_1$, $t=6$, $t=t_2$, and $t=15$, respectively. The parameters are $u_l=-1$, $u_m=-1.5$, $u_r=0.5$, $D=50$, $\alpha=-1$. (a) Riemann invariants $r$; (b) Combination of Riemann invariants $R$; (c) Analytical solutions of Eq.~(1) based on the Whitham modulation theory; (d) Numerical solutions of Eq.~(1).}
\label{Figs.~9.}
\end{figure*}
\begin{figure}
\includegraphics[scale=0.39]{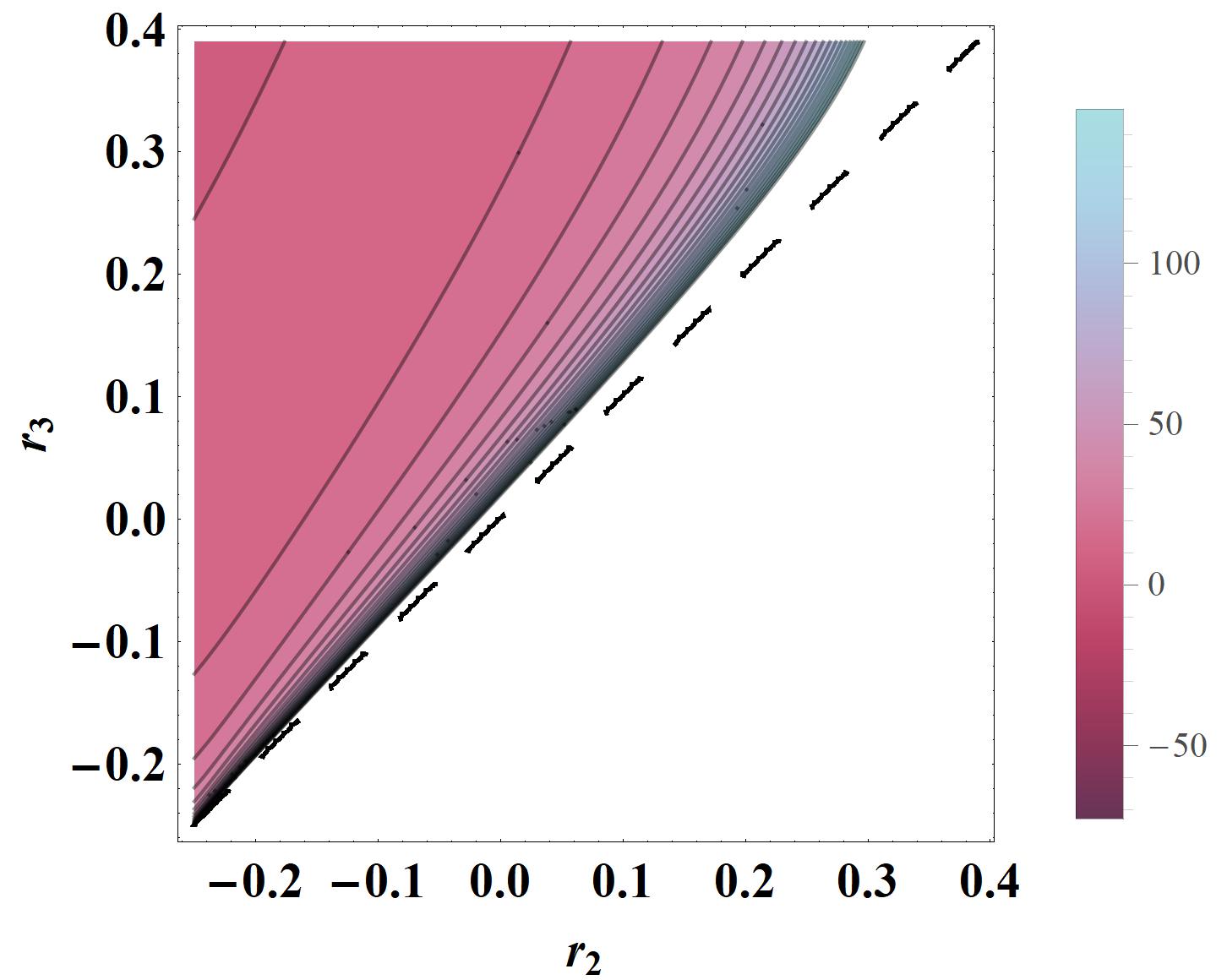}\hfill
\includegraphics[scale=0.29]{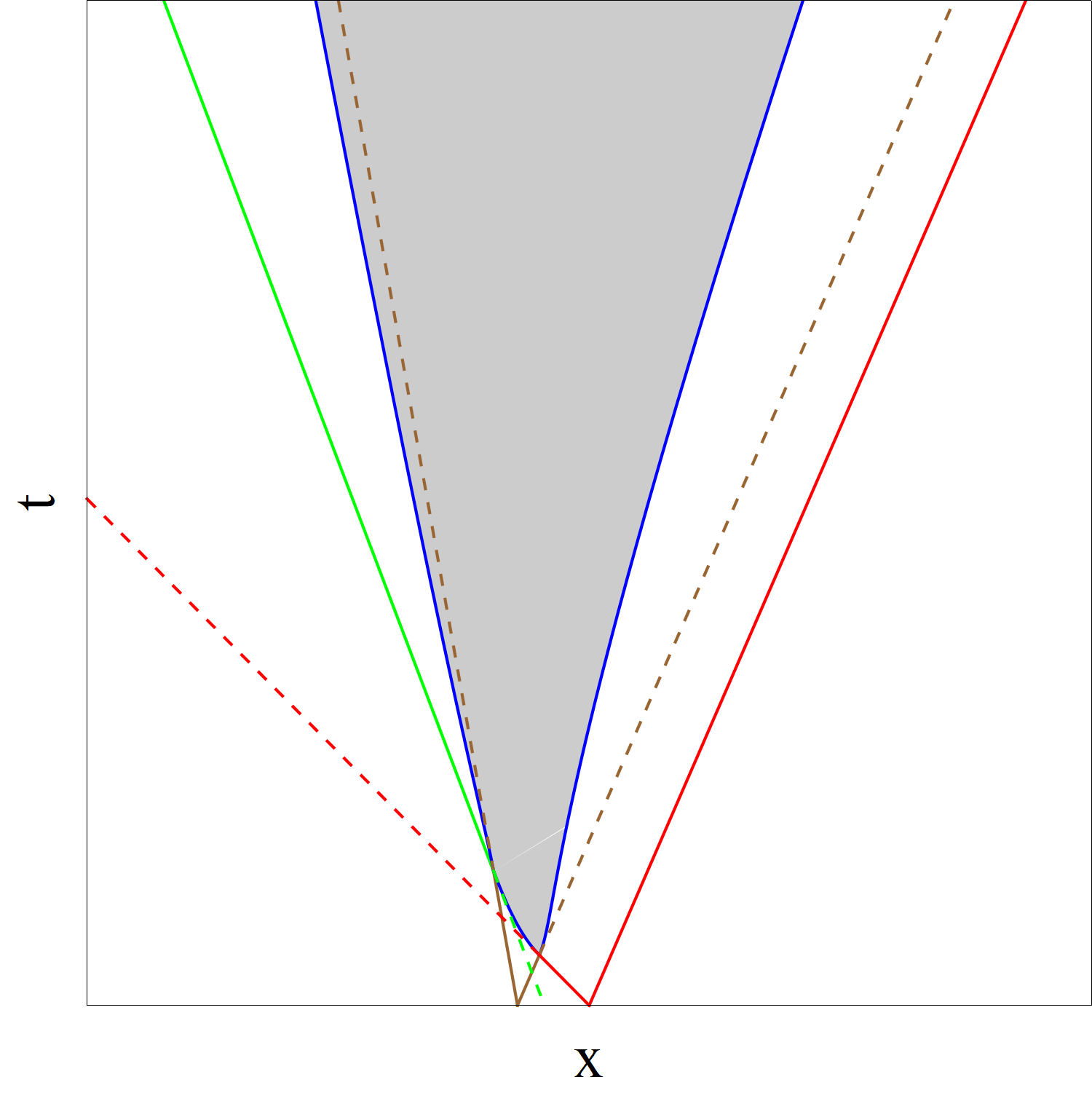}\hfill\\
{\footnotesize\hspace{0.0cm}(a)\hspace{4cm}(b)}
\caption{(a) Contour plot of $\frac{\tilde{w}_3(r_2,r_3)-\tilde{w}_2(r_2,r_3)}{v_2(r_2,r_2,r_3)-v_3(r_2,r_2,r_3)}$ on the hodograph $(r_2,r_3)$ plane; (b) Evolution on the $(x,t)$ plane. The brown, red and blue lines are edges of the RW, CDSW and interaction region, respectively, and the dashed lines represents original trajectories if there had been no collision. The green line is the trailing edge of the CDSW existing from the interaction region.}
\label{Figs.~10.}
\end{figure}
\begin{figure*}
\includegraphics[scale=0.3]{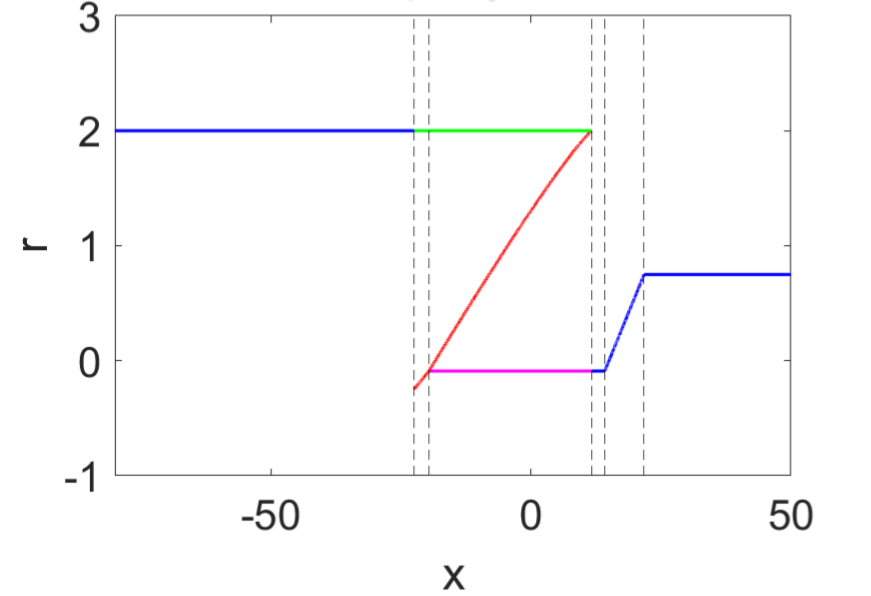}\hfill
\includegraphics[scale=0.3]{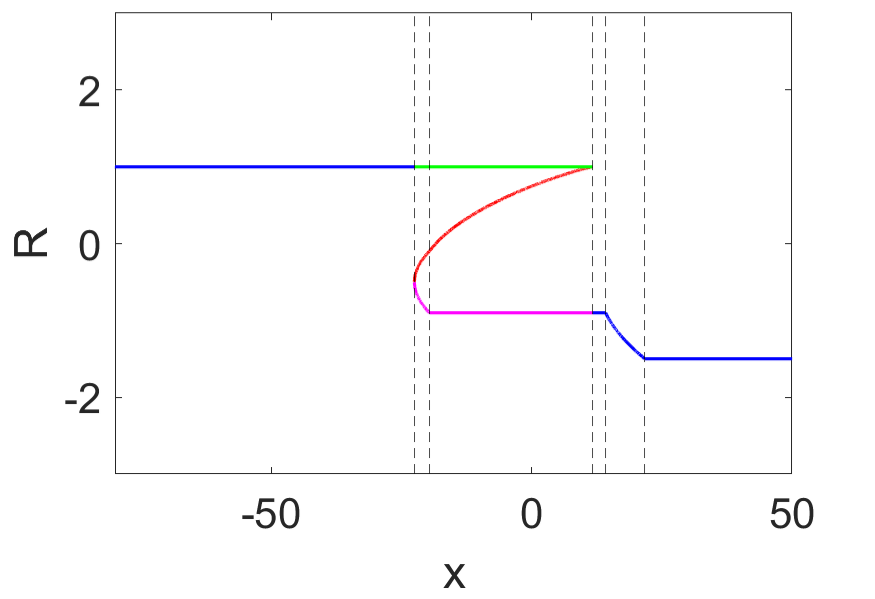}\hfill
\includegraphics[scale=0.3]{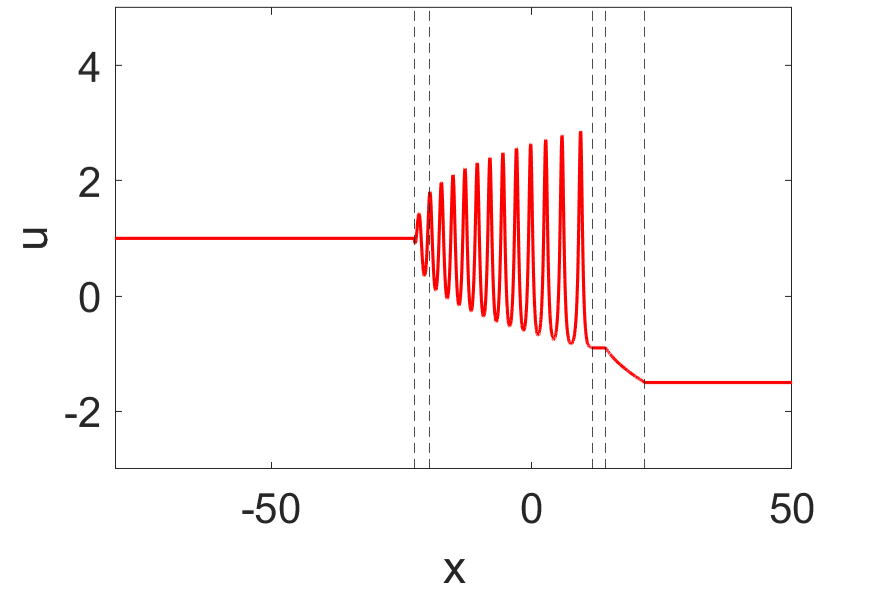}\hfill
\includegraphics[scale=0.3]{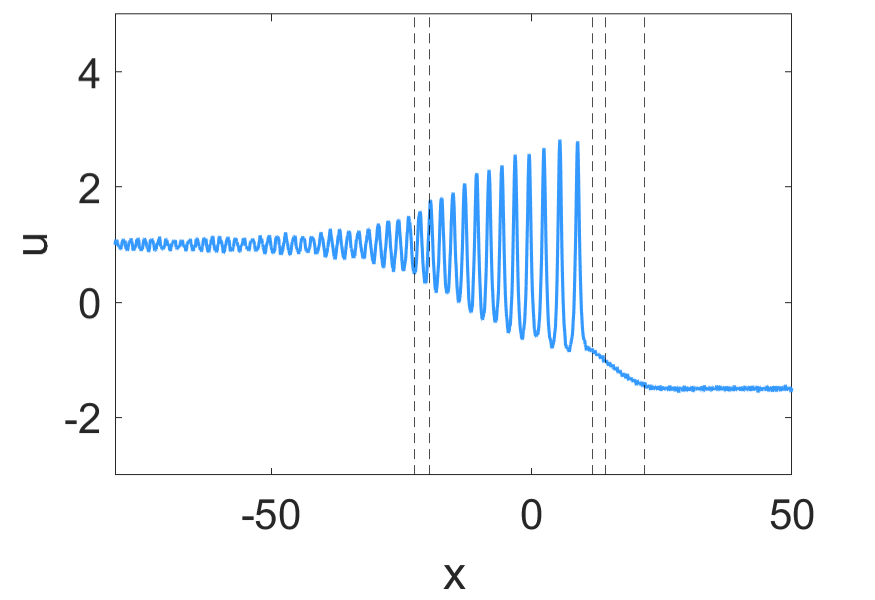}\hfill\\
\includegraphics[scale=0.3]{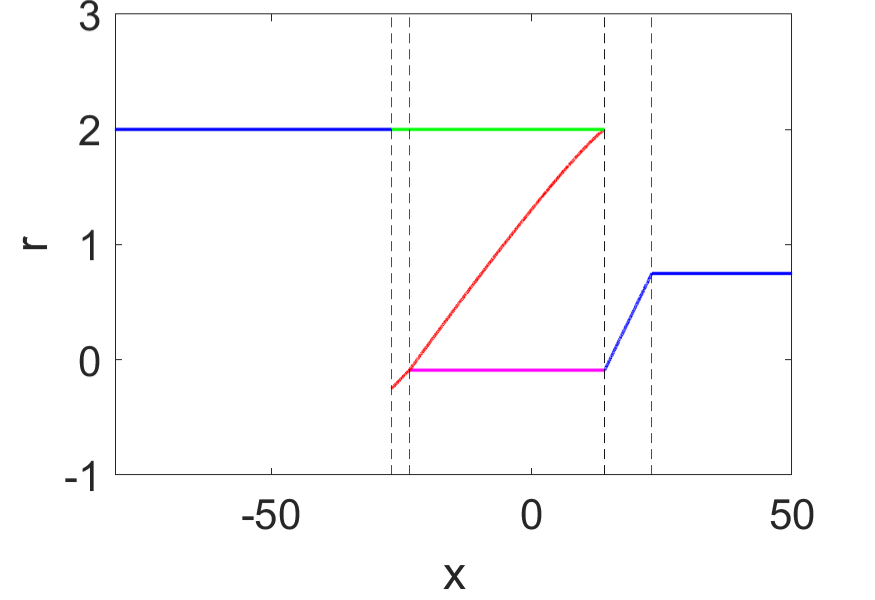}\hfill
\includegraphics[scale=0.3]{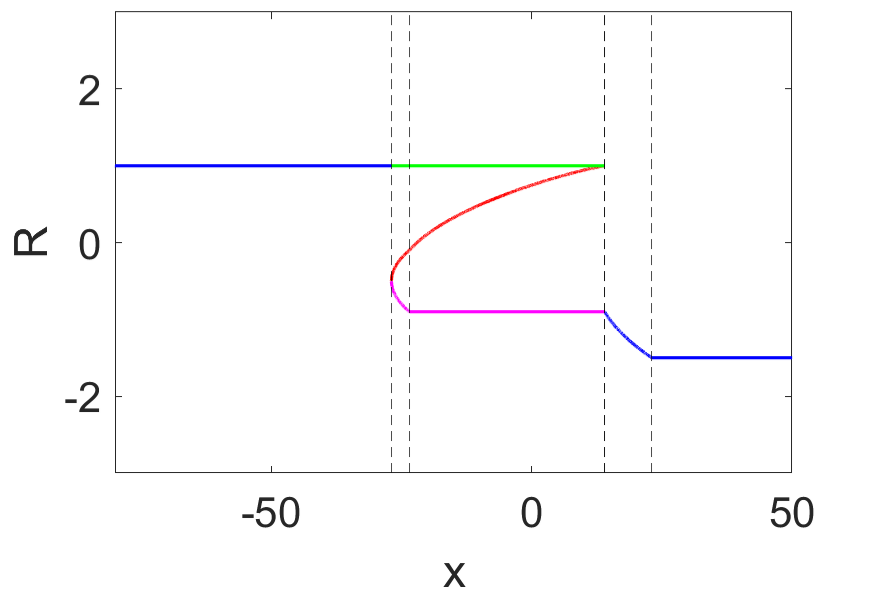}\hfill
\includegraphics[scale=0.3]{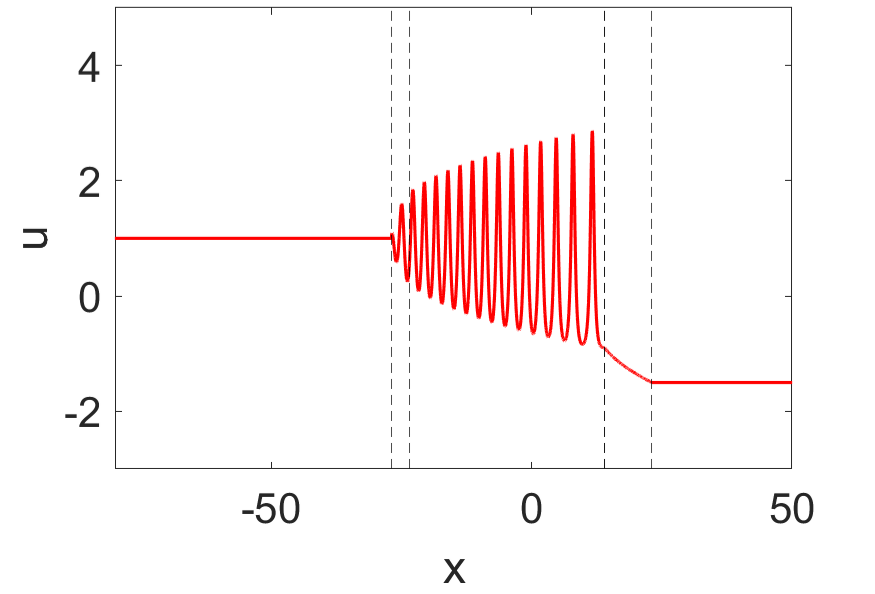}\hfill
\includegraphics[scale=0.3]{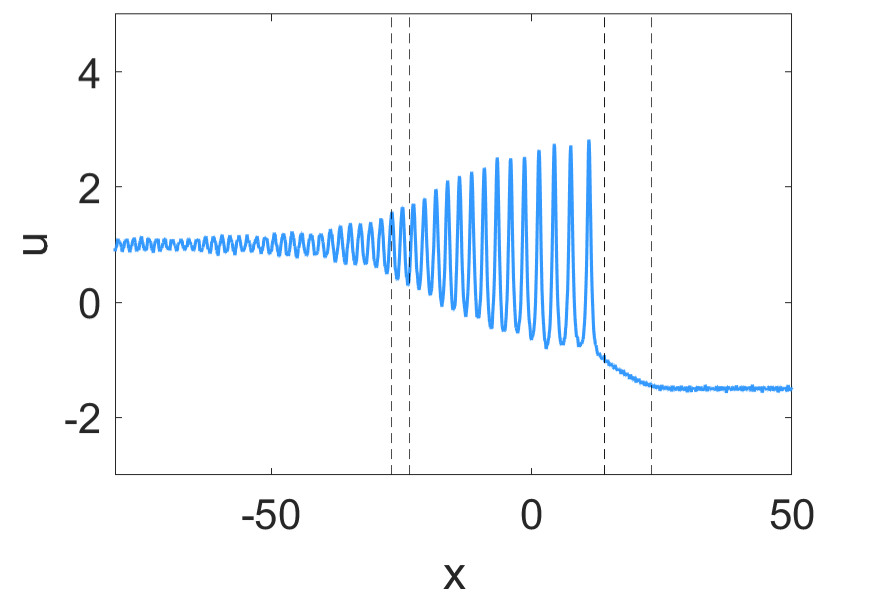}\hfill\\
\includegraphics[scale=0.3]{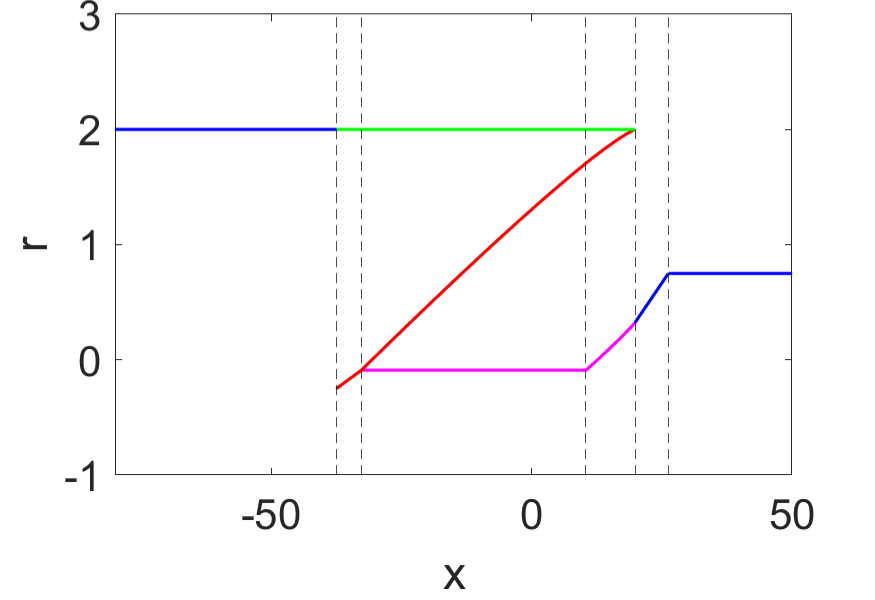}\hfill
\includegraphics[scale=0.3]{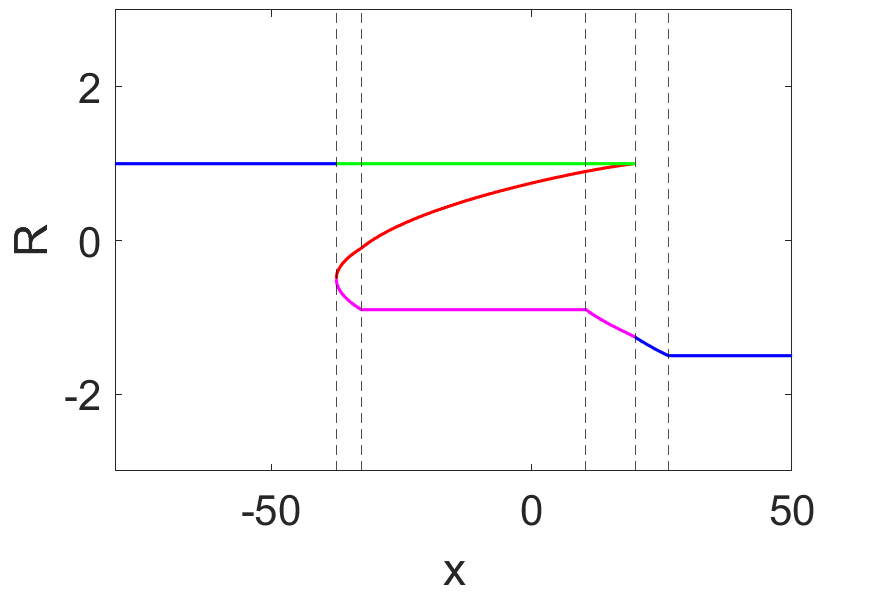}\hfill
\includegraphics[scale=0.3]{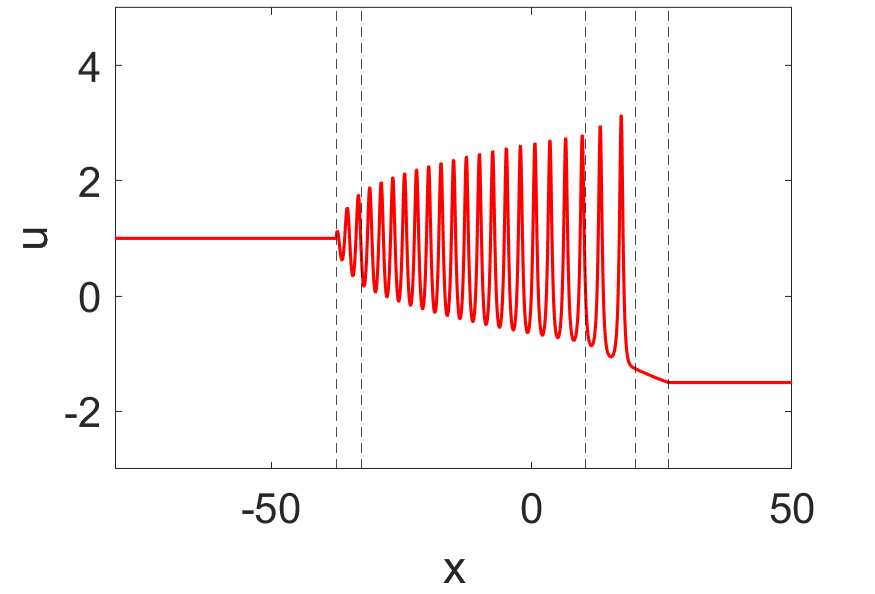}\hfill
\includegraphics[scale=0.3]{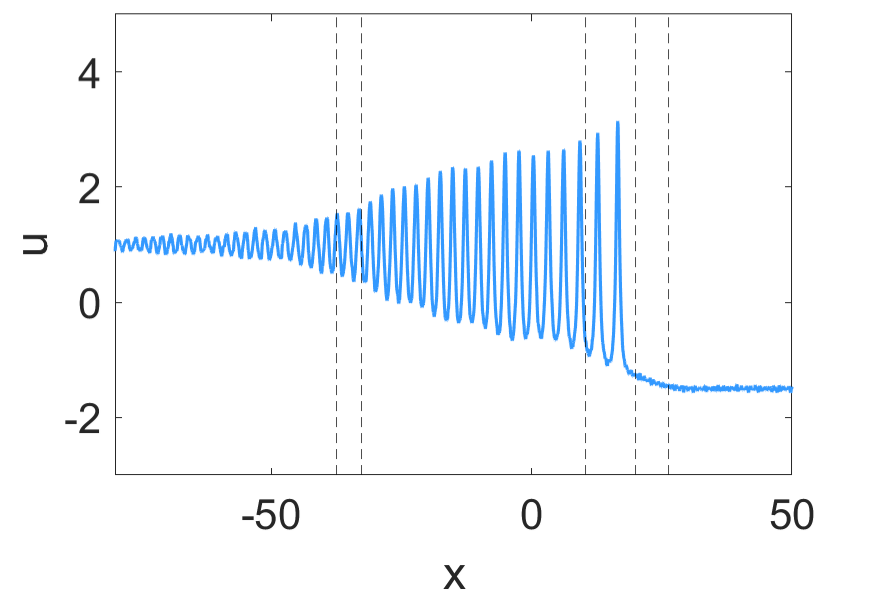}\hfill\\
\includegraphics[scale=0.3]{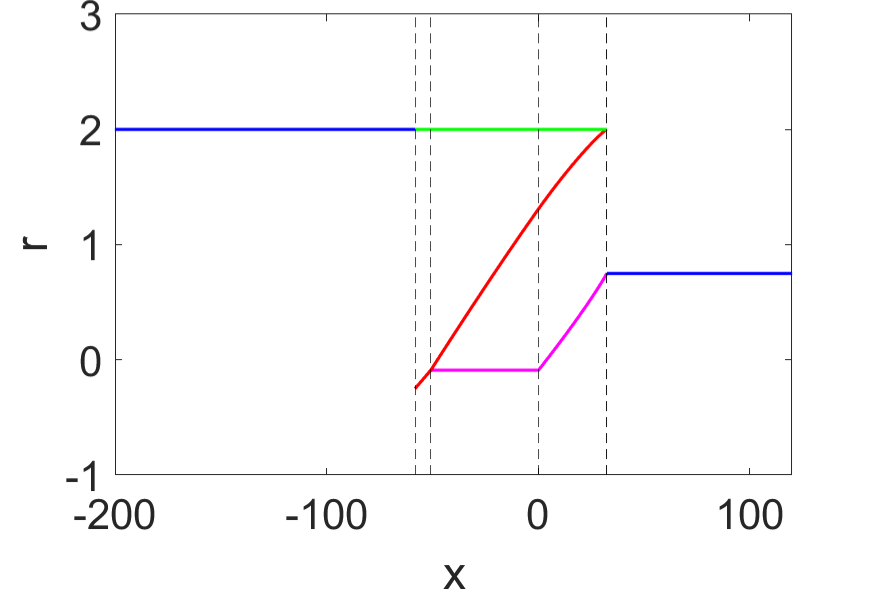}\hfill
\includegraphics[scale=0.3]{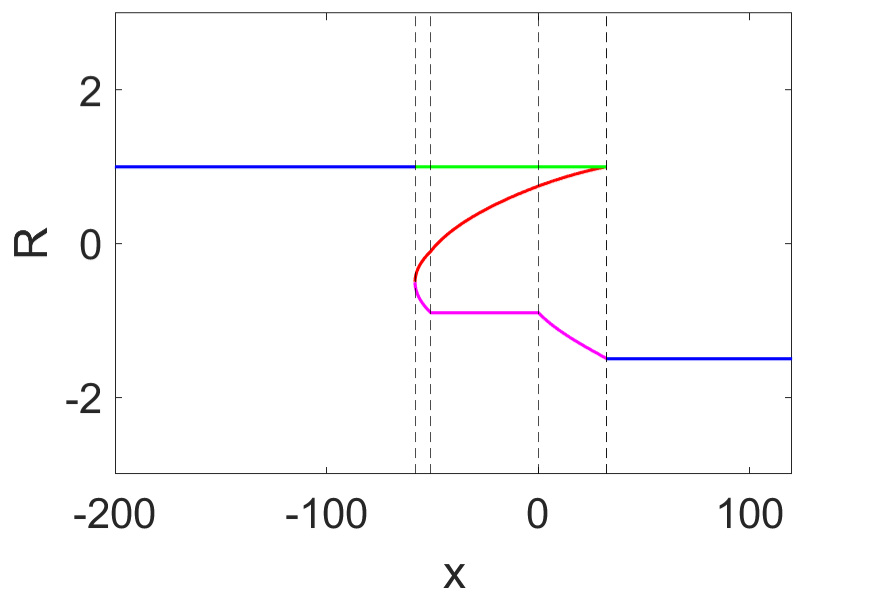}\hfill
\includegraphics[scale=0.3]{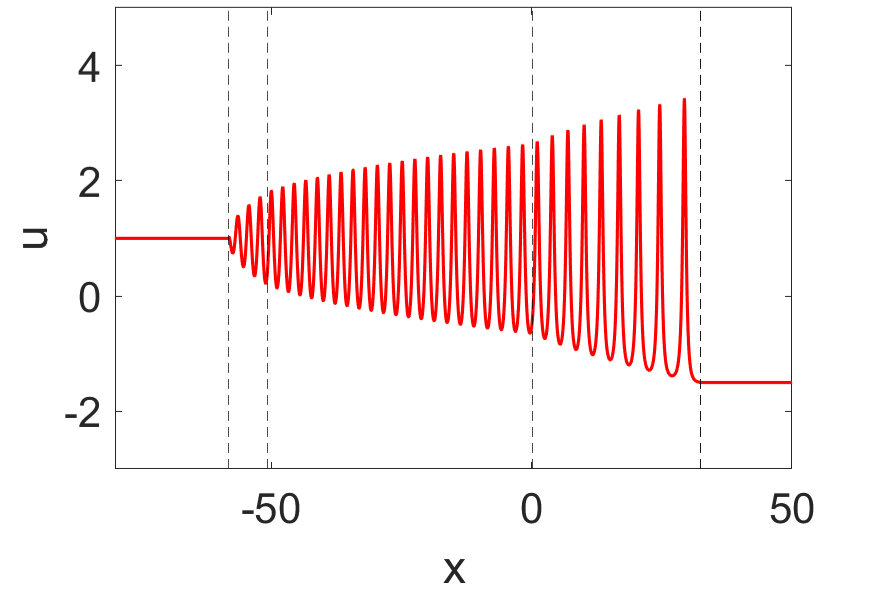}\hfill
\includegraphics[scale=0.3]{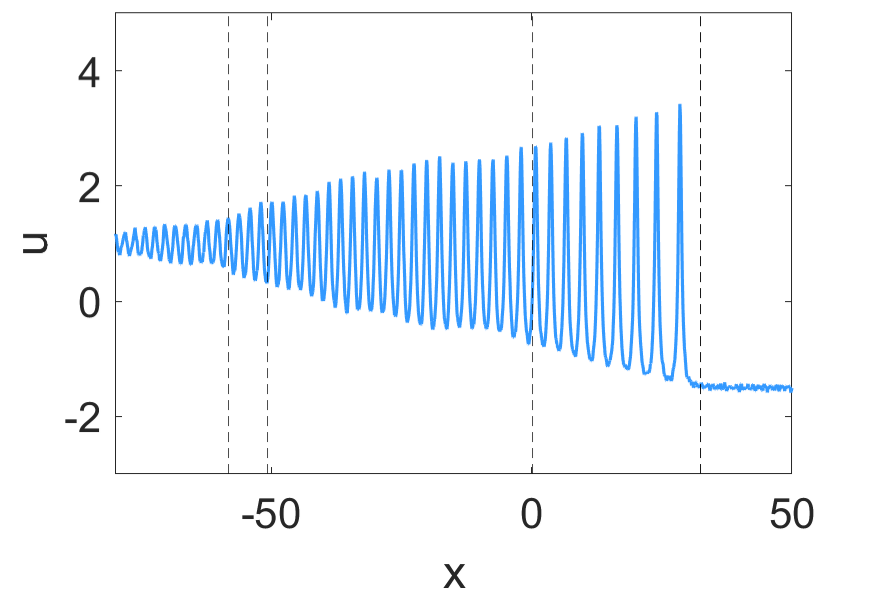}\hfill\\
\includegraphics[scale=0.3]{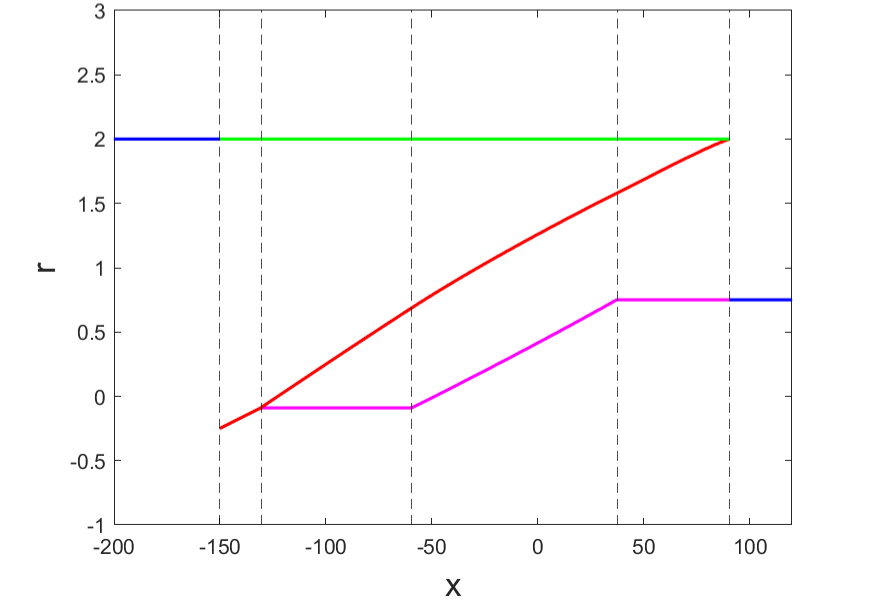}\hfill
\includegraphics[scale=0.3]{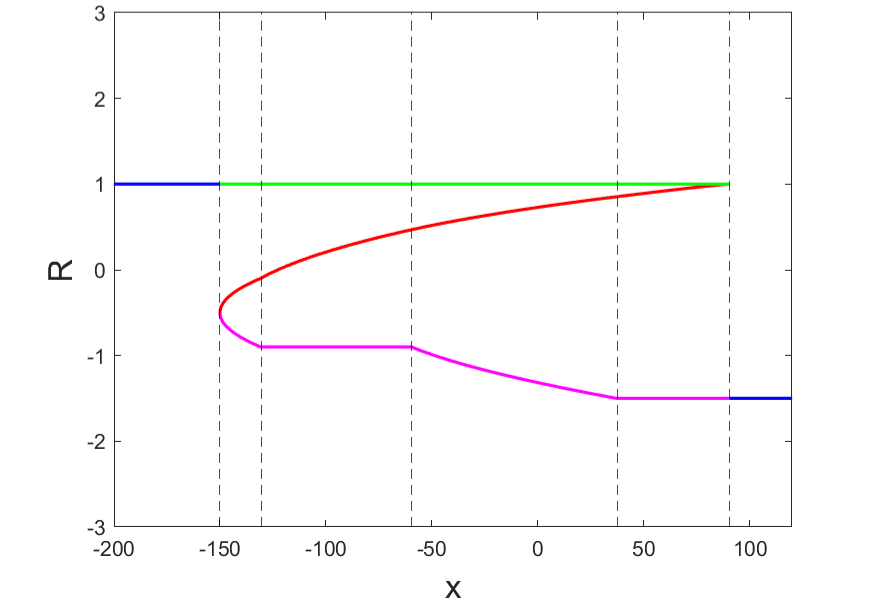}\hfill
\includegraphics[scale=0.3]{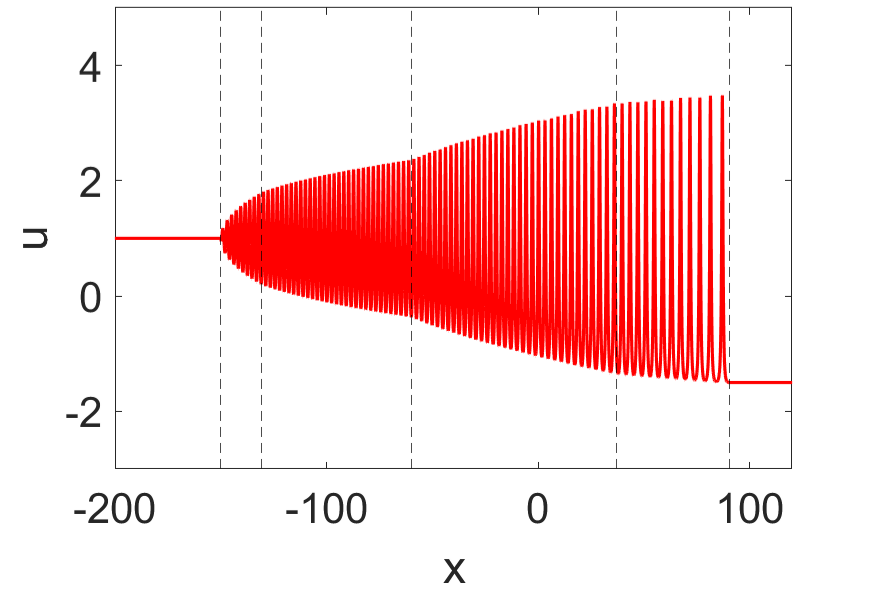}\hfill
\includegraphics[scale=0.3]{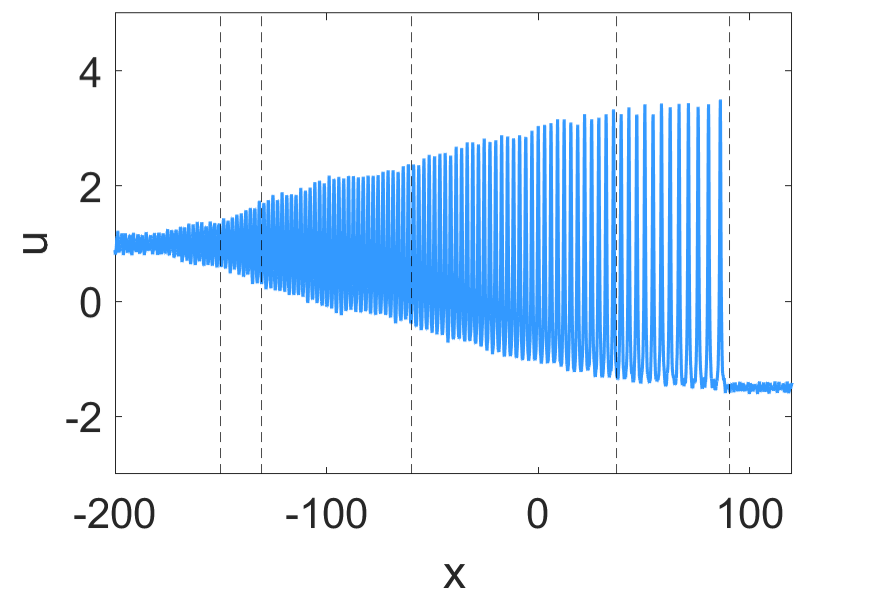}\hfill\\

{\footnotesize\hspace{0.0cm}(a)\hspace{4cm}(b)\hspace{4cm}(c)\hspace{4cm}(d)}
\caption{Evolution of DSW$|$CDSW-RW interaction with transmitted DSW, at $t=1.5$, $t=t_1$, $t=2.5$, $t=t_2$, and $t=10$, respectively. Parameters are $u_l=1$, $u_m=-0.9$, $u_r=-1.5$, $D=15$, $\alpha=-1$. (a) Riemann invariants $r$; (b) Combination of Riemann invariants $R$; (c) Analytical solutions of Eq.~(1) based on the Whitham modulation theory; (d) Numerical solutions of Eq.~(1).}
\label{Figs.~11.}
\end{figure*}
\begin{figure*}
\includegraphics[scale=0.3]{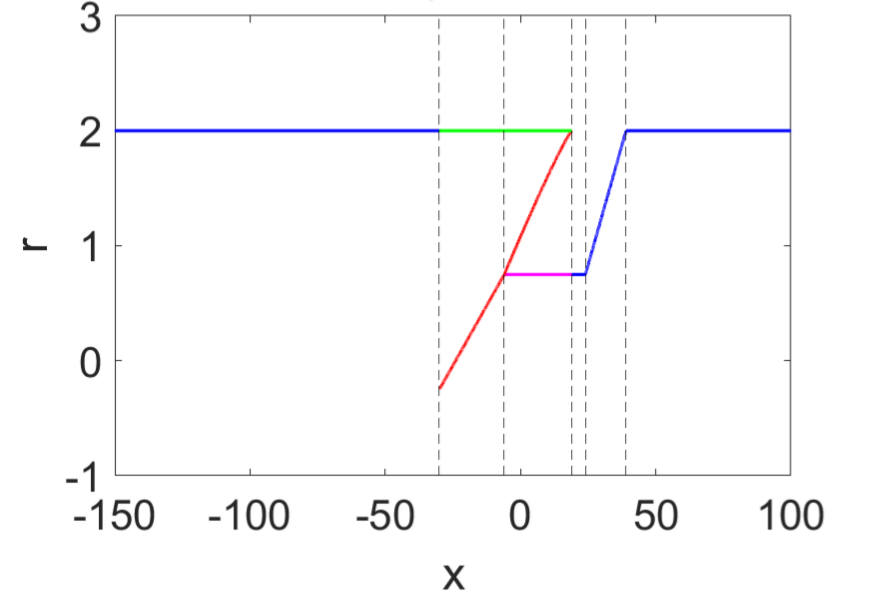}\hfill
\includegraphics[scale=0.3]{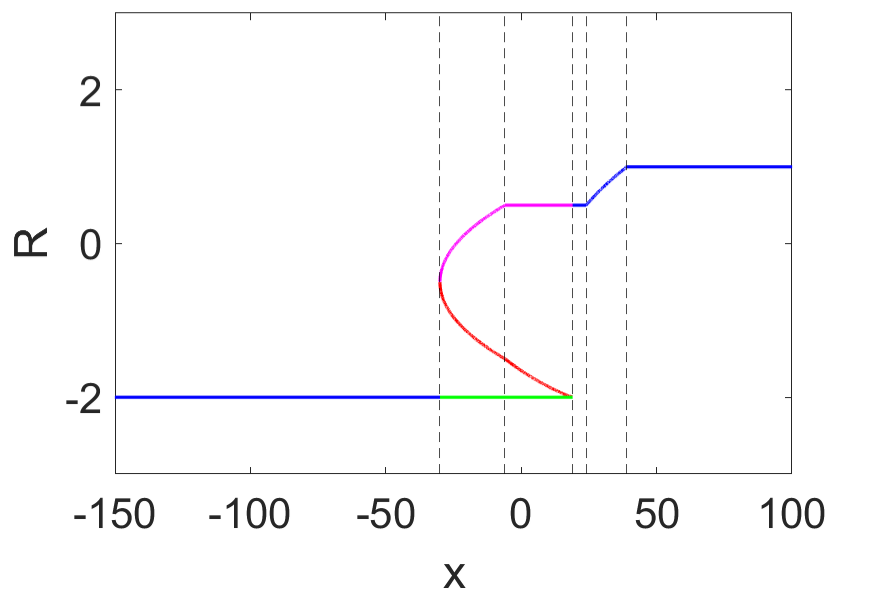}\hfill
\includegraphics[scale=0.3]{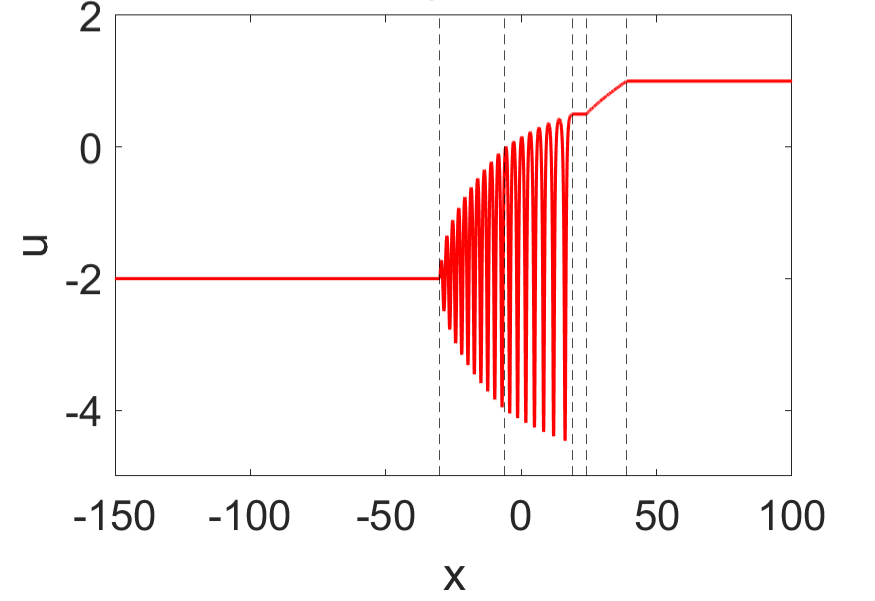}\hfill
\includegraphics[scale=0.3]{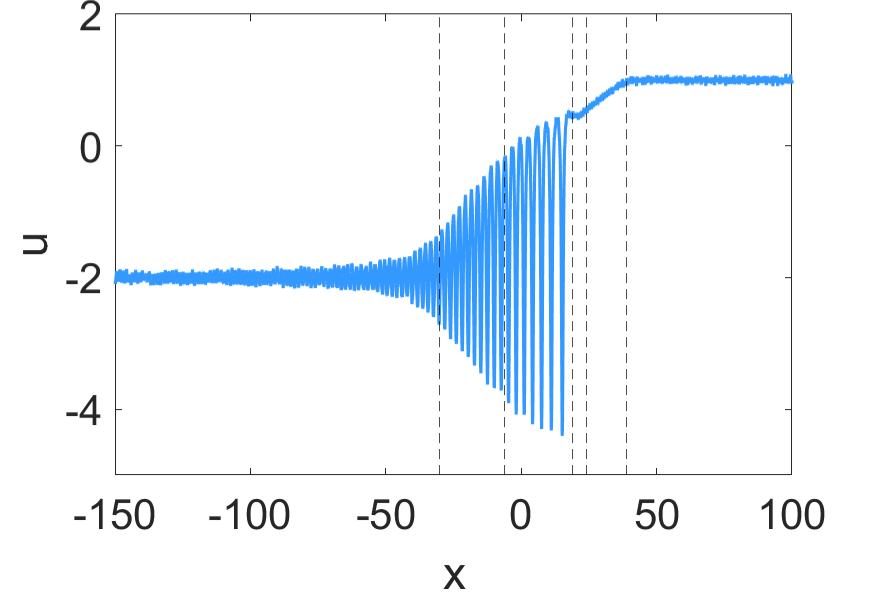}\hfill\\
\includegraphics[scale=0.3]{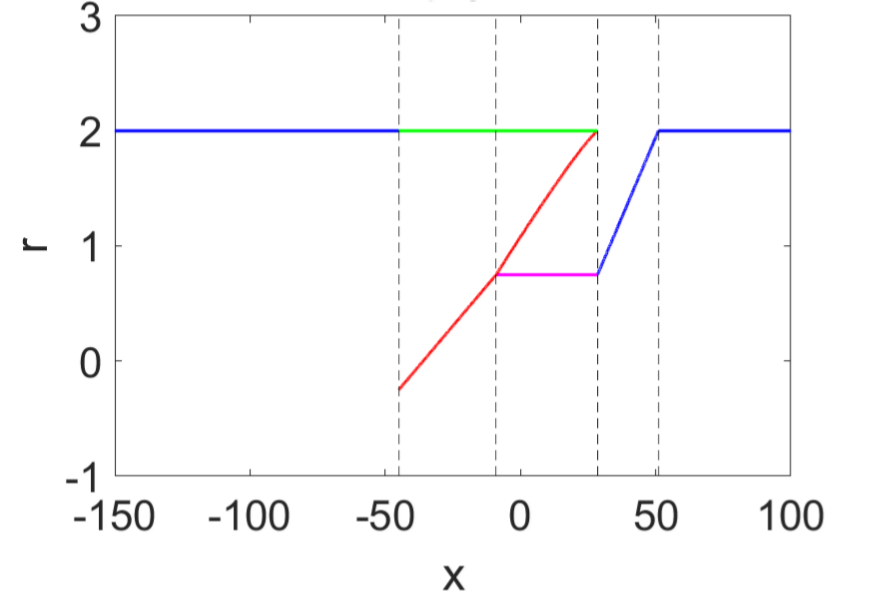}\hfill
\includegraphics[scale=0.3]{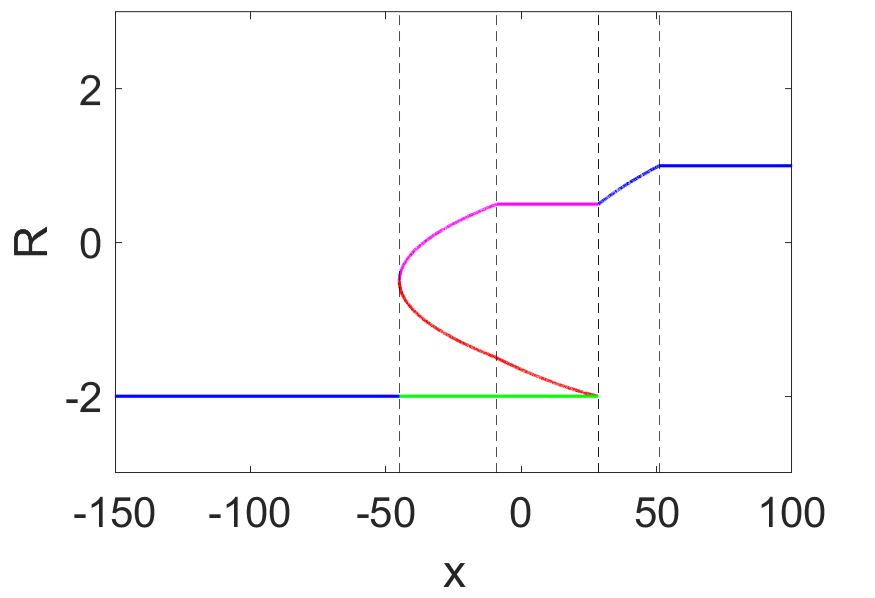}\hfill
\includegraphics[scale=0.3]{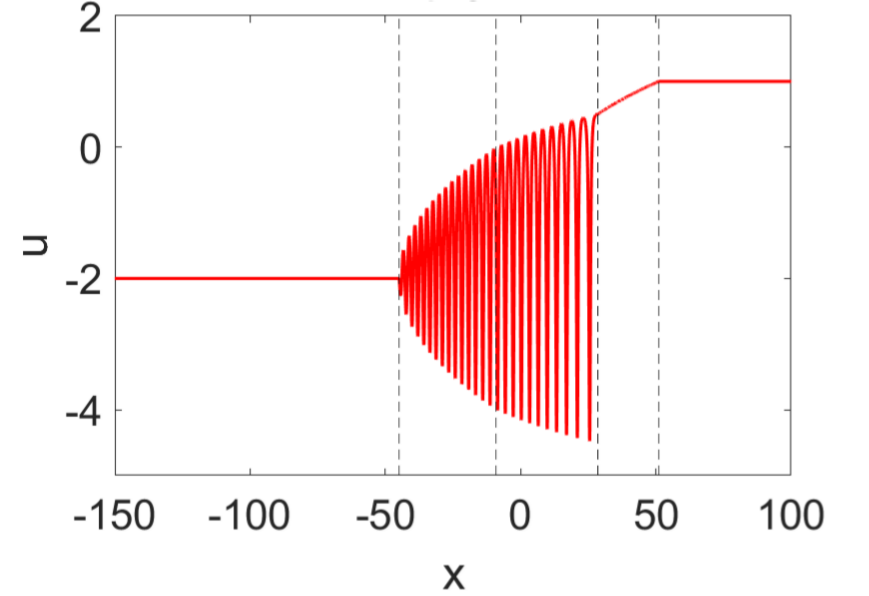}\hfill
\includegraphics[scale=0.3]{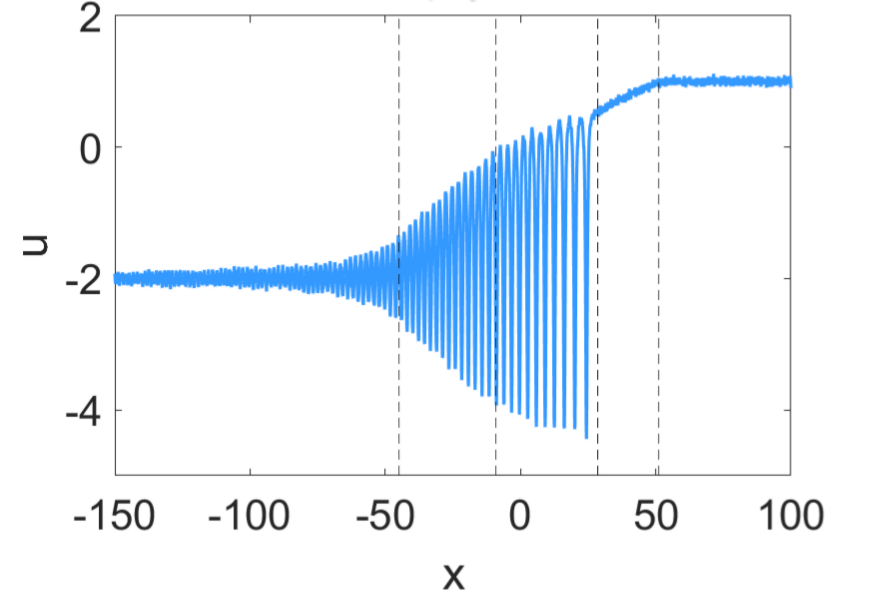}\hfill\\
\includegraphics[scale=0.3]{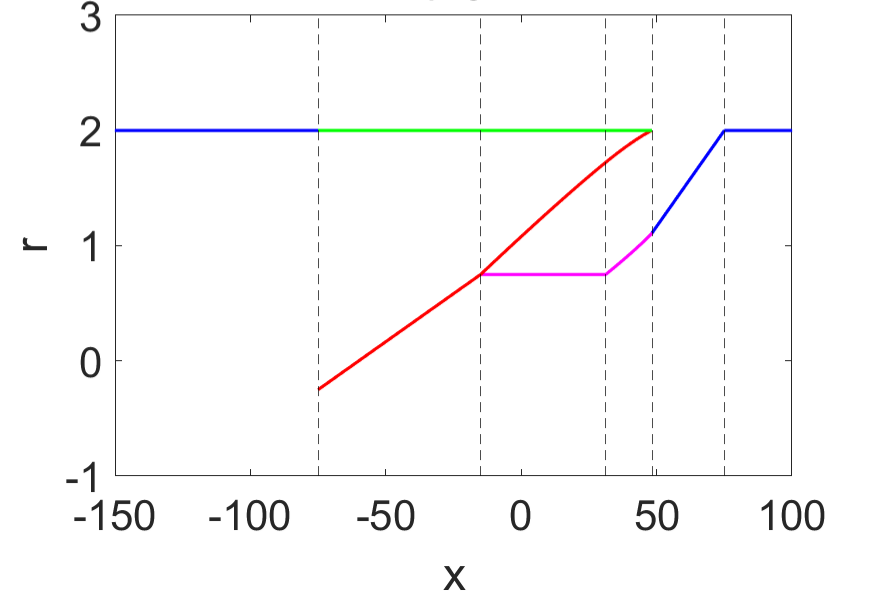}\hfill
\includegraphics[scale=0.3]{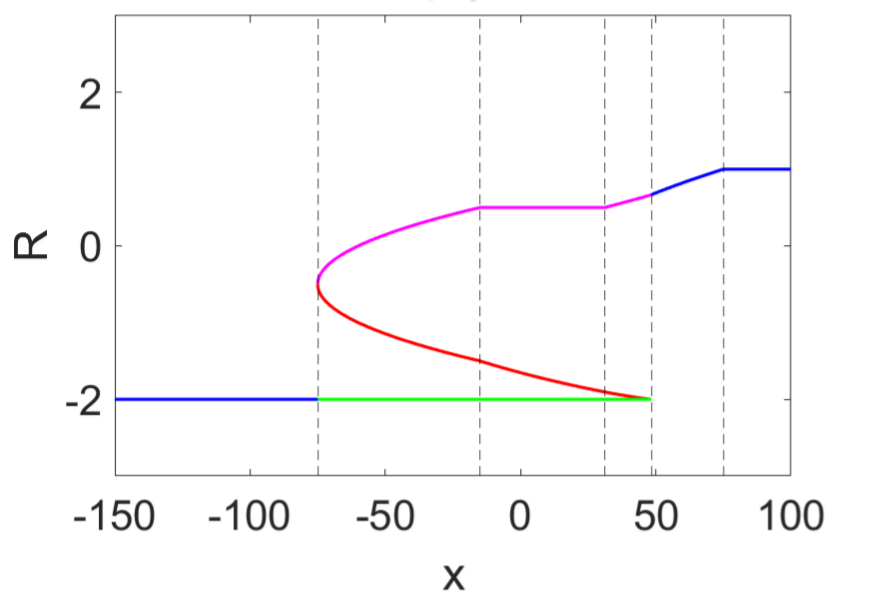}\hfill
\includegraphics[scale=0.3]{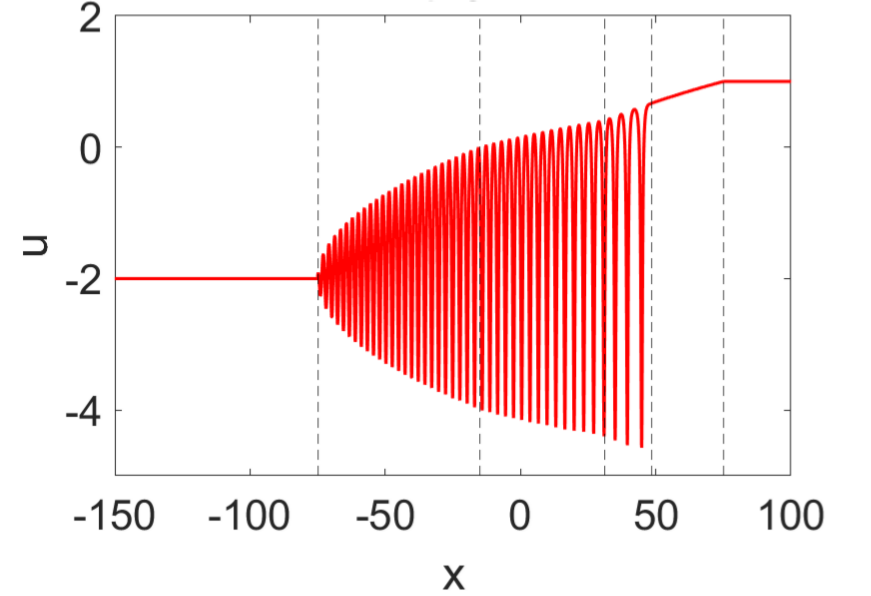}\hfill
\includegraphics[scale=0.3]{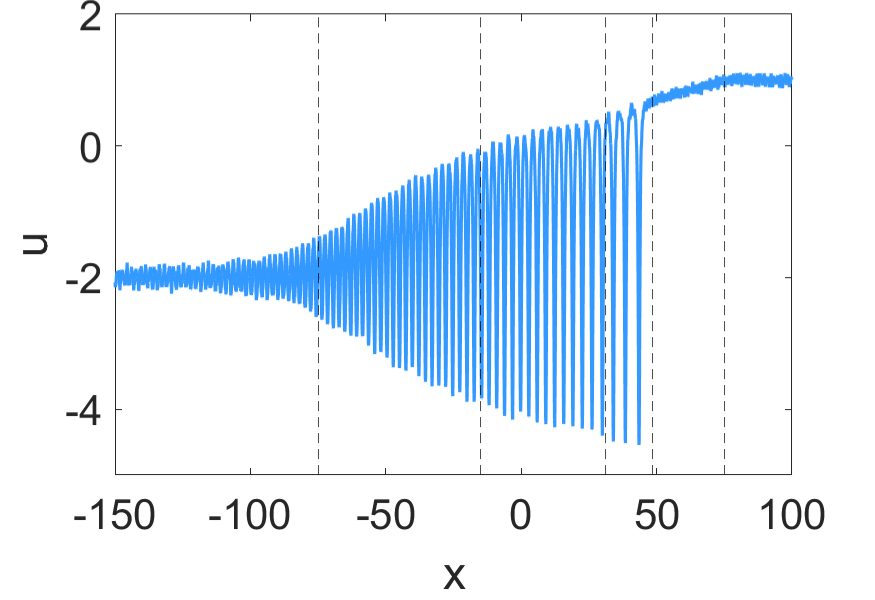}\hfill\\

{\footnotesize\hspace{0.0cm}(a)\hspace{4cm}(b)\hspace{4cm}(c)\hspace{4cm}(d)}
\caption{Evolution of DSW$|$CDSW-RW interaction with retained RW, at $t=2$, $t=t_1$, and $t=5$, respectively. Parameters are $u_l=-2$, $u_m=0.5$, $u_r=1$, $D=15$, $\alpha=-1$. (a) Riemann invariants $r$; (b) Combination of Riemann invariants $R$; (c) Analytical solutions of Eq.~(1) based on the Whitham modulation theory; (d) Numerical solutions of Eq.~(1).}
\label{Figs.~12.}
\end{figure*}
\begin{figure*}
\includegraphics[scale=0.3]{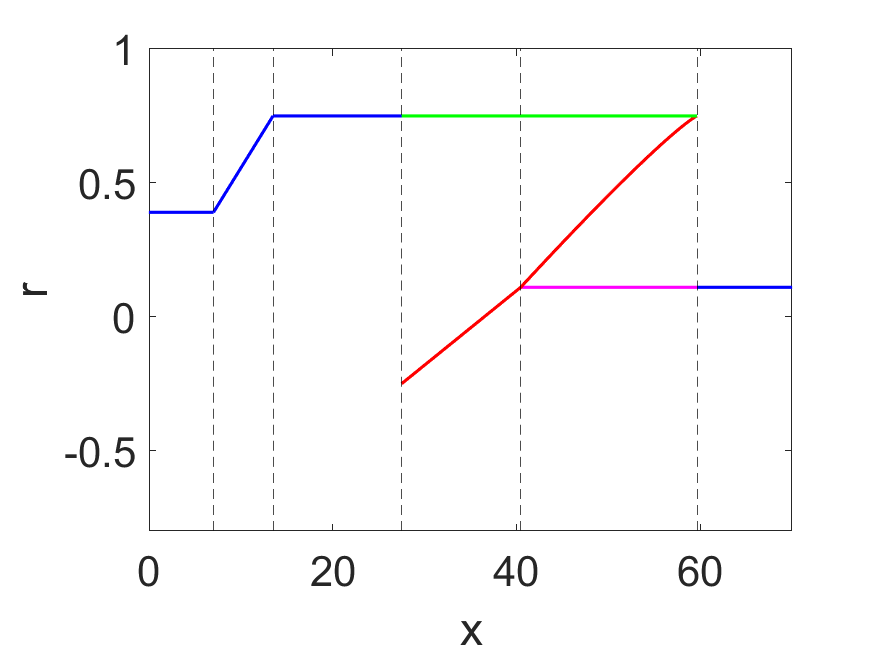}\hfill
\includegraphics[scale=0.3]{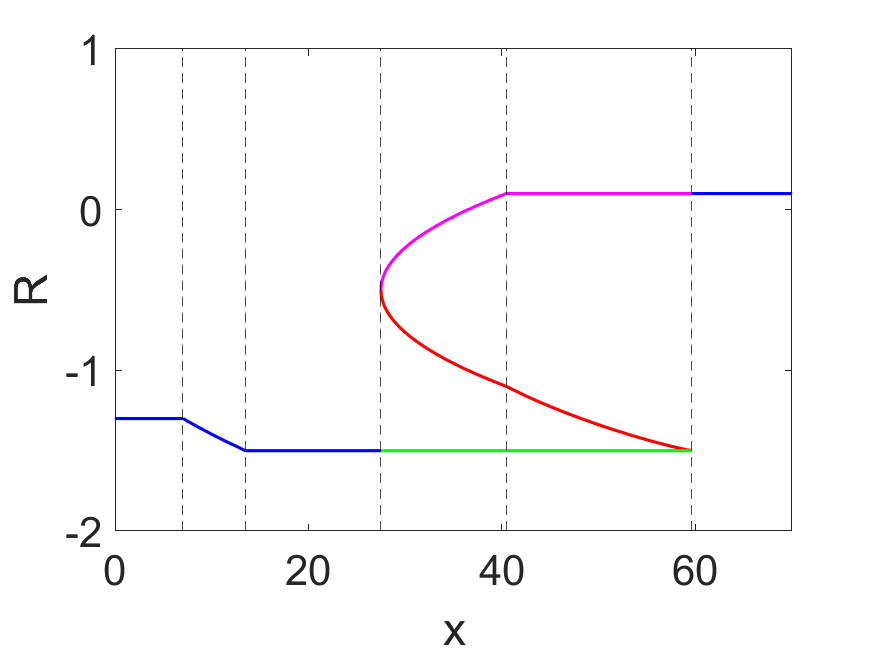}\hfill
\includegraphics[scale=0.3]{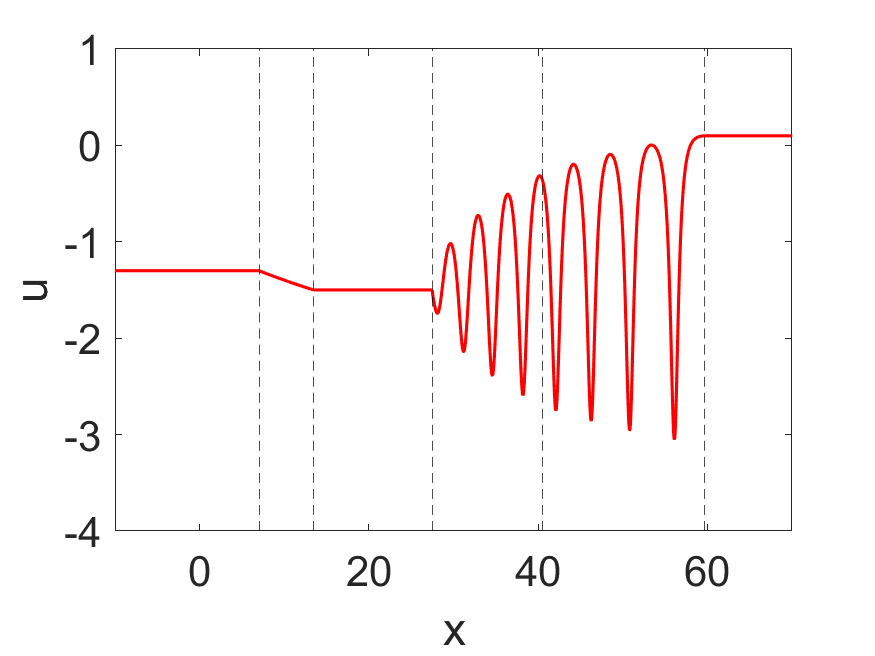}\hfill
\includegraphics[scale=0.3]{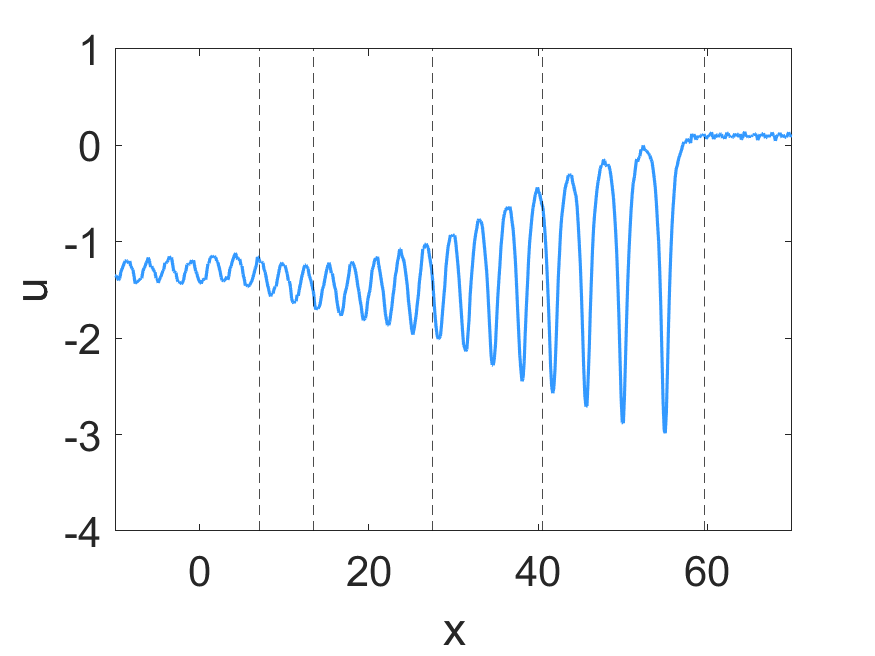}\hfill\\
\includegraphics[scale=0.3]{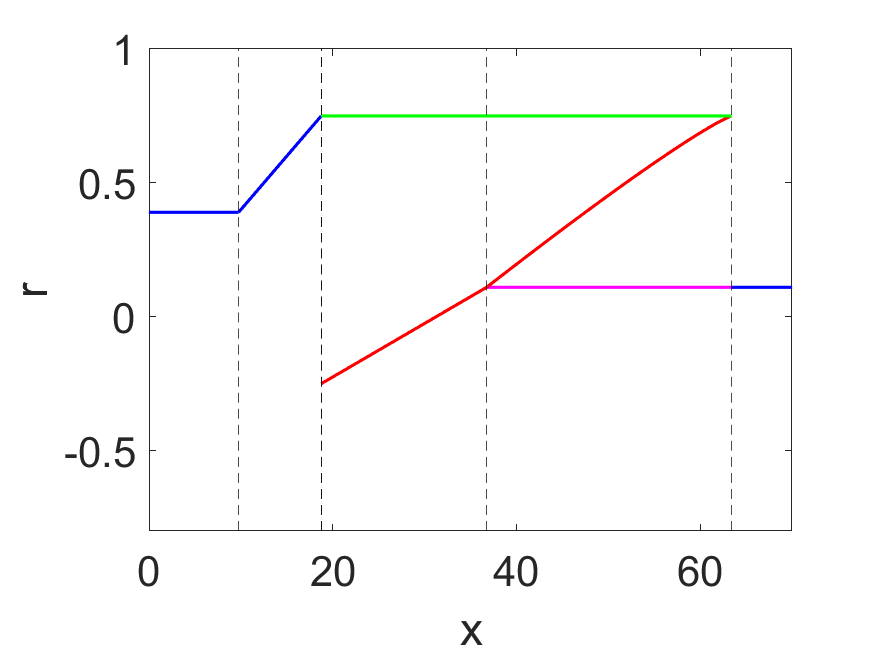}\hfill
\includegraphics[scale=0.3]{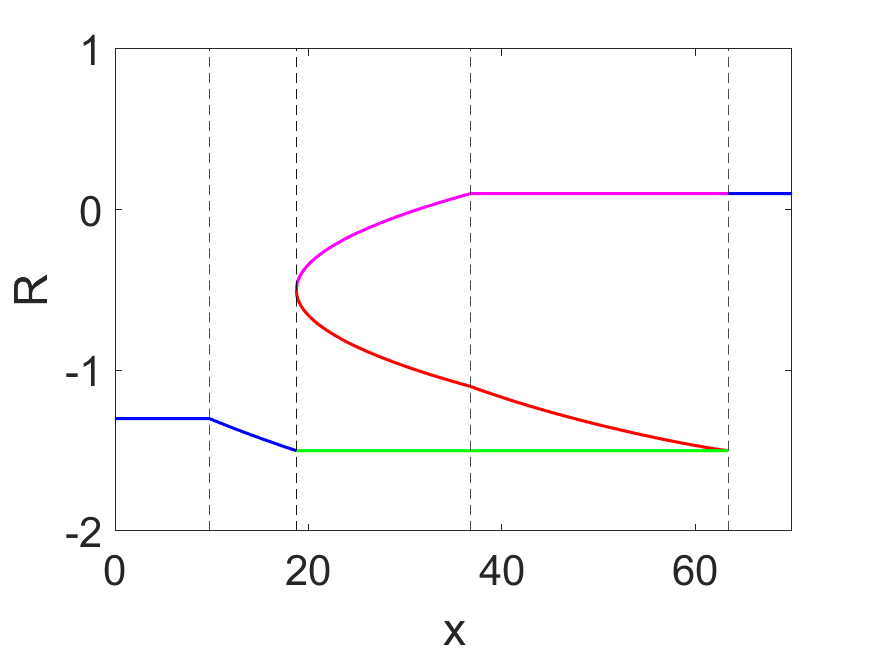}\hfill
\includegraphics[scale=0.3]{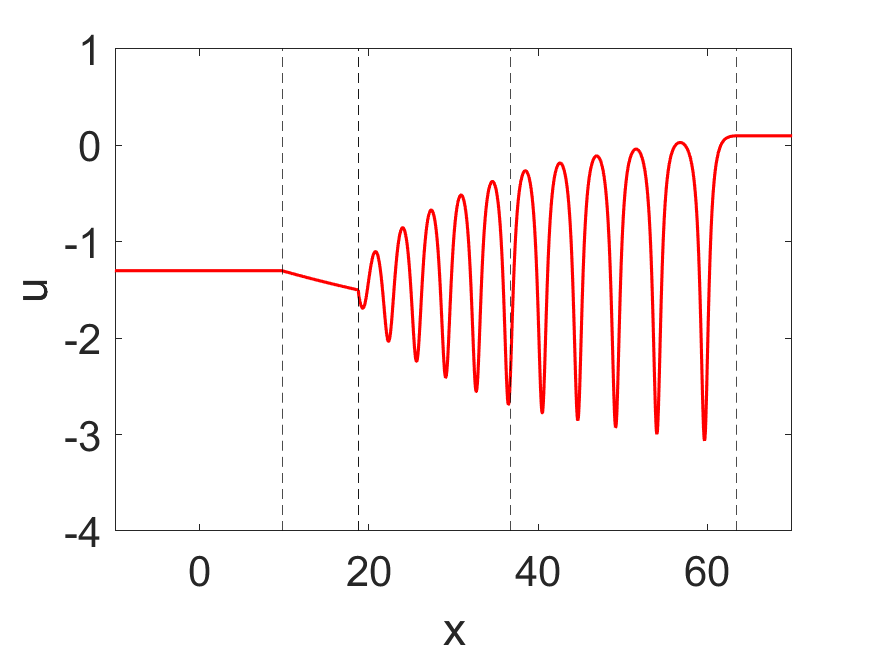}\hfill
\includegraphics[scale=0.3]{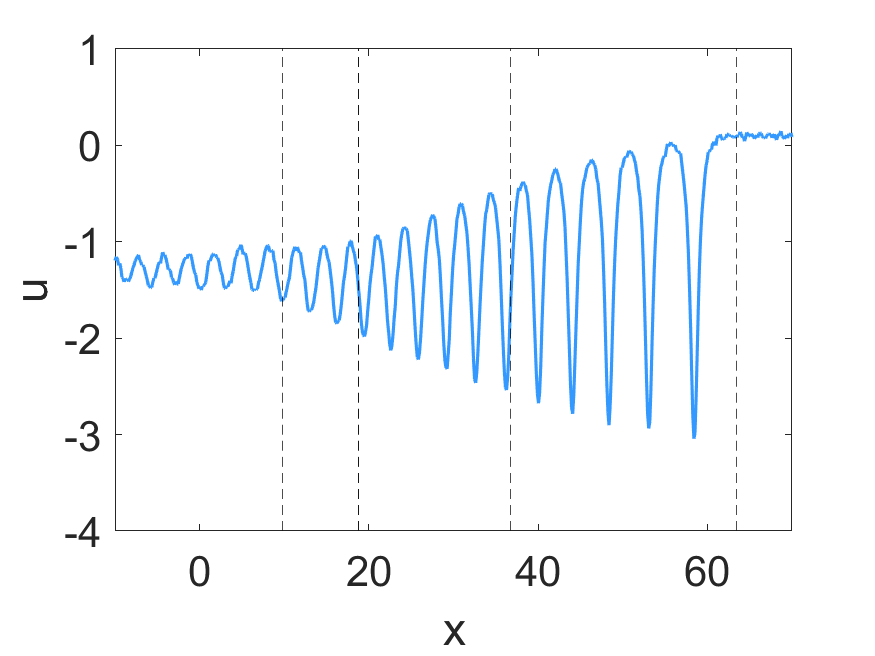}\hfill\\
\includegraphics[scale=0.3]{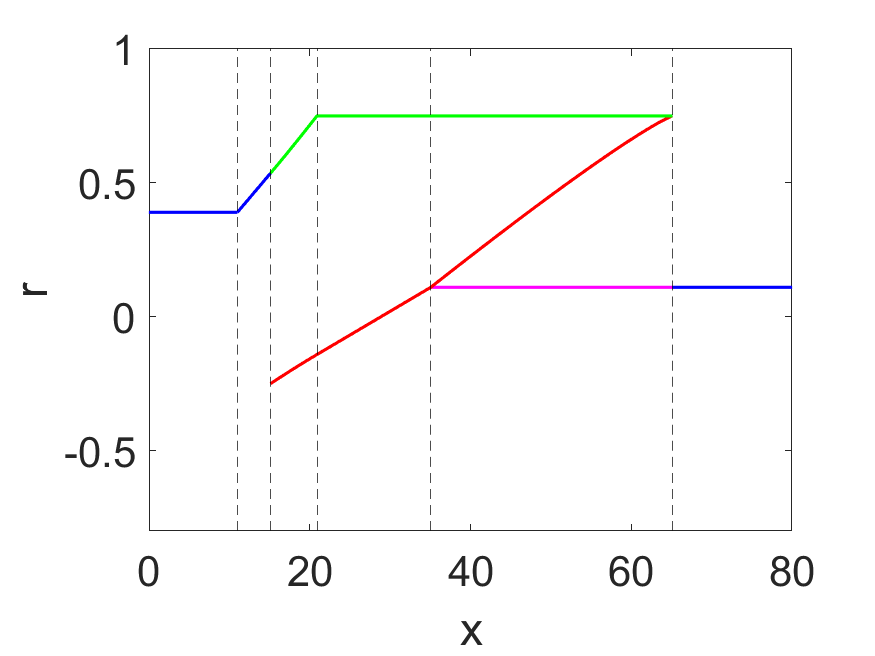}\hfill
\includegraphics[scale=0.3]{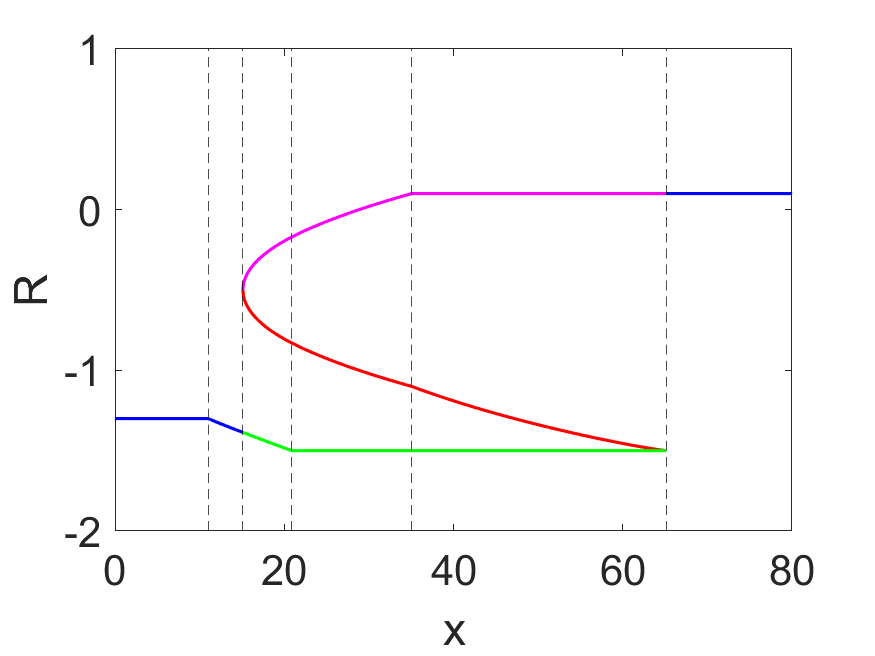}\hfill
\includegraphics[scale=0.3]{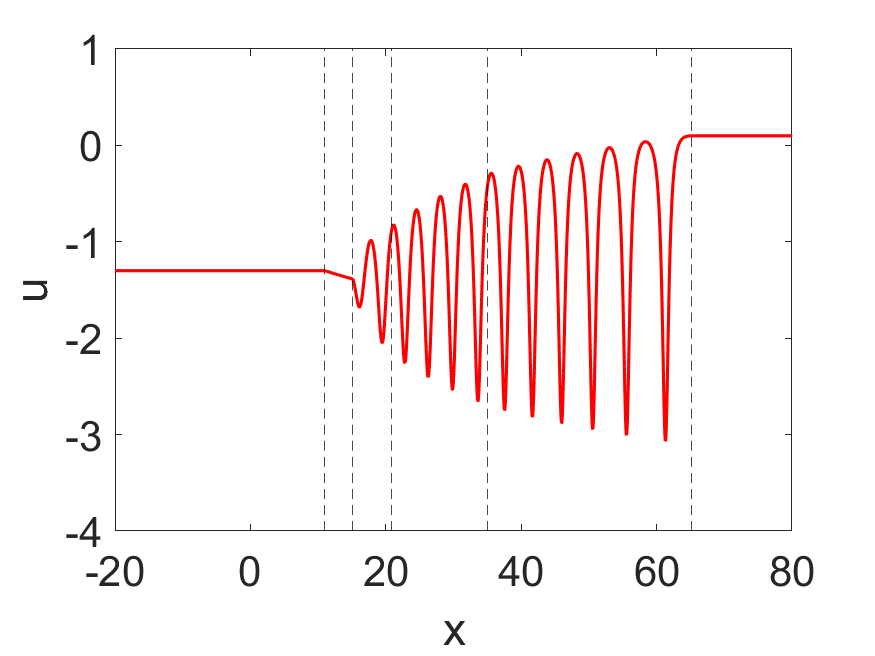}\hfill
\includegraphics[scale=0.3]{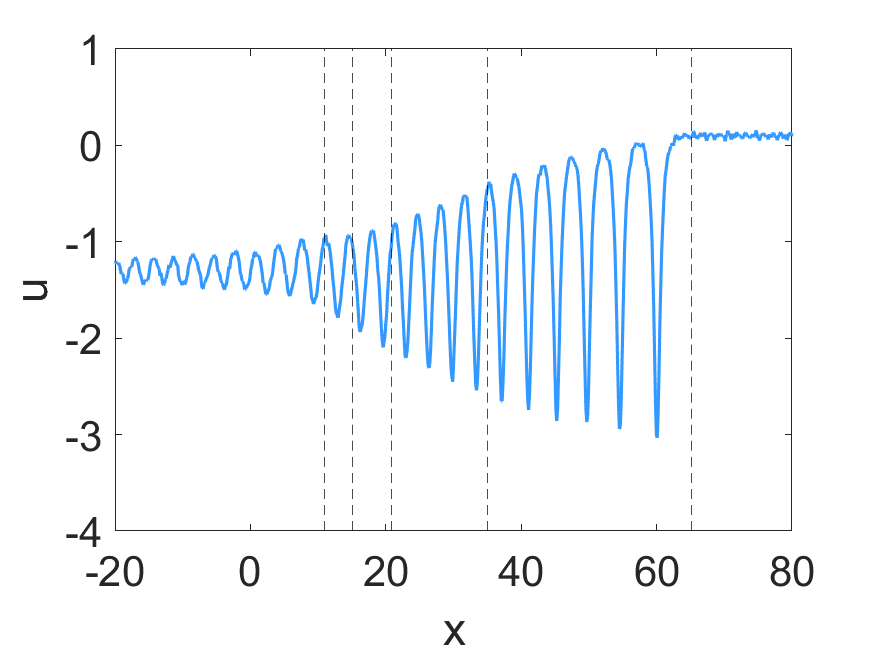}\hfill\\
\includegraphics[scale=0.3]{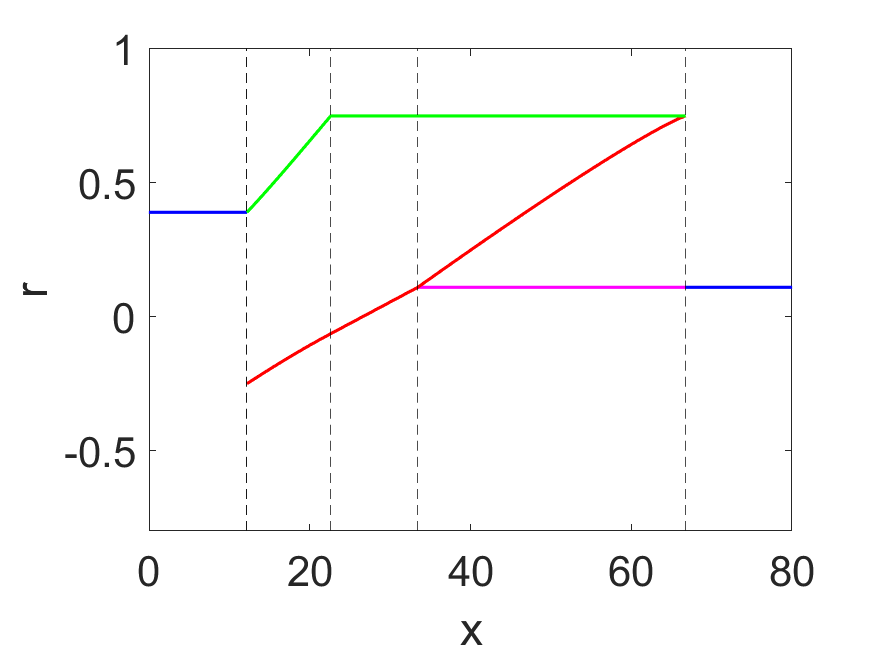}\hfill
\includegraphics[scale=0.3]{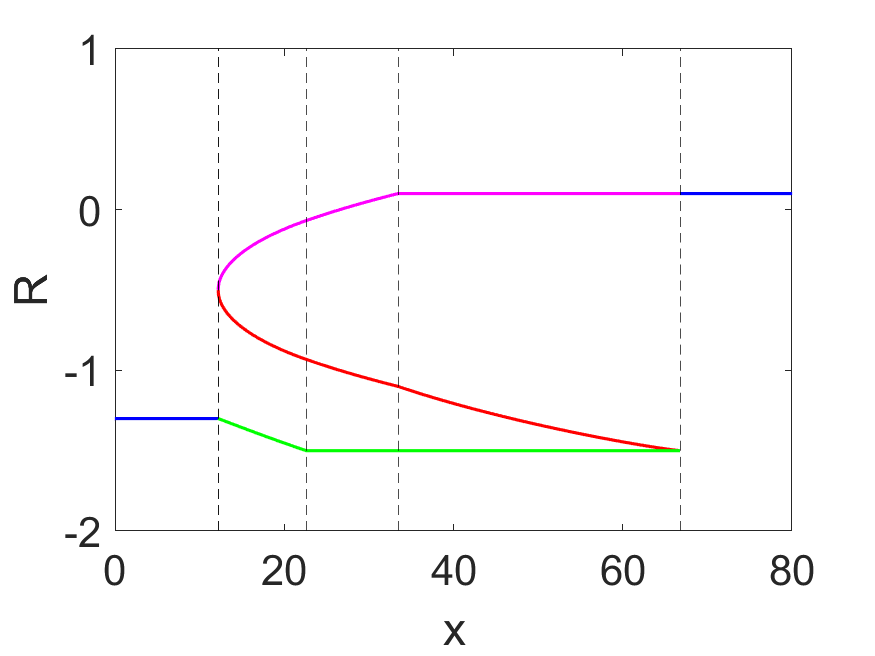}\hfill
\includegraphics[scale=0.3]{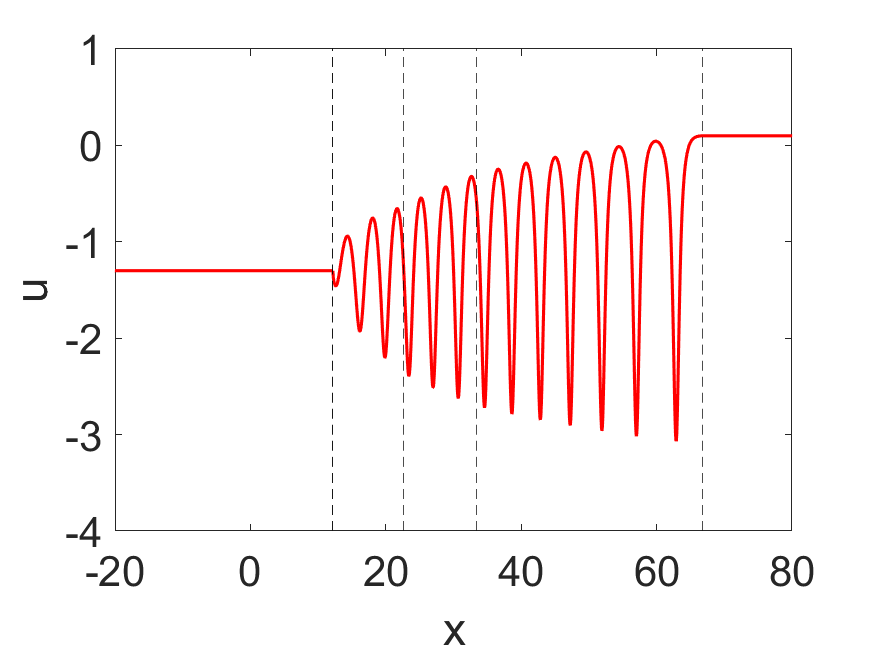}\hfill
\includegraphics[scale=0.3]{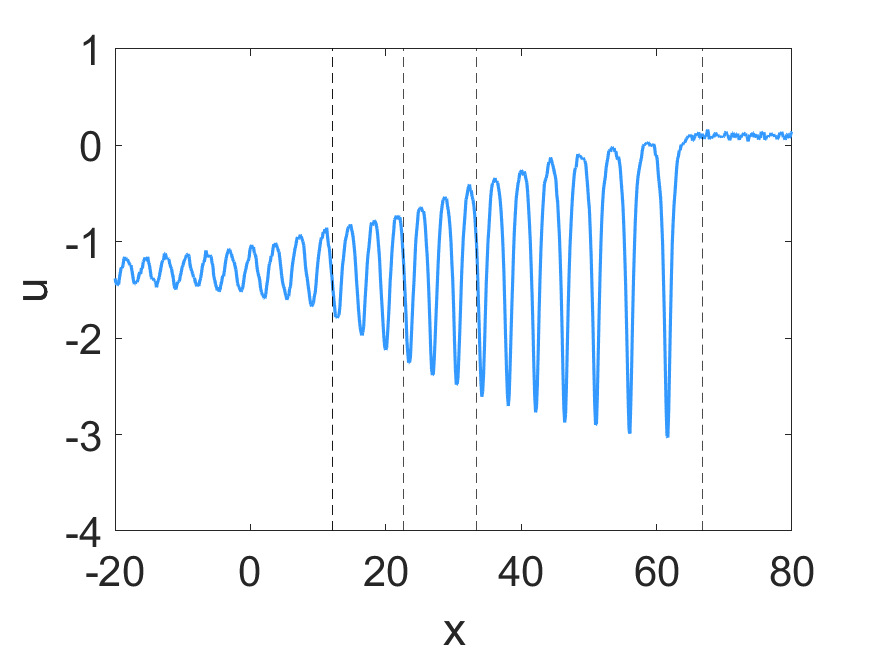}\hfill\\
\includegraphics[scale=0.3]{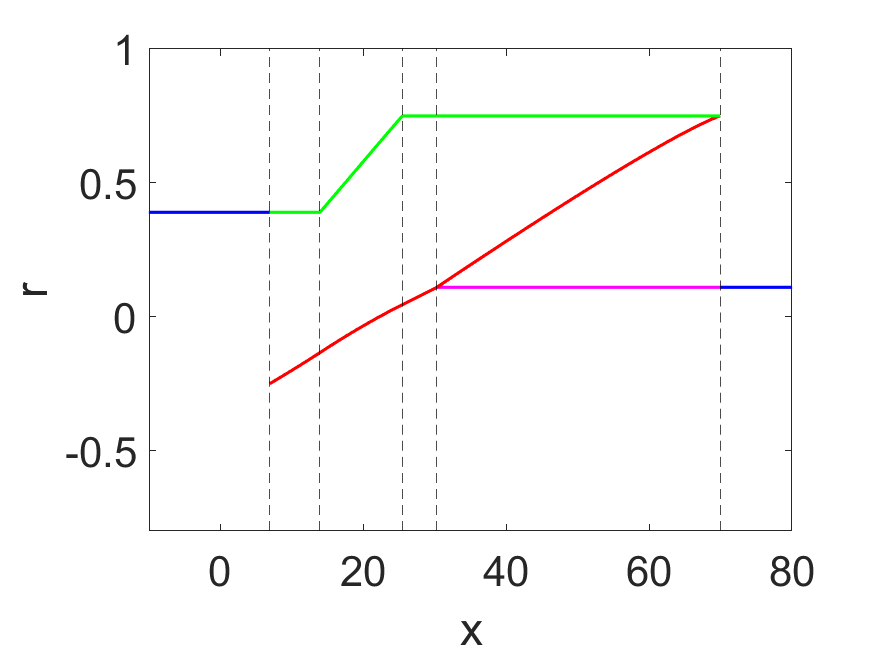}\hfill
\includegraphics[scale=0.3]{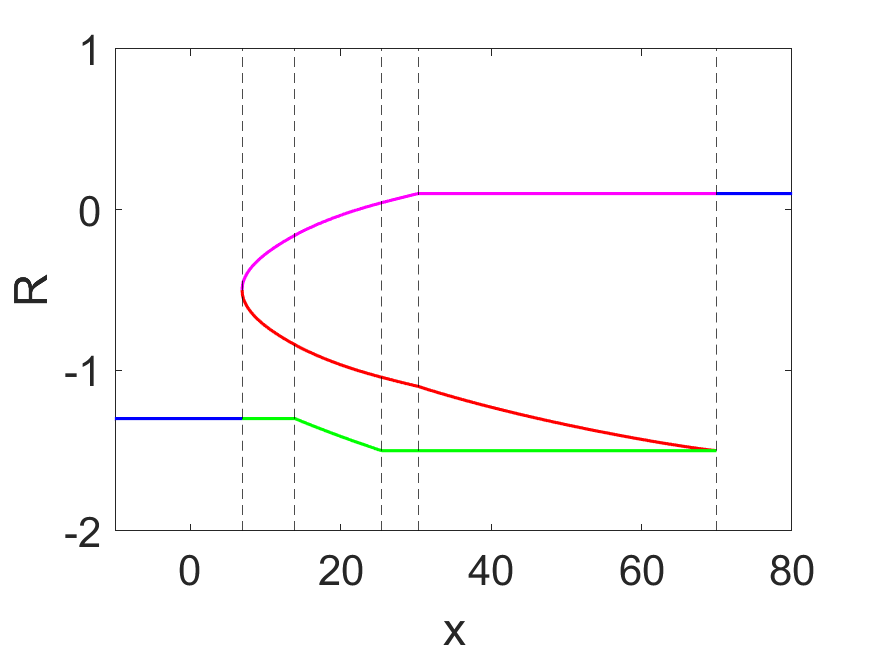}\hfill
\includegraphics[scale=0.3]{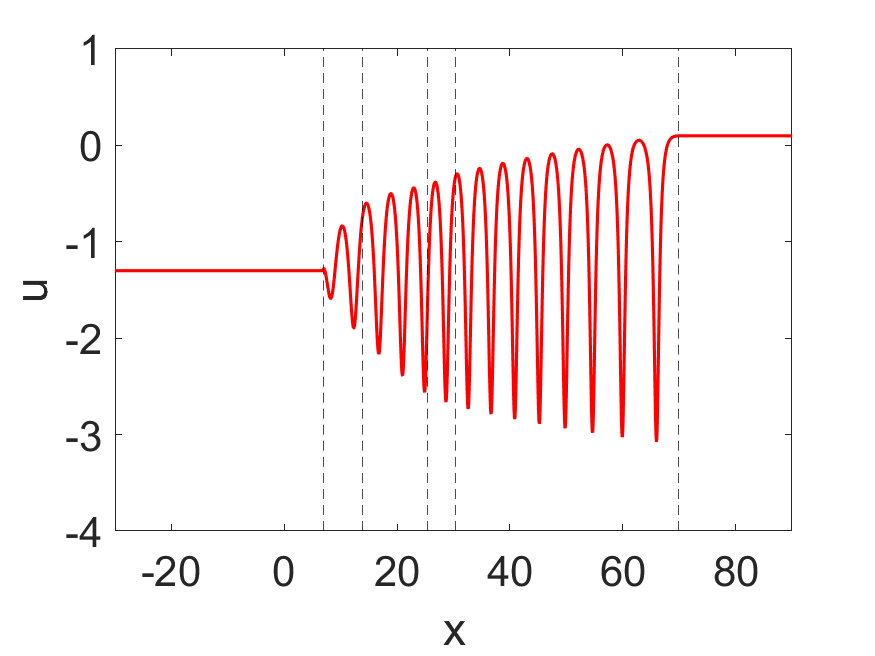}\hfill
\includegraphics[scale=0.3]{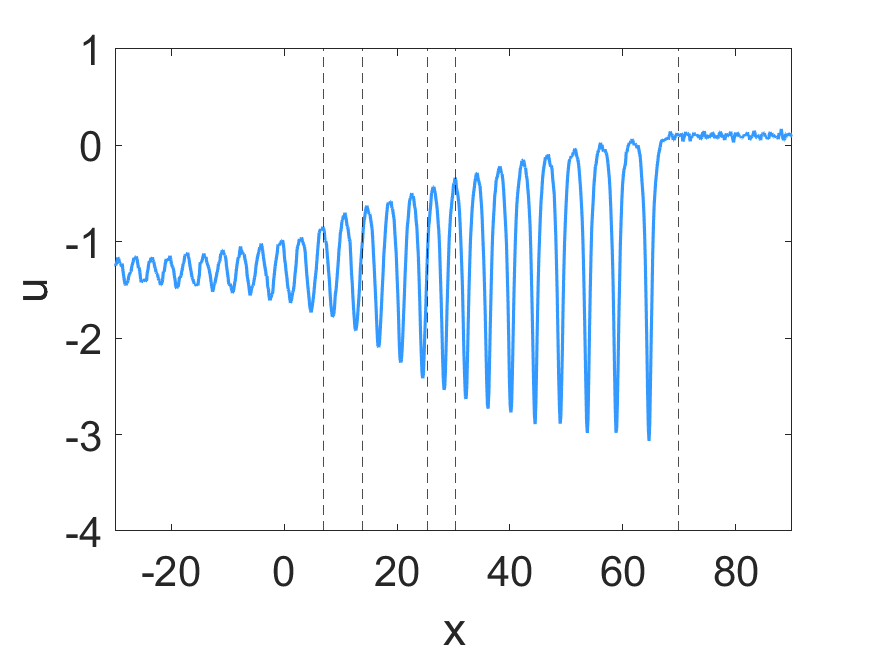}\hfill\\
\includegraphics[scale=0.3]{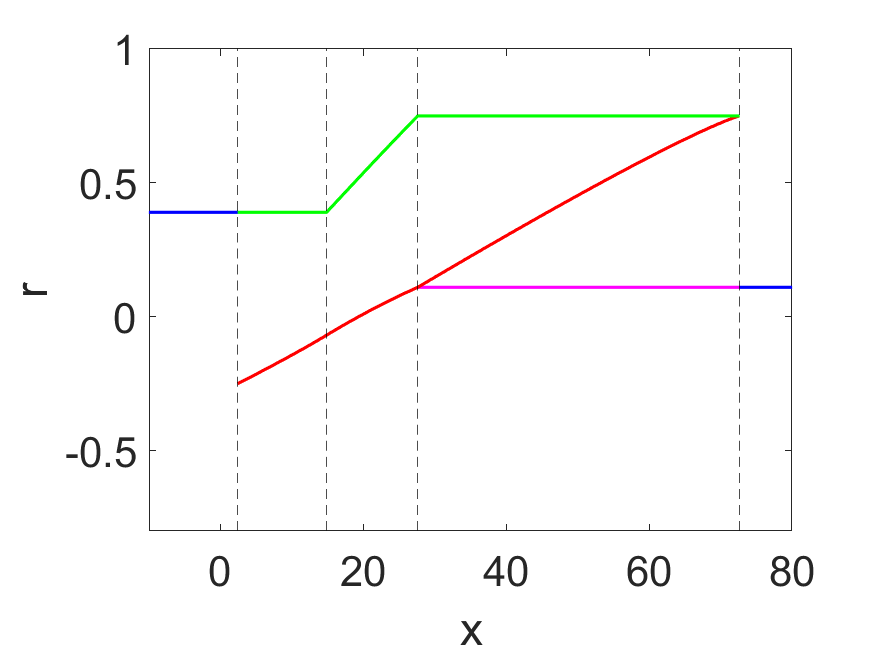}\hfill
\includegraphics[scale=0.3]{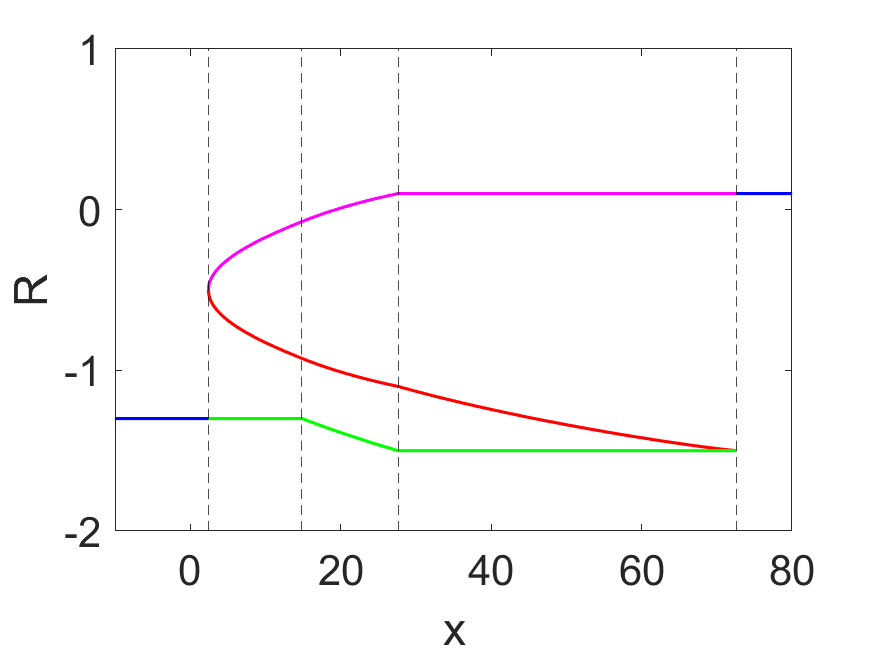}\hfill
\includegraphics[scale=0.3]{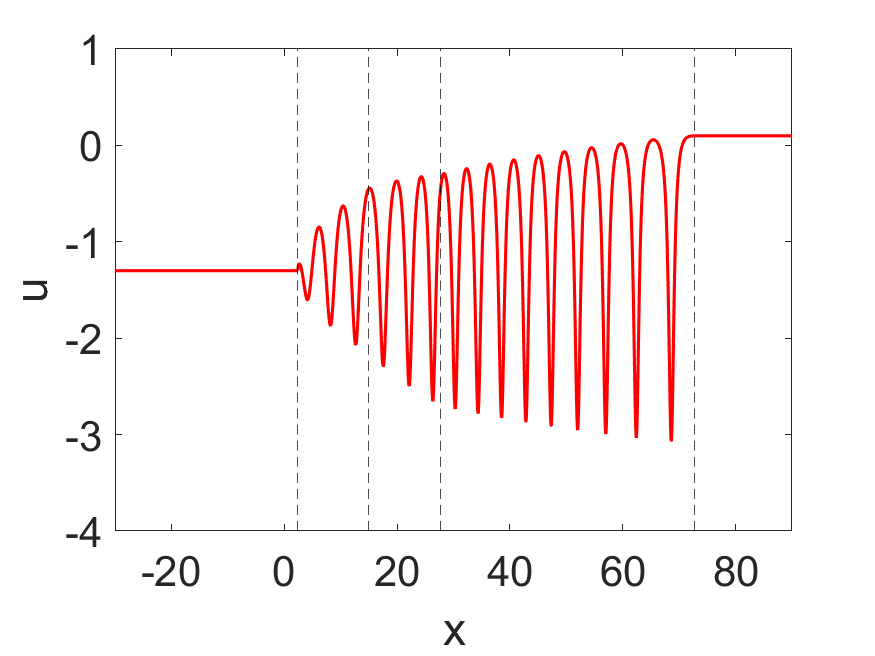}\hfill
\includegraphics[scale=0.3]{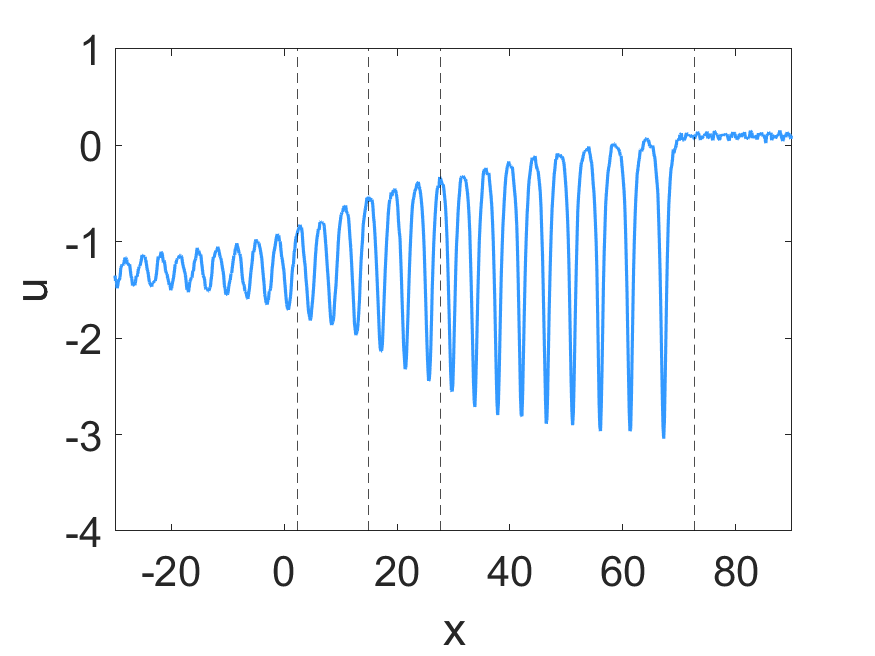}\hfill\\
\includegraphics[scale=0.3]{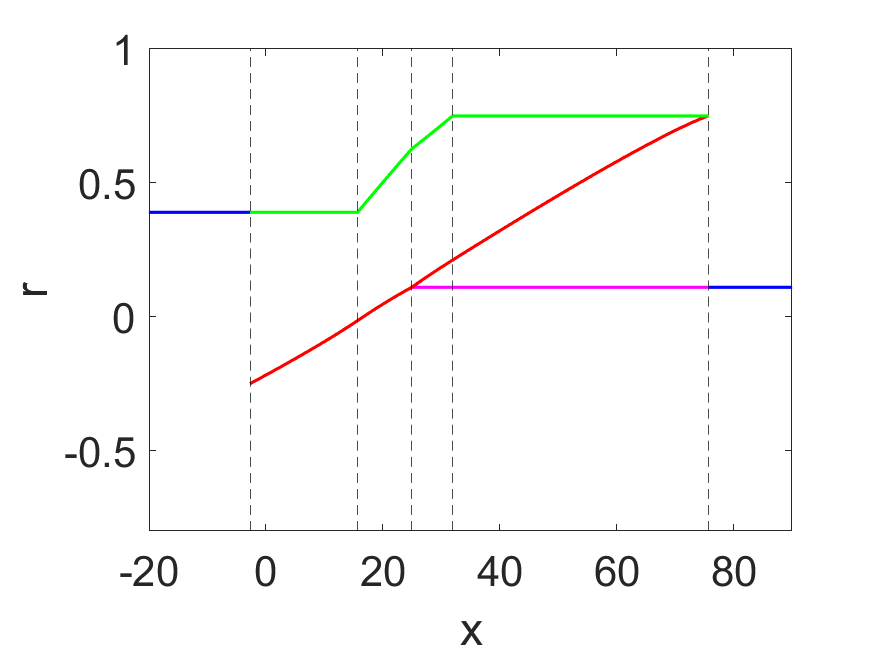}\hfill
\includegraphics[scale=0.3]{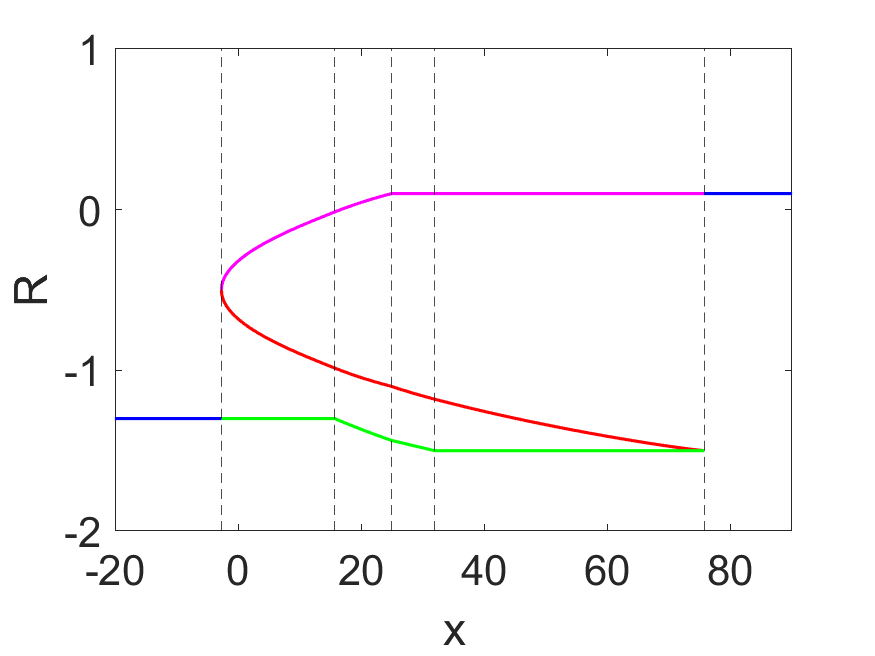}\hfill
\includegraphics[scale=0.3]{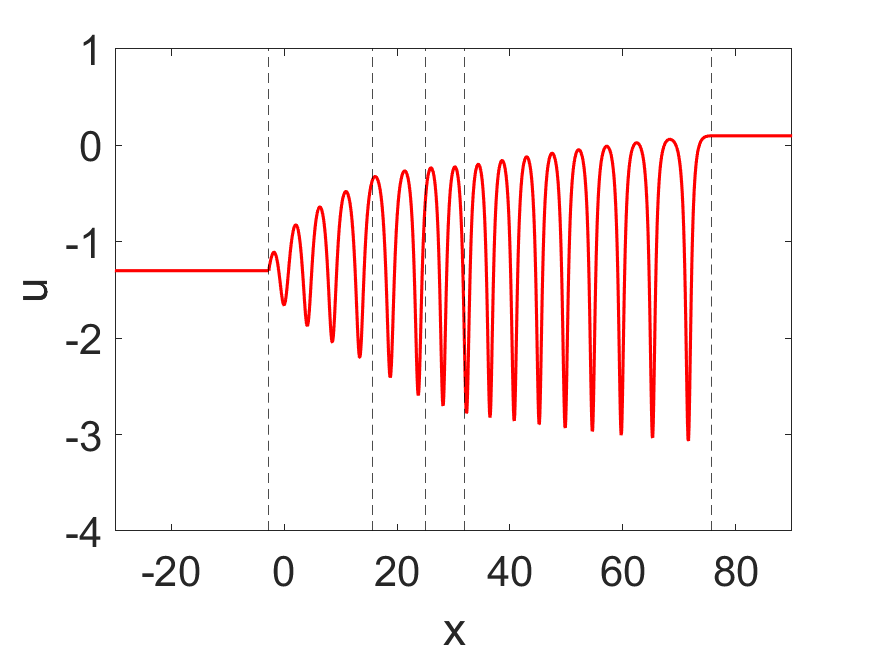}\hfill
\includegraphics[scale=0.3]{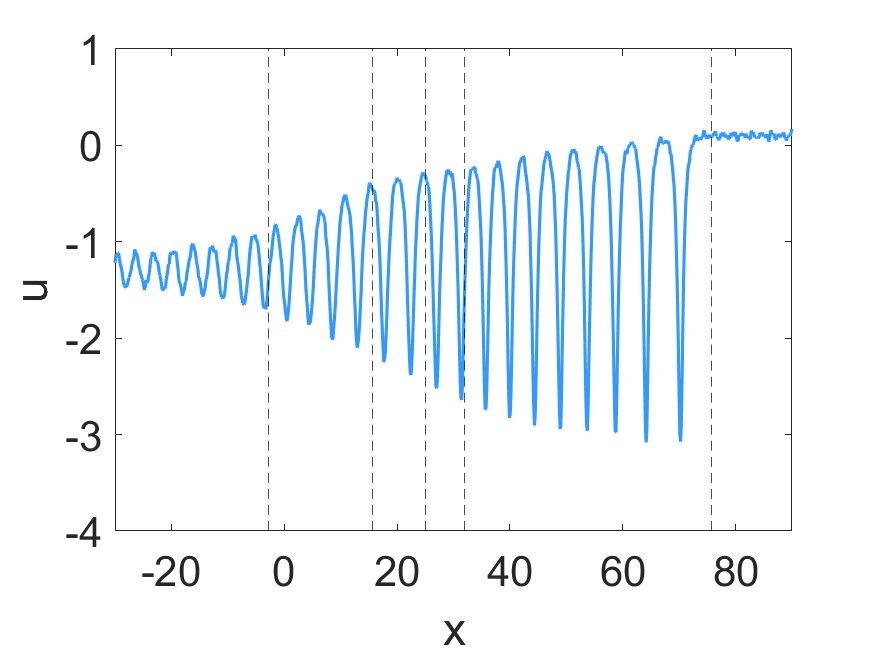}\hfill\\
\end{figure*}
\begin{figure*}
\includegraphics[scale=0.3]{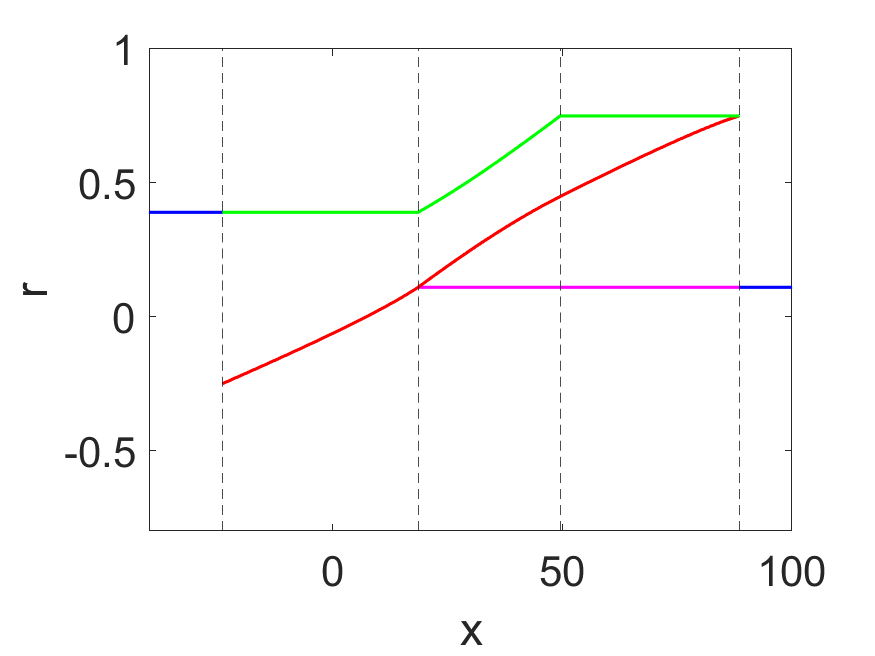}\hfill
\includegraphics[scale=0.3]{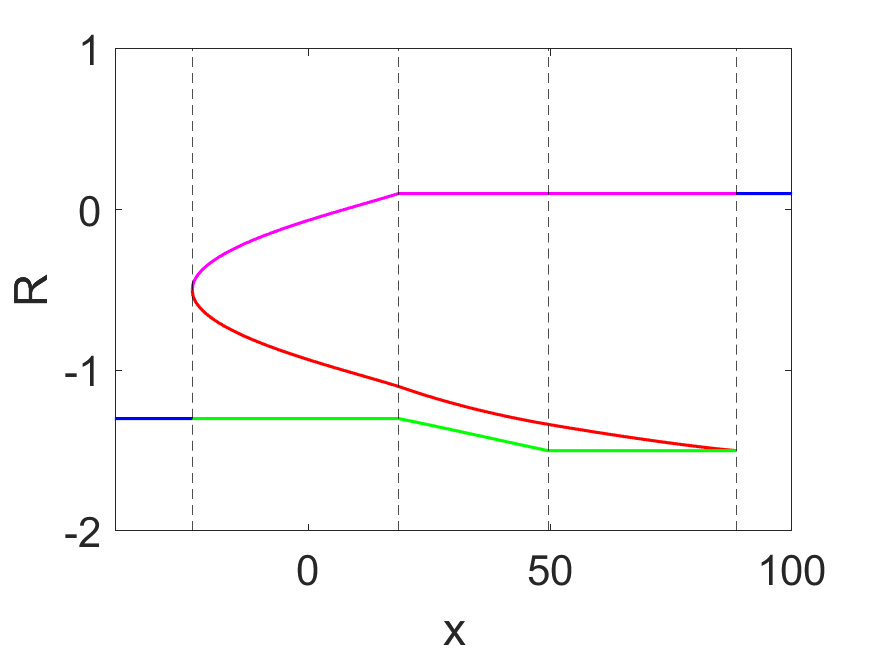}\hfill
\includegraphics[scale=0.3]{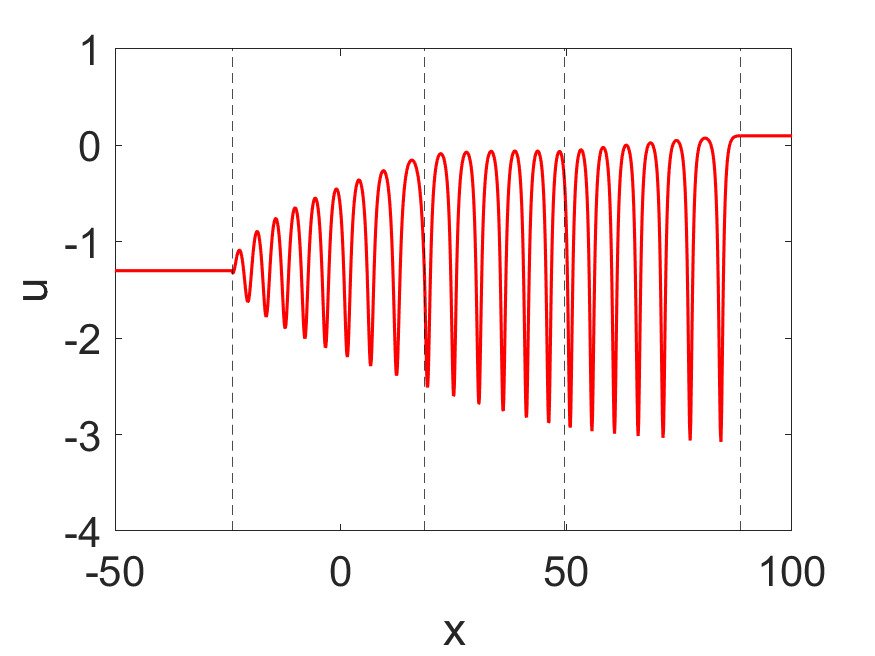}\hfill
\includegraphics[scale=0.3]{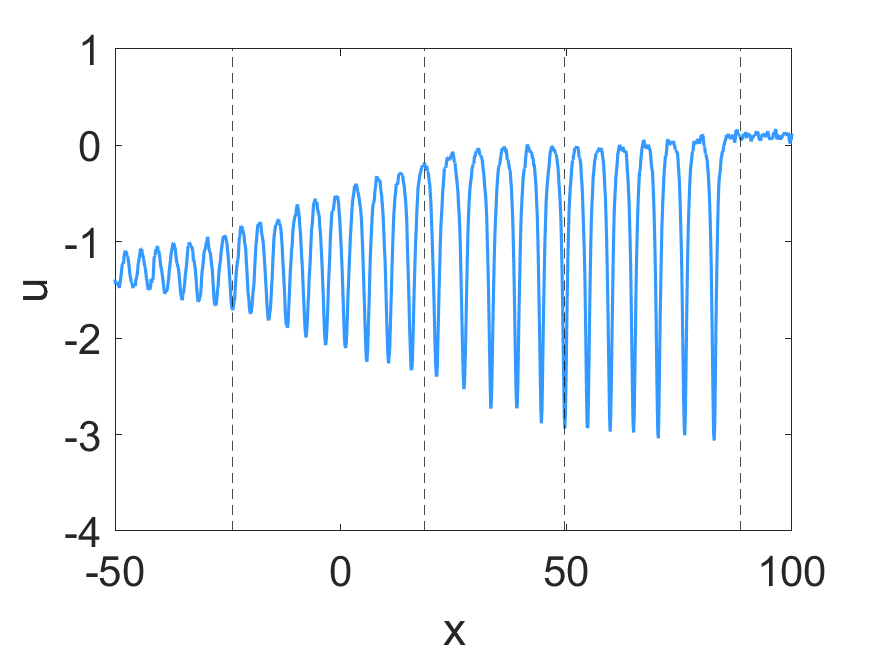}\hfill\\
\includegraphics[scale=0.3]{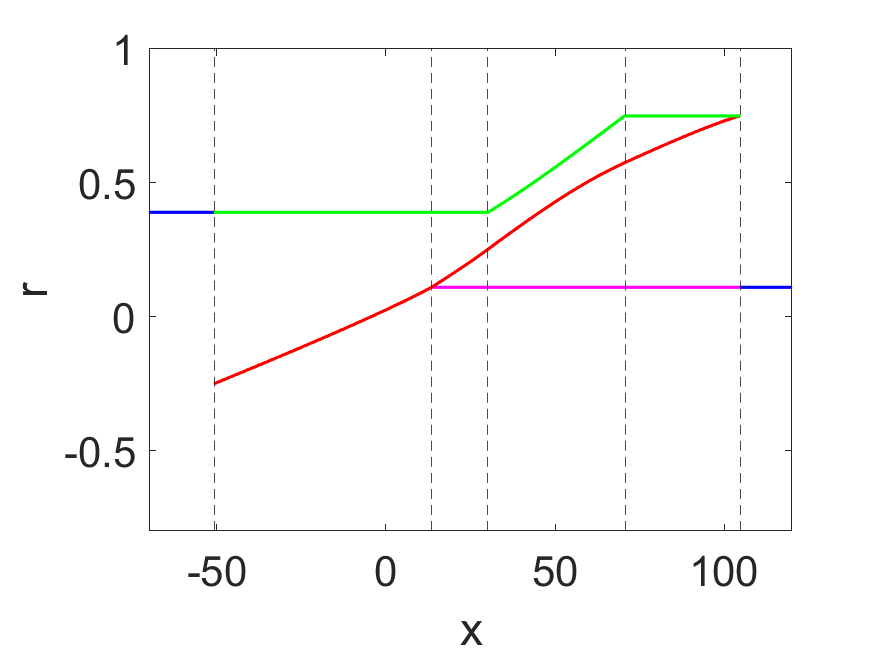}\hfill
\includegraphics[scale=0.3]{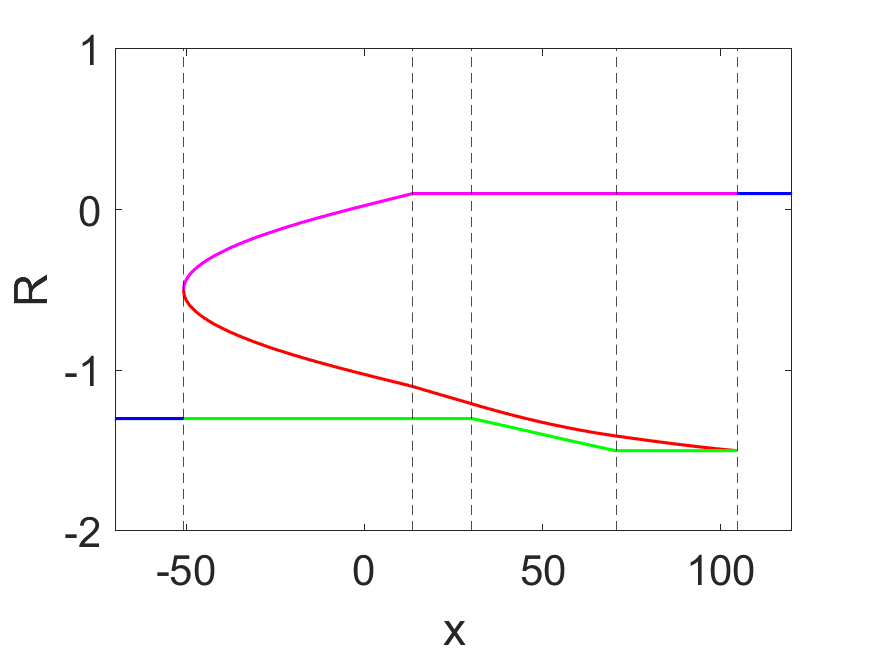}\hfill
\includegraphics[scale=0.3]{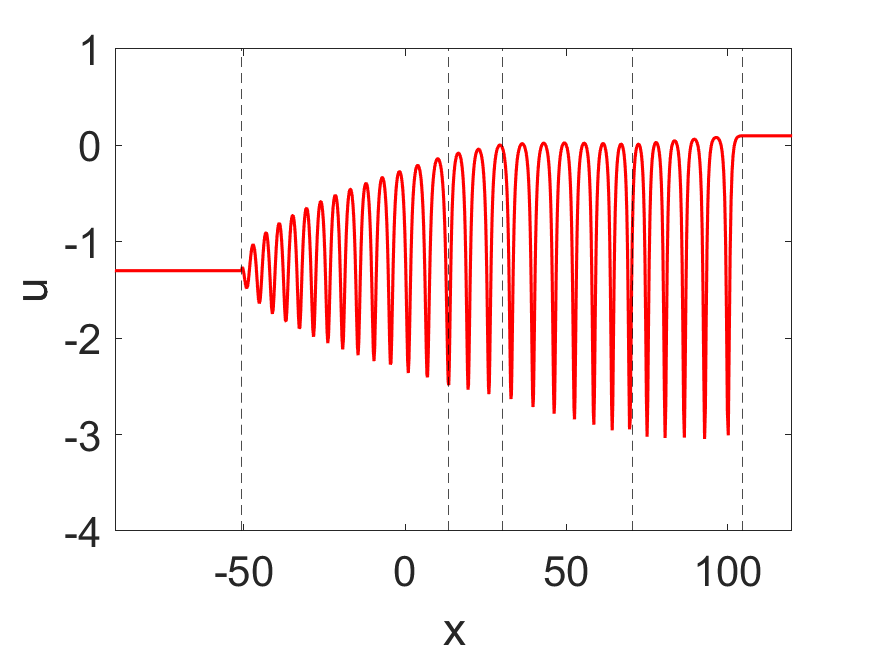}\hfill
\includegraphics[scale=0.3]{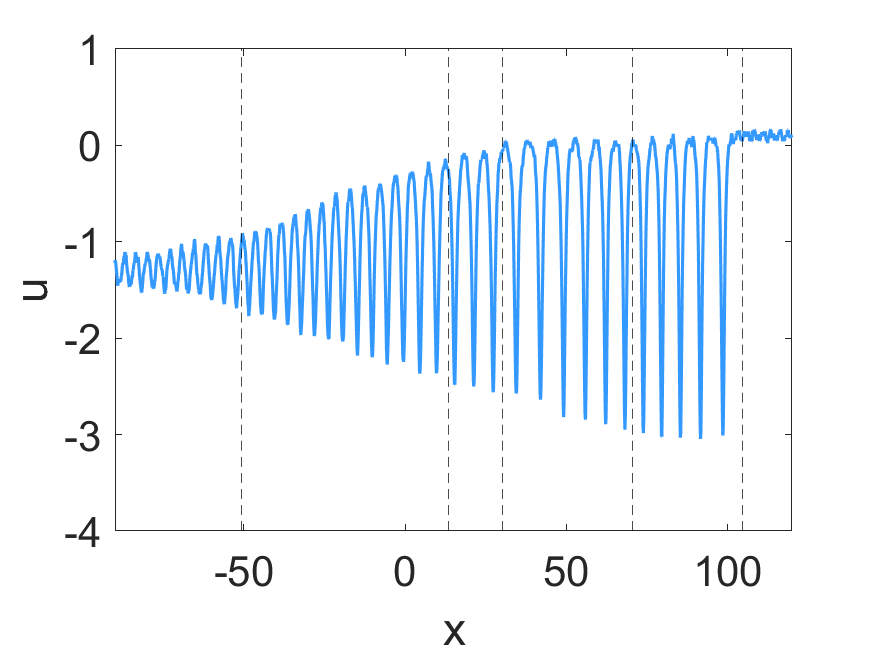}\hfill\\
{\footnotesize\hspace{0.0cm}(a)\hspace{4cm}(b)\hspace{4cm}(c)\hspace{4cm}(d)}
\caption{Evolution of RW-CDSW$|$DSW interaction at $t=3$, $t=t_1$, $t=4.7$, $t=t_2$, $t=6.2$, $t=t_3$, $t=8$, $t=t_4$ and $t=17$, respectively. Parameters are $u_l=-1.3$, $u_m=-1.5$, $u_r=0.1$, $D=50$, $\alpha=-1$. (a) Riemann invariants $r_i$; (b) Combination of Riemann invariants $R_i$; (c) Analytical solutions of Eq.~(1) based on the Whitham modulation theory; (d) Numerical solutions of Eq.~(1).}
\label{Figs.~13.}
\end{figure*}

Since the two interaction processes, namely, RW-normal CDSW and RW-reversed CDSW, are analogous, differing only in their opposite polarities, we will take the former as an example to discuss. For the initial data (3) such that $\frac{1}{\alpha}<u_l<u_m=\frac{1}{\alpha}-u_r$, there are a RW located on $x=0$ and a normal CDSW on $x=D$ in the early stage of the evolution, and then they intersect when $x_{\text{RW}}^+=x_{\text{CDSW}}^-$ at $t_1=\frac{D}{12\chi_m-\frac{3}{\alpha}}$. As time goes on, an interaction region occurs, in which all Riemann invariants are not constant, so we still need the hodograph transform to solve the more general Gardner--Whitham modulation equations.

At the left boundary of the interaction region,
\begin{equation}
R_1=R_2=\frac{1}{2\alpha},\ \ \ \
r_1=r_2=\frac{1}{4\alpha},
\end{equation}
while at the right boundary of the interaction region,
\begin{align}
R_1+R_2=\frac{1}{\alpha},\ \ \
	&R_3=u_m,\notag\\
r_1=r_2,\ \ \ \ &r_3=\chi_m.
\end{align}
Due to all Riemann invariants vary in the interaction region, we must consider the following equations
\begin{align}
x-v_1(r_1,r_2,r_3)t&=\tilde{w}_1(r_1,r_2,r_3),\notag\\
x-v_2(r_1,r_2,r_3)t&=\tilde{w}_2(r_1,r_2,r_3),\\
x-v_3(r_1,r_2,r_3)t&=\tilde{w}_3(r_1, r_2,r_3),\notag
\end{align}
on the hodograph plane. Eqs.~(75)-(76) can be transformed to boundary conditions for phase $\tilde{g}(r_1, r_2,r_3)$. On the left edge, the interaction region is connected to the RW, so
\begin{align}
\tilde{w}_3&(\frac{1}{4\alpha},\frac{1}{4\alpha},r_3)=\notag\\
&\tilde{g}(\frac{1}{4\alpha},\frac{1}{4\alpha},r_3)-\frac{L(\frac{1}{4\alpha},\frac{1}{4\alpha},r_3)}{\partial_3L(\frac{1}{4\alpha},\frac{1}{4\alpha},r_3)}\partial_3\tilde{g}(\frac{1}{4\alpha},\frac{1}{4\alpha},r_3)=0,
\end{align}
integrating the above equation yields
\begin{equation}
\tilde{g}(\frac{1}{4\alpha},\frac{1}{4\alpha},r_3)=\frac{\tilde{c}_1}{\sqrt{1-4\alpha r_3}}.
\end{equation}
On the right edge, the interaction region is connected to the CDSW, so
\begin{align}
\tilde{w}_1&(r_1,r_2,\chi_m)=\tilde{g}(r_1,r_2,\chi_m)\notag\\
&-\frac{L(r_1,r_2,\chi_m)}{\partial_1L(r_1,r_2,\chi_m)}\partial_1\tilde{g}(r_1,r_2,\chi_m)=D,\notag\\
\tilde{w}_2&(r_1,r_2,\chi_m)=\tilde{g}(r_1,r_2,\chi_m)\notag\\
&-\frac{L(r_1,r_2,\chi_m)}{\partial_2L(r_1,r_2,\chi_m)}\partial_2\tilde{g}(r_1,r_2,\chi_m)=D,
\end{align}
integrating the above equation yields
\begin{equation}
\tilde{g}(r_1,r_2,\chi_m)=D+\tilde{c}_2L(r_1,r_2,\chi_m).
\end{equation}
For simplicity, we set $\tilde{c}_2=0$ in subsequent discussions. The general solution for $\tilde{g}(r_1,r_2,r_3)$ can be setting as
\begin{align}
\tilde{g}(r_1,r_2,r_3)=&\int_{\frac{1}{4\alpha}}^{r_1}\frac{\tilde{\phi}_1(r)}{\sqrt{(r_1-r)(r_2-r)(r_3-r)}}dr\notag\\
&+\int_{\frac{1}{4\alpha}}^{r_2}\frac{\tilde{\phi}_2(r)}{\sqrt{(r_1-r)(r_2-r)(r_3-r)}}dr\notag\\
&+\int_{\chi_m}^{r_3}\frac{\tilde{\phi}_3(r)}{\sqrt{(r-r_1)(r-r_2)(r-r_3)}}dr.
\end{align}
We assume that $\tilde{\phi}_1(r)=0$ and the second integral converges to zero as $r_2\rightarrow \frac{1}{4\alpha}$ (to be justified). Then Eq.~(79) can be rewritten as
\begin{equation}
\tilde{g}(\frac{1}{4\alpha},\frac{1}{4\alpha},r_3)=\int_{\chi_m}^{r_3}\frac{\tilde{\phi}_3(r)}{\sqrt{r-r_3}(r-\frac{1}{4\alpha})}dr=\frac{\tilde{c}_1}{\sqrt{1-4\alpha r_3}}.
\end{equation}
By means of the Abel transform,
\begin{align}
\tilde{\phi}_3(r_3)&=-i\frac{(r_3-\frac{1}{4\alpha})}{\pi}\frac{d}{dr_3}\int_{\chi_m}^{r_3}\frac{\tilde{c}_1}{\sqrt{1-4\alpha r}\sqrt{r-r_3}}dr\notag\\
&=\frac{c_1}{4\alpha\pi}\frac{\sqrt{1-4\alpha\chi_m}}{\sqrt{\chi_2-r_3}}.
\end{align}
And then
\begin{equation}
\int_{\chi_m}^{r_3}\frac{\tilde{\phi}_3(r)}{\sqrt{(r-r_1)(r-r_2)(r-r_3)}}dr=-\frac{c_1\sqrt{1-4\alpha \chi_m}}{2\pi\alpha}\text{K}(\tilde{k}),
\end{equation}
with $\tilde{k}=\frac{(r_2-r_1)(\chi_m-r_3)}{(r_3-r_1)(\chi_m-r_2)}$. Considering the above equation tends to $-\frac{\tilde{c}_1\sqrt{1-4\alpha \chi_m}}{4\alpha\sqrt{\chi_m-r_1}\sqrt{\chi_m-r_2}}$ as $r_3\rightarrow \chi_m$, the boundary condition (81) can be transformed to
\begin{align}
\int_{\frac{1}{4\alpha}}^{r_2}&\frac{\tilde{\phi}_2(r)}{\sqrt{(r_1-r)(r_2-r)(\chi_m-r)}}dr\notag\\
&=-\frac{\tilde{c}_1\sqrt{1-4\alpha \chi_m}}{4\alpha\sqrt{\chi_m-r_1}\sqrt{\chi_m-r_2}}+D.
\end{align}
By the Abel transform,
\begin{align}
\tilde{\phi}_2(r_2)=&\frac{\sqrt{(r_1-r_2)(\chi_m-r_2)}}{\pi}\Bigg(\frac{D}{\sqrt{r_2-\frac{1}{4\alpha} }}\Bigg.\notag\\
&\Bigg.+\frac{\tilde{c}_1(1-4\alpha \chi_m)}{4\alpha(\chi_m-r_2)\sqrt{1-4\alpha r_2}\sqrt{\chi_m-r_1}} \Bigg),
\end{align}
and then, the second integral in Eq.~(82) can be expressed in terms of the complete elliptic integrals
\begin{align}
&\int_{\frac{1}{4\alpha}}^{r_2}\frac{\tilde{\phi}_2(r)}{\sqrt{(r_1-r)(r_2-r)(r_3-r)}}dr=\notag\\
&\frac{\chi_m-\frac{1}{4\alpha}}{\pi\sqrt{r_3-\frac{1}{4\alpha}}\sqrt{\chi_m-r_2}}\Bigg(2D\Pi(\tilde{n}',\tilde{k}')+\frac{\tilde{c}_1\text{K}(\tilde{k}')}{\sqrt{\alpha(r_1-\chi_m)}}\Bigg)
\end{align}
with
$$
\tilde{n}'=-\frac{r_2-\frac{1}{4\alpha}}{\chi_m-r_2},\ \ \tilde{k}'=\frac{(r_2-\frac{1}{4\alpha})(\chi_2-r_3)}{(\chi_m-r_2)(r_3-\frac{1}{4\alpha})}.
$$
Note that the previous assumption gives $\tilde{c}_1=D\sqrt{1-4\alpha \chi_m}$. Finally, we obtain the expression of $\tilde{g}(r_2,r_3)$ as follows
\begin{align}
\tilde{g}&(r_1,r_2,r_3)=\frac{D(1-4\alpha \chi_m)}{2\alpha \pi\sqrt{\chi_m-r_2}}\notag\\
&\Bigg(\frac{\sqrt{\chi_m-\frac{1}{4\alpha}}\text{K}(\tilde{k}')}{\sqrt{\chi_m-r_1}\sqrt{r_3-\frac{1}{4\alpha}}}-\frac{\Pi(\tilde{n}',\tilde{k}')}{\sqrt{r_3-\frac{1}{4\alpha}}}-\frac{\text{K}(\tilde{k})}{\sqrt{r_3-r_1}}\Bigg).
\end{align}

Different from the RW-DSW interaction, there always exists a moment $t_2$ when the trailing edge of the RW catches up with the left boundary of the interaction region, and tends to exceed it due to $u_l>\frac{1}{2\alpha}$. At the left boundary of the interaction region
\begin{equation}
r_1=r_2=\frac{1}{4\alpha},\ r_3=\chi_l,
\end{equation}
so
\begin{equation}
t_2=\frac{\tilde{w}_2(\frac{1}{4\alpha},\frac{1}{4\alpha},\chi_l)}{12(\chi_l-\frac{1}{4\alpha})}.
\end{equation}
When $t>t_2$, the RW disappears, and both sides of the interaction region are connected by incomplete CDSWs, where the left CDSW exhibits a small-amplitude edge without a soliton edge, while the right one possesses a soliton edge without a small-amplitude edge. Due to the continuous matching at $t_2$ of the two stages before and after the disappearance of the RW, the degenerate form of Eq.~(89) still applies to the transmitted CDSW with $r_3=\chi_l$ and two varying Riemann invariants governed by
\begin{equation}
x=\underset{r_1\rightarrow r_2}{\lim}\big(v_2(r_1,r_2,\chi_l)t+\tilde{w}_2(r_1,r_2,\chi_l)\big).
\end{equation}
At the trailing edge, $r_1=r_2=\frac{1}{4\alpha}$, and the trajectory is
\begin{equation}
x=(\frac{3}{\alpha}-6\chi_l)t+\tilde{w}_2(\frac{1}{4\alpha},\frac{1}{4\alpha},\chi_l).
\end{equation}
Compared with the original CDSW before the collision, we can find that not only the velocity changes, but there is also a phase shift $\Delta=\tilde{w}_2(\frac{1}{4\alpha},\chi_l)-D$ (see the intersection of the green dashed line with the $x$-axis in Fig.~10(b)) at the trailing edge.

We show the evolution for $r_i$, $R_i$ and $u$ in Figs.~8-9. Although the evolution of Riemann invariants is similar, the evolution of $R_i$ differs completely for the RW-normal CDSW and the RW-reversed CDSW, owing to Eqs.~(73)-(74). As the interaction region expands, the variation range of Riemann invariants gradually increases, reflected in the upper bound (at the right boundary) of $r_2$ approaching $\chi_m$ and the lower bound (at the left boundary) of $r_3$ approaching $\frac{1}{4\alpha}$. As shown in Fig.~10(a), $r_3$ first reaches the possible minimum lower bound, which is determined by $u_l$ in the initial data (3). As time continues to evolve, the limit $r_2\rightarrow r_3$ is first achieved at the left boundary of the interaction region, indicating that algebraic solitons initially emerge at the left boundary.
\section{INTERACTION BETWEEN RWS AND COMPOSITE STRUCTURES FOR $\alpha<0$}
Based on above analyses, we will discuss the interaction between a RW and a composite structure consisting of contact and cnoidal parts.
\subsection{CDSW$|$DSW-RW interaction}
We focus on the initial data (3) satisfying $u_r<u_m<\frac{1}{2\alpha}<u_l$ and $u_l+u_m>\frac{1}{\alpha}$, resulting $\text{min}(\chi_l,\chi_r)>\chi_m$. Whether the RW is completely absorbed and the DSW escapes is determined by the relative magnitude of $\chi_l$ and $\chi_r$, the evolutions are shown in Figs.~11-12. The discontinuity $x=0$ evolves into a composite structure consisting of a normal incomplete CDSW and a DSW. And the discontinuity $x=D$ evolves into a RW expanding between $x_{\text{RW}}^-=6\chi_mt+D$ and $x_{\text{RW}}^+=6\chi_rt+D$.
Obviously, the soliton edge of the DSW intersects with the trailing edge of the RW at $(x_1,t_1)$ following Sec.~\uppercase\expandafter{\romannumeral3} A, and then the solution of the Gardner--Whitham equation in the interaction region can be described by Eqs.~(B2)-(B3) and (21). If $\chi_l>\chi_r$, there exists $t_2$ defined to be the time when the RW disappears, such that the degree of interaction reaches its maximum, and $r_1=\chi_r$ at the right boundary of the interaction region. Subsequently, the DSW escapes from the interaction region with the following changed parameters
$$
r_1=\chi_r,\ \ \ \ x=v_2(\chi_r,r_2,\chi_l)t+\hat{w}_2(\chi_r,r_2),\ \ \ \ r_3=\chi_l.
$$
If $\chi_l<\chi_r$, the degree of interaction increases until the final asymptotic state is reached. Noting that the contact part in the composite structure does not take part in the interaction due to $\chi_r>\chi_m$.

Following Sec.~\uppercase\expandafter{\romannumeral3} A, we have $r_2\rightarrow r_1$ as $t\rightarrow \infty$. However, the difference is that this limit corresponds to a normal CDSW, instead of a small amplitude wave train (see Eq.~(73) and Appendix A). For the interaction between the composite structure reversed CDSW$|$DSW and the RW, the analysis is similar besides the polarity.
\subsection{RW-CDSW$|$DSW interaction}
For the initial data satisfying $\chi_m>\chi_l$ and $\chi_m>\chi_r$, there are a RW with edges
$$
x_{\text{RW}}^-=6\chi_lt,\ \ \ x_{\text{RW}}^+=6\chi_mt,
$$
and a composite structure consisting of reversed CDSW and DSW parts with edges
$$
x^-=\left(\frac{3}{\alpha}-6\chi_m\right)t+D,\ \ x^+=(2\chi_r+4\chi_m)t+D,
$$
and the match point
$$
x_m=6(2\chi_r-\chi_m)t+D.
$$
The leading edge of the RW intersects with the trailing edge of the composite structure at
\begin{equation}
t_1=\frac{D}{12\chi_m-\frac{3}{\alpha}},\ \ x_1=\frac{D}{2\chi_m-\frac{1}{2\alpha}}.
\end{equation}
And then the interaction region is the same as Sec.~\uppercase\expandafter{\romannumeral5}, namely, the solution of the Gardner--Whitham system is implicitly expressed by Eqs.~(77) and (89), until the entire CDSW part enters the interaction region, resulting in the right boundary of the interaction region intersects with the match point, i.e., $r_1=r_2=\chi_r$ and $r_3=\chi_m$. The coordinates of this key point in the $(x,t)$ plane is
\begin{align}
&t_3=-\frac{D}{96\alpha(\chi_m-\chi_r)^2}\Bigg(1-4\alpha \chi_r+\frac{16\big(\alpha(\chi_r-\chi_m)\big)^\frac{3}{2}}{\sqrt{1-4\alpha \chi_m}} \Bigg),\notag\\
&x_3=\frac{D(\chi_m-2\chi_r)}{\sqrt{\chi_m-\chi_r}}\left(\frac{1-4\alpha \chi_r}{16\alpha(\chi_m-\chi_r)^\frac{3}{2}}-\frac{\sqrt{-\alpha}}{1-4\alpha \chi_m} \right)+D.
\end{align}
Similar to Eqs.~(90)-(91), the RW is completely absorbed into the interaction region at $t=t_2$, which may be either larger or smaller than $\bar{t}_3$, contingent upon the exact value of $\chi_l$.

When $t>t_3$, the interaction region consists of two subregions, namely the left RW-CDSW part and the right RW-DSW part. The latter inherits the boundary condition
\begin{equation}
\bar{g}(r_2,r_3)=\tilde{g}(\chi_l,r_2,r_3),
\end{equation}
at the boundary of the two parts. And the right boundary is connected with the DSW, satisfying
\begin{equation}
\bar{g}(r_2,\chi_m)=D+\bar{c}\,\text{K}\left(\frac{r_2-\chi_r}{\chi_m-\chi_r}\right),
\end{equation}
with a constant $\bar{c}$. Noting that $\tilde{g}(\chi_r,r_2,\chi_m)=D$, then one yields $\bar{g}(r_2,r_3)=\tilde{g}(\chi_r,r_2,r_3)$, which accompanied by Eqs.~(B2)-(B3) constitutes the solution of the Gardner--Whitham equation.

If $\chi_l>\chi_r$, the CDSW part escapes from the interaction region at $t_4$ as follows
\begin{equation}
t_4=\frac{\bar{w}_3(\chi_r,\chi_r,\chi_l)-\bar{w}_2(\chi_r,\chi_r,\chi_l)}{v_2(\chi_r,\chi_r,\chi_l)-v_3(\chi_r,\chi_r,\chi_l)},
\end{equation}
after which, the interaction region returns to a state with only one subregion RW-DSW. The entire evolution, featuring the four critical instants, is shown in the Figs.~13.
\section{CONCLUSIONS}
Within the framework of Eq.~(1), we have discussed the interaction between RWs and DSWs for both convex and non-convex cases by means of the Whitham modulation theory, with particular emphasis on the influence of non-convexity. By the hodograph transform, we have derived the analytic expression of the oscillation in the interaction region for several interaction types, including RW-DSW, DSW-RW, RW-CDSW, and the interaction between composite structures.

For the interaction between convex RWs and DSWs, where the initial data don't cross the inflection point, there are two situations where the DSW is to the left of RW and the RW is to the left of DSW, and they collide at $t=t_1$, which is determined by the leading edge velocity of the left one, the trailing edge velocity of the right one, and the initial interval between them. After the formation of the interaction region, two of Riemann invariants change and the other is still a constant. We have obtained analytical solutions for $r_i$, which can describe the oscillation in the interaction region. When $\chi_l>\chi_r$, there exists $t_2$ such that the RW completely enters the interaction region, and then a new DSW exits from the right or left side, corresponding to above different situations, accompanied by a phase shift and the changed constant Riemann invariant, compared to the DSW before collision. When $t\rightarrow\infty$, two varying Riemann invariants tend toward each other, which leads to the oscillation degenerating into small amplitude waves or a soliton train with different polarity depending on the monotonicity of $\chi(u)$.

Considering the non-convexity, Eq.~(1) admits special DSWs: the pure kink ($\alpha>0$) and CDSW ($\alpha<0$), the former's interaction with a RW has been studied previously~\cite{sol-mean}, whereas the latter's is investigated in this work. We found that the pure CDSW can only collide with the RW located on the left side, as its leading edge velocity is the same as the trailing one of the RW on the same background. In this scenario, the constant Riemann invariant ceases to exist within the interaction region; instead, three $r_i$ are all varying, with two of them maintaining equality. We have solved the more general Gardner--Whitham equations. With the progressive intensification of the interaction, the RW vanishes and is fully absorbed into the interaction region, in contrast to the convex case where a portion could be preserved under specific initial data. This behavior stems from the fact that $r_1=r_2$ on the trailing edge of the CDSW attains the minimum of $\chi(u)$. Subsequently, a CDSW which does not contain an algebraic soliton at the leading edge, exits from the left side accompanied by a phase shift and changed $r_3$. It should be noted that although the variations of $r_i$ associated with the interactions of normal or reversed CDSW are qualitatively similar, evolutions of their combination $R_i$ are fundamentally different. As $t\rightarrow\infty$, $r_2$ tends toward $r_3$, signifying a degeneration into an algebraic soliton train; notably, this limit is achieved first on the left boundary of the interaction region.

Based on interactions between fundamental structures, we turn to the case of composite structures. For $\alpha>0$, the RW and DSW exist coupled with a kink if the initial data cross the inflection point. The kink will switch the polarity of the RW or DSW colliding with it, without affecting the interaction. As for $\alpha<0$, it can be divided into two situations. For the CDSW$|$DSW-RW interaction, the CDSW part does not participate in the interaction. The difference from the convex case is that although $r_2$ still tends towards $r_1$ as $t\rightarrow\infty$, this limit corresponds to a CDSW, rather than a small amplitude wave train. For the second situation where RW is on the left of CDSW$|$DSW, the RW interacts with the CDSW part until $t=t_3$. Then the interaction region splits into two subregions, designated as RW-CDSW and RW-DSW. If $\chi_l>\chi_r$, it consists solely of the RW-DSW subregion after $t=t_4$, implying that the CDSW part has completely exited from the region, accompanied by modified parameters and a phase shift. At this stage, the interaction region undergoes a transition from one subregion to two and then back to one. Subsequently, a part of the DSW escapes from the left boundary of the interaction region, likewise accompanied by altered parameters and a phase shift.

We have exhaustively discussed all possible interaction scenarios between a RW and DSW for Eq.~(1) under the prescribed double-step initial data, covering both the convex and non-convex cases. This analysis is beneficial for studying analogous problems in other equations that possess non-convex fluxes. However, the DSW-DSW interactions pertinent to multi-phase modulation theory have not yet been investigated, and this will constitute a direction for future research.
\begin{acknowledgments}
We express our sincere thanks to each member of our discussion group for their suggestions. This work has been supported by the National Natural Science Foundation of China under Grant No. 12575005, the Shanxi Province Science Foundation under Grant No. 202303021221031, and the Research Project Supported by Shanxi Scholarship Council of China under Grant No. 2024-033.
\end{acknowledgments}
\appendix
\section{FINITE GENUS SOLUTIONS AND GARDNER--WHITHAM EQUATIONS}
In Appendix A, we provide brief information on the relevant results of finite genus solutions and the Gardner--Whitham equations~\cite{pre2012}.

The dispersionless limit of Eq.~(1) admits the self-similar solution
\begin{equation}
\frac{x}{t}=6\chi(u),
\end{equation}
which describe rarefaction waves.

For the case of genus-1, the Gardner traveling solutions $u=u(\xi)$, $\xi=x-Vt-\xi_0$ are described by the ordinary differential equation
\begin{equation}
u^2_\xi=\alpha(u-u_1)(u-u_2)(u-u_3)(u-u_4),
\end{equation}
where the phase velocity is given by $V=\alpha(u_1u_2+u_1u_3+u_1u_4+u_2u_3+u_2u_4+u_3u_4)$. Noting that we only focus on the modulationally stable case, where four real roots related by the condition $\sum_{i=1}^4{u_i}=\frac{2}{\alpha}$ have an order of $u_1\leq u_2\leq u_3\leq u_4$.

(i) For $\alpha>0$, the periodic solution
\begin{align}
u_p&=u_2+\frac{\left( u_3-u_2 \right) \text{cn}^2\left( \theta ,m_1 \right)}{1-\frac{u_3-u_2}{u_4-u_2}\text{sn}^2\left( \theta ,m_1 \right)}\notag\\
&=u_4-\frac{u_4-u_3}{1-\frac{u_3-u_2}{u_4-u_2}\text{sn}^2(\theta,m_1)}
\end{align}
oscillates between $u_2$ and $u_3$, with
\begin{align}
&\theta =\sqrt{|\alpha| \left( u_3-u_1 \right) \left( u_4-u_2 \right)}\xi /2,\notag\\
&m_1=\frac{\left( u_3-u_2 \right) \left( u_4-u_1 \right)}{\left( u_4-u_2 \right) \left( u_3-u_1  \right)}.\notag
\end{align}

If $m\rightarrow1$, the periodic solution (A4) allows for two different types of degradation, namely, the bright soliton $
u=u_1+\frac{u_3-u_1}{\cosh^2\theta -\frac{u_3-u_1}{u_4-u_1}\sinh^2\theta}
$
propagating against a constant background $\bar{u}=u_1<\frac{1}{2\alpha}$ as $u_2\rightarrow u_1$, and the dark soliton
$
u=u_4-\frac{u_4-u_2}{\cosh^2\theta -\frac{u_4-u_2}{u_4-u_1}\sinh^2\theta}
$
on the constant background $\bar{u}=u_4>\frac{1}{2\alpha}$ as $u_3\rightarrow u_4$.

When $u_3\rightarrow u_2$ ($m_1\rightarrow 0$), the periodic wave asymptotically transforms into a linear harmonic wave $u\cong u_2+\frac{1}{2}(u_3-u_2)\text{cos}(2\theta)$.

(ii) For $\alpha<0$, the positive values of $Q(u)$ occurs in the two intervals
$$
u_1\leq u\leq u_2,\ \ \ \ u_3\leq u \leq u_4.
$$

If $u_1\leq u\leq u_2$, the periodic solution has the following expression
\begin{align}
u_p&=u_2-\frac{\left( u_2-u_1 \right) \text{cn}^2\left( \theta ,m_2 \right)}{1+\frac{u_2-u_1}{u_4-u_2}\text{sn}^2\left( \theta ,m_2 \right)}\notag\\
&=u_4-\frac{u_4-u_1}{1+\frac{u_2-u_1}{u_4-u_2}\text{sn}^2(\theta,m_2)},
\end{align}
with $
m_2=\frac{\left( u_4-u_3 \right) \left( u_2-u_1 \right)}{\left( u_4-u_2 \right) \left( u_3-u_1 \right)}
$.

As $u_3\rightarrow u_2$ ($m_2\rightarrow 1$), the periodic solution reduces to a dark soliton
$
u=u_2-\frac{u_2-u_1}{\cosh^2\theta +\frac{u_2-u_1}{u_4-u_2}\sinh^2\theta}
$
on the constant background $\bar{u}=u_2$. When $u_2\rightarrow u_1$ ($m_2\rightarrow 0$), the periodic wave asymptotically transforms into a harmonic limit $u\cong u_2-\frac{1}{2}(u_2-u_1)\text{cos}(2\theta)$. Interestingly, when $u_4=u_3$, but $u_1\neq u_2$, one yields the special DSW, i.e., the reversed CDSW $
u=u_2-\frac{\left( u_2-u_1 \right) \cos ^2\theta}{1+\frac{u_2-u_1}{u_3-u_2}\sin ^2\theta}$.

If we consider the case of $u_3\leq u \leq u_4$, the periodic solution can be derived as
\begin{align}
u_p&=u_3+\frac{\left( u_4-u_3 \right) \text{cn}^2\left( \theta ,m_2 \right)}{1+\frac{u_4-u_3}{u_3-u_1}\text{sn}^2\left( \theta ,m_2 \right)}\notag\\
&=u_1+\frac{u_4-u_1}{1+\frac{u_4-u_3}{u_3-u_1}\text{sn}^(\theta,m_2)}.
\end{align}
In the soliton limit $u_3\rightarrow u_2\ (m_2\rightarrow 1)$, one can get the bright soliton
$
u=u_2+\frac{u_4-u_2}{\cosh^2\theta +\frac{u_4-u_2}{u_2-u_1}\sinh^2\theta}
$
propagating on a constant background $\bar{u}=u_2$. In the limit $u_3\rightarrow u_4$ $(m_2\rightarrow 0)$, the periodic wave asymptotically transforms into a harmonic limit $u\cong u_3+\frac{1}{2}(u_4-u_3)\text{cos}(2\theta)$. If $u_2\rightarrow u_1$, but $u_3\neq u_4$,  one can get the normal CDSW $
u=u_3+\frac{\left( u_4-u_3 \right) \cos ^2\theta}{1+\frac{u_4-u_3}{u_3-u_1}\sin ^2\theta}$.

As is well known, the analytical description of DSW is established within the framework of Whitham modulation theory~\cite{GP1974}. Due to the fact that Eq.~(1) is a completely integrable equation, the Gardner--Whitham system can be expressed as the Riemann diagonal form
\begin{equation}
\frac{\partial r_i}{\partial t}+\upsilon _i\frac{\partial r_i}{\partial x}=0,\ \ \ i=1,2,3,
\end{equation}
with the Whitham characteristic velocities
\begin{equation}
\upsilon _i=\left( 1-\frac{L}{\partial _iL}\partial _i \right) V,\ \ \ \ \partial _i=\frac{\partial}{\partial r_i}.
\end{equation}
Due to the invariance of Eq.~(1) with respect to the following transformation
\begin{equation}
u\rightarrow \frac{1}{\alpha}-u,
\end{equation}
the two parameter families $r_i$ and $u_i$ do not correspond one-to-one, but are linked by the following equations
\begin{align}
r_1=\frac{1}{2}\left( u_1+u_2 \right) -\frac{\alpha}{4}\left( u_1+u_2 \right) ^2,\notag \\
r_2=\frac{1}{2}\left( u_1+u_3 \right) -\frac{\alpha}{4}\left( u_1+u_3 \right) ^2,\\ \notag
r_3=\frac{1}{2}\left( u_2+u_3 \right) -\frac{\alpha}{4}\left( u_2+u_3 \right) ^2,
\end{align}
for $\alpha>0$, and
\begin{align}
r_1=\frac{1}{2}\left( u_2+u_3 \right) -\frac{\alpha}{4}\left( u_2+u_3 \right) ^2,\notag\\
r_2=\frac{1}{2}\left( u_1+u_3 \right) -\frac{\alpha}{4}\left( u_1+u_3 \right) ^2,\\
r_3=\frac{1}{2}\left( u_1+u_2 \right) -\frac{\alpha}{4}\left( u_1+u_2 \right) ^2,\notag
\end{align}
for $\alpha<0$, with ordering $r_1<r_2<r_3$. Based on above relations, the modulus of elliptic function for both signs of $\alpha$ can be unified as
\begin{equation}
m_1=m_2=m=\frac{r_2-r_1}{r_3-r_1},
\end{equation}
and the wavelength $L=\frac{2K(m)}{\sqrt{(r_3-r_1)}}$. Then the Whitham characteristic velocities can be written as
\begin{align}
&v_1=2(r_1+r_2+r_3)+\frac{4(r_2-r_1)\text{K}(m)}{\text{E}(m)-\text{K}(m)},\notag\\
&v_2=2(r_1+r_2+r_3)-\frac{4(r_2-r_1)(1-m)\text{K}(m)}{\text{E}(m)-(1-m)\text{K}(m)},\\
&v_3=2(r_1+r_2+r_3)+\frac{4(r_3-r_2)\text{K}(m)}{\text{E}(m)},\notag
\end{align}
where $\text{K(m)}$ and $\text{E(m)}$ are the complete elliptic integrals of the first and second kind, respectively. On the two edges of DSW, Eq.~(A6) reduces to systems in which a formulae is consistent with the dispersionless limit of Eq.~(1). Specifically, in the small amplitude edge $m=0$, i.e., $r_1=r_2$, the Whitham characteristic velocities reduce to
\begin{equation}
v_1=v_2=12r_1-6r_3,\ \ \ v_3=6r_3.
\end{equation}
And in the soliton edge $m=1$, i.e., $r_2=r_3$, we have
\begin{equation}
v_1=6r_1,\ \ \ v_2=v_3=2r_1+4r_3.
\end{equation}
\section{HODOGRAPH TRANSFORM AND EULER-POISSON-DARBOUX EQUATION}
In the interaction region, the solution of the Gardner--Whitham system is no longer self-similar. In order to obtain the general solutions of Eq.~(A6), we need to introduce the hodograph transform. If $w_i$ solve the Tsarev equation
\begin{equation}
\frac{\partial_j w_i}{w_i-w_j}=\frac{\partial_j v_i}{v_i-v_j},\ \ i\neq j,
\end{equation}
then the solution $(r_i, r_j)$ of
\begin{equation}
x-v_i t=w_i,\ \ \ \ x-v_j t=w_j,
\end{equation}
satisfies Eq.~(A6). Due to the symmetry (B1), we can assume $w_i$ has the following form
\begin{equation}
w_i=(1-\frac{L}{\partial_i L})g,\ \ i=1,2,3,
\end{equation}
similar to Eq.~(A7). In this way, the original problem is simplified to solving for a single function $g(r_i,r_j)$.

Substituting Eqs.~(A7) and (B3) into (B1), the unknown function $g(r_i,r_j)$ satisfies the EPD equation
\begin{equation}
2(r_i-r_j)\partial_{ij}g=\partial_ig-\partial_jg,\ \ i\neq j.
\end{equation}
Following Eisenhart~\cite{Eisenhart}, the general solutions of Eq.~(B4) can be expressed as
\begin{align}
g(r_i,r_j)=&\int_{a_i}^{r_i}\frac{\phi_i}{\sqrt{(r-r_i)(r-r_j)}}dr\notag\\
&+\int_{a_j}^{r_j}\frac{\phi_j}{\sqrt{(r-r_i)(r-r_j)}}dr,
\end{align}
where $a_{i,j}$ are arbitrary constant and $\phi_{i,j}$ are determined by the matching conditions on the boundaries separating different regions.
\section{MODULATED PHASE IN SINGLE-PHASE PERIODIC WAVES}
Taking into account the Whitham modulation theory, single-phase periodic waves in Appendix A can describe DSWs, but the dependence of the initial phase $\xi_0$ on $x$ and $t$ has not been constructed which is the task here.

The angular phase
\begin{equation}
\Xi=k\xi=k(x-Vt-\xi_0),
\end{equation}
gives
\begin{equation}
\Xi_x=k,\ \ \ \ \ \ \Xi_t=-kV=-\omega,
\end{equation}
in which $k$ is the wavenumber, and $\omega$ is the frequency. The generalised phase relationships (C2) implies $\frac{\partial _i\omega}{\partial _ik}=v_i$, and should also be satisfied in modulated periodic waves.

Differentiating Eq.~(C1) with respect to $t$, one yields
\begin{align}
\varXi _t&=-\omega +\sum_{i=1}^3{\frac{\partial r_i}{\partial t}\left( \frac{\partial k}{\partial r_i}x-\frac{\partial \omega}{\partial r_i}t-\frac{\partial k}{\partial r_i}\xi _0-\frac{\partial \xi _0}{\partial r_i}k \right)}\notag\\
&=-\omega +\sum_{i=1}^3{\frac{\partial r_i}{\partial t}\frac{\partial k}{\partial r_i}\left( x-v_it-\xi _0-\frac{\partial _i\xi _0}{\partial _ik}k \right)}\notag\\
&=-\omega +\sum_{i=1}^3{\frac{\partial r_i}{\partial t}\frac{\partial k}{\partial r_i}\left( x-v_it-\xi _0+\frac{L}{\partial _iL}\partial _i\xi _0 \right)},
\end{align}
where we have considered $\frac{\partial _i\omega}{\partial _ik}=v_i$ and $k=\frac{2\pi}{L}$. Comparing the second one of Eq.~(C2) with (C3), one gets
\begin{equation}
x-v_it=(1-\frac{L}{\partial _iL}\partial _i)\xi _0,
\end{equation}
in which initial phase $\xi_0$ is an analogue of $g$ in Eq.~(B2), so
\begin{equation}
\xi=g+zL,
\end{equation}
with constant $z=n+\frac{1}{2}$ solved below.

At the soliton edge, i.e., $m\rightarrow1$, we have $\frac{L}{\partial_2L}=\frac{L}{\partial_3L}=0$. From Eqs.~(A7) and (B3), one yields
\begin{equation}
x-Vt-g=-\frac{L}{\partial_iL}(\partial_ig+2t),
\end{equation}
and then
\begin{equation}
x-Vt-g=0.
\end{equation}
For the modulated periodic solution (A3) oscillating between $u_2$ and $u_3$, the leading edge is either connected to $u_2$ corresponding to the normal DSW, or connected to $u_3$ corresponding to the reversed DSW. This implies that $\text{sn}^2(\theta,m_1)\rightarrow 1$ for the normal DSW, so
\begin{equation}
x-Vt-g-zL=(n-\frac{1}{2})L,\ \ \ \ \ n=\pm1,\pm2,\pm3,...,
\end{equation}
which together with Eq.~(C7) gives $z=\frac{1}{2}-n$. But for the reversed DSW, $\text{sn}^2(\theta,m_1)\rightarrow 0$ at the soliton edge, and then
\begin{equation}
x-Vt-g-zL=nL,
\end{equation}
so $z=-n$. Similarly, for Eqs.~(A4)-(A5), the soliton edge is connected to $u_2$ or $u_3$, implying $\text{sn}^2(\theta,m_2)\rightarrow 1$ as $m\rightarrow1$, and then $z=\frac{1}{2}-n$.
\bibliography{ref}

@BOOK{Nlw1,
   author       = {R. H. J. Grimshaw},
   year         = 2005,
   title        = {Nonlinear waves in fluids: Recent advances and modern applications},
   publisher    = {Springer},
   address      = {Berlin}
}

@ARTICLE{Nlw2,
   author       = "P. Bandyopadhyay and G. Prasad and A. Sen and P. K. Kaw",
   journal      = "Phys. Rev. Lett.",
   title        = "Experimental study of nonlinear dust acoustic solitary waves in a dusty plasma",
   year         = "2008",
   volume       = "101",
   number       = "6",
   pages        = "065006",
   doi          = "10.1103/physrevlett.101.065006"
}

@BOOK{Nlw3,
   author       = {N. Akhmediev and A. Ankiewicz},
   year         = 1997,
   title        = {Solitons: Nonlinear Pulses and Beams},
   publisher    = {Chapman and Hall},
   address      = {London}
}

@ARTICLE{Nlw4,
   author       = "D. V. Shaykin and A. M. Kamchatnov",
   journal      = "Phys. Rev. A",
   title        = "Theory of dispersive shock waves induced by the {Raman} effect in optical fibers",
   year         = "2025",
   volume       = "112",
   number       = "6",
   pages        = "063511",
   doi          = "10.1103/94zy-78my"
}

@ARTICLE{Nlw5,
   author       = "S. Mossman and S. I. Mistakidis and G. C. Katsimiga and A. Romero-Ros and G. Biondini and P. Schmelcher and P. Engels and P. G. Kevrekidis",
   journal      = "Phys. Rev. Lett.",
   title        = "Nonlinear stage of modulational instability in repulsive two-component {Bose-Einstein} condensates",
   year         = "2025",
   volume       = "135",
   number       = "11",
   pages        = "113401",
   doi          = "10.1103/6jsr-f8q1"
}

@ARTICLE{Scholarpedia2009,
   author       = "M. Hoefer and M. Ablowitz",
   journal      = "Scholarpedia",
   title        = "Dispersive shock waves",
   year         = "2009",
   volume       = "4",
   number       = "11",
   pages        = "5562",
}

@ARTICLE{physicad2016,
   author       = "G. A. El and M. A. Hoefer",
   journal      = "Physica D",
   title        = "Dispersive shock waves and modulation theory",
   year         = "2016",
   volume       = "333",
   pages        = "11-65",
   doi          = "10.1016/j.physd.2016.04.006"
}

@ARTICLE{pre2026mkdv,
   author       = "L. F. Calazans de Brito and A. Gammal and A. M. Kamchatnov",
   journal      = "Phys. Rev. E",
   title        = "Evolution of localized pulses in the defocusing modified {Korteweg--de Vries} equation theory",
   year         = "2026",
   volume       = "113",
   number       = "6",
   pages        = "064208",
   doi          = "10.1103/sh7l-lxck"
}

@ARTICLE{pre2026FPU,
   author       = "S. Yang",
   journal      = "Phys. Rev. E",
   title        = "Quasicontinuum descriptions of rarefaction and dispersive shock waves in {Fermi-Pasta-Ulam} lattices with {Hertzian} potentials",
   year         = "2026",
   volume       = "113",
   number       = "5",
   pages        = "054204",
   doi          = "10.1103/wjjy-qt58"
}

@ARTICLE{prl2025MR,
   author       = "G. Biondini and A. Bivolcic and M. A. Hoefer",
   journal      = "Phys. Rev. Lett.",
   title        = "Mach reflection and expansion of two-dimensional dispersive shock waves",
   year         = "2025",
   volume       = "135",
   number       = "6",
   pages        = "067201",
   doi          = "10.1103/cdvf-xnfw"
}

@ARTICLE{GP1974,
   author       = "A. V. Gurevich and L. P. Pitaevski\v{i}",
   journal      = "Sov. Phys. JETP",
   title        = "Nonstationary structure of a collisionless shock wave",
   year         = "1974",
   volume       = "38",
   number       = "2",
   pages        = "291-297",
}

@ARTICLE{whitham1965,
   author       = "G. B. Whitham",
   journal      = "Proc. R. Soc. London, Ser. A",
   title        = "Non-linear dispersive waves",
   year         = "1965",
   volume       = "283",
   number       = "1393",
   pages        = "238-261",
   doi          = "10.1098/rspa.1965.0019"
}

@BOOK{2000,
   author       = {A. M. Kamchatnov},
   year         = 2000,
   title        = {Nonlinear periodic waves and their modulations: An introductory course},
   publisher    = {World Scientific},
   address      = {Singapore}
}

@ARTICLE{chaos2005nonintegrable,
   author       = "G. A. El",
   journal      = "Chaos",
   title        = "Resolution of a shock in hyperbolic systems modified by weak dispersion",
   year         = "2005",
   volume       = "15",
   number       = "3",
   pages        = "037103",
   doi          = "10.1063/1.1947120"
}

@ARTICLE{pre2019nonintegrable,
   author       = "A. M. Kamchatnov",
   journal      = "Phys. Rev. E",
   title        = "Dispersive shock wave theory for nonintegrable equations",
   year         = "2019",
   volume       = "99",
   number       = "1",
   pages        = "012203",
   doi          = "10.1103/PhysRevE.99.012203"
}

@ARTICLE{physicad1995,
   author       = "G. A. El and V. V. Geogjaev and A. V. Gurevich and A. L. Krylov",
   journal      = "Physica D",
   title        = "Decay of an initial discontinuity in the defocusing {NLS} hydrodynamics",
   year         = "1995",
   volume       = "87",
   number       = "1-4",
   pages        = "186-192",
}

@ARTICLE{siam2017,
   author       = "G. A. El and M. A. Hoefer and M. Shearer",
   journal      = "SIAM Rev.",
   title        = "Dispersive and diffusive-dispersive shock waves for nonconvex conservation laws",
   year         = "2017",
   volume       = "59",
   number       = "1",
   pages        = "3-61",
   doi          = "10.1137/15M1015650"
}

@ARTICLE{pre2012,
   author       = "A. M. Kamchatnov and Y.-H. Kuo and T.-C. Lin and T.-L. Horng and S.-C. Gou and R. Clift and G. A. El and R. H. J. Grimshaw",
   journal      = "Phys. Rev. E",
   title        = "Undular bore theory for the {Gardner} equation",
   year         = "2012",
   volume       = "86",
   number       = "3",
   pages        = "036605",
   doi          = "10.1103/PhysRevE.86.036605"
}

@ARTICLE{jpc2018,
   author       = "A. M. Kamchatnov",
   journal      = "J. Phys. Commun.",
   title        = "Evolution of initial discontinuities in the {DNLS} equation theory",
   year         = "2018",
   volume       = "2",
   number       = "2",
   pages        = "025027",
   doi          = "10.1088/2399-6528/aaae12"
}

@ARTICLE{pra2020,
   author       = "S. K. Ivanov",
   journal      = "Phys. Rev. A",
   title        = "Riemann problem for the light pulses in optical fibers for the generalized {Chen-Lee-Liu} equation",
   year         = "2020",
   volume       = "101",
   number       = "5",
   pages        = "053827",
   doi          = "10.1103/PhysRevA.101.053827"
}

@ARTICLE{pre2023,
   author       = "H. Gao and D. S. Wang",
   journal      = "Phys. Rev. E",
   title        = "Optical undular bores in {Riemann} problem of photon fluid with quintic nonlinearity",
   year         = "2023",
   volume       = "108",
   number       = "2",
   pages        = "024222",
   doi          = "10.1103/PhysRevE.108.024222"
}

@ARTICLE{PRL2017,
   author       = "G. Xu and M. Conforti and A. Kudlinski and A. Mussot and S. Trillo",
   journal      = "Phys. Rev. Lett.",
   title        = "Dispersive Dam-Break Flow of a Photon Fluid",
   year         = "2017",
   volume       = "118",
   number       = "25",
   pages        = "254101",
   doi          = "10.1103/physrevlett.118.254101"
}

@ARTICLE{pradbe,
   author       = "S. Chandramouli and S. I. Mistakidis and G. C. Katsimiga and P. G. Kevrekidis",
   journal      = "Phys. Rev. A",
   title        = "Dispersive shock waves in a one-dimensional droplet-bearing environment",
   year         = "2024",
   volume       = "110",
   number       = "2",
   pages        = "023304",
   doi          = "10.1103/PhysRevA.110.023304"
}

@ARTICLE{DRS,
   author       = "H. Kim and E. Kim and C. Chong and P. G. Kevrekidis and J. Yang",
   journal      = "Phys. Rev. Lett.",
   title        = "Demonstration of dispersive rarefaction shocks in hollow elliptical cylinder chains",
   year         = "2018",
   volume       = "120",
   number       = "19",
   pages        = "194101",
   doi          = "10.1103/PhysRevLett.120.194101"
}

@ARTICLE{pradsoliton,
   author       = "A. Romero-Ros and G. C. Katsimiga and P. G. Kevrekidis and B. Prinari and G. Biondini and P. Schmelcher",
   journal      = "Phys. Rev. A",
   title        = "On-demand generation of dark soliton trains in {Bose-Einstein} condensates",
   year         = "2021",
   volume       = "103",
   number       = "2",
   pages        = "023329",
   doi          = "10.1103/PhysRevA.103.023329"
}

@ARTICLE{pradbsoliton,
   author       = "A. Romero-Ros and G. C. Katsimiga and P. G. Kevrekidis and B. Prinari and G. Biondini and P. Schmelcher",
   journal      = "Phys. Rev. A",
   title        = "On-demand generation of dark-bright soliton trains in {Bose-Einstein} condensates",
   year         = "2022",
   volume       = "105",
   number       = "2",
   pages        = "023325",
   doi          = "10.1103/PhysRevA.105.023325"
}

@ARTICLE{cpam1980,
   author       = "H. Flaschka and M. G. Forest and D. W. McLaughlin",
   journal      = "Comm. Pure Appl. Math.",
   title        = "Multiphase averaging and the inverse spectral solution of the {Korteweg--de Vries} equation",
   year         = "1980",
   volume       = "33",
   number       = "6",
   pages        = "739-784",
}

@ARTICLE{jns2006,
   author       = "G. Biondini and Y. Kodama",
   journal      = "J. Nonlinear Sci.",
   title        = "On the {Whitham} equations for the defocusing nonlinear {Schr\"{o}dinger} equation with step initial data",
   year         = "2006",
   volume       = "16",
   number       = "5",
   pages        = "435-481",
   doi          = "10.1007/s00332-005-0733-2"
}

@ARTICLE{pre2009,
   author       = "M. J. Ablowitz and D. E. Baldwin and M. A. Hoefer",
   journal      = "Phys. Rev. E",
   title        = "Soliton generation and multiple phases in dispersive shock and rarefaction wave interaction",
   year         = "2009",
   volume       = "80",
   number       = "1",
   pages        = "016603",
   doi          = "10.1103/PhysRevE.80.016603"
}

@ARTICLE{pre2013,
   author       = "M. J. Ablowitz and D. E. Baldwin",
   journal      = "Phys. Rev. E",
   title        = "Dispersive shock wave interactions and asymptotics",
   year         = "2013",
   volume       = "87",
   number       = "2",
   pages        = "022906",
   doi          = "10.1103/PhysRevE.87.022906"
}

@ARTICLE{pla2013,
   author       = "M. J. Ablowitz and D. E. Baldwin",
   journal      = "Phys. Lett. A",
   title        = "Interactions and asymptotics of dispersive shock waves---{Korteweg--de Vries} equation",
   year         = "2013",
   volume       = "377",
   number       = "7",
   pages        = "555-559",
   doi          = "10.1016/j.physleta.2012.12.040"
}

@ARTICLE{physicad2007,
   author       = "M. A. Hoefer and M. J. Ablowitz",
   journal      = "Physica D",
   title        = "Interactions of dispersive shock waves",
   year         = "2007",
   volume       = "236",
   number       = "1",
   pages        = "44-64",
   doi          = "10.1016/j.physd.2007.07.017"
}

@ARTICLE{chaos2002,
   author       = "G. A. El and R. H. J. Grimshaw",
   journal      = "Chaos",
   title        = "Generation of undular bores in the shelves of slowly-varying solitary waves",
   year         = "2002",
   volume       = "12",
   number       = "4",
   pages        = "1015-1026",
   doi          = "10.1063/1.1507381"
}

@ARTICLE{physicad2012,
   author       = "G. A. El and V. V. Khodorovskii and A. M. Leszczyszyn",
   journal      = "Physica D",
   title        = "Refraction of dispersive shock waves",
   year         = "2012",
   volume       = "241",
   number       = "18",
   pages        = "1567-1587",
   doi          = "10.1016/j.physd.2012.06.002"
}

@ARTICLE{siamj2017,
   author       = "P. Sprenger and M. A. Hoefer",
   journal      = "SIAM J. Appl. Math.",
   title        = "Shock waves in dispersive hydrodynamics with nonconvex dispersion",
   year         = "2017",
   volume       = "77",
   number       = "1",
   pages        = "26-50",
   doi          = "10.1137/16M1082196"
}

@ARTICLE{waterwave2025,
   author       = "S. Baqer and T. P. Horikis and D. J. Frantzeskakis",
   journal      = "Water Waves",
   title        = "On shallow water non-convex dispersive hydrodynamics: The extended KdV model",
   year         = "2025",
   volume       = "7",
   number       = "2",
   pages        = "225-262",
   doi          = "10.1007/s42286-025-00114-9"
}

@ARTICLE{CPL2026,
   author       = "Y. H. Ye and J. J. Wu",
   journal      = "Chin. Phys. Lett.",
   title        = "The well and box types initial value problems of the defocusing {mKdV} equation",
   year         = "2026",
   volume       = "43",
   number       = "4",
   pages        = "040001",
   doi          = "10.1088/0256-307X/43/4/040001"
}

@ARTICLE{jpsj,
   author       = "M. Wadati",
   journal      = "J. Phys. Soc. Jpn.",
   title        = "Wave propagation in nonlinear lattice. {II}",
   year         = "1975",
   volume       = "38",
   number       = "3",
   pages        = "681-686",
   doi          = "10.1143/JPSJ.38.681"
}

@ARTICLE{bg1,
   author       = "W. M. Moslem and S. Rezk and U. M. Abdelsalam and S. K. El-Labany",
   journal      = "Adv. Space Res.",
   title        = "Shocklike soliton because of an impinge of protons and electrons solar particles with {Venus} ionosphere",
   year         = "2018",
   volume       = "61",
   number       = "8",
   pages        = "2190-2197",
   doi          = "10.1016/j.asr.2018.01.023"
}

@ARTICLE{bg2,
   author       = "G. M. Coclite and F. Maddalena and G. Puglisi and M. Romano and G. Saccomandi",
   journal      = "SIAM J. Appl. Math.",
   title        = "The {Gardner} equation in elastodynamics",
   year         = "2021",
   volume       = "81",
   number       = "6",
   pages        = "2346-2361",
   doi          = "10.1137/21M1407537"
}

@ARTICLE{bg3,
   author       = "T. Talipova and O. Kurkina and A. Kurkin and E. Didenkulova and E. Pelinovsky",
   journal      = "Microgravity Sci. Technol.",
   title        = "Internal wave breathers in the slightly stratified fluid",
   year         = "2020",
   volume       = "32",
   number       = "1",
   pages        = "69-77",
   doi          = "10.1007/s12217-019-09738-2"
}

@ARTICLE{bg4,
   author       = "E. Demler and A. Maltsev",
   journal      = "Ann. Phys.",
   title        = "Semiclassical solitons in strongly correlated systems of ultracold bosonic atoms in optical lattices",
   year         = "2011",
   volume       = "326",
   number       = "7",
   pages        = "1775-1805",
   doi          = "10.1016/j.aop.2011.04.001"
}

@BOOK{Hyperbolicsystems,
   author       = {P. D. Lax},
   year         = 1973,
   title        = {Hyperbolic systems of conservation laws and the mathematical theory of shock waves},
   publisher    = {SIAM},
   address      = {Philadelphia}
}

@ARTICLE{sol-mean,
   author       = "J. X. Niu and R. Guo and J. W. Zhang",
   journal      = "Nucl. Phys. B",
   title        = "Soliton--mean flow interaction problem for a non-convex system with quadratic-cubic nonlinearity in the stratified fluid",
   year         = "2026",
   volume       = "1025",
   pages        = "117379",
   doi          = "10.1016/j.nuclphysb.2026.117379"
}

@ARTICLE{Eisenhart,
   author       = "L. P. Eisenhart",
   journal      = "Ann. Math.",
   title        = "Triply conjugate systems with equal point invariants",
   year         = "1919",
   volume       = "20",
   number       = "4",
   pages        = "262-273",

}

\end{document}